\documentclass[12pt,preprint]{aastex}
\usepackage{hyperref}
\usepackage{epsfig}
\usepackage{natbib}                
\usepackage{lscape}
\usepackage{verbatim}              
\usepackage{graphicx}              
\usepackage{amsmath,amsthm,amsfonts,amsopn,amssymb} 
\usepackage{longtable,pdflscape}
\usepackage{afterpage}
\usepackage{rotating}					
\usepackage{ulem}

\shorttitle{Stellar Diameters and Temperatures II. Main Sequence K \& M Stars}
\shortauthors{Boyajian \& von Braun et al.}
\begin{document}


\title{Stellar Diameters and Temperatures \\
II. Main Sequence K \& M Stars}

\author{Tabetha S. Boyajian\altaffilmark{1,2}, 
Kaspar von Braun\altaffilmark{3},
Gerard van Belle\altaffilmark{4}, \\ 
Harold A. McAlister\altaffilmark{1}, 
Theo A. ten Brummelaar\altaffilmark{5}, 
Stephen R. Kane\altaffilmark{3},
Phil Muirhead\altaffilmark{9},
Jeremy Jones\altaffilmark{1},
Russel White\altaffilmark{1},
Gail Schaefer\altaffilmark{5}, 
David Ciardi\altaffilmark{3}, 
Todd Henry\altaffilmark{1}, 
Mercedes L\'{o}pez-Morales\altaffilmark{6,7},
Stephen Ridgway\altaffilmark{8}, 
Douglas Gies\altaffilmark{1},
Wei-Chun Jao\altaffilmark{1},
B\'{a}rbara Rojas-Ayala\altaffilmark{10},
J. Robert Parks\altaffilmark{1},
Laszlo Sturmann\altaffilmark{5}, 
Judit Sturmann\altaffilmark{5},  
Nils H. Turner\altaffilmark{5},
Chris Farrington\altaffilmark{5}, 
P. J. Goldfinger\altaffilmark{5}, 
David H. Berger\altaffilmark{11}}

\altaffiltext{1}{Center for High Angular Resolution Astronomy and Department of Physics and Astronomy, Georgia State University, P. O. Box 4106, Atlanta, GA 30302-4106} 
\altaffiltext{2}{Hubble Fellow} 
\altaffiltext{3}{NASA Exoplanet Science Institute, California Institute of Technology, MC 100-22, Pasadena, CA 91125} 
\altaffiltext{4}{Lowell Observatory, Flagstaff, AZ 86001}
\altaffiltext{5}{The CHARA Array, Mount Wilson Observatory, Mount Wilson, CA 91023}
\altaffiltext{6}{Institut de Ci\`{e}ncies de L'Espai (CSIC-IEEC), Spain}
\altaffiltext{7}{Department of Terrestrial Magnetism, Carnegie Institution of Washington, 5241 Broad Branch Road, NW, Washington, DC 20015}
\altaffiltext{8}{National Optical Astronomy Observatory, P.O. Box 26732, Tucson, AZ 85726-6732}
\altaffiltext{9}{Department of Astrophysics, California Institute of Technology, MC 249-17, Pasadena, CA 91125}
\altaffiltext{10}{Department of Astrophysics, Division of Physical Sciences, American Museum of Natural History, Central Park West at 79th Street, New York, NY 10024}
\altaffiltext{11}{System Planning Corporation, 3601 Wilson Blvd, Arlington, VA  22201}



\begin{abstract}


We present interferometric angular diameter measurements of 21 low-mass, K- and M- dwarfs made with the CHARA Array.  This sample is enhanced by adding a collection of radii measurements published in the literature to form a total data set of 33 K-M dwarfs with diameters measured to better than 5\%.  We use these data in combination with the $Hipparcos$ parallax and new measurements of the star's bolometric flux to compute absolute luminosities, linear radii, and effective temperatures for the stars. We develop empirical relations for $\sim$K0 to M4 main-sequence stars that link the stellar temperature, radius, and luminosity to the observed ($B-V$), ($V-R$), ($V-I$), ($V-J$), ($V-H$), ($V-K$) broad-band color index and stellar metallicity [Fe/H]. These relations are valid for metallicities ranging from [Fe/H] $= -0.5$ to $+0.1$~dex, and are accurate to $\sim$2\%, $\sim$5\%, and $\sim$4\% for temperature, radius, and luminosity, respectively.  Our results show that it is necessary to use metallicity dependent transformations in order to properly convert colors into stellar temperatures, radii, and luminosities. Alternatively, we find no sensitivity to metallicity on relations we construct to the global properties of a star omitting color information e.g., temperature-radius and temperature-luminosity.  Thus, we are able to empirically quantify to what order the star's observed color index is impacted by the stellar iron abundance.  In addition to the empirical relations, we also provide a representative look-up table via stellar spectral classifications using this collection of data. Robust examinations of single star temperatures and radii compared to evolutionary model predictions on the luminosity - temperature and luminosity - radius planes reveals that models overestimate the temperatures of stars with surface temperatures $<5000$~K by $\sim 3$\%, and underestimate the radii of stars with radii $<0.7$~R$_{\odot}$ by $\sim 5$\%.  These conclusions additionally suggest that the models over account for the effects that the stellar metallicity may have on the astrophysical properties of an object. By comparing the interferometrically measured radii for the single star population to those of eclipsing binaries, we find that for a given mass, single and binary star radii are indistinguishable. However, we also find that for a given radius, the literature temperatures for binary stars are systematically lower compared to our interferometrically derived temperatures of single stars by $\sim 200$ to 300~K. The nature of this offset is dependent on the validation of binary star temperatures; where bringing all measurements to a uniform and correctly calibrated temperature scale is needed to identify any influence stellar activity may have on the physical properties of a star.  Lastly, we present a empirically determined HR diagram using fundamental properties presented here in combination with those in \citet{boy12} for a total of 74 nearby, main-sequence, A- to M-type stars, and define regions of habitability for the potential existence of sub-stellar mass companions in each system.

\end{abstract}



\section{Introduction}                              
\label{sec:introduction}


Direct size measurements of low-mass stars represent vital tests of theoretical models of stellar evolution, structure, and atmospheres.  As seen in the results of \citet{ber06}, notable disagreements exist between interferometrically determined radii and those calculated in low-mass stellar models such as those of \citet{cha97} or \citet{sie97} in the sense that interferometrically obtained values for the stellar diameters are systematically larger by more than 10\% than those predicted from models. A similar trend exists for the linear radii and temperatures of low-mass stars obtained from eclipsing binaries (EBs), which are systematically larger and cooler (respectively) for given mass (see \citealt{irw11}, and references therein). Additional evidence for this ongoing discrepancy can be found in the works of, e.g., \citet{lop05}, \citet{von08}, and \citet{boy08}, providing motivation for adjustments to models in order to match the observations \citep{cha07,dot08} and fulfilling the obvious need for more directly determined diameters of late-type dwarfs to provide further constraints to stellar models \citep{boy10}.

The emergence of the observed radius discrepancy over the years (e.g. \citealt{pop97, tor02a}) for low-mass EBs has triggered different theoretical scenarios to reconcile model predictions with observations. \citet{rib08} identify that the elevated activity levels in binary stars must be accounted for in defining the properties of low-mass stars. In addition, \citet{mor10} show that observational analysis, and thus the derived parameters of a system, may be biased if the activity (manifested as star spots) is not properly accounted for in EB light curves.  Strong magnetic fields within the stellar interior have been shown to inhibit convection in fully convective M-type stars \citep{mul01}. Other contributing factors to the EB radii offsets with respect to models such as the orbital period and the rotation rate have been considered and refuted \citep{kra11, irw11}. 

While such proposed scenarios alleviate the discrepancy between models and observations of binary stars, they are not capable of explaining the presence of the same shortcomings observed in predictions of single star properties. However, for the case of single stars, the interferometric observations in \citet{ber06} suggest that theoretical models for low-mass stars may be missing some opacity source in order to properly reproduce observed radii. This result was challenged by the conclusions in \citet{dem09}, which reveal no such correspondence. In many instances in the literature, the effective temperatures of stars have been shown to be overestimated by models, also alluding to a possible connection with stellar abundances (e.g., see \citealt{boy08}, and references therein). 

The first measurement of a stellar diameter for a star other than our Sun was that of Betelgeuse in 1921 \citep{mic21}.  Since then, several hundred stars have been observed at very high spatial resolution to measure their angular sizes.  The most difficult to target are stars on the low-mass end of the main sequence. Although plentiful and nearby, these stars are intrinsically small and faint, and require sensitive high angular resolution techniques to resolve them adequately\footnote{Their linear sizes amount to only several Jupiter radii, and several orders of magnitude less luminous than the Sun. In comparison, Betelgeuse is $\sim 11,000$ times the size of Jupiter and $\sim 140,000$ the luminosity of our Sun!}. The current number of single K- and M- dwarfs whose angular sizes have been obtained with long-baseline interferometry to better than 5\% precision is 17 \citep{lan01,seg03,dif04,ber06,boy08,ker08a,dem09,van09,von11a,von12} and fewer than half of these are M-dwarfs.
Empirical radius and temperature measurements toward the low-mass end of the main-sequence are sparsely available, and any placement of low-mass stars onto an HR diagram presently relies upon model atmosphere results that are in need of observational verification and constraints.  We also identify that with the limited amount of the data available in the literature presently, it is difficult to discern whether or not the discrepancy between models and observations can be explained by a unified theory to encompass issues observed in both binary and single stars.  

In the context of the present paper, we stress the observational components of an extensive interferometric survey to determine the fundamental properties of nearby stars.  The results of our recent paper focus on main-sequence A-, F-, and G-type stars \citep{boy12}.  Here, we present high-precision interferometric observations to determine their angular diameters of 21 late-type K- and M-type dwarfs (\S~\ref{sec:observations}).  In order to characterize these systems, we use trigonometric parallax values and literature photometry to calculate physical diameters and stellar bolometric fluxes.  This allows us to determine their effective temperatures and luminosities (\S~\ref{sec:stellar_params}). In \S~\ref{sec:discussion}, we take our stellar parameters and literature data to establish relations between stellar astrophysical parameters and observables. Throughout the discussion, we elaborate on the agreement of the data to predictions of stellar atmosphere and evolutionary models.  We address the aforementioned radius and temperature discrepancy with models in \S~\ref{sec:global_relations}. We provide motivation to validate estimates of binary star temperatures in \S~\ref{sec:teff_validation}, as they are found to be several hundred Kelvin lower than those of single stars, and we conclude in \S~\ref{sec:conclusion}.


\section{Observations with the CHARA Array: Angular Diameters}
\label{sec:observations}


We acquired interferometric observations at the CHARA Array with the Classic beam combiner in single-baseline mode in the near-infrared $K^{\prime}$ and $H$ bands \citep{ten05} for 21 K- and M-type dwarfs (Table~\ref{tab:observations}). Targets were selected to be of northern declination ($-10^{\circ} <$~dec~$< 90^{\circ}$), to have no known companion with separation $< 2^{\prime\prime}$, to be bright enough for the instrument ($H,K^{\prime} < 6.5, V<11$), and to have estimated angular sizes large enough to resolve the stellar disk to better than 5\% precision ($\theta > 0.4$~mas for our configuration).  The majority of the data were collected from 2008 through 2011 with the longest of CHARA's baselines $B_{\rm Telescope~\#1/Telescope~\#2}$: $B_{\rm S1/E1}=331$~m, $B_{\rm E1/W1}=313$~m, $B_{\rm S2/E1}=302$~m, $B_{\rm S1/E2}=279$~m, $B_{\rm S1/W1}=279$~m, $B_{\rm S2/E2}=248$~m, and $B_{\rm E1/W2}=218$~m.  Additionally, we made use of CHARA archived data for GJ~687 taken in 2004, and for GJ~526 from 2006. 

We employ the standard observing strategy for interferometric observations where each science target is observed in bracketed sequences with calibrator stars \citep[for a detailed discussion see][]{boy09a, boy12}.  In short, we were cautious that every science target was observed with (1) at least two calibrators, each chosen to be unresolved point sources in close proximity to the object on the sky, (2) on a minimum of two nights, and (3) with more than one baseline configuration\footnote{The exception to item \#3 is GJ~33, only observed on S1/E1}. We collected an average of 22 bracketed observations (ranging from $n=2$ to 51) for each star, ideal for reducing noise and increasing the reliability of the final diameter fits. The low number of observations for GJ~702B ($n=2$) is a consequence of technical difficulties we had to acquire and lock tip-tilt on the secondary star which lies $\sim 4.5$ arcseconds from GJ~702A, the much brighter primary in the visual binary system\footnote{The interferometric field of view is on the order of 2 arcseconds, and thus the individual component visibilities are not affected by incoherent light on the detector}. However, the observations of GJ~702B still meet the aspects 1-3 outlined above. 
In order to estimate the calibrator star's angular diameter, $\theta_{\rm SED}$, we fit flux calibrated photometry to a Kurucz model spectral energy distribution (for details, see \citealt{boy12}). As noted in item 1 above, an ideal calibrator is an unresolved point source in order to minimize any biases that may arise through the assumed calibrator star diameter. Similar to the strategy employed in Paper~I, we follow the example in \citet{van05}, aiming to select calibrators with $\theta_{\rm SED} < 0.45$~mas.  In this way, the errors on the object's calibrated visibilities are mainly a product of measurement error, and the relative errors introduced by calibrator angular size prediction error are minimized.  All targets observed in this work have {\it at least} one calibrator that meets this criteria. In several cases however, we use partially resolved calibrator stars with angular diameters larger than $\theta_{\rm SED} = 0.45$~mas. All calibrated points are cross-checked for consistency with respect to one another in order to identify any biases that stem from using partially resolved calibration sources, as well as any signs of duplicity (i.e., if a selected calibrator star is a previously unidentified binary star.)  A log of our observations can be found in Table~\ref{tab:observations}, and a list of our calibrators is presented in Table~\ref{tab:calibs}.

To solve for the angular diameter of our science star, we used a non-linear least-squares fitting routine in IDL (MPFIT, \citealt{mar09}) to fit the calibrated data points to the uniform disk and limb darkened visibility function for a single star using the equations in \citet{han74} (see also \citealt{von11a}).  To correct for limb darkening, we use the linear limb-darkening coefficients $\mu_{\lambda}$ from PHOENIX models presented in \citet{cla00}, where $\lambda$ represents the wavelength of observation ($H$ or $K$, where here we assume $K \approx K^{\prime}$). The determination of a limb-darkened angular diameter is very forgiving in the infrared, where the change from $\theta_{\rm UD}$ to $\theta_{\rm LD}$ is typically on the order of $2-3$~\%.  In addition to a wavelength dependence, the coefficients are also dependent on the star's effective temperature, surface gravity, and metallicity. 

When adopting limb-darkening coefficients, we find that modest errors on the assumed stellar parameters of temperature and gravity ($\Delta T_{\rm EFF} = 200$~K, $\Delta \log g = 0.5$~dex) contribute to under a tenth of a percent to the overall error budget of the limb-darkened angular diameter.  The dependence on the coefficients with respect to metallicity is negligible throughout the range of metallicity values in our sample, so for all objects, the respective limb-darkening coefficients assume solar metallicity.  In order to assign the appropriate limb-darkening coefficient to an object, we perform an iterative procedure, beginning with an initial guess of the temperature based on spectral type from \citet{cox00}. For all stars, we adopt a surface gravity value for lower main-sequence stars to be $\log g = 4.5$ \citep{cox00}.  The resulting fit for the limb-darkened angular diameter using this coefficient is then combined with the measured bolometric flux (see Section \ref{sec:stellar_diameters}) to derive a temperature.  This temperature is then used to identify a new limb darkening coefficient, if needed. All temperatures are simply rounded to the nearest temperature in the \citet{cla00} grid, which is in 200~K increments.

\begin{figure}										
\centering
\begin{tabular}{cc}
\epsfig{file=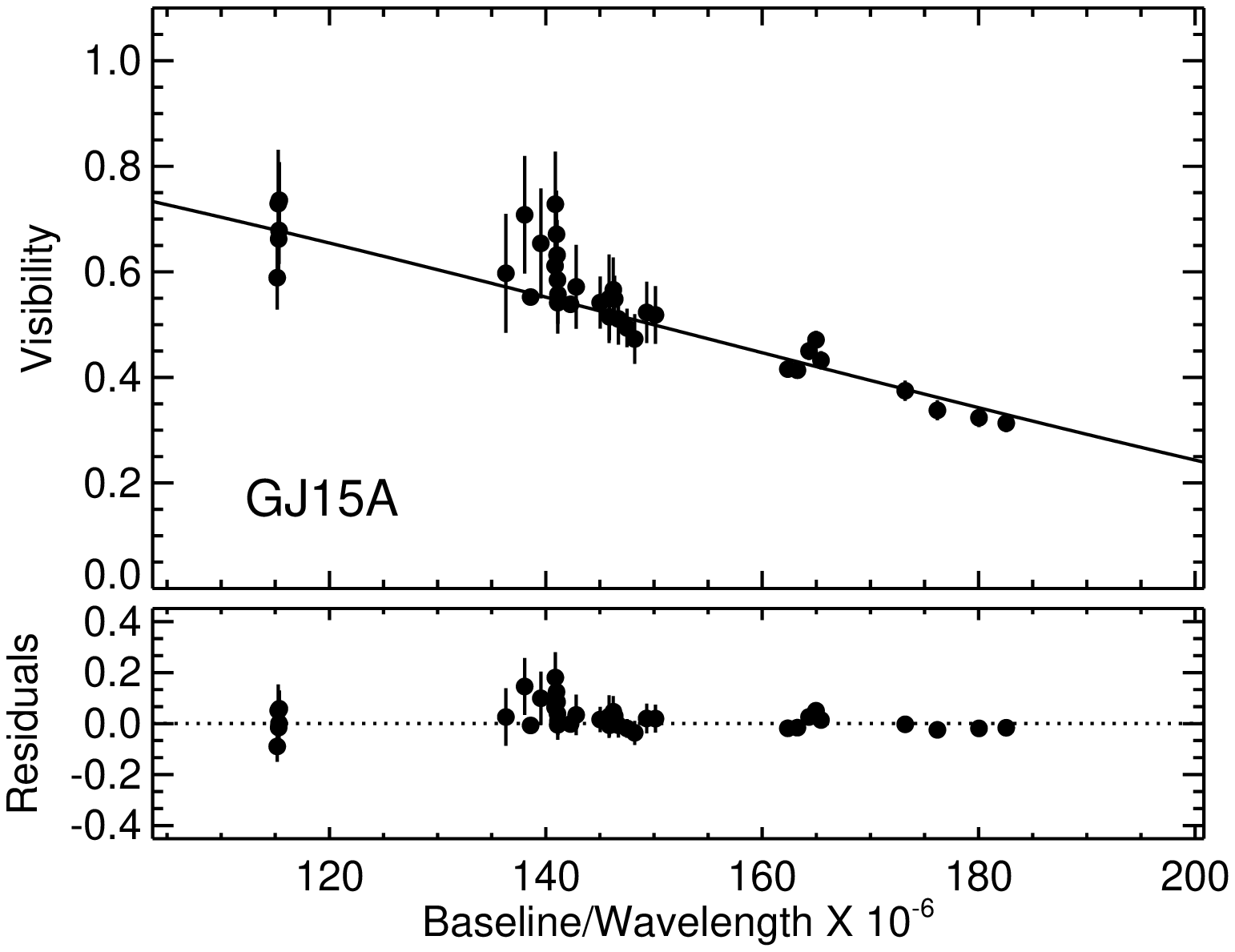, width=0.5\linewidth, clip=} &
\epsfig{file=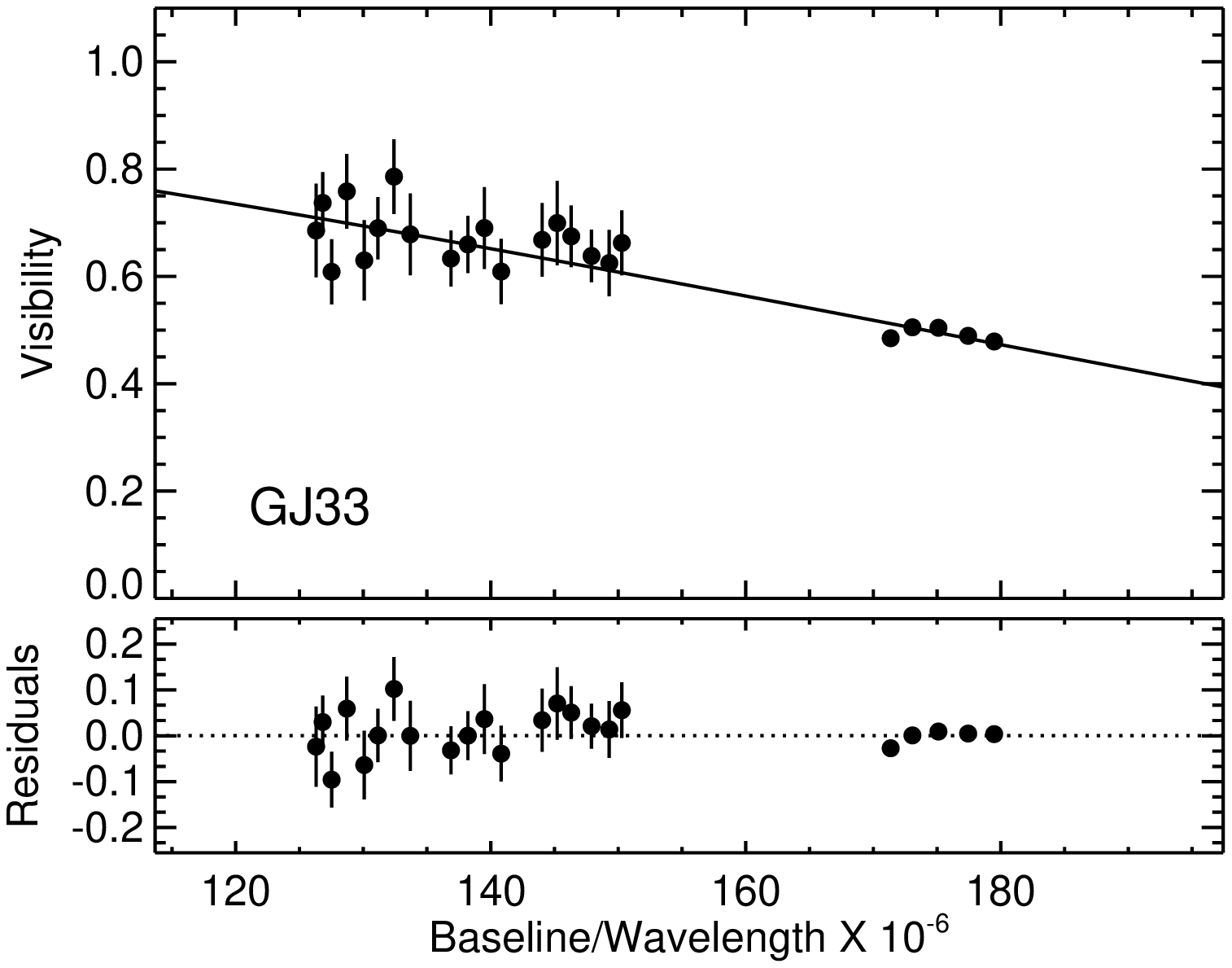, width=0.5\linewidth, clip=} \\
\epsfig{file=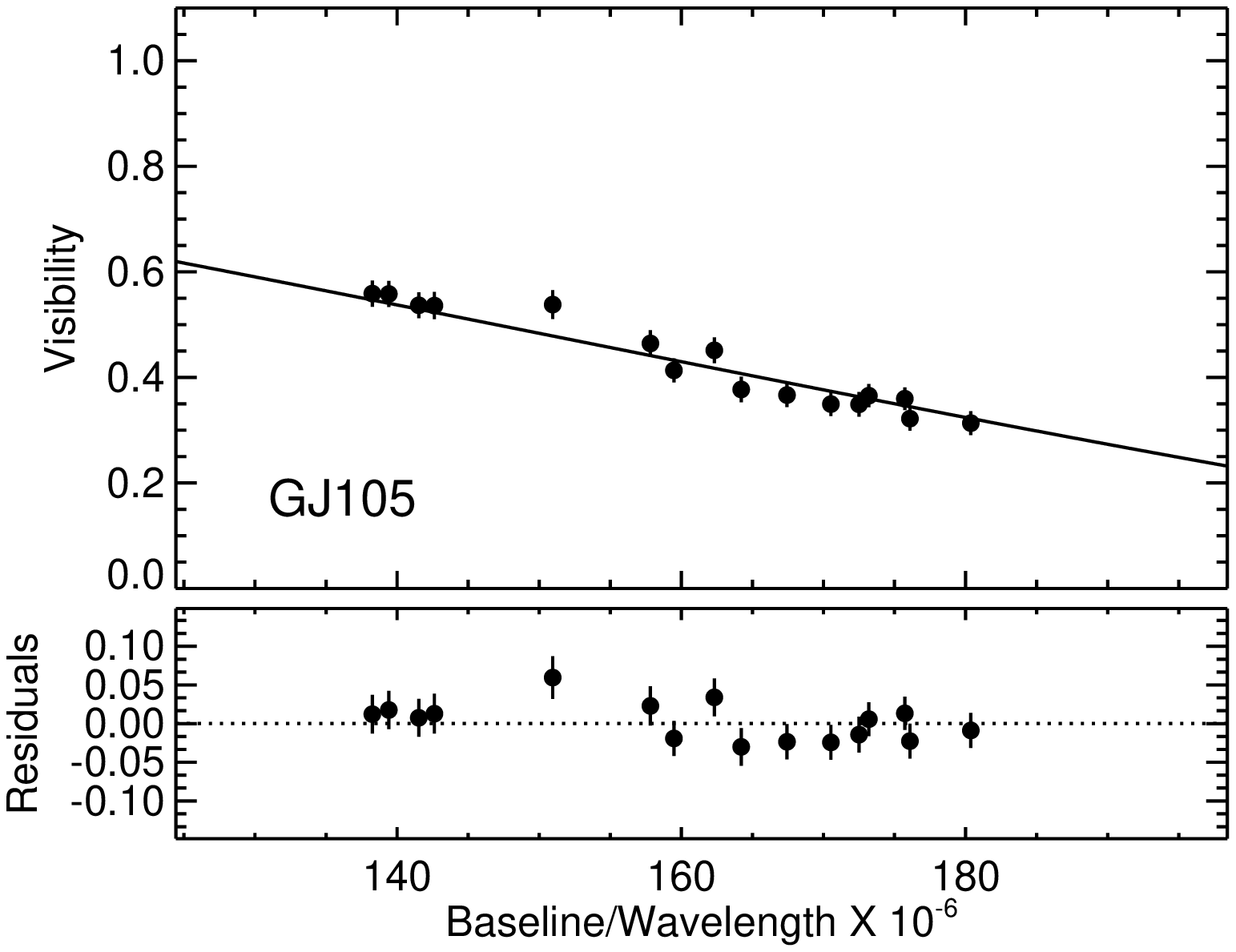, width=0.5\linewidth, clip=} &
\epsfig{file=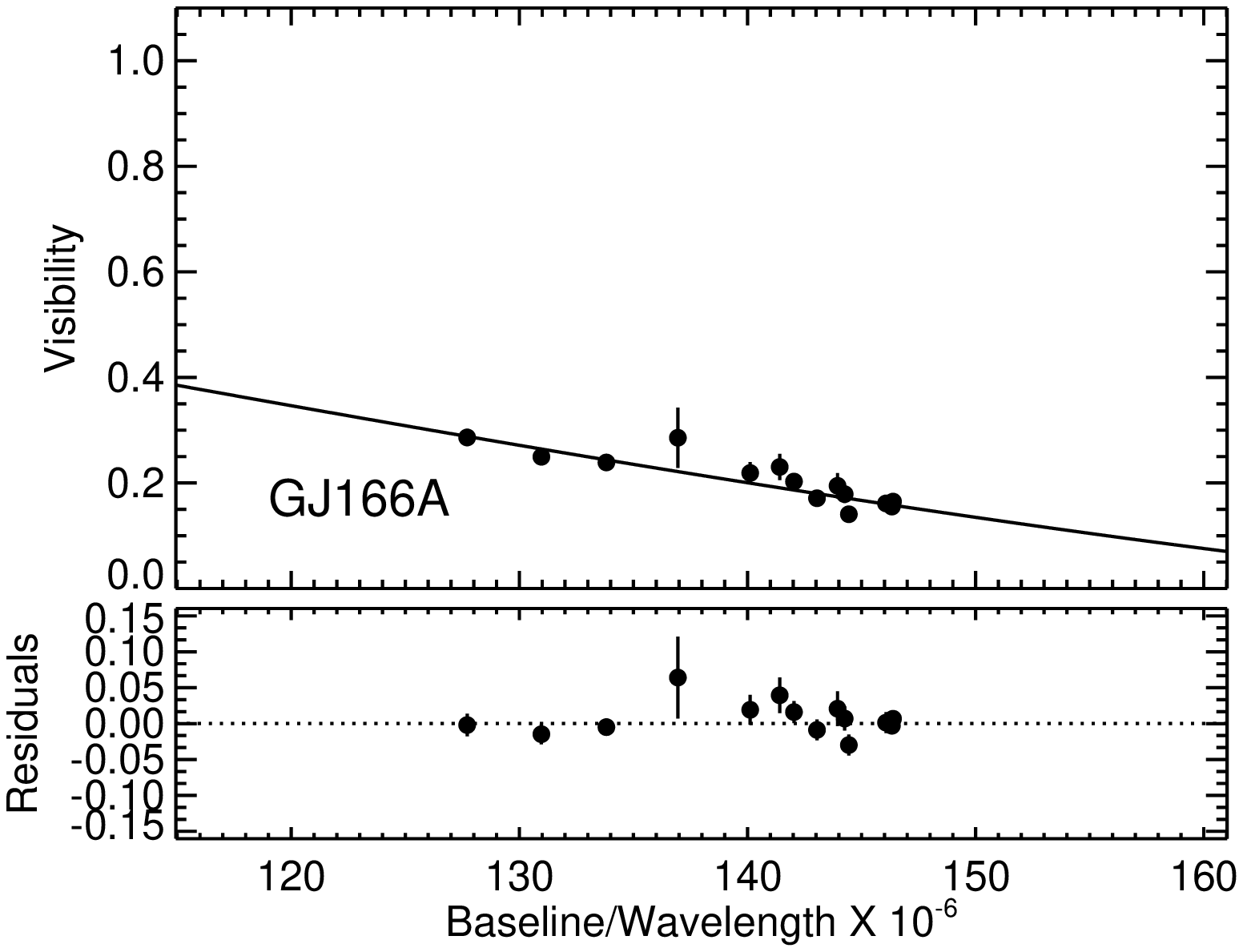, width=0.5\linewidth, clip=} \\
\epsfig{file=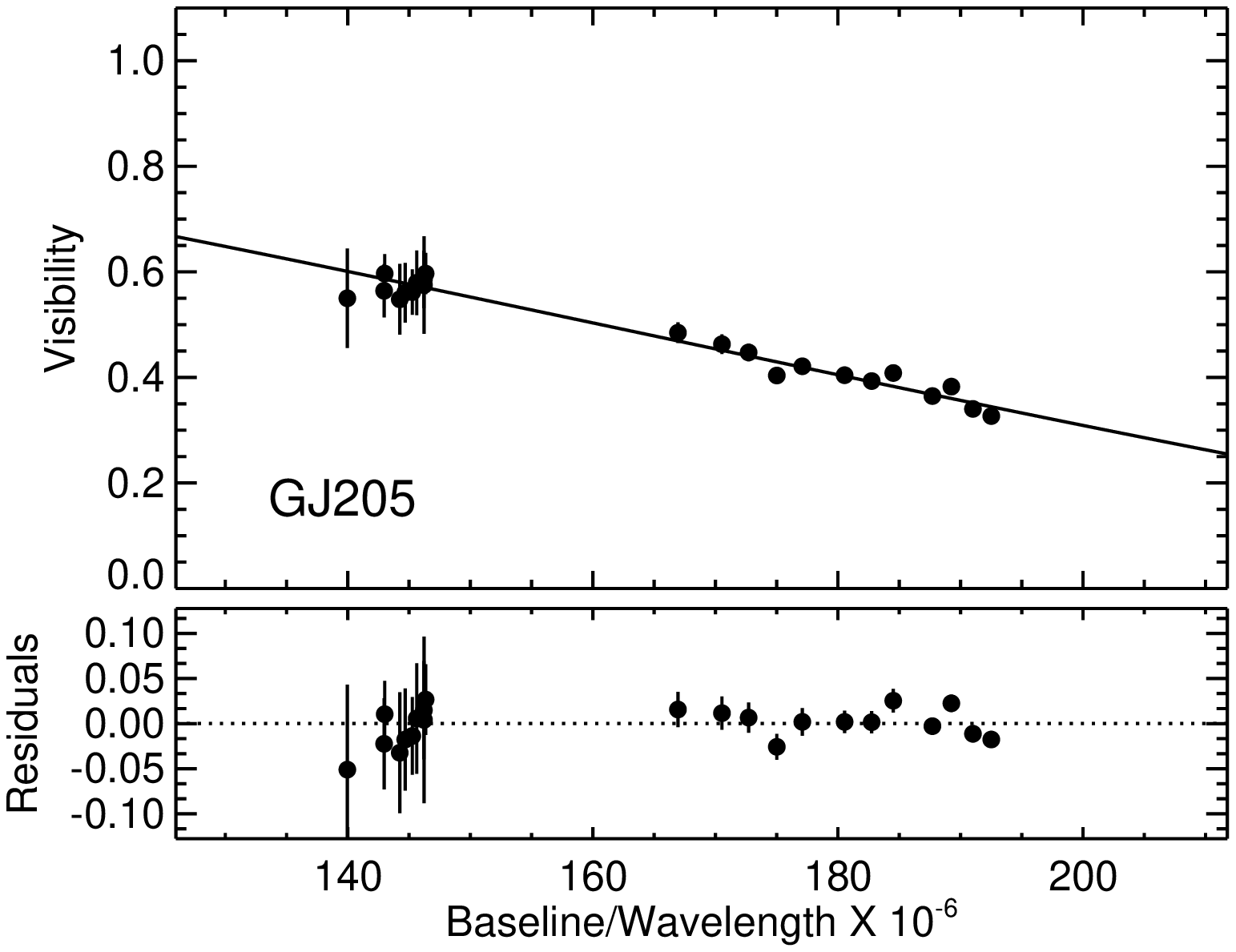, width=0.5\linewidth, clip=} &
\epsfig{file=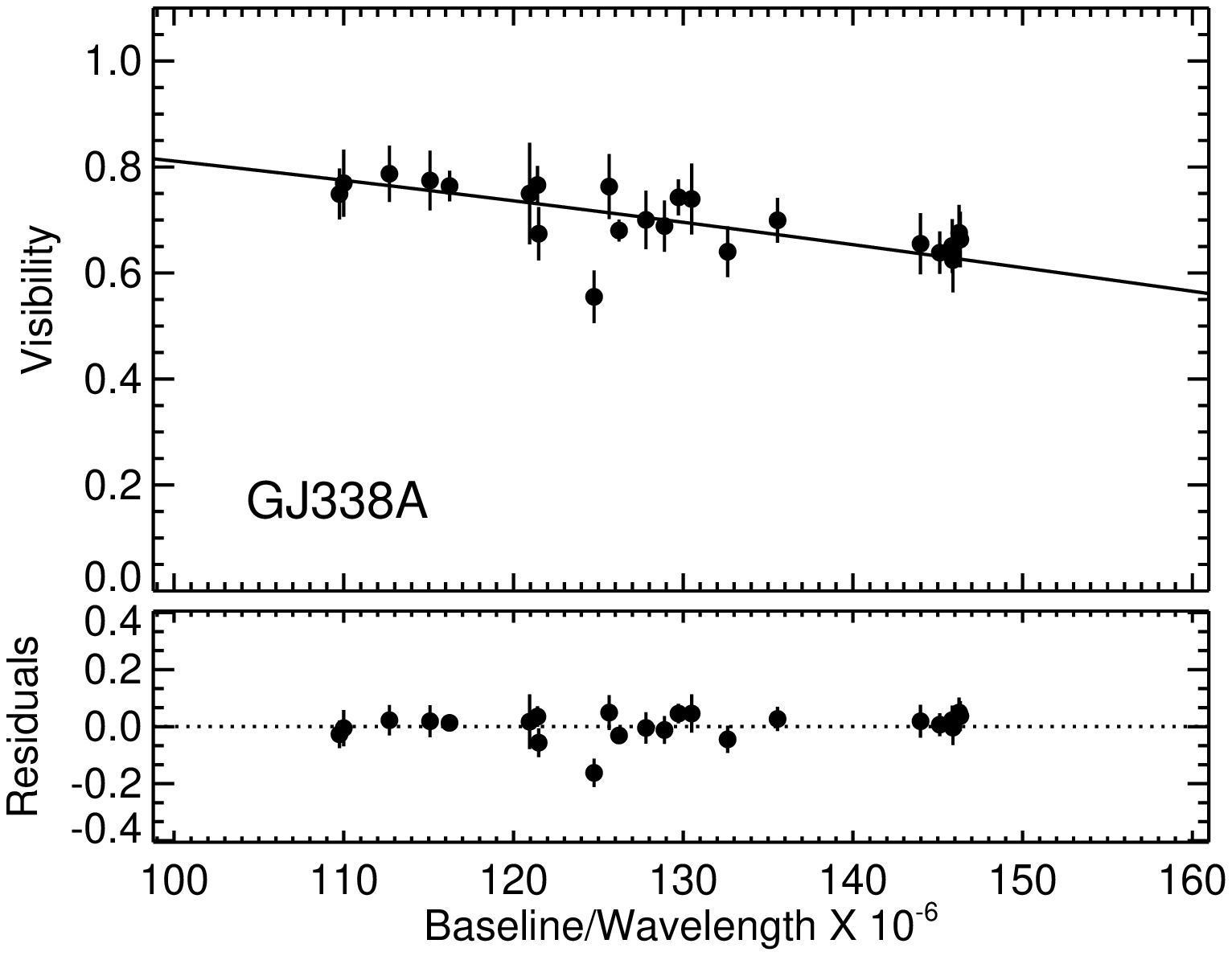, width=0.5\linewidth, clip=} 
 \end{tabular}
 \caption[Angular Diameters] {Calibrated observations plotted with the limb-darkened angular diameter fit for each star.}
 \label{fig:diameters1}
 \end{figure}

\begin{figure}										
 \centering
 \begin{tabular}{cc} 
\epsfig{file=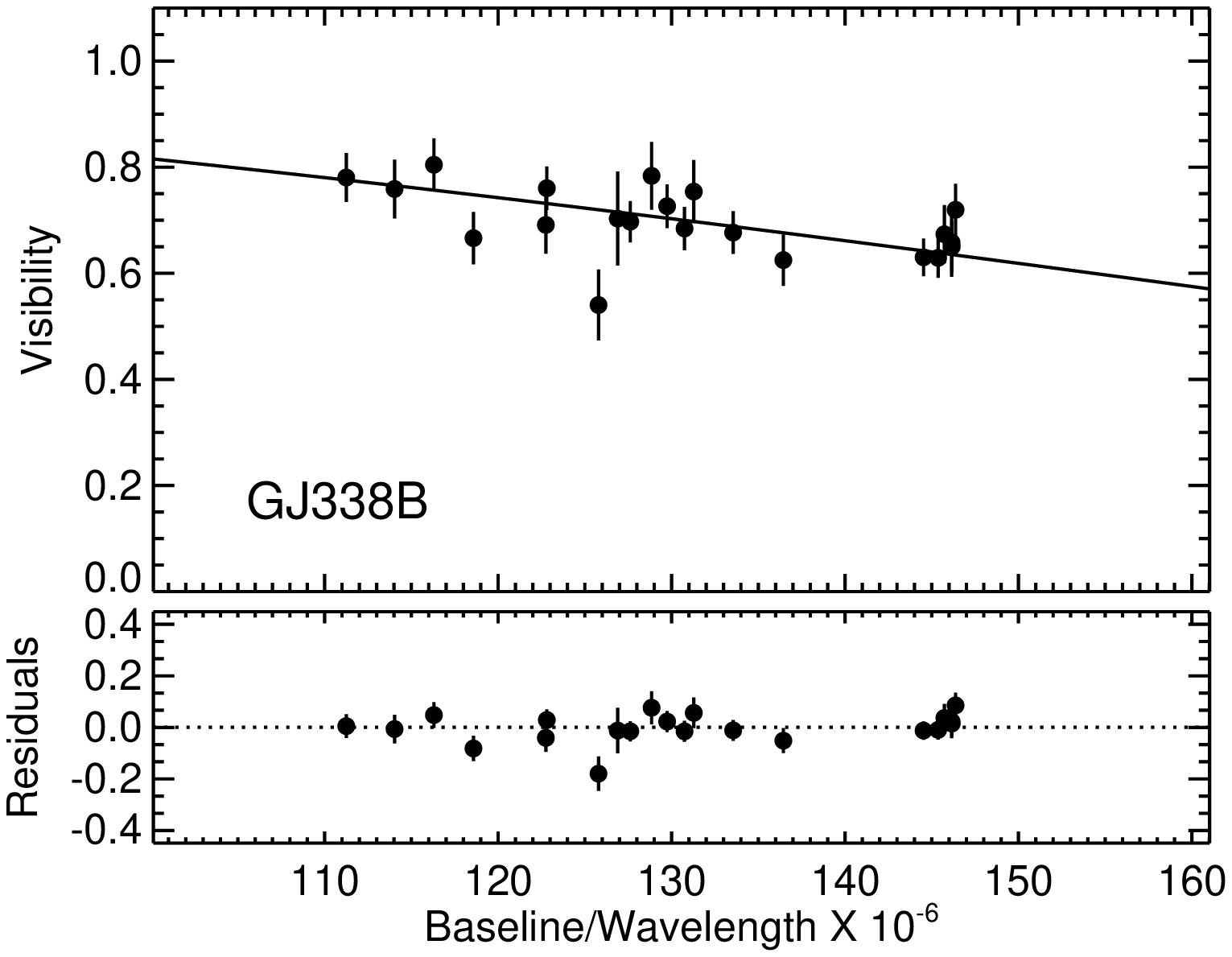, width=0.5\linewidth, clip=} &
\epsfig{file=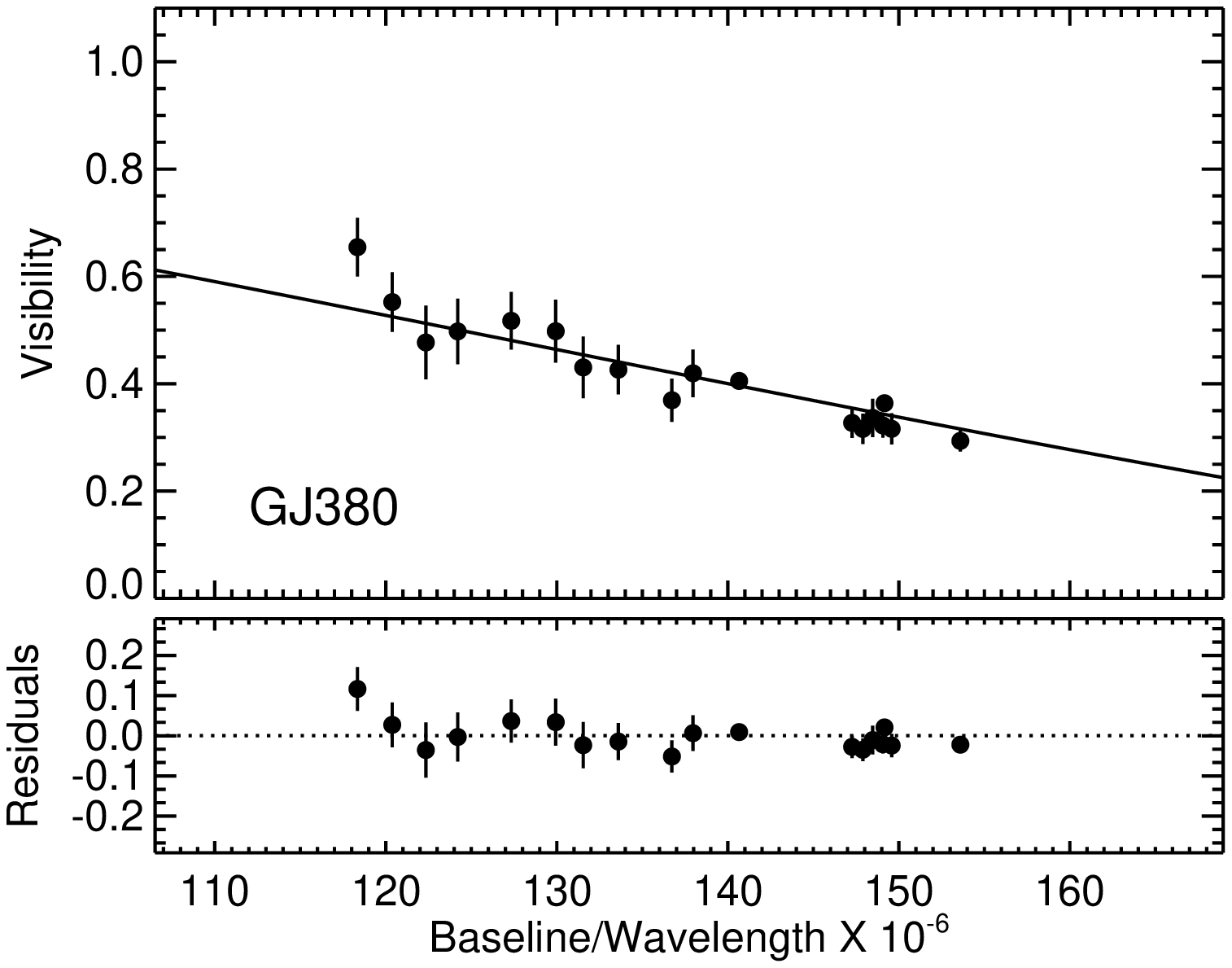, width=0.5\linewidth, clip=} \\
\epsfig{file=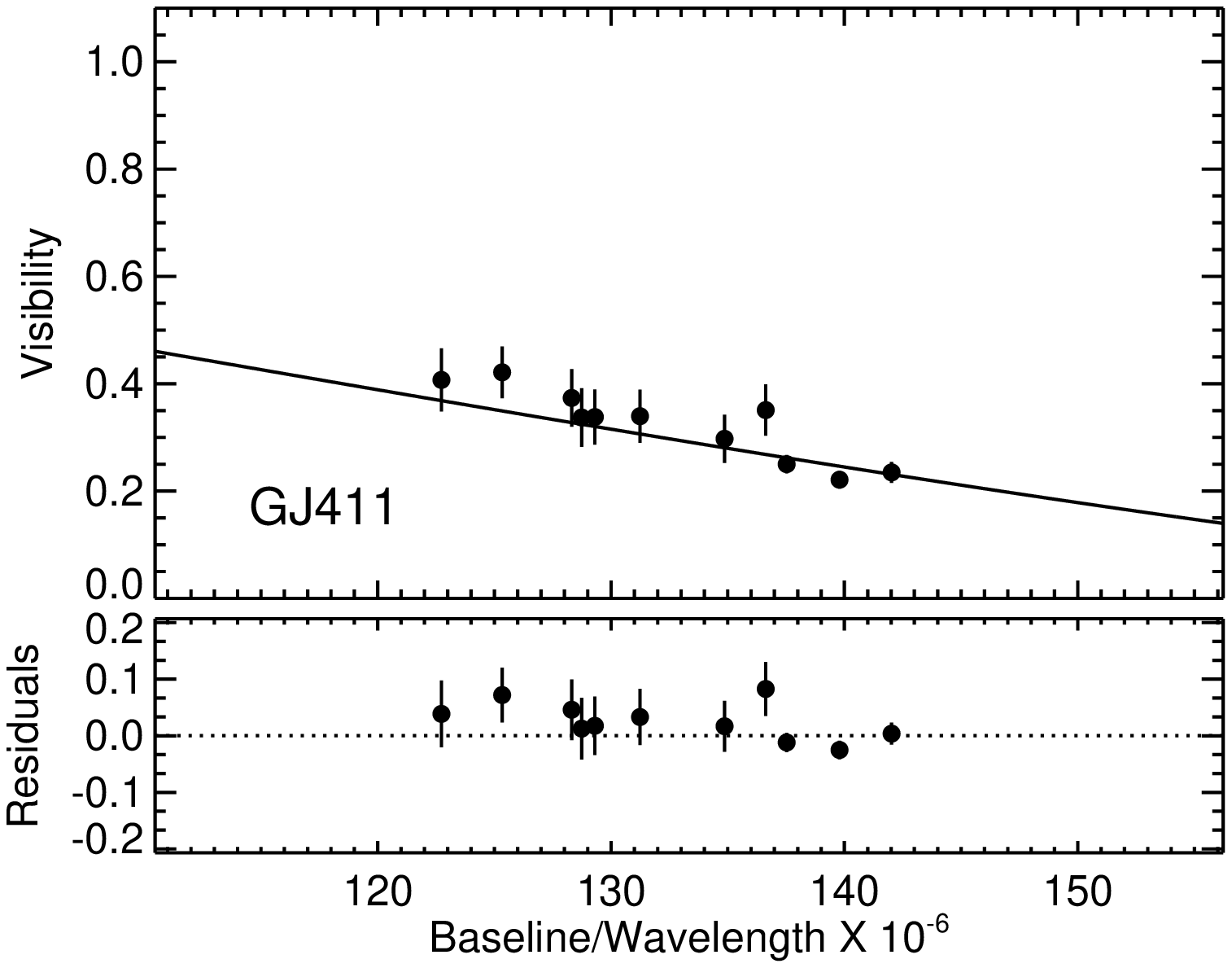, width=0.5\linewidth, clip=} &
\epsfig{file=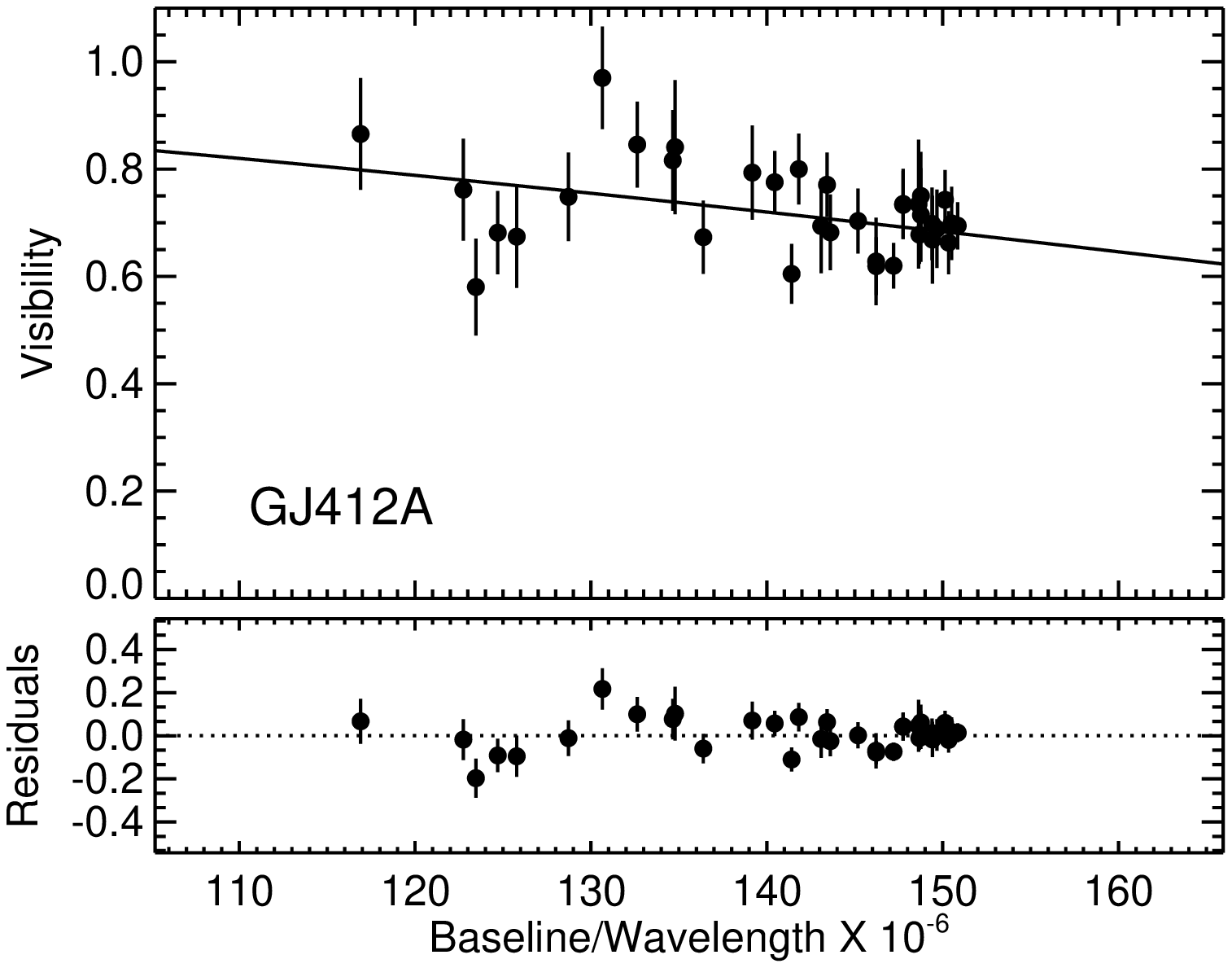, width=0.5\linewidth, clip=} \\
\epsfig{file=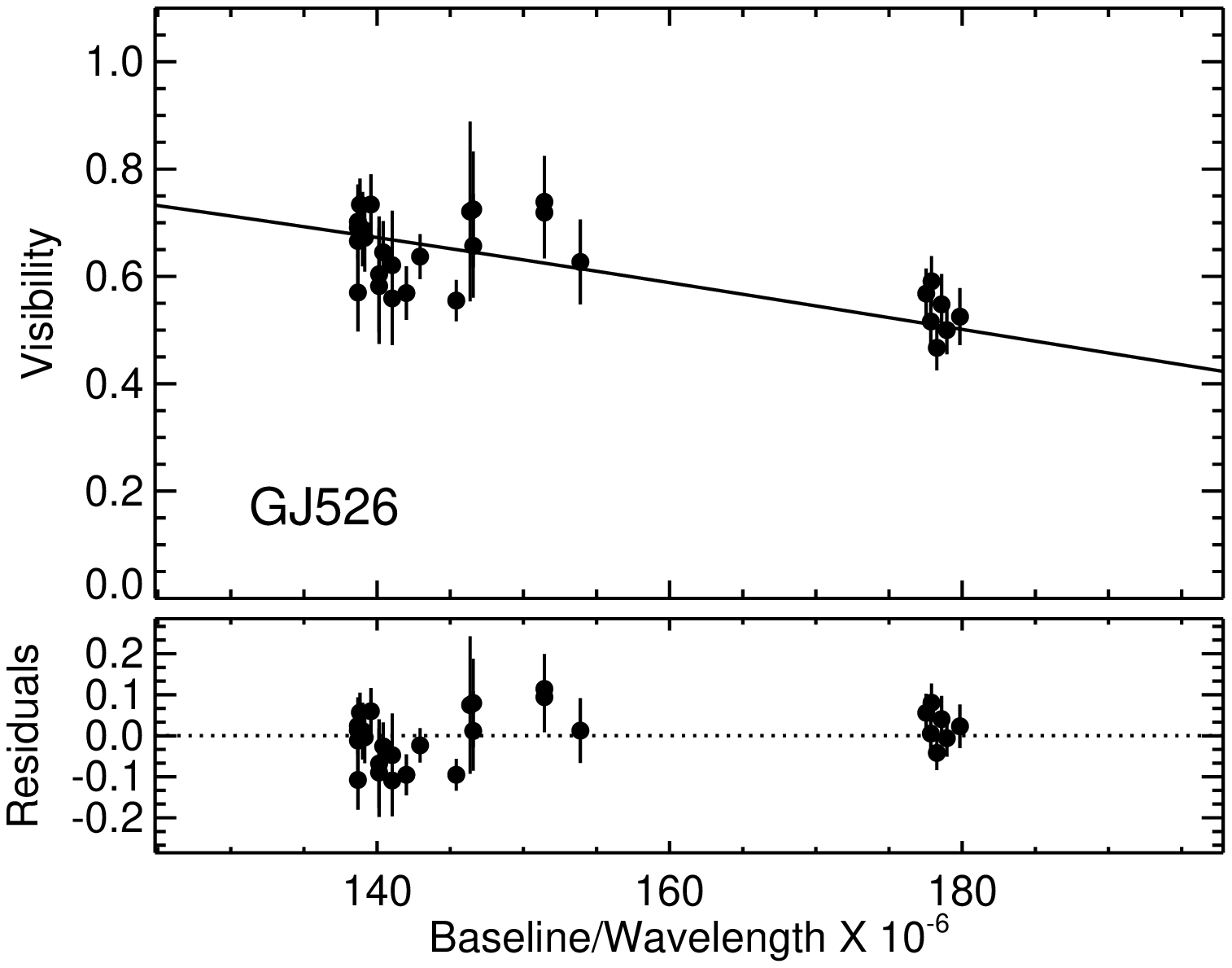, width=0.5\linewidth, clip=} &
\epsfig{file=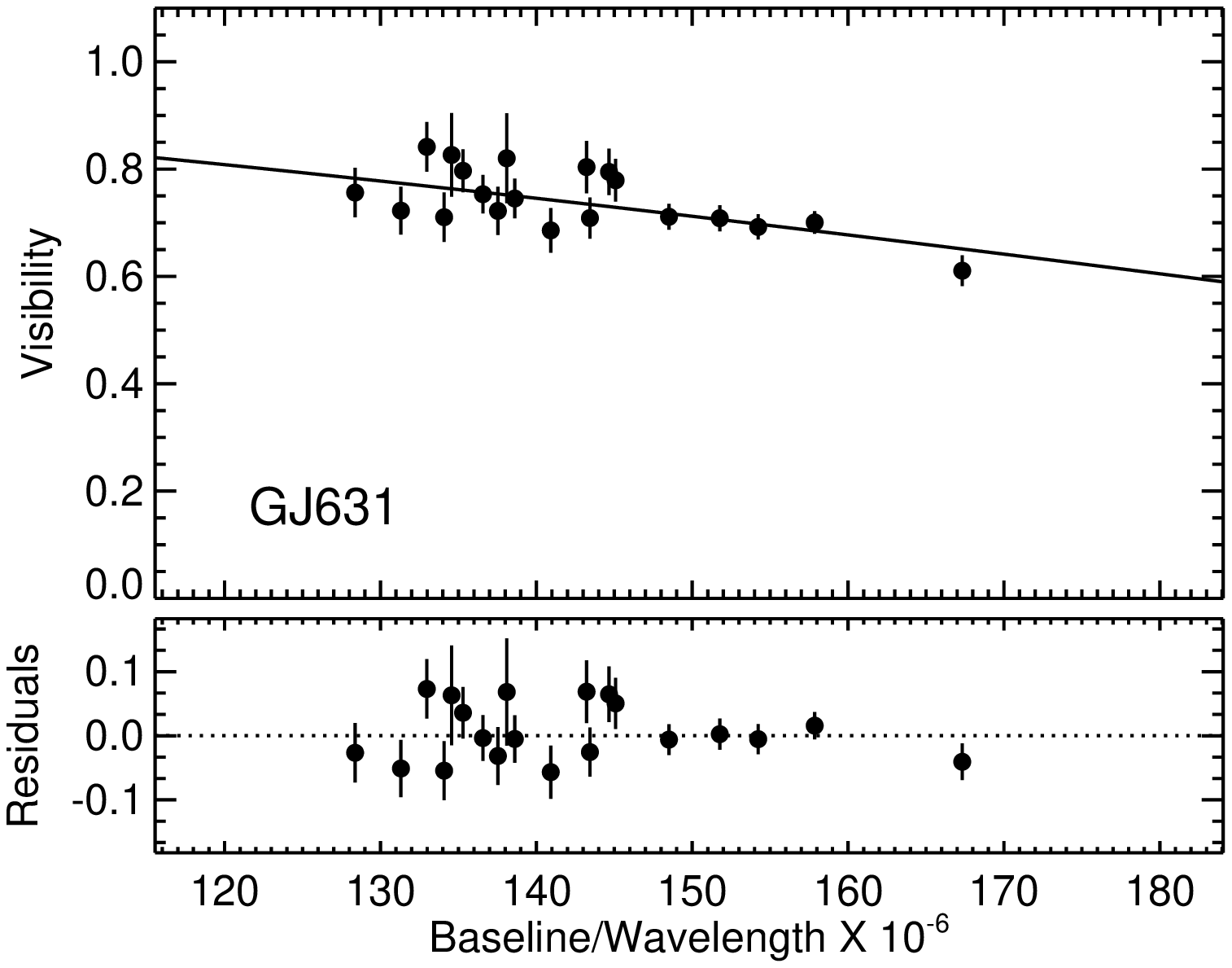, width=0.5\linewidth, clip=} 
 \end{tabular}
 \caption[Angular Diameters] {Calibrated observations plotted with the limb-darkened angular diameter fit for each star.}
 \label{fig:diameters2}
 \end{figure}

\begin{figure}										
 \centering
 \begin{tabular}{cc}
\epsfig{file=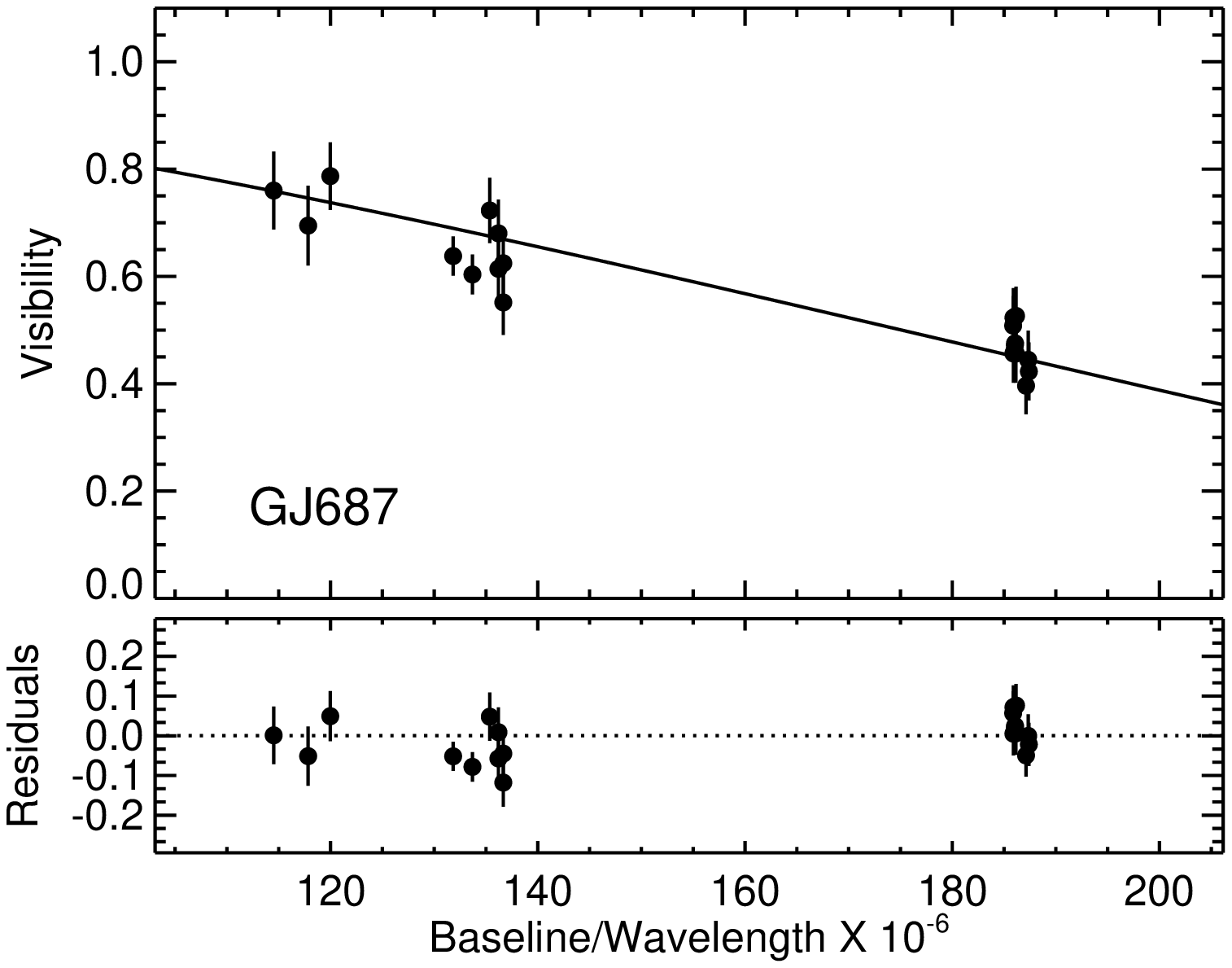, width=0.5\linewidth, clip=} &
\epsfig{file=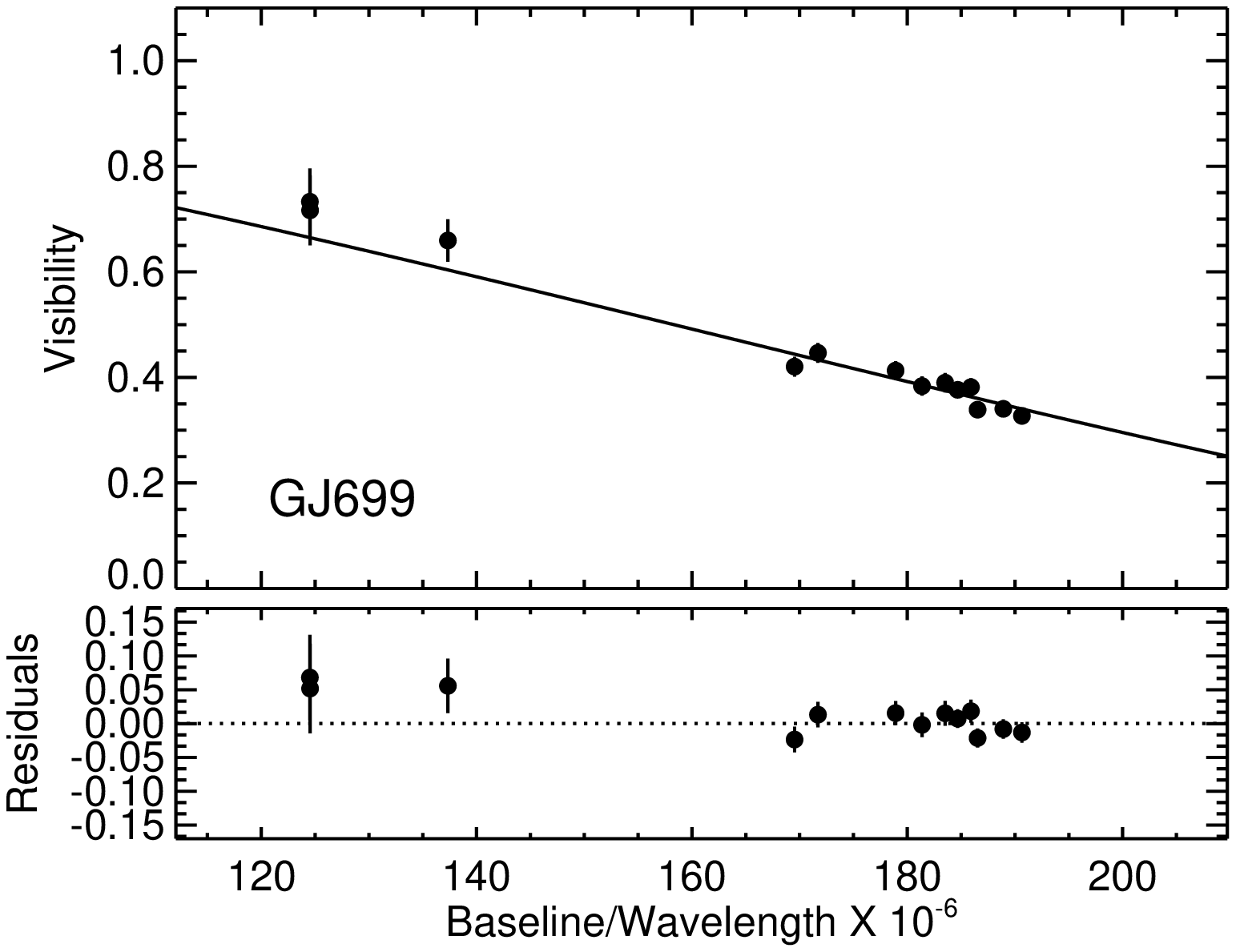, width=0.5\linewidth, clip=} \\
\epsfig{file=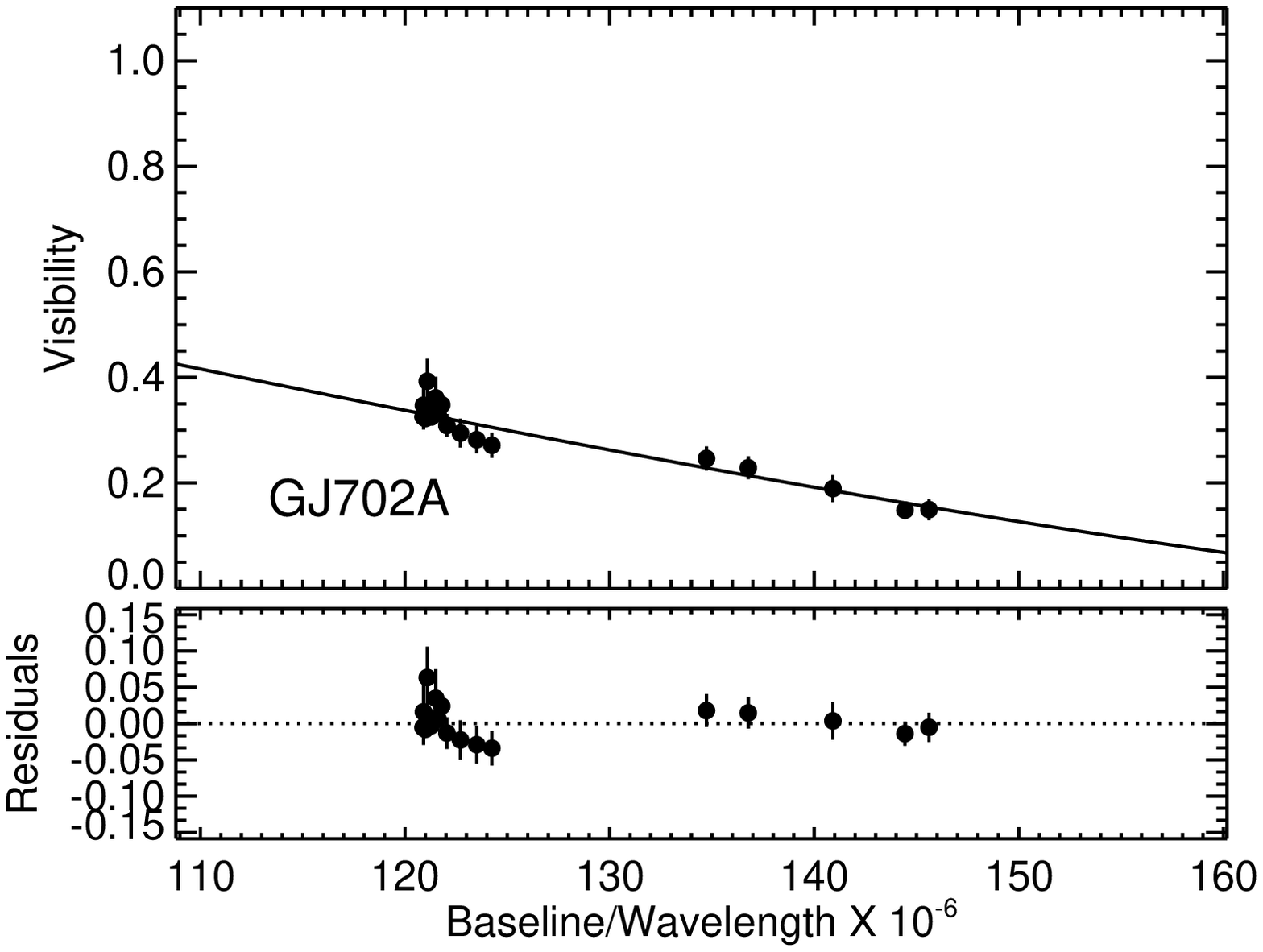, width=0.5\linewidth, clip=} &
\epsfig{file=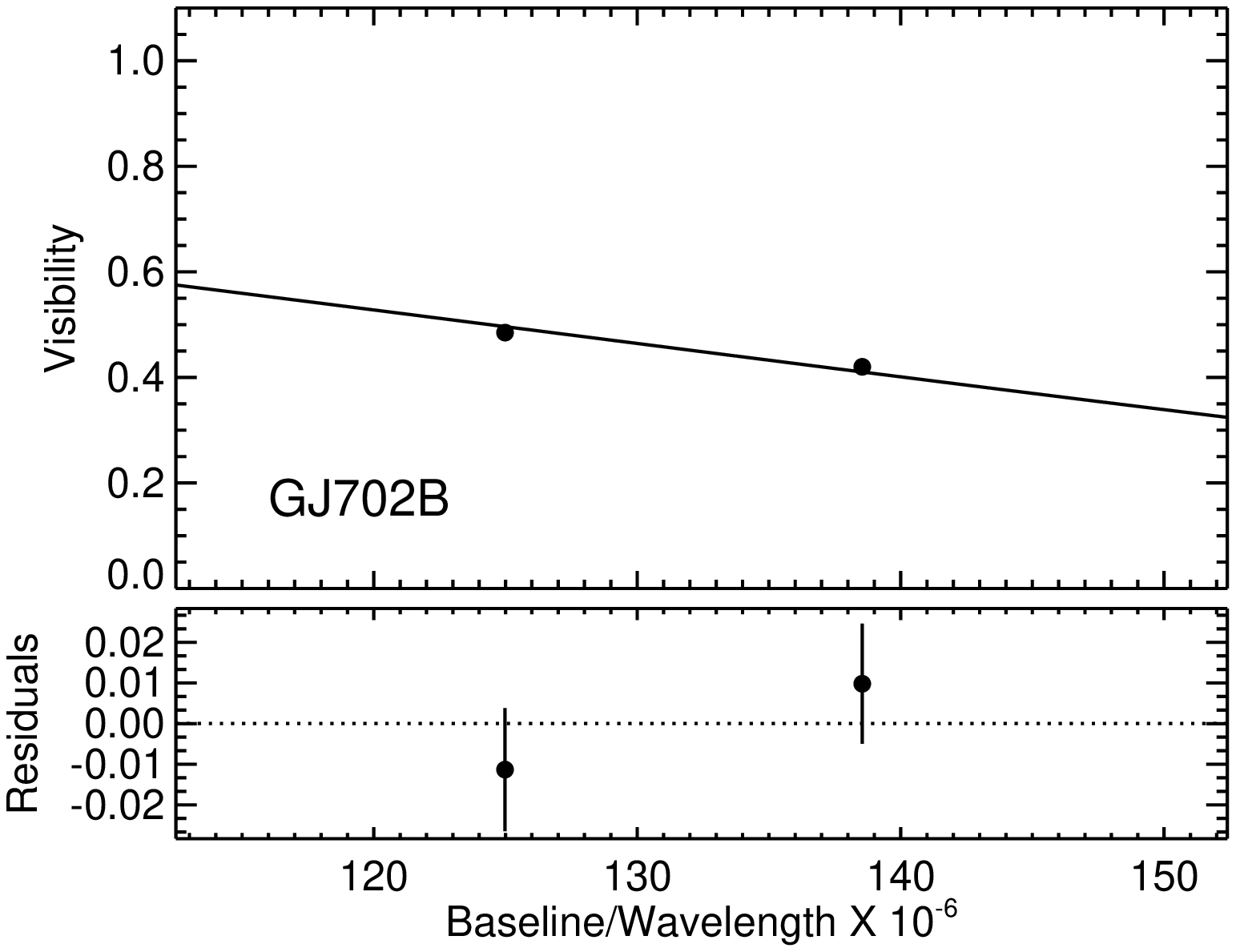, width=0.5\linewidth, clip=} \\
\epsfig{file=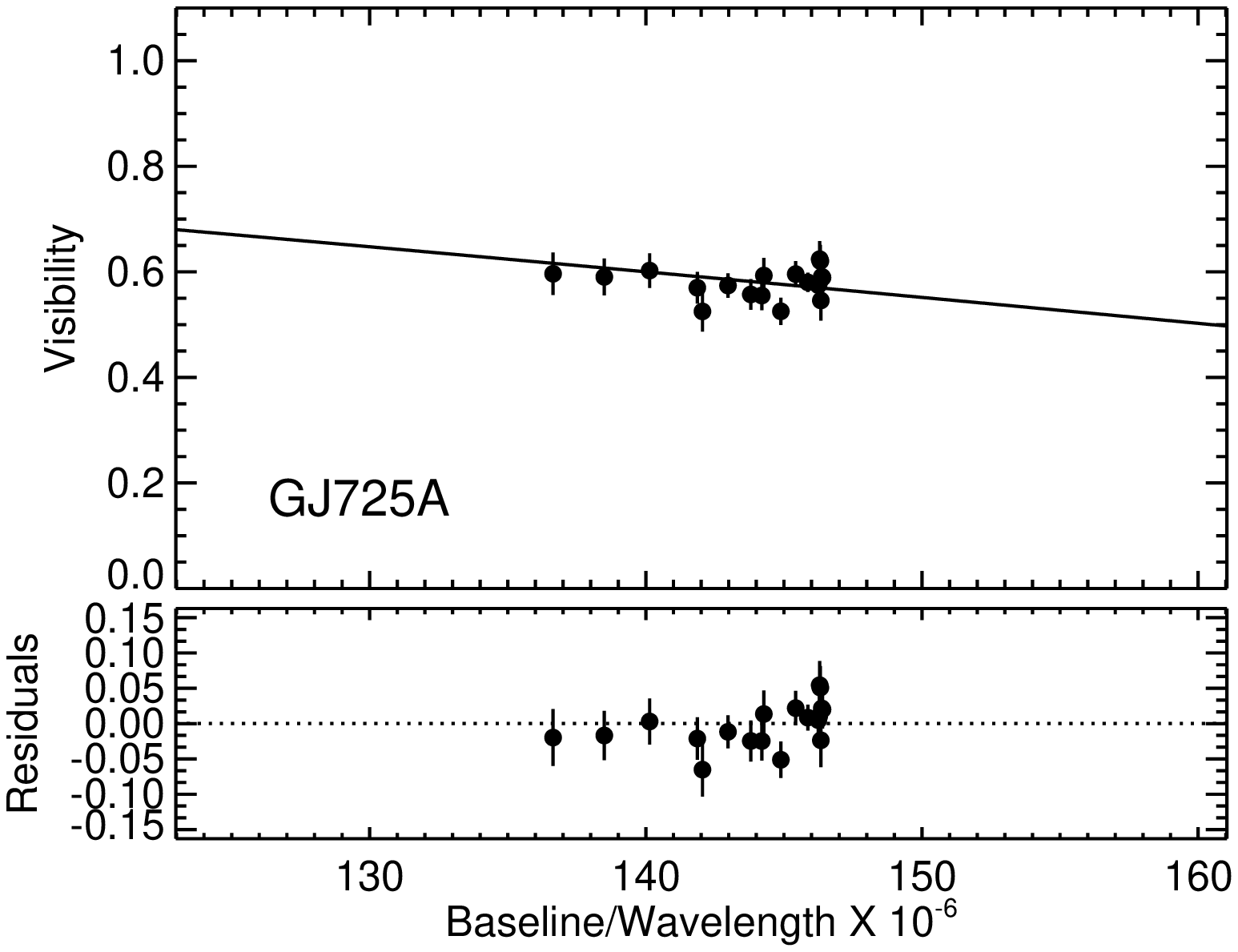, width=0.5\linewidth, clip=} &
\epsfig{file=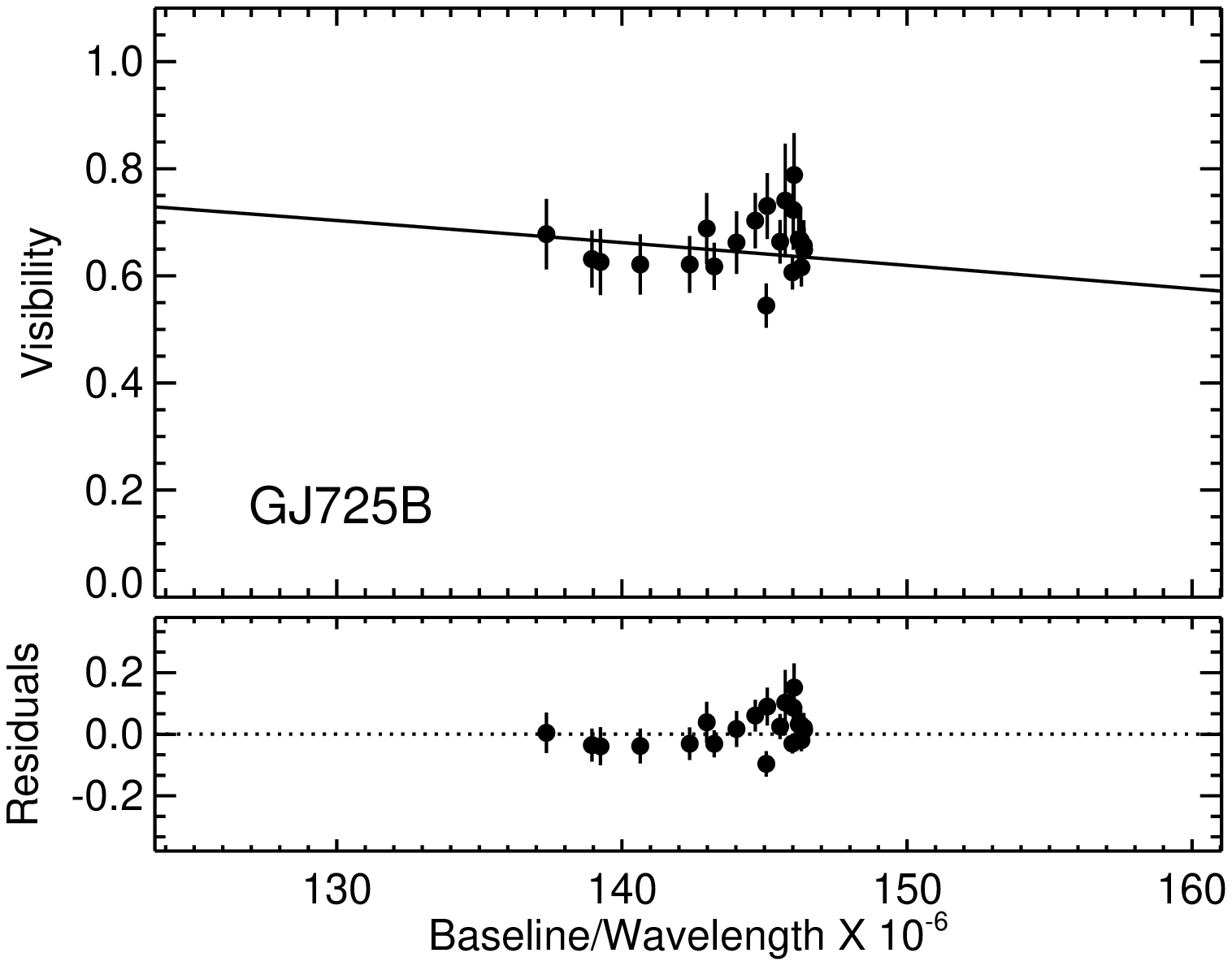, width=0.5\linewidth, clip=} 
 \end{tabular}
 \caption[Angular Diameters] {Calibrated observations plotted with the limb-darkened angular diameter fit for each star.}
 \label{fig:diameters3}
 \end{figure}

\begin{figure}										
 \centering
 \begin{tabular}{c}
\epsfig{file=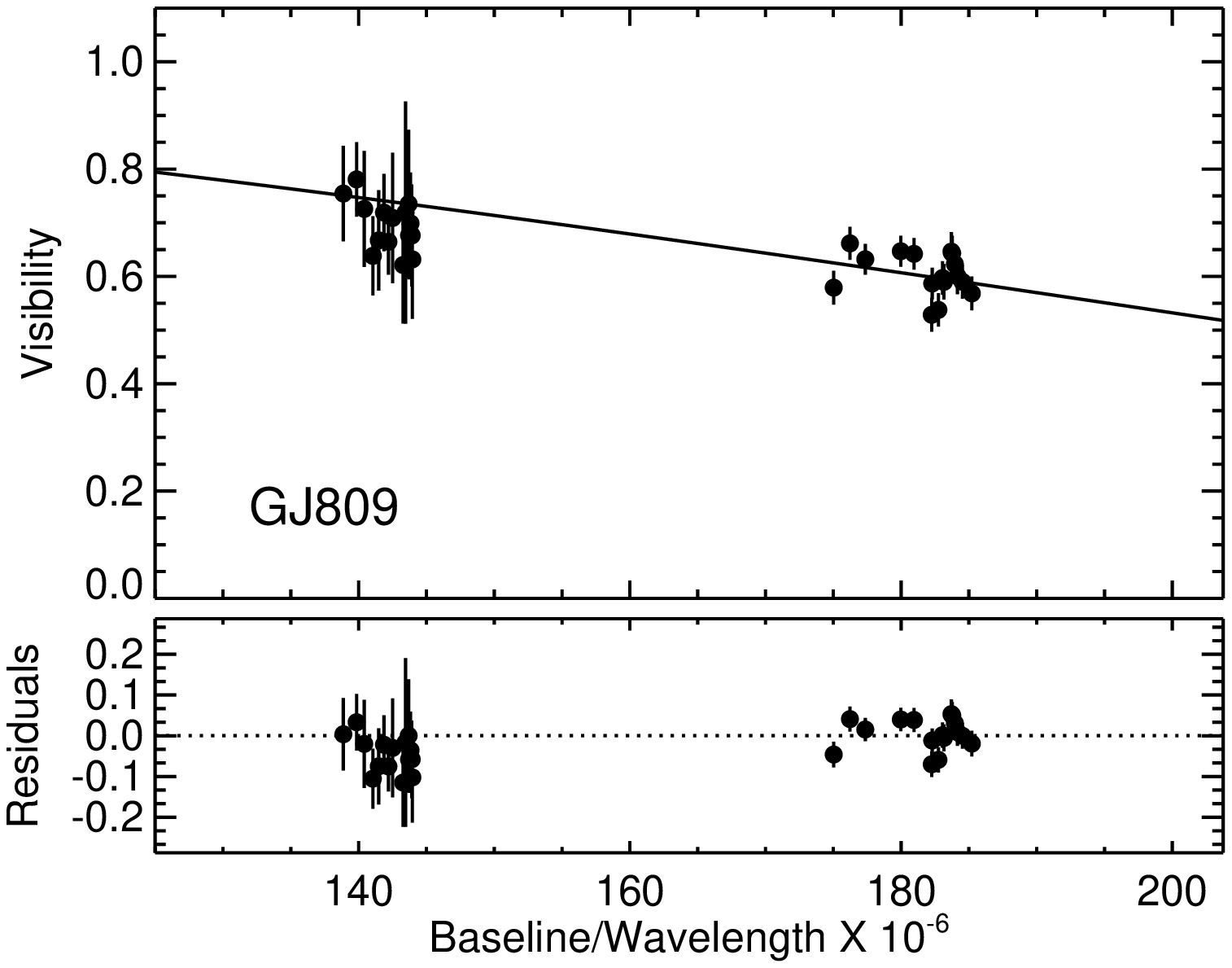, width=0.5\linewidth, clip=} \\
\epsfig{file=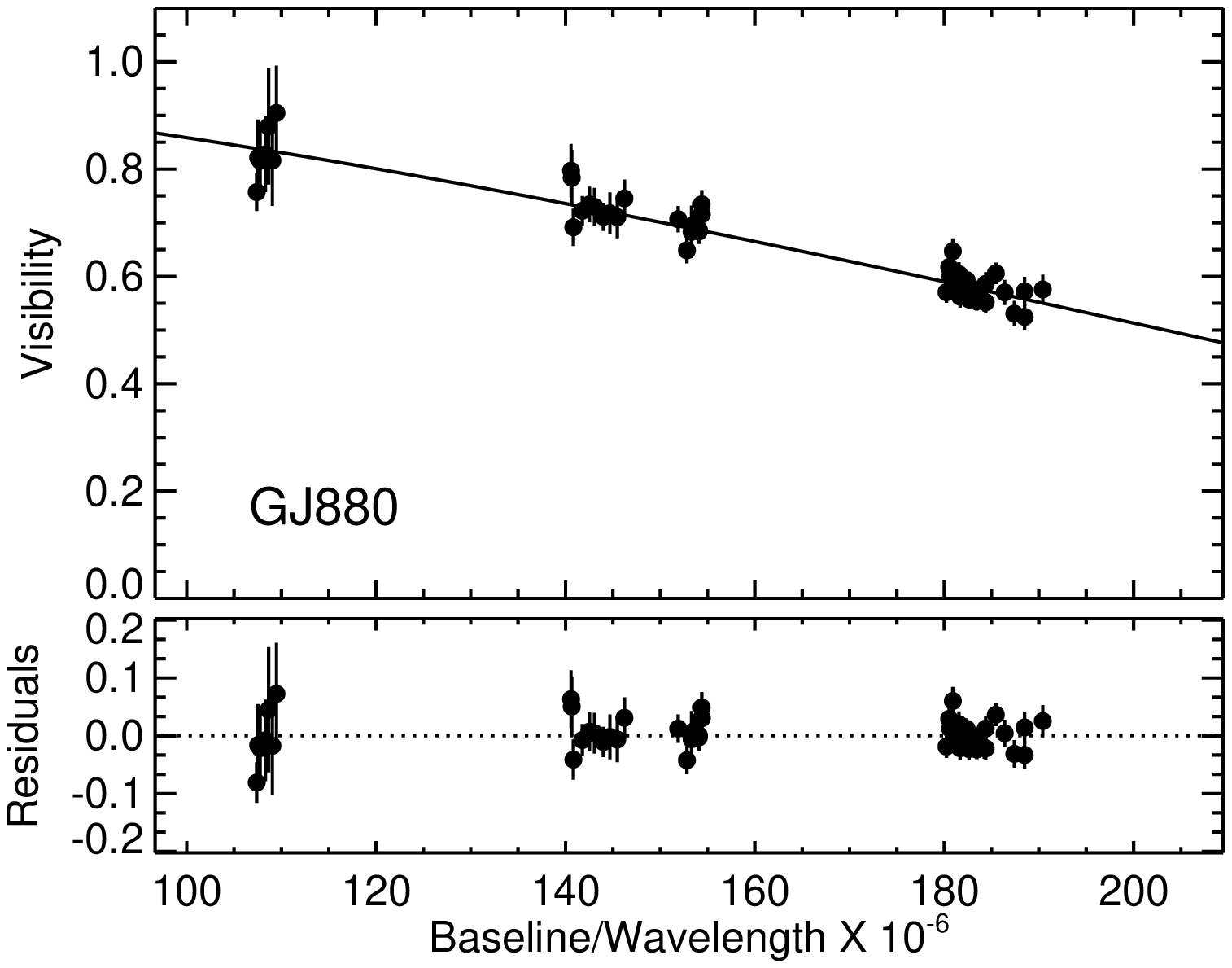, width=0.5\linewidth, clip=} \\
\epsfig{file=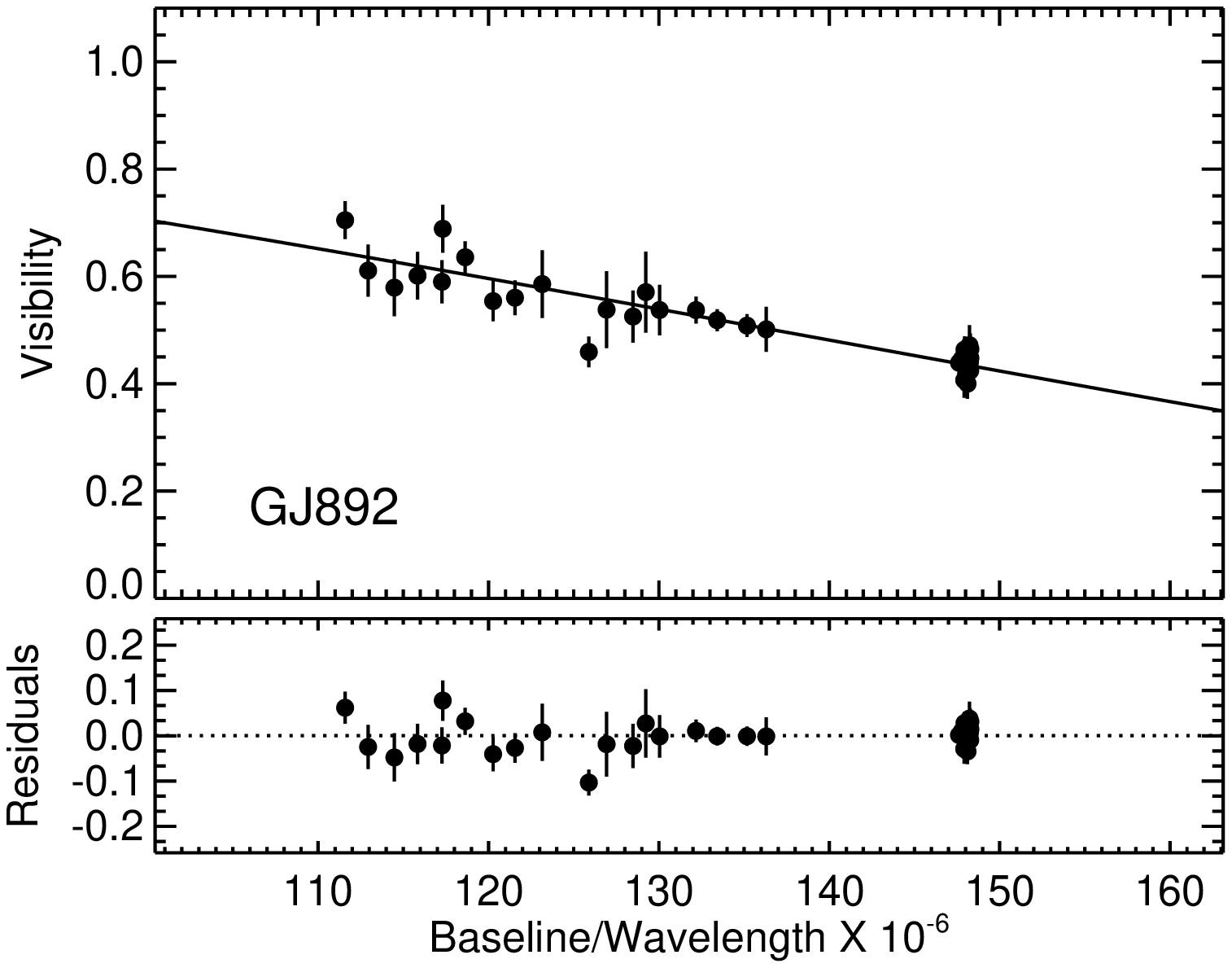, width=0.5\linewidth, clip=} 
 \end{tabular}
\caption[Angular Diameters] {Calibrated observations plotted with the limb-darkened angular diameter fit for each star.}
\label{fig:diameters4}
\end{figure}

Our results and associated uncertainties are given in Table~\ref{tab:diameters}, and Figures~\ref{fig:diameters1}, \ref{fig:diameters2}, \ref{fig:diameters3}, and \ref{fig:diameters4} display the fits to the data.  In summary, we measured the angular diameters of 9 K stars and 12 M stars with a precision of better than $2.5$\%. 

\begin{figure}										
  \centering
 \epsfig{file=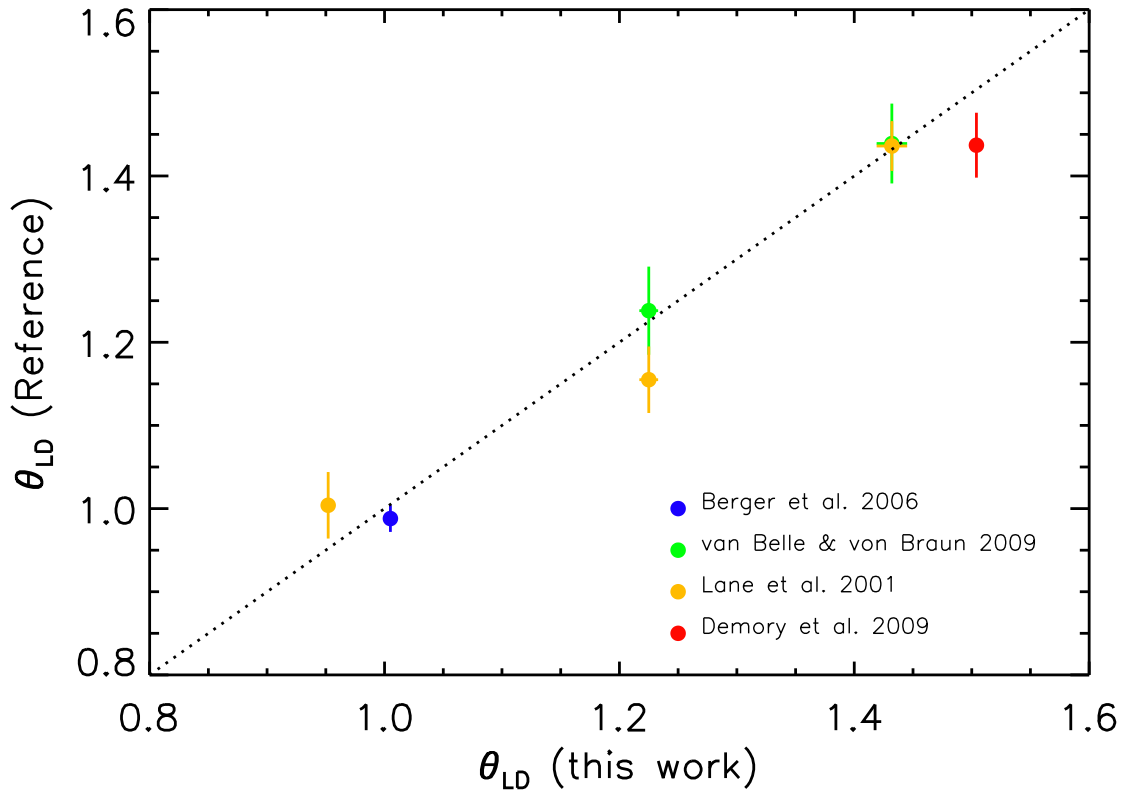, width=1.0\linewidth, clip=} 
 \caption[ ] {Limb-darkened diameters of the stars observed in common with the works of \citet{lan01, seg03, ber06, van09, dem09} compared to the values measured in this work (left to right: GJ~699, GJ~15A, GJ~380, GJ~411, GJ~166A).  Results of this consistency check leads to an agreement of $\theta_{\rm LD, this~work}/\theta_{\rm LD, reference} = 1.008$. See Section~\ref{sec:observations} for details.}
 \label{fig:comp_stars_5p}
 \end{figure}

We intentionally chose several target stars that have previously been observed by other interferometers and beam combiners to check whether our measurements are consistent with other published values and/or to improve on previous measurements with larger uncertainties. Table~\ref{tab:comp_stars} shows this comparison for 12 of the stars presented here along with spectral type, angular diameter value, and instrument used.  The table further provides the relative error of each measurement as well as the statistical difference between our measurement and the previously published value.  We indicate by marking a measurement with a $c$ where the previously published diameter error is greater than 5\%, leaving only 5 stars (7 literature measurements) that have $<$5\% uncertainty for comparison with our results (Figure~\ref{fig:comp_stars_5p}).  Literature measurements with large errors are consistent with CHARA measurements of diameters $< 1$~mas.  This is a direct consequence of the high angular resolution available with CHARA's long baselines.

An exception is the work in \citet{ber06}, where they also use the CHARA Array to measure angular diameters of stars., although their errors are still quite large. However, the \citet{ber06} results are nonetheless different from our work, possibly from their measurement errors being underestimated. For instance, as listed above, our observing strategy is to use more than one calibrator, more than one baseline, and observe on more than one night in order to minimize calibration issues that might arise. In \citet{ber06} this is not the case for GJ~15A, GJ~687, or GJ~880, and although GJ~526 is observed on more than one occasion, it lacks the additional calibrator and baselines.  We also make use of the $H$-band filter for our observations, improving the interferometers resolution by a factor of $\sim 1.5$. Upgrades to the instrument have moreover improved sensitivity and minimized calibration effects arising in data acquisition and reduction \citep{stu10}. 

The other measurements compared in Table~\ref{tab:comp_stars} come from the Palomar Testbed Interferometer (PTI) \citep{lan01, van09} and the Very Large Telescope Interferometer (VLTI) \citep{seg03, dem09} with much shorter baselines than CHARA.  We see no statistical difference between the measurements at PTI, VLTI and our own, and because of CHARA's larger baselines, we improve the precision of these measurements by no less than a factor of three for all previous results, and in one case by up to a factor of 10. Note that \citet{dem09} state that beam combiners such as CLASSIC that do not use spatial filtering are prone to systematic calibration errors. While it is true that beam combiners without spatial filtering do indeed have larger errors, we have found no evidence either in our own measurements or in the literature that these errors are systematic. Furthermore, we point out that for many years now, the CLASSIC beam combiner has included spatial filtering optimized for the faint targets \citep{ten08}.

In conclusion, we present new interferometric diameter measurements of 21 low-mass stars with an average precision of 1\%.  Nine of these (five M-stars and four K-stars) have not been measured before this work.  Of the 12 previously observed stars, our new data improve the precision for seven of these objects to better than 5\%.  The five remaining stars in our sample (GJ~15A, GJ~166A, GJ~380, GJ~411, and GJ~699) are used as consistency checks across different interferometers and instruments.  We find that the average ratio  $\theta_{\rm LD, this~work}/\theta_{\rm LD, reference} = 1.008$ proves that there is excellent agreement among the different data sets.


\section{Stellar Parameters}     
\label{sec:stellar_params}


In this section, we use our angular diameters in combination with literature data to determine stellar parameters for the 21 stars in 
Table~\ref{tab:diameters}. We use these same methods and procedures to determine properties of stars in the literature with interferometrically determined radii (see Section~\ref{sec:literature}).


\subsection{Stellar Diameters, Effective Temperatures, and Luminosities}     
\label{sec:stellar_diameters}


All of the observed stars are bright enough to have precise trigonometric parallaxes measured by Hipparcos \citep{van07}.  These values are converted to distances $d$, which range from $1.82$ to $9.75$~parsecs, and are known with an average precision of 0.5\%.  The physical stellar radii $R$ are trivially calculated via 

\begin{equation}
\label{eq:radius}
\theta_{\rm LD} =\frac{2 R}{d}.	
\end{equation}

We produce stellar energy distribution (SED) fits for all targets using flux calibrated photometry published in the literature (See Table\ref{tab:SED_phot}). We fit the photometry to spectral templates in the \citet{pic98} library, providing us with a robust and empirical approach to measuring the bolometric flux $F_{\rm BOL}$ (average errors of $\sim 1.3$\%). General details of this method are given in \citet{van08}, although contrary to the approach in \citet{van08} where interstellar reddening is a free variable in the SED fitting, this work sets the value of the interstellar extinction identical to zero since all our targets are nearby. This ensures consistency across all targets in our survey. However, we note that this approach does not attempt to take into account any potential circumstellar reddening that may be present in individual systems. Thus, it is possible that previously published stellar parameters for stars may be slightly different (within $1 \sigma$), even though they are based on largely the same data (e.g, GJ~581; \citealt{von11a}).   We were not able to measure a bolometric flux for visual binary GJ~702A and GJ~702B because the photometry for these two objects is blended due to their close separation.  Thus for these two stars, we use the luminosity values computed from \citet{egg08} using bolometric corrections.

With knowledge of $\theta_{\rm LD}$, and $F_{\rm BOL}$, we calculate stellar effective temperature $T_{\rm EFF}$ via the rewritten Stefan-Boltzmann Law

\begin{equation} \label{eq:temperature}
T_{\rm EFF} ({\rm K}) = 2341 (F_{\rm BOL}/\theta_{\rm LD}^2)^{\frac{1}{4}},
\end{equation}

\noindent where $F_{\rm BOL}$ is in units of $10^{-8}$~erg~cm$^{-2}$~s$^{-1}$ and $\theta_{\rm LD}$ is in units of mas.   The total luminosity given by $L = 4 \pi d^2 F_{\rm BOL}$ is also derived for the stars in the sample. We present these results and associated errors in Table~\ref{tab:fund_params_combined}.


\subsection{Photometry, Metallicities, Masses, and X-ray Brightnesses}
\label{sec:metallicities}


One of the main contributions of this paper is to provide empirical relations linking observables, namely broad-band photometry and metallicity, to the measured physical properties of late-type stars.  Due to the fact that these objects are very bright in the infrared, we are unable to get reliable photometry from the $2MASS$ Catalog \citep{cut03}. Alternatively, we list in Table~\ref{tab:Object_phot_forfits} the collection of broad-band Johnson $BVRIJHK$ photometry for all the program stars, along with the reference for each measurement.

Metallicities for M-dwarfs are difficult to calibrate accurately due to their complex spectral features and molecular bands contaminating the continuum. [Fe/H] can be used for diagnostic purposes to understand the effects of a star's metallicity on relations between other stellar parameters.  This is a crucial point of interest when comparing the stars to models, in which opacities in the atmospheres of M-dwarfs may not have properly been accounted for. The metallicity values for the stars in this paper were preferentially collected from a uniform source - so that we could minimize any systematic offsets with respect to different methods/references in the metallicity scales.  Unfortunately, a single catalog of stellar metallicities does not contain all of the stars in this survey.

For this study, we cite the metallicities of M-dwarfs from the calibration presented in \citet{roj12}, wherever available. Alternatively, we quote the metallicity values from \citet{nev11}, e.g., for GJ~15A and GJ~880.  We find that all of the K-dwarfs have entries in \citet{and11}, who provide average metallicity values of several available literature references. While the \citet{and11} metallicities are an average - and thus not of uniform origin - we cross check these averaged metallicities with those in the SPOCS catalog \citep{val05}, where entries are available for the majority of our stars, and we find that the values are within 0.04$-$0.05~dex each other, revealing no systematic offset.  The [Fe/H] values used in this study and references can be found in Table~\ref{tab:fund_params_combined}.

In order to compute stellar masses for the stars in the sample, we employ an absolute $K$-band mass-luminosity relation.  While the relation from \citet{del00} is commonly used for M-type stars, it does not extend to be valid for K-type stars.  Due to this, we chose to use the relation from \citet{hen93} because it is good throughout the whole mass range of the sample presented in this paper.  We note that figure~1 in \citet{del00} illustrates consistency in masses derived via the piece-wise solution derived in \citet{hen93} and the polynomial form of their solution in \citet{del00}.  We apply the conversion from \citet{leg92} to transform the Johnson $K$ magnitudes in Table~\ref{tab:Object_phot_forfits} into the CIT system \citep{1982AJ.....87.1029E} so the relations in \citet{hen93} could be applied. These corrections are on the order of 0.02~magnitudes.  The resulting estimates of the stellar mass are included in Table~\ref{tab:fund_params_combined} and assume a 10\% uncertainty.

Lastly, we use equation~1 in \citet{lop07} to convert hardness ratios from the ROSAT All-Sky Survey Bright Source Catalogue \citep{vog99} to X-ray flux for each object and compute the X-ray to bolometric luminosity ratio $L_{\rm X}/L_{\rm BOL}$, a useful diagnostic to the activity levels in late-type stars. These values are presented in the last column of Table~\ref{tab:fund_params_combined}.


\subsection{Literature Angular Diameters and Fundamental Properties}
\label{sec:literature}


At present, there are 10 published papers that present interferometrically determined radii of K and M dwarf stars with precision to better than 5\%: \citet{lan01, seg03, ber06, dif07, ker08, boy08, van09, dem09, von11a, von12}.  These results within these papers amount to 22 measurements of 17 unique sources, and are compiled in Table~\ref{tab:fund_params_combined}, which also contains the values and references for the stellar metallicities (\S~\ref{sec:metallicities}). 

We perform a SED fit for each of these stars to determine the bolometric flux, luminosity, and effective temperature in the same manner as we did for our targets in Section~\ref{sec:stellar_diameters}. The stellar mass is again derived from the $K$-band mass-luminosity relation in \citet{hen93}. Lastly, we also list the derived values of the derived bolometric to X-ray luminosity ratio $L_{\rm X}/L_{\rm BOL}$.

\subsection{Stars with Multiple Measurements}
\label{sec:averages}

Table~\ref{tab:fund_params_combined} lists each individual measurement in our survey, as well as in the literature in the top two partitions. We also mark in Table~\ref{tab:fund_params_combined} stars that have multiple measurements with a $^{\dagger}$ or a $^{\dagger\dagger}$.  The $^{\dagger}$ marked stars include the stars discussed in Section~\ref{sec:observations} that we used for consistency checks across different instruments and interferometers: GJ~15A, GJ~166A, GJ~380, GJ~411, and GJ~699.  Note that both GJ~380 and GJ~411 have two measurements in the literature. The $^{\dagger\dagger}$ marked stars indicate stars that have multiple literature measurements, but are not included in our survey: GJ~820A, GJ~820B, and GJ~887.  We determine the weighted mean of these stars' radii and temperatures, and list them in the third portion of Table~\ref{tab:fund_params_combined}.  The other properties of these stars, i.e., the spectral type, metallicity, bolometric flux, luminosity, mass, and $L_{\rm X}/L_{\rm BOL}$, are not affected by the combination of multiple measurements, and we do not reprint them in this section of Table~\ref{tab:fund_params_combined} for clarity.

In summary, there are 17 late-type dwarfs in the literature with high-precision (better than 5\%), direct measurements their radii, of which 10 are K-dwarfs and 7 are M-dwarfs. This publication adds equivalent (or higher precision) measurements of 16 additional late type stars (7 K-dwarfs and 9 M-dwarfs), plus 5 additional measurements of the 17 dwarfs in the literature for the purpose of consistency checks across facilities and techniques. The total number of directly measured radii of late-type dwarfs with precision better than 5\% thus stands at 33: 17 K-dwarfs and 16 M-dwarfs.


\section{Color - Metallicity - [Temperature, Radius, Luminosity] Relations for K- and M-Dwarfs}
\label{sec:discussion}


In this Section, we determine relations between the observable properties of stars: broad-band color index and metallicity, to our empirically measured quantities: effective temperature, radius, luminosity.  To achieve this, the observables are fit to a multi-parameter, 2nd order polynomial function, and solutions are found using the non-linear least-squares fitting routine MPFIT \citep{mar09}.  

The following analysis uses our results plus the compilation of results from the literature, i.e., data presented in Table~\ref{tab:fund_params_combined}.  We note that GJ~551 (M5.5~V, \citealt{haw96}) is the latest spectral type star in the sample, departing from the rest of the bunch by $\sim 1.7$ magnitudes in its ($V-K$) color. This places it toward the extreme boundaries in the relations, and thus introduces a bias in the fit. Although we include it in our color-metallicity-temperature fits (Equation~\ref{eq:temp_relation}), we caution that the properties of such late-type stars are not properly accounted for in this work. The coolest spectral type covered in these relations is M4~V.  As for the color-metallicity-radius, color-metallicity-luminosity relations (Equations~\ref{eq:radius_metallicity_relation} and \ref{eq:luminosity_metallicity_relation}), we do not include GJ~551 in the analysis because the adopted functional form of the relation is not appropriate over such a long baseline.  This is covered in more detail in the sections to follow.  The metallicity of the sample ranges from [Fe/H]~$= -0.68$ to $+0.35$, where the median metallicity value is sub-solar at [Fe/H]~$=-0.19$.  The full sample of data is used in the fitting procedures regardless of metallicity, however we impose limits to where the relations hold true to where there is uniform sampling within the data set: stars with metallicity values ranging from $-0.5$ to $\sim +0.1$~dex.

The specific form of each function (i.e., color-metallicity-temperature, color-metallicity-radius, color-metallicity-luminosity.) is defined in the sections to follow. For the temperature, radius, and luminosity relations, we find that adding the metallicity as an additional parameter improved the fit up to a factor of two, where the strongest influence is exhibited on the redder (M-type) stars in the sample. Our choice of polynomial functions is based merely on the way the data appeared to present themselves, and has no physical relevance.  For the radius and luminosity relations, the use of higher order ($>2$) polynomials to fit the data was tested, but did not model the data any better.  We stress that the functional form of the polynomials used for the radius and luminosity relations are strictly valid only for the range of stars specified.  Beyond this range on both the hot and cool ends, the structure of the curve modelling the stellar properties will exhibit inflection points (see discussion in the sections to follow), requiring the use of higher order functions.
We find that the major weakness in using a higher order polynomial is the lack of a more uniform sampling the data at the endpoints of the relations with respect to an object's color and metallicity, leading to poor constraints on the fits. For this reason, we do not incorporate these solutions in this work. However, we are able to ameliorate this aspect on the blue end of the fits by adding data for G-type stars to establish realistic (and empirical) boundary conditions in the color-metallicity-temperature relations, while still using a 2nd order polynomial.  This is discussed more in Section~\ref{sec:color_temperature}.

The form of the function relating stellar temperature to other parameters varies in the literature. For instance, color-temperature relations (without metallicity) have been expressed ranging from a 2nd order polynomial (e.g, see \citealt{bla91}) to a 6th order polynomial (e.g., see \citealt{gra92, boy12}), and even a power function \citep{van09}.  When including the color and metallicity as variables, the preferred form is the one we adopt here, although it is parametrized to contain the reciprocal temperature, $\theta_{\rm EFF} = 5040/T_{\rm EFF}$, as the fitting constant (e.g., see \citealt{alo96}). Additional corrections have been applied by fitting a 6th order polynomial to the residuals of the 2-variable color-metallicity function, e.g., in the work of \citet{ram05}.  We anticipate that in the near future, new data published on the fundamental properties of these types of stars will remedy this shortfalling and allow for robust testing to the true form of these relations (See Section~\ref{sec:conclusion}). We conclude this discussion by introducing a look-up table presenting the average temperature and radius for each spectral type in the sample (Section~\ref{sec:spectral_type_relations}). 

In each of the following sections defining the empirical color-metallicity-temperature (Section~\ref{sec:color_temperature}) , color-metallicity-radius (Section~\ref{sec:empirical_radius_relations}), and color-metallicity-luminosity (Section~\ref{sec:empirical_lumin_relations}) relations, we overlay model predictions to compare with our results.  However, we caution that a straightforward comparison of our solutions with models is indirect when regarded on the observational (color) plane, for the physical solutions of models' predictions must be transformed to the observational plane through color tables.  Errors introduced in color table conversions, whether synthetic or empirical based, are still under scrutiny \citep{dot08, van10}.  Due to this, we refrain on discussing in great detail any incongruity within the observations and models on the color planes.  In Section~\ref{sec:global_relations} however, we bypass the color planes and relate the physical quantities of stars directly.  In these such cases, we are able to fully describe and quantify any anomalies in the model predictions compared to our observations.


\subsection{Color-Metallicity-Temperature}
\label{sec:color_temperature}


In this Section, we derive empirical color-metallicity-temperature relations with the combined data set (Table~\ref{tab:fund_params_combined}), using the indices of the Johnson photometric system: ($B-V$), ($V-R$), ($V-I$), ($V-J$), ($V-H$), and ($V-K$) (Table~\ref{tab:Object_phot_forfits}). In order to properly model the relations at the blue end of our data range, we add an additional seven G-type stars from our Paper~I \citep{boy12}, whose properties were determined in the same manner as this work.  We provide a table of the G-stars from Paper~I in Table~\ref{tab:DT1_stars}. While other G-type star angular diameters are available in Paper~I, we find that only the seven listed in Table~\ref{tab:DT1_stars} have a complete photometric collection of all $BVRIJHK$ magnitudes. So in order to ensure consistency of the improvement that the inclusion of additional measurements of G-stars will have on the color-metallicity-temperature relations we derive, we selected only these seven G-stars that will contribute equally in the relations for all color indices applied in this work. 

\begin{figure}										
  \centering
   \begin{tabular}{cc}
 \epsfig{file=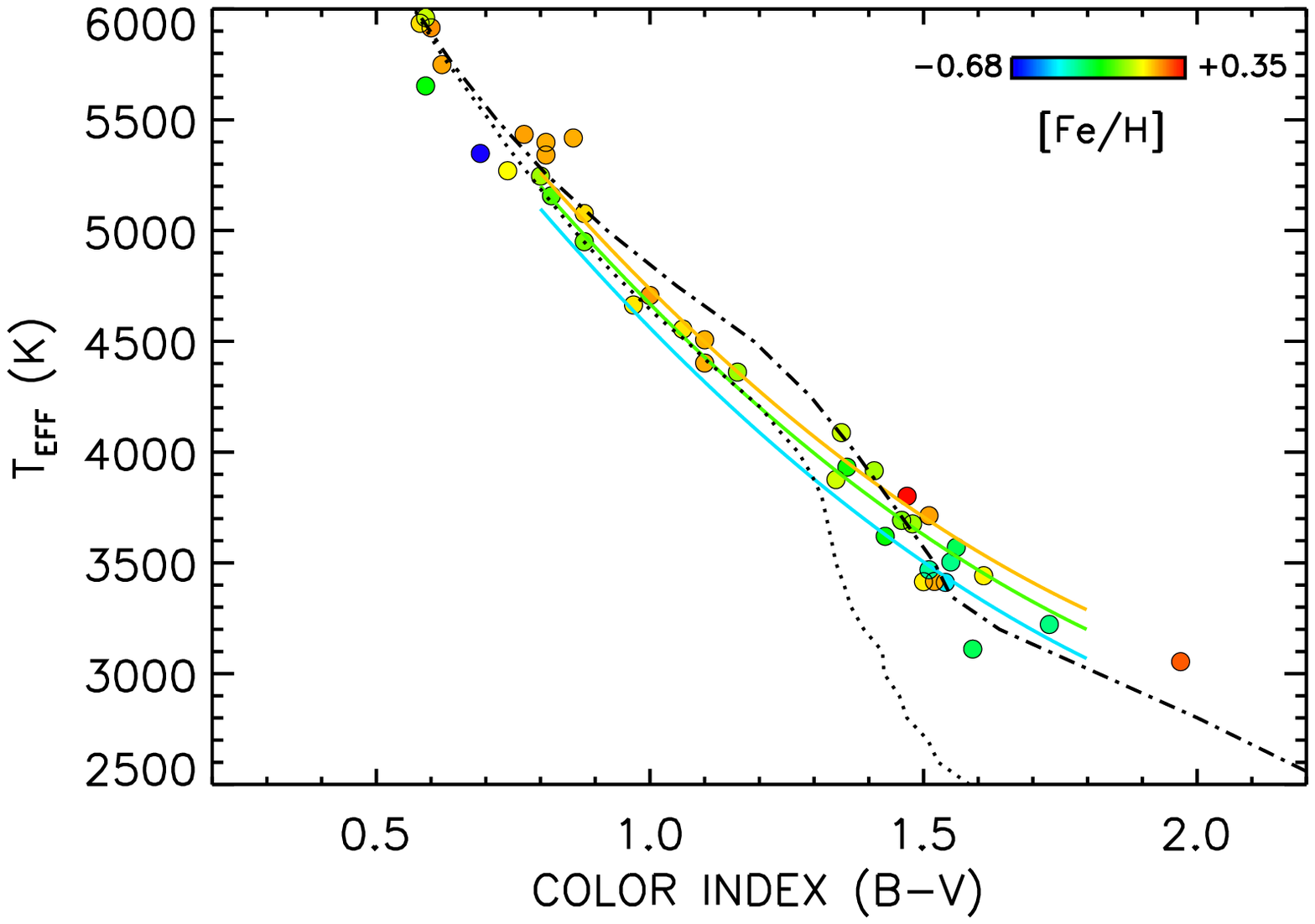, width=0.4\linewidth, clip=}  &
 \epsfig{file=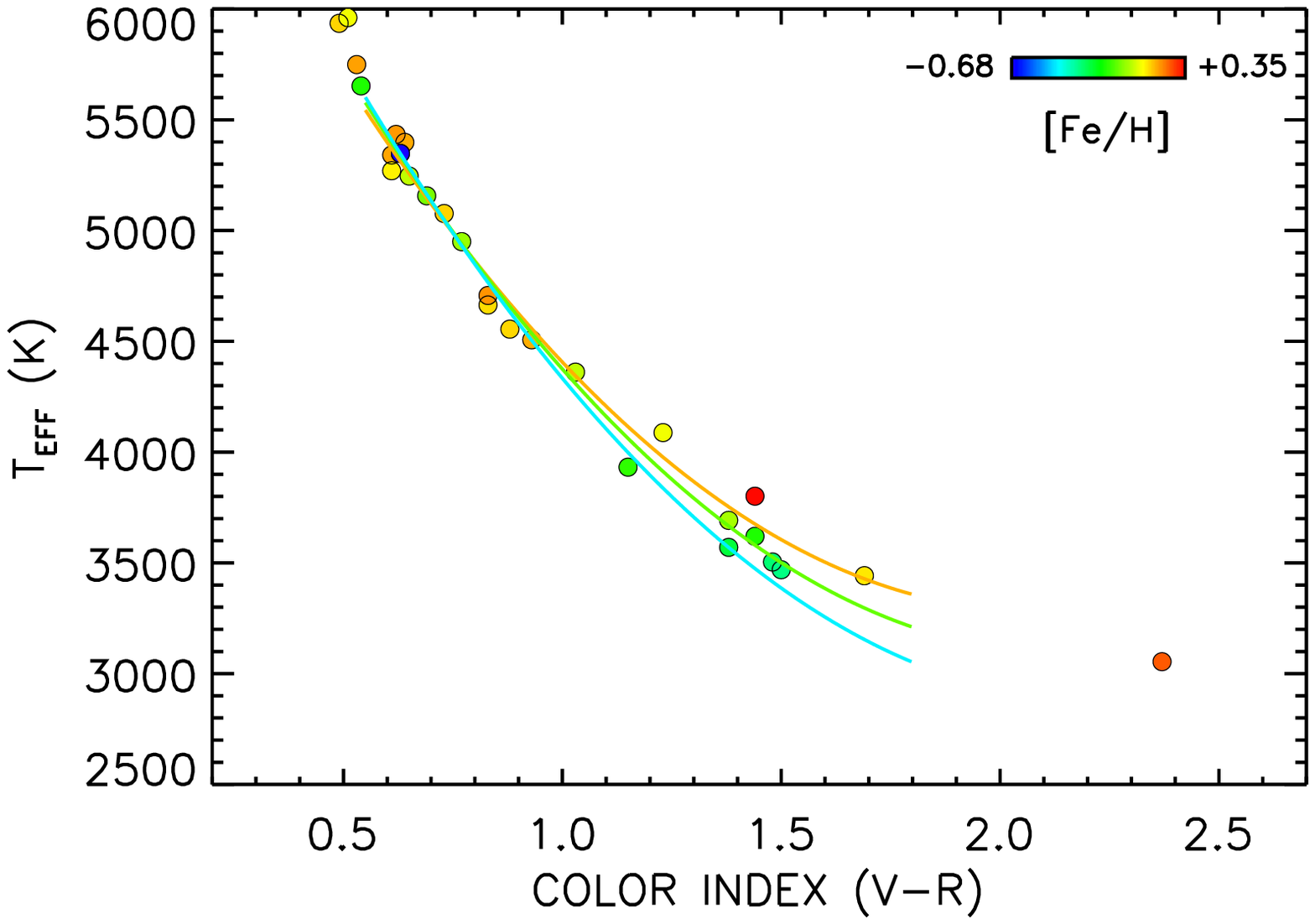, width=0.4\linewidth, clip=}  \\
 \epsfig{file=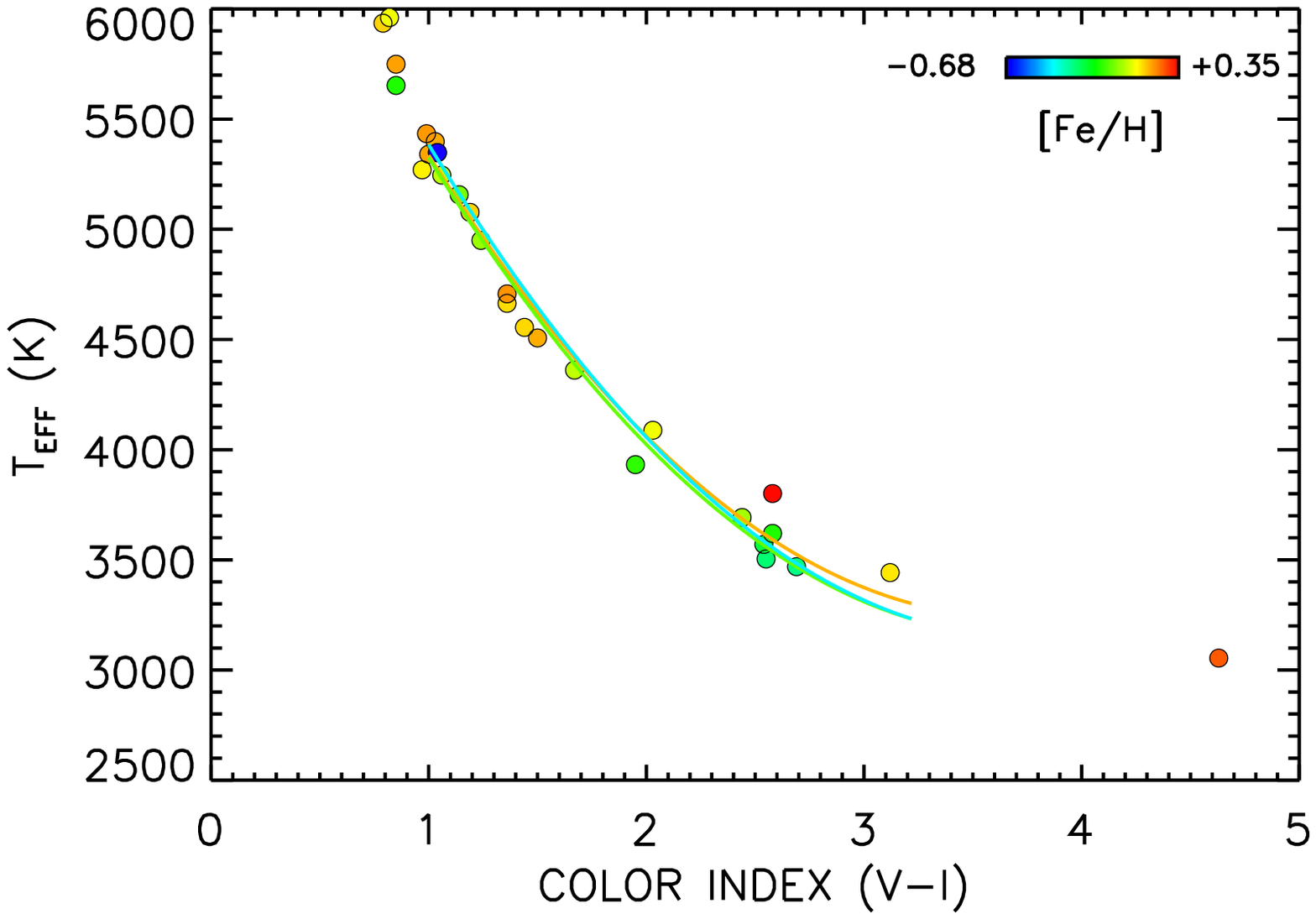, width=0.4\linewidth, clip=}  &
     \epsfig{file=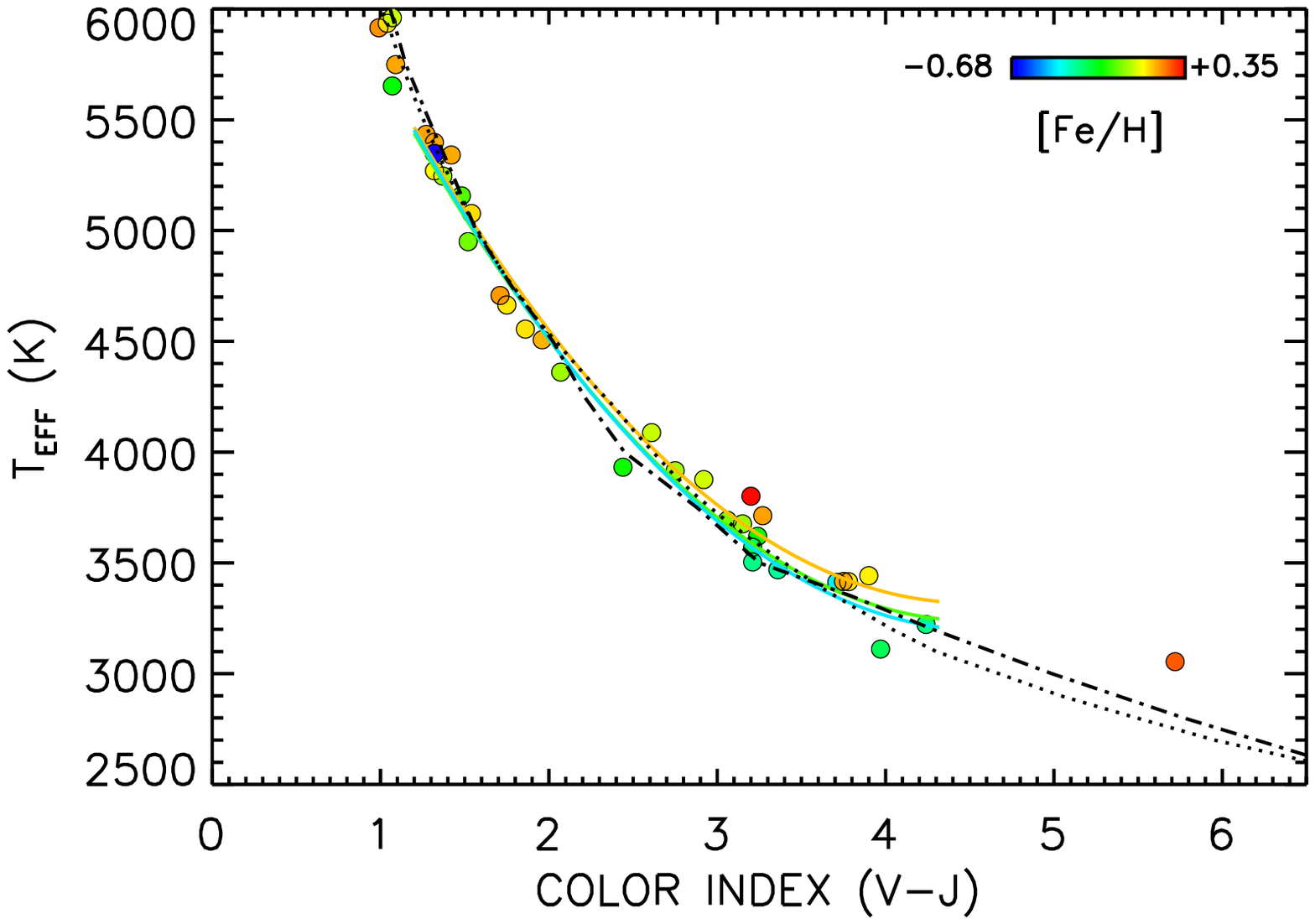, width=0.4\linewidth, clip=} \\
          \epsfig{file=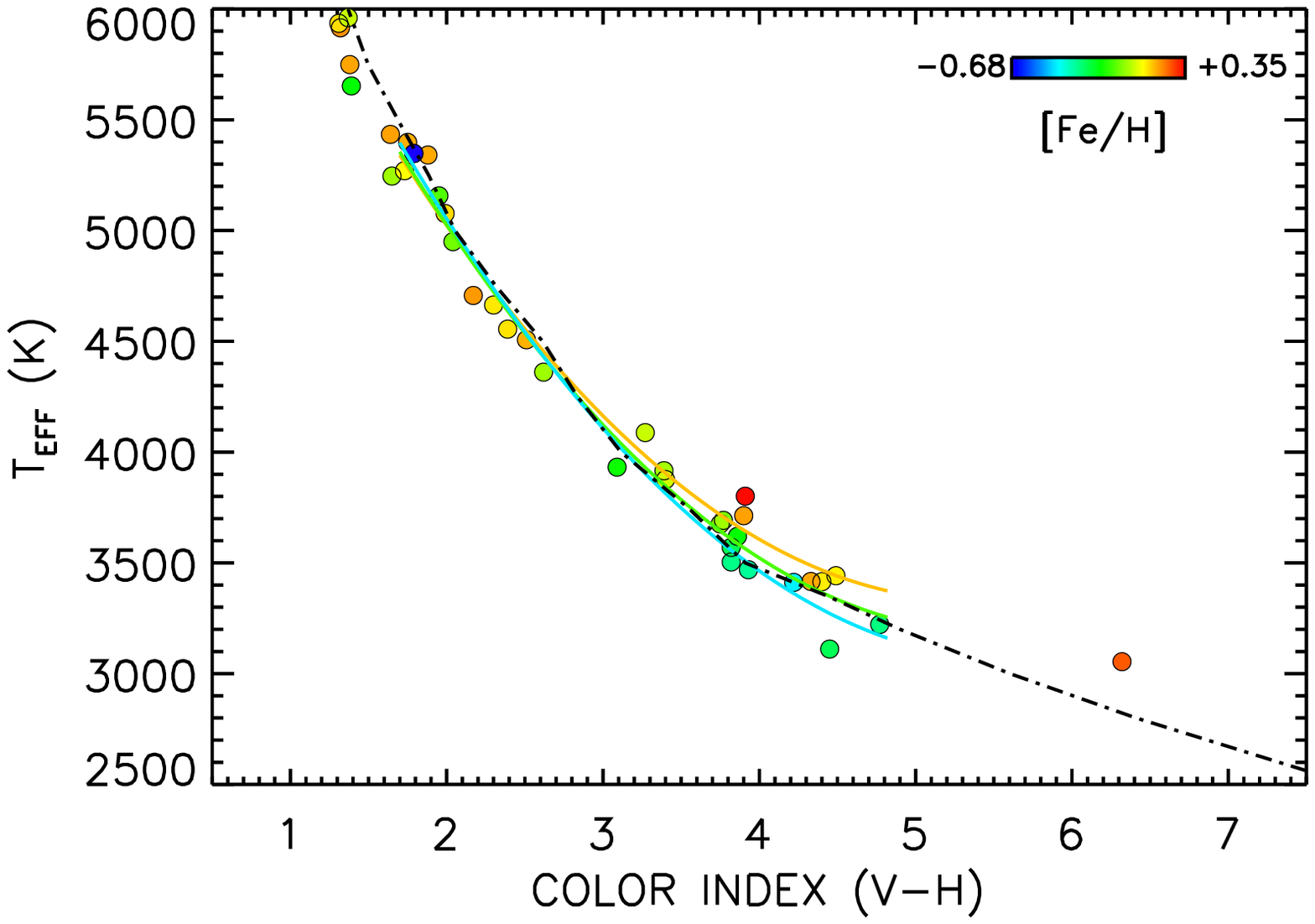, width=0.4\linewidth, clip=} 	&
            \epsfig{file=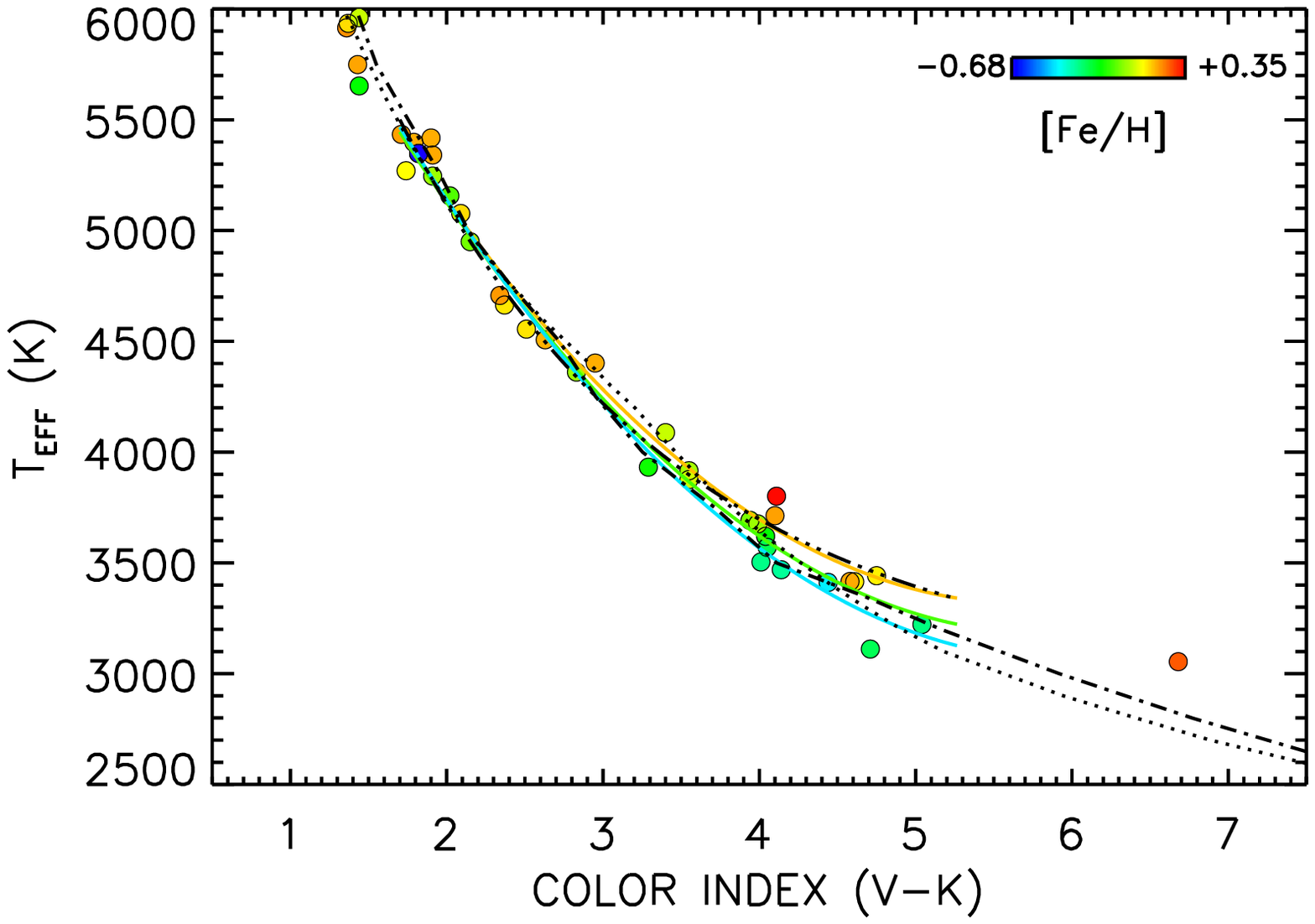, width=0.4\linewidth, clip=}  
    \end{tabular}
  \caption[] {Empirical color-metallicity-temperature relations presented in Section~\ref{sec:color_temperature}. The color of the data point reflects the metallicity of the star as depicted in the legend. The colored lines are solutions to the metallicity dependent fits, where the line color (orange, green, teal) represents our solution for an iso-metallicity line to [Fe/H]~$=0.0,-0.25,-0.5$.  The dash-dotted lines are the solutions from \citet{lej98}. Solutions from the empirical relation established for dwarfs via interferometric measurements in \citet{van09} is shown as a dashed-triple-dotted line.  Color-temperature curves from the BT-Settl PHOENIX model atmospheres at a [Fe/H]~$=0$ are illustrated as a dotted line. See Equation~\ref{eq:temp_relation} for the form of the equation and Table~\ref{tab:poly_solutions} for the coefficients and statistical overview for each color-metallicity-temperature solution in this work.}
 \label{fig:temp_VS_colors_a}
 \end{figure}

The solution for temperature as a function of color and metallicity is in the form: 

\begin{eqnarray}
\label{eq:temp_relation}
T_{\rm EFF} ({\rm K}) &	= 	& a_0 + a_1 X + a_2 X^2 + a_3 X Y + a_4 Y + a_5 Y^2
\end{eqnarray}

\noindent where variable $X$ is the color index and $Y$ is the stellar metallicity [Fe/H].  The coefficients $a_0$-$a_5$ and a statistical overview containing the number of data points used in the fit ($n$), the approximate color range over which the fit is valid, and the median absolute deviation about the best fit for each of the relations are presented in Table~\ref{tab:poly_solutions}\footnote{Note that some of the K and M stars do not have published measurements in all photometric bands, and can be seen when the total number of points used in the fit $n$ is less than 33.}.  The data and solutions are plotted in Figure~\ref{fig:temp_VS_colors_a}.  We color code the data plotted in these figures to reflect the metallicity of the star: blue indicating metal poor and red metal rich (a [Fe/H] range of $-0.68$ to $+0.35$).   We also show the solution to the polynomial fit as colored solid lines illustrating a range of solutions for fixed metallicity values.

All solutions give a median absolute deviation in temperature between 43~K and 70~K. The tightest of these relations are the $V$-band to infrared color-indices (($V-H$), and ($V-K$)), perhaps because they provide the most sensitivity as to where the peak of the stellar energy distribution is located, and hence a better determination of the effective temperature.  However, we find that the ($V-J$) color gives the largest scatter.  We are uncertain about the source of the deficiency in this bandpass index, but speculate that the systematics seen for $1.5<(V-J)<2.0$, where all the points fall beneath the curve, and $1.2<(V-J)<1.5$, where the points all fall above the curve, is indicitave of the fact that a higher order function might be needed to properly model the rapidly changing slope of the data.  

In Figure~\ref{fig:temp_VS_colors_a} we show our solution as iso-metallicity lines for fixed values of metallicity in 0.25~dex increments. It can be seen that each of these iso-metallicity curves converge for the earlier-type stars. We find that adding metallicity as an extra parameter in the color-metallicity-temperature relations only improves the median absolute deviation of the data to the fit by $\sim 30$\%, with the most prominent effect between 3500 and 4000~K, where the solution for [Fe/H]~$=-0.5$ is about 100~K cooler than a solar abundance of [Fe/H]~$=0$. This is true even for the color-indices with long baselines, (i.e., ($V-J$), ($V-H$), and ($V-K$)), which we find to be just as sensitive to metallicity as the shorter baseline color-indices examined here. This could be caused by the redder apex of the SED in these types of stars and the formation of molecular features in the atmosphere that contribute to larger flux variations in the IR band fluxes.  Although this is a weak detection to the influence of the metallicity to color and temperature, it can perhaps be refined with a better sampled range of metallicities, especially for high and low values of metallicity at the cool endpoint of the relations.

We include the ($B-V$) color-metallicity-temperature relation but caution against using this relation for stars with peculiar abundances since ($B-V$) color is known to be strongly affected by stellar metallicity. In fact, Figure~\ref{fig:temp_VS_colors_a} shows one of the most extreme outliers toward the bluest of ($B-V$) colors and several sigma under the curve is the metal-poor star GJ~53A ($\mu$~Cas~A).  We note that GJ~53A does not fall on the ($B-V$)-metallicity-temperature relation, but it is not an outlier in any of the other color-metallicity-temperature relations (see Figure \ref{fig:temp_VS_colors_a}).  We chose to include it in our color-metallicity-temperature analysis to be complete, and we note that it does not change our results when removing it and re-fitting the data. In fact, all solutions except for the ($B-V$) color-metallicity-temperature relation are shown to converge at temperatures hotter than $\sim 4000$~K.
 
In Table~\ref{tab:poly_solutions} we show that the median absolute deviation of the fits using the extended GKM sample is within 5~K of the KM sample. This is shown graphically in Figure~\ref{fig:temp_VS_colors_a}, where we plot the G-dwarfs used to constrain the fit on the hotter side, along with the K- and M-dwarf sample, showing that there is a smooth transition of temperature extending to the bluer colors.   Thus, the extrapolation of the solutions on the hotter side is valid through spectral type G0.

On the other hand, extrapolation to stars redder than the ranges specified in Table~\ref{tab:poly_solutions} is not recommended. Figure~\ref{fig:temp_VS_colors_a} shows that past this specified range, there is a large gap in color index before the reddest endpoint, the star GJ~551, is encountered. So, although we use the data for GJ~551 in the fitting procedure, the paucity of data in this range makes the results unreliable for such red colors, as for the relations presented here, the position of GJ~551 deviates several hundred Kelvin from the modelled curve. 


In Figure~\ref{fig:temp_VS_colors_a} we show the ($V-K$)-temperature solution for dwarfs (expressed as a power function) derived from interferometric measurements in \citet{van09}, displaying excellent agreement with our fit. Also for comparison, we show the model based temperature scale for the solar metallicity ($B-V$), ($V-J$), ($V-H$), and ($V-K$) color-temperature solution from \citet{lej98} in Figure~\ref{fig:temp_VS_colors_a}. While the temperatures from \citet{lej98} for the earlier type stars show excellent agreement with our own, the ($V-J$), ($V-H$), and ($V-K$) solution diverges at temperatures $<4500$~K (approximately the K5 spectral type) to produce temperatures $\sim 200$ Kelvin cooler than the ones that we measure with interferometry.  For the ($B-V$) colors, the \citet{lej98} scale diverges to predict higher temperatures from $5000 < T_{\rm EFF} < 4000$~K.  Below this temperature, the data are modelled more satisfactorily with the \citet{lej98} scale, although with a slope much different than our own.  The difference in slope is no surprise, for the low-order polynomial we use in this work is not capable of modelling any rapidly changing slope and/or kinks in the data. Lastly, we show the BT-Settl PHOENIX model atmosphere color-temperature curves \citep{all12} for solar metallicity in the ($B-V$), ($V-J$), and ($V-K$) plots, expressing by far the best agreement with the temperatures we derive here for all ranges of temperatures.  It is only in the ($B-V$)-temperature solution that for stellar effective temperatures less than 4000~K, the BT-Settl model predicts a steep drop in temperature at colors too red by about $(B-V) \sim 0.2$~magnitudes.  This is due to troubles in determining the $B$ magnitudes from the synthetic PHOENIX model grid (see the discussion in \citealt{dot08}). 

\begin{figure}										
  \centering
     \epsfig{file=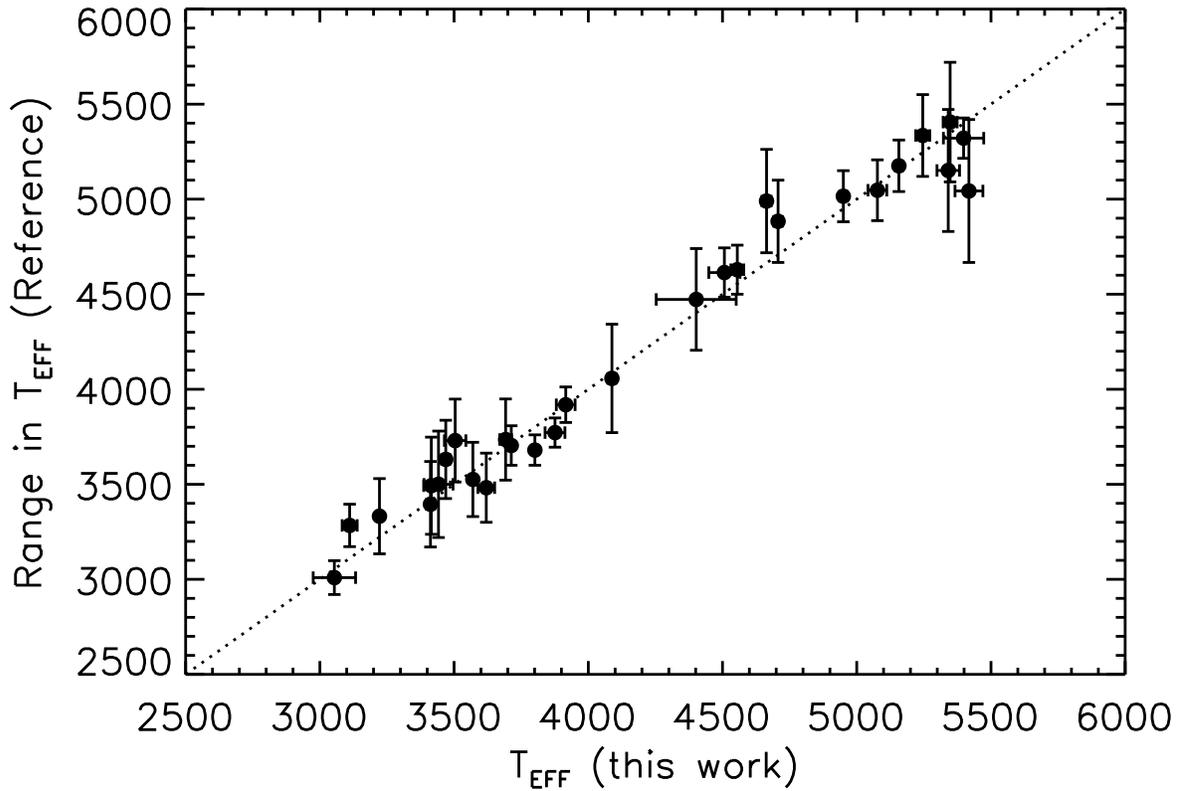, width=1.0\linewidth, clip=} 	 
  \caption[] {Temperatures determined in this work compared to those found in the literature for each star.  The position in the x-direction is our measured temperature and associated 1-$\sigma$ error on temperature.  The extent of the y-error bars indicate the range in temperature that can be found in the literature for each star. See Section~\ref{sec:color_temperature} for details.}
 \label{fig:temp_compare}
 \end{figure}

Specific references citing temperatures of these stars in Table~\ref{tab:fund_params_combined} are plentiful.  For instance, for any given target in the sample, there are, on average, tens of references to the effective temperature in the literature, estimated using a number of different techniques. As such, a quantitative comparison of our temperature values to all temperature measurements in the literature is beyond the scope of this paper.  Instead, for each star we prepare an overall qualitative assessment of how our temperatures compare to previously obtained ones by examining the full range of temperature estimates listed in Vizier\footnote{Vizier provides access to the most complete library of published astronomical catalogues and data tables available on-line.  For a description of the catalog, see \citet{och00}}.  Figure~\ref{fig:temp_compare} shows the results of this exercise, where the $x$-errors indicate our measurement precision, and the $y$-errors indicate the range in temperatures found for each object. Note that a search in Vizier may not gather all references to the temperature in the literature, but rather a collection of available on-line material. In fact, a search in Vizier resulted in no references for temperature estimates of GJ~702B, and only one for GJ~551.  We thus performed a more thorough literature search to gather a range of temperature estimates for these stars to make this comparison (the additional references include \citealt{egg08} and \citealt{luc05} for GJ~702B, and \citealt{mor08, dem09, wri11} for GJ~551).


\subsection{Color-Metallicity-Radius}
\label{sec:empirical_radius_relations}


Similar to  \S \ref{sec:color_temperature}, we use the data in Table~\ref{tab:fund_params_combined} to derive empirical color-metallicity-radius relations in the form of Equation~\ref{eq:radius_metallicity_relation}, where $X$ is the color index, and $Y$ is the stellar metallicity [Fe/H].  

\begin{eqnarray}
\label{eq:radius_metallicity_relation}
R ({\rm R_{\odot}}) &	= 	&	a_0 + a_1 X + a_2 X^2 + a_3 X Y + a_4 Y + a_5 Y^2
\end{eqnarray}

However, as opposed to the temperature relations, we refrain from including the G-type type stars in Paper~I to constrain the bluer end of the relations, as their radius may be influenced by stellar evolution (we assume that a star with radius less than 0.8 R$_{\odot}$ has not had time to evolve off the main-sequence).  The inclusion of GJ~551 is also omitted from this analysis, for the adopted form of the equation we use does not allow for extrapolation of the curve to such red colors.  

\begin{figure}										
  \centering
  \begin{tabular}{cc}
         \epsfig{file=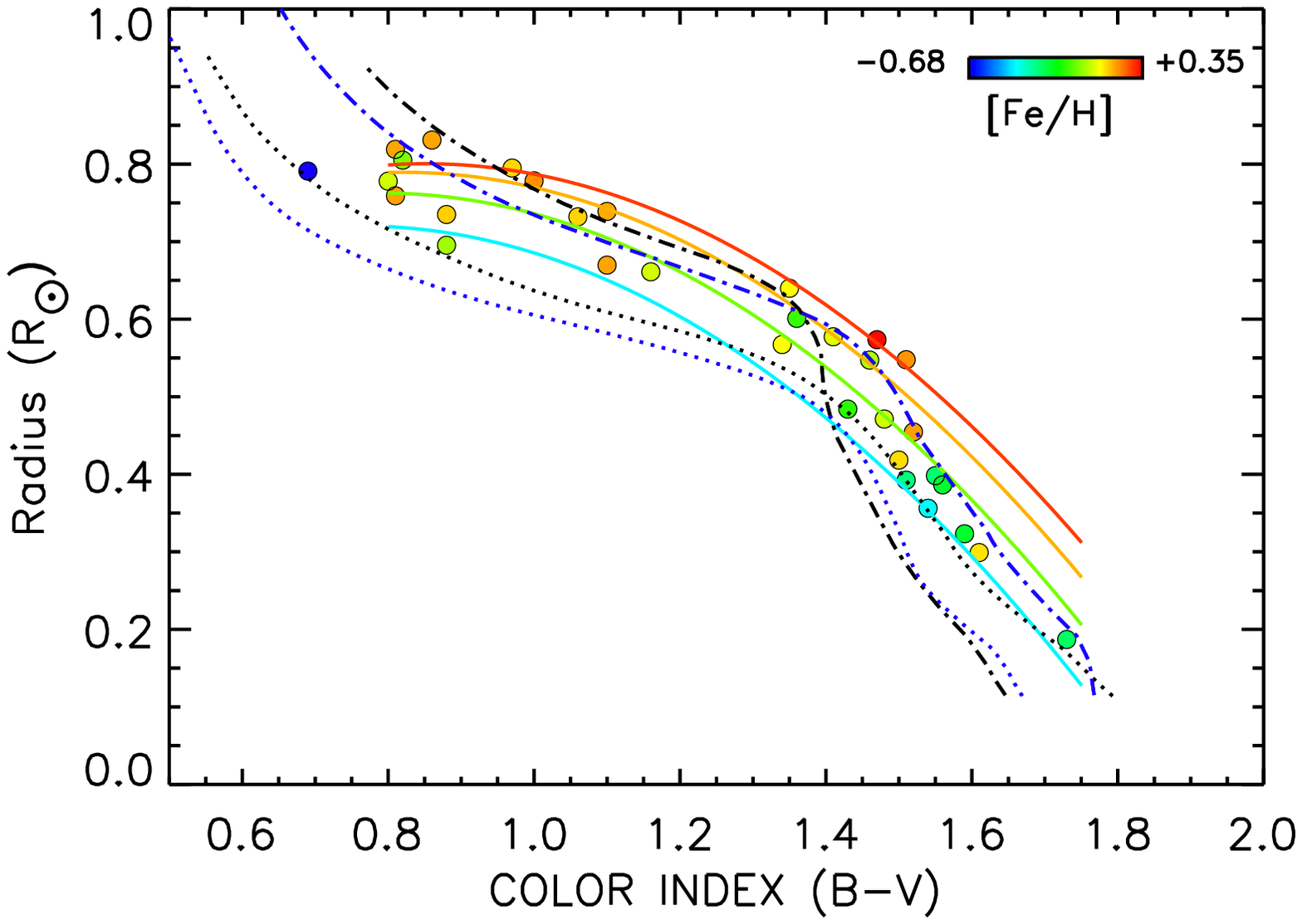, width=0.5\linewidth, clip=} &
         \epsfig{file=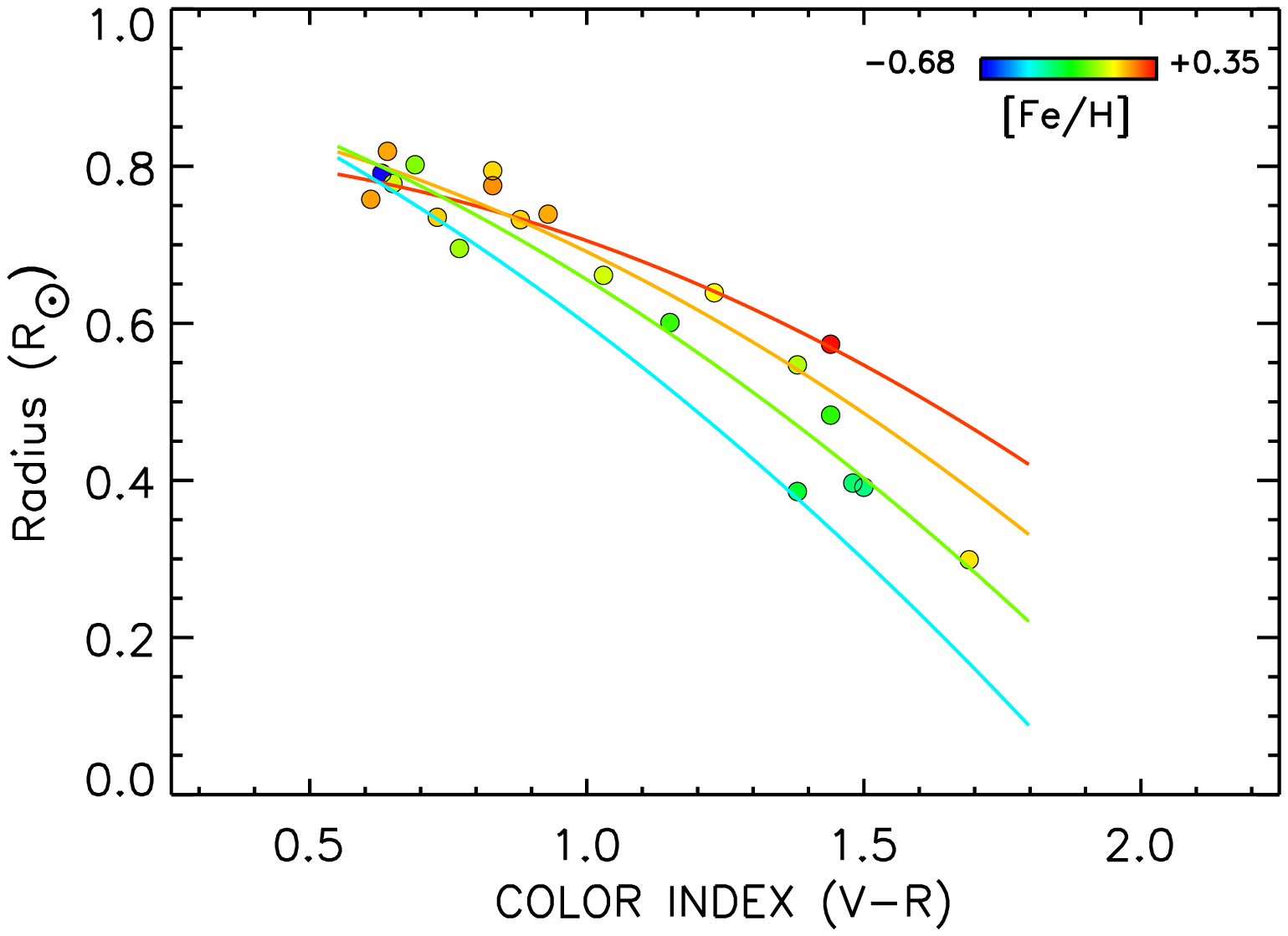, width=0.5\linewidth, clip=} \\
         \epsfig{file=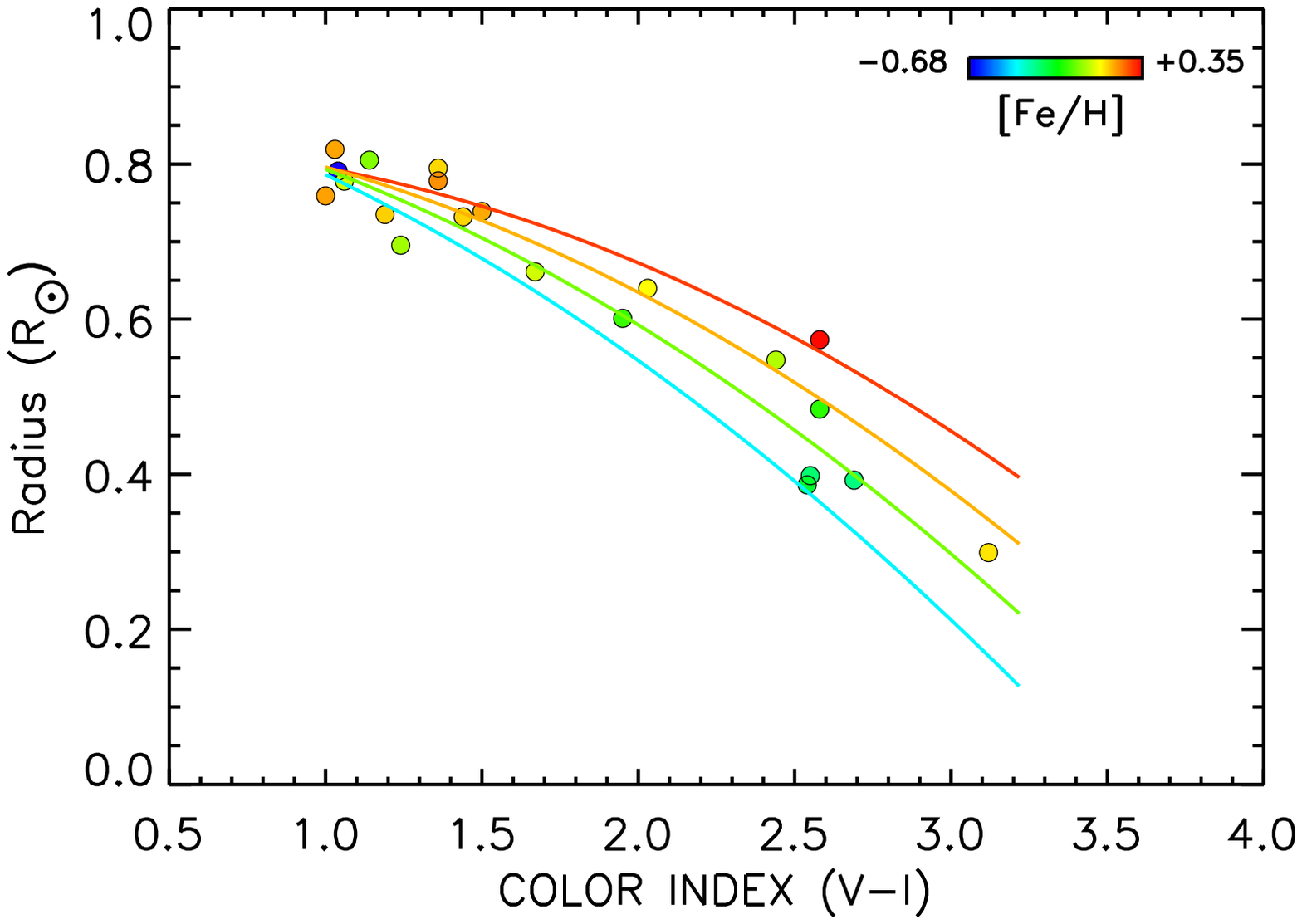, width=0.5\linewidth, clip=} &
         \epsfig{file=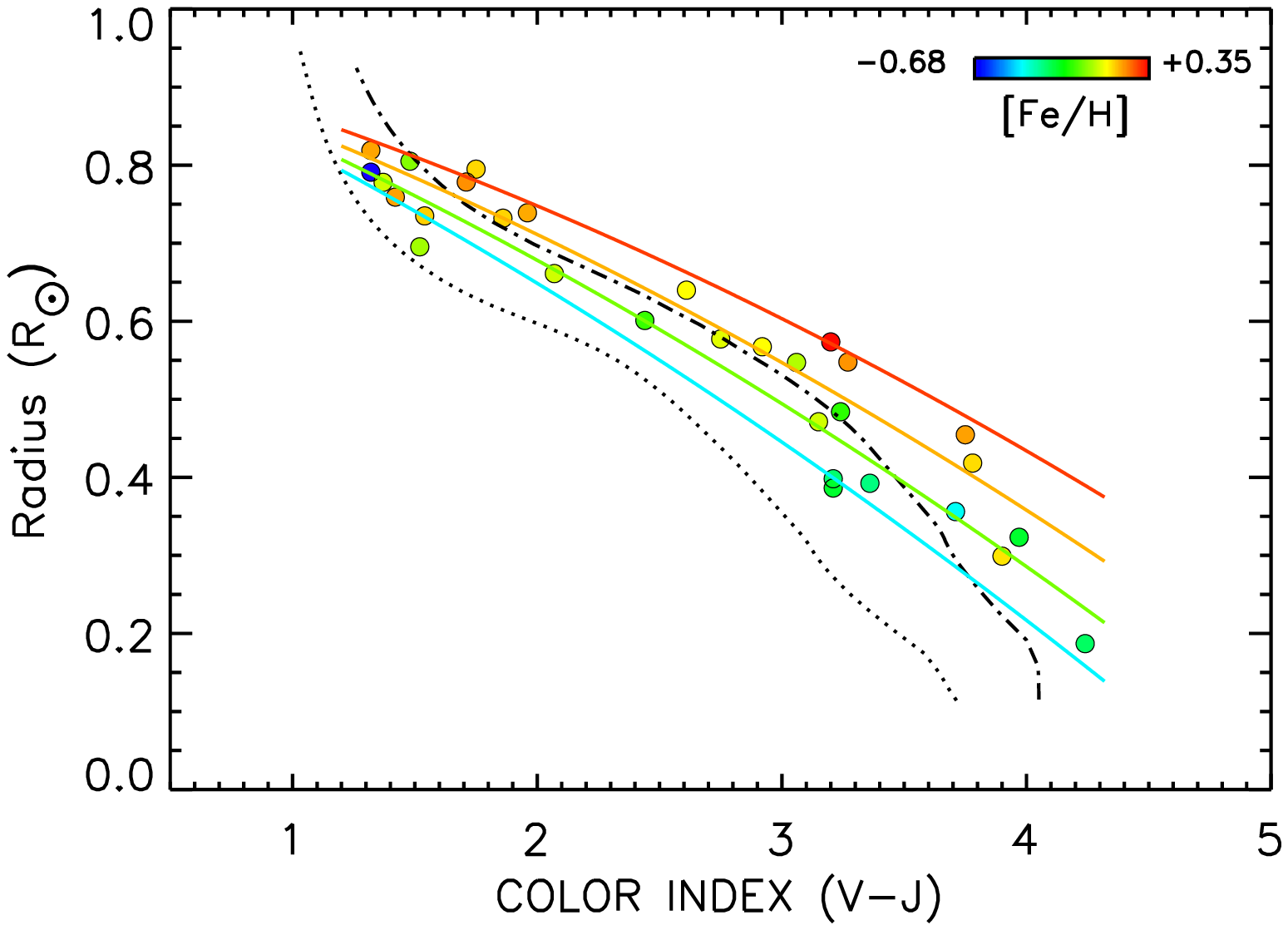, width=0.5\linewidth, clip=} \\
         \epsfig{file=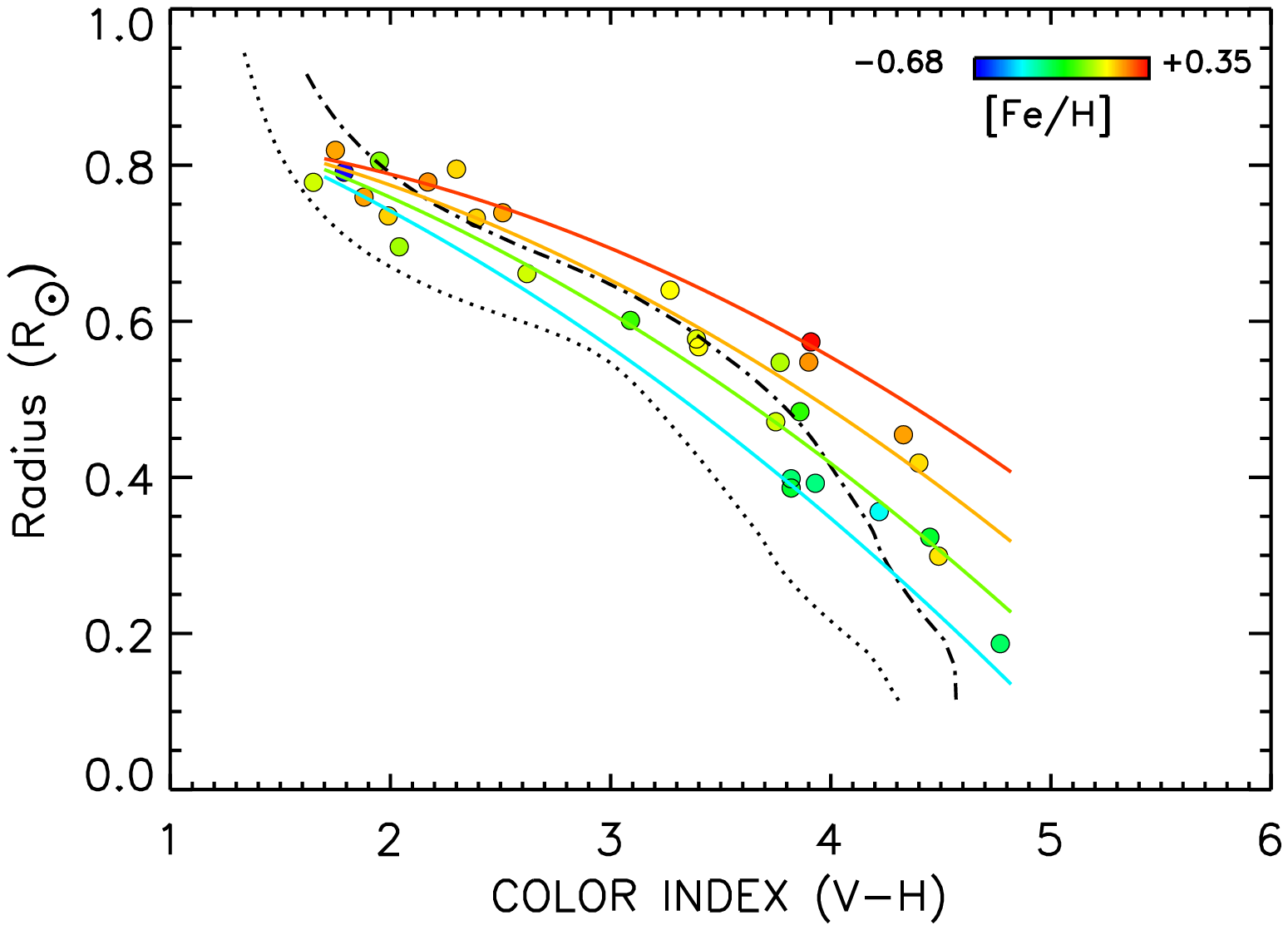, width=0.5\linewidth, clip=} &
         \epsfig{file=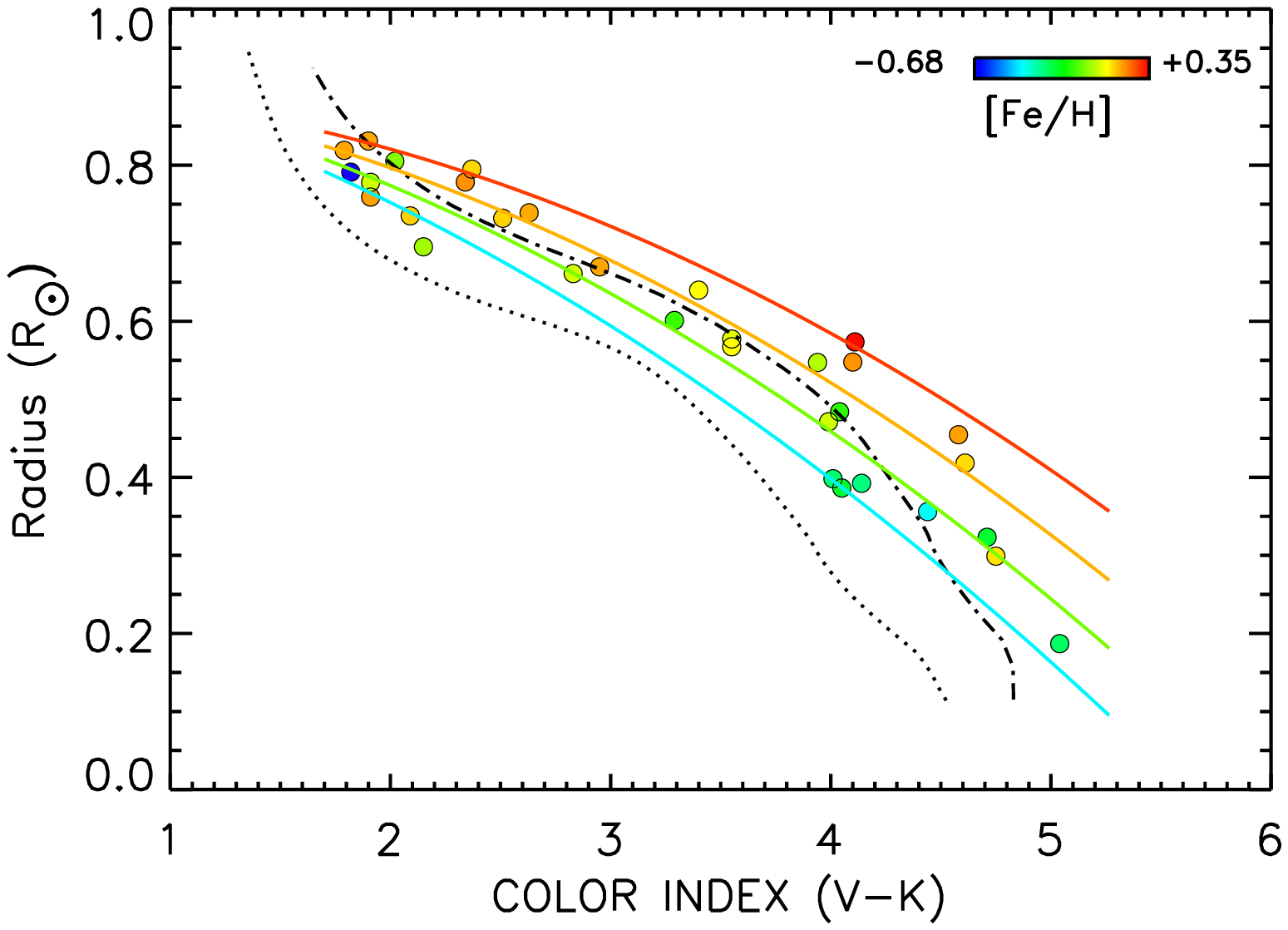, width=0.5\linewidth, clip=}           
    \end{tabular}
  \caption[] {Empirical color-metallicity-radius relations presented in Section~\ref{sec:empirical_radius_relations}.  The color of the data point reflects the metallicity of the star. The colored lines are solutions to the metallicity dependent fits, where the line color (red, orange, green, teal) represents our solution for an iso-metallicity line to [Fe/H]~=~$+0.25, 0.0,-0.25,-0.5$. The panels displaying the ($B-V$), ($V-J$), ($V-H$), and ($V-K$) relations also include Dartmouth 5~Gyr isochrones as dash-dotted ([Fe/H]~$= 0.0$) and dotted lines ([Fe/H]~$=-0.5$).  Within the ($B-V$) panel, we include solutions with the same Dartmouth model specifications, but using the semi-empirical color transformation (blue lines) to derive the ($B-V$) colors. See Equation~\ref{eq:radius_metallicity_relation} for the form of the equation and Table~\ref{tab:poly_solutions_radii} for the coefficients and statistical overview for each color-metallicity-radius solution.} 
 \label{fig:radius_VS_colors_b}
 \end{figure}

The coefficients and statistical overview of the solutions are in Table~\ref{tab:poly_solutions_radii}.  We find that, when adding metallicity as an extra parameter adds significant improvement to the significance of our fits, cutting the median absolute deviation of the observed versus calculated radius in half.  Figure~\ref{fig:radius_VS_colors_b} displays the data and our fit using the same color scheme as identified in the temperature relations to illustrate the metallicity. The data show a large spread in radii for a given color index toward the coolest stars (e.g., for ($B-V$)~$\sim 1.5$ and ($V-K$)~$\sim 4$, the radii range from $0.4 - 0.6 R_{\odot}$). We discuss this effect in more detail in Section~\ref{sec:empirical_MR_relations}.

The ranges of color indices where the relations are valid are listed in Table~\ref{tab:poly_solutions_radii}.  The ranges listed are hard limits, and do not allow for extrapolation. There are several reasons for these strict boundaries.  On the blue end, we are not able to observationally constrain the form of the upper part of this curve due to stellar evolution.  This causes the unphysical behavior of the curve to 1) fold over as a parabola, and 2) display cross-overs among fiducials of different metallicities (see Figure~\ref{fig:radius_VS_colors_b}).  In actuality, a zero-age main sequence radius curve would have an inflection point near this upper boundary we impose on the ranges in Table~\ref{tab:poly_solutions_radii}.  Past this point, the stellar radius will continue to rise with decreasing color index (but unfortunately also changes with increasing age). The use of the color-metallicity-radius relation (Equation~\ref{eq:radius_metallicity_relation}) on the blue end, despite the cross-over, will still lead to results within the stated errors, so long as the object in question is non-evolved. The red ends of the relations are strictly hard limits as well, where, by visual inspection of Figure~\ref{fig:radius_VS_colors_b}, extrapolation would result in null radii estimates approaching the late M-type star regime.

We show 5~Gyr Dartmouth isochrones \citep{dot08} based on the $UBV(RI)_{\rm C}JHK_{\rm S}$ synthetic color transformations \citep{bes90, cut03} within the panels displaying the ($B-V$), ($V - J$), ($V - H$), and ($V - K$) data and relations in Figure~\ref{fig:radius_VS_colors_b}\footnote{Since conversions between $K$ and $K_{\rm S}$ of these stars are on the order of the magnitude errors, we assume $K_{\rm S}\approx K$ in these plots.}. These isochrones are for two metallicities, [Fe/H]$=0$ and $-0.5$, roughly encompassing the metallicity range of the data included in this paper.  The ($B-V$)-radius plot also shows the Dartmouth isochrones under the same conditions, but with the semi-empirical $BV(RI)_{C}$ color-transformations (\citealt{van03}, blue lines). Dartmouth models for Johnson $R$ and $I$ colors are not provided, so we are unable to compare these color indices to our observations.  

There is acceptable agreement of the models and data with the earlier type stars, and the isochrones also show the sharp change in slope to increasing radii when extending beyond our data set to bluer colors (G-type stars). Extending to the later-type stars however, we find that for stars with radii around 0.6R$_{\odot}$ and below, the models predict a drop in radius at colors too blue by several tenths of a magnitude compared to what we observe in the ($V - J$), ($V - H$), and ($V - K$) colors. The ($B-V$) panel shows that at this point, the solar metallicity model unexpectedly drops and crosses over the model $-0.5$~dex curve.  This is a well documented failure of the synthetic color grid's ability to reproduce accurate colors for short wavelengths (e.g., see discussion in \citealt{dot08} for the accuracies in the color conversions). For instance, the use of the semi-empirical color transformations in the ($B-V$) panel (blue lines) do not show this flaw, demonstrating the contrast of using different color-table conversions. 

Overall, the models show a pattern of divergence with the observations in the color-radius plane. The reason behind this disagreement is not easily identified, for it could be a consequence of under-predicted radii for a given color index, or an offset in the color-tables used to calculate the colors of the later-type stars (see Section~\ref{sec:global_relations}).


\subsection{Color-Metallicity-Luminosity}
\label{sec:empirical_lumin_relations}


In this Section we use the data in Table~\ref{tab:fund_params_combined} to derive relations between a star's color and metallicity to its luminosity, which can be inverted to solve for the distance, if unknown, so long as one can assume that the star is on the main sequence. In the same way as the radius and temperature relations, we express luminosity as a function of color and metallicity using a polynomial in the form of:

\begin{eqnarray}
\label{eq:luminosity_metallicity_relation}
\log L ({\rm L_{\odot}}) &	= 	&	a_0 + a_1 X + a_2 X^2 + a_3 X Y + a_4 Y + a_5 Y^2,
\end{eqnarray}

\noindent where the variable $X$ is the color index and $Y$ is the stellar metallicity [Fe/H]. Table~\ref{tab:poly_solutions_lumin} lists the coefficients and overview of each relation, and Figure~\ref{fig:lumin_VS_colors_b} show the data and solutions for our fits.  Similar to the radius relations in Section~\ref{sec:empirical_radius_relations}, we find that the addition of metallicity as an extra parameter reduces the median absolute deviation by a factor of two. The plots in Figure~\ref{fig:lumin_VS_colors_b} show that the solutions converge for the earlier type stars regardless of the metallicity. However, for the later-type stars there is a spread in luminosity for a given color index, confirming that the metallicity is a necessary factor in estimating a star's luminosity using color relations.  

We impose the same hard limits for the range of color indices over which the color-metallicity-luminosity relations are valid, just as we do in the radius relations.  That is, no extrapolation of the curve is warranted for stars bluer or redder than the indicated color ranges listed in Table~\ref{tab:poly_solutions_lumin}.  Once again there is a tendency for some of the iso-metallicity solutions displayed in Figure~\ref{fig:lumin_VS_colors_b} to cross over each other.  This effect is non-physical, and is a product of the metallicity dependent solutions converging at the blue end of the relations. Despite the appearance of the cross over, any application of the relations with real data will still lead to reliable results with the quoted errors - so long as the boundary conditions are carefully regarded (Table~\ref{tab:poly_solutions_lumin}).

The luminosity versus color plots can be interpreted as empirical isochrones. Thus, the panels displaying the ($B-V$), ($V-J$), ($V-H$), and ($V-K$) relations in Figure~\ref{fig:lumin_VS_colors_b} also show Dartmouth isochrones \citep{dot08} for an age of 5~Gyr and two metallicities 0.0 and $-0.5$, roughly bracketing the metallicity range of the data included in this paper. Within the ($B-V$) panel, the Dartmouth model solutions using both the synthetic (black) and semi-empirical (blue) color conversions are shown, as described above in Section~\ref{sec:empirical_radius_relations}.  Two items are immediately clear based on the comparison of the models to our data.  First, the models reproduce the properties of the bluer stars (i.e., stars with $\log L/L_{\odot} > -1$).  Below this limit, the models predict an observed steep drop in luminosity, however, this drop is predicted at colors bluer than what we observe. This is very similar to our color-radius results (Section~\ref{sec:empirical_radius_relations}; Table~\ref{tab:poly_solutions_radii}; Figure~\ref{fig:radius_VS_colors_b}), and provides substantial motivation for modellers to identify whether or not the source of this disagreement is within the computed astrophysical properties, or in the tabulated colors of the later-type stars.  We discuss this more in Section~\ref{sec:global_relations}.


\subsection{Relations to Spectral Type}
\label{sec:spectral_type_relations}

For each spectral type observed, we use the data in Table~\ref{tab:fund_params_combined} in order to build a look-up table referring a spectral type to the star's temperature and radius.  These results are in Table~\ref{tab:spectral_type_table}. The quantity $n$ indicates the number of stars that fall into that spectral type bin. The value of the parameter given is the average value of all stars within the spectral type bin, and the $\sigma$ is the standard deviation of the parameter uncertainties for each spectral type bin.  The spectral types with only one measurement ($n = 1$) simply lists the individual value and the measured error of that measurement.


\section{Relations to the Global Properties [Luminosity, Temperature, Radius, Mass] for K- and M-Dwarfs}
\label{sec:global_relations}


The Sections~\ref{sec:empirical_LT_relations}, \ref{sec:empirical_LR_relations}, \ref{sec:empirical_TR_relations}, and \ref{sec:empirical_MR_relations} link the global stellar properties to each other: luminosity-temperature, luminosity-radius, temperature-radius, and mass-radius. In each Section, we derive empirical relations based on the data of the 33 stars in Table~\ref{tab:fund_params_combined}.  Following the observational results, we introduce and discuss model predictions of each star, as well as compare overall agreement to our relations and model relations of the physical quantities. 
 
In the following analysis, we use the Dartmouth models in the interpretation of our data because of the flexibility and user-friendly interface of the web-based interpolator that generates customized isochrones for users\footnote{http://stellar.dartmouth.edu/models/isolf.html}. We note that there are several evolutionary models for low-mass stars available to the community.   Figure~\ref{fig:logL_vs_T_and_R_phil} shows our data and four models: Padova, Dartmouth, BCAH, and Yonsei-Yale \citep{gir00,dot08,bar98,dem04}. In the luminosity-temperature, luminosity-radius, and temperature-radius plots, the dashed line corresponds to the solar metallicity model (see color legend within plots), and the hashed region spans a metallicity of 0 to $-0.5$~dex.  In the mass-radius plot, the solid and dashed lines mark a change in metallicity from [Fe/H]~$=0$ to $-0.5$.  We restrict the metallicity range plotted to conserve consistency across models (BCAH98 models only cover this range in metallicity).  Note that for the BCAH98 models we use mixing length of $\alpha = 1.0$ for stars $<0.6$~M$_{\odot}$ and the solar calibrated mixing length of $\alpha = 1.9$ for stars $>0.6$~M$_{\odot}$. Also, only a solar metallicity solution with the BCAH98 models is shown for the stars $>0.6$~M$_{\odot}$ for it is the only model available.  Of these four models, the Dartmouth and BCAH show to reproduce the trends of our data the best.  While each model is approximately consistent for stars brighter than 0.16~L$_{\odot}$ (earlier than $\sim$K5), the conclusions here are specific to the Dartmouth models, and should reflect the best case scenario to the observed offset with models for stars below this mark where the agreement is poor across model predictions (See Figure~\ref{fig:logL_vs_T_and_R_phil}).


\subsection{Stellar Luminosity versus Temperature}
\label{sec:empirical_LT_relations}


We use the 33 stars in Table~\ref{tab:fund_params_combined} to derive a relation between the luminosity as a function of temperature expressed as:

\begin{eqnarray}
\label{eq:temp_lumin_relation}
\log L ({\rm L_{\odot}}) &	= 	&	-5960.5710 (\pm 207.7541) + 4831.6912 (\pm 172.6408) X \nonumber \\
						&		&	- 1306.9966 (\pm 47.8104) X^2 + 117.9716 (\pm 4.4125) X^3  
\end{eqnarray}

\noindent where the variable $X$ stands for the logarithmic effective temperature $\log (T_{\rm EFF})$, and is valid for non-evolved objects within a temperature range of 3200 to 5500~K. A metallicity dependent solution did not show any improvement to the fit, therefore this relation is metallicity independent. The median absolute deviation of this fit is 0.0057~L$_{\odot}$, similar to the abundance dependent color-metallicity-luminosity relations (Equation~\ref{eq:luminosity_metallicity_relation}, Table~\ref{tab:poly_solutions_lumin}). The top plot in Figure~\ref{fig:Lumin_VS_Temp} shows the data and the solution to this fit. The middle plot in Figure~\ref{fig:Lumin_VS_Temp} shows the fractional deviation of observed luminosities compared to the empirical relation expressed in Equation~\ref{eq:temp_lumin_relation}.

The fractional residuals in the middle panel of Figure~\ref{fig:Lumin_VS_Temp} (calculated as $(L_{\rm Obs} - L_{\rm Fit}) / L_{\rm Fit}$, where $L_{\rm Fit}$ is the luminosity of the empirical relation at the star's observed temperature) show that for stars with temperatures above $\sim 3500$~K, the observed luminosities lie within $\pm 10$\% of our empirical solution. However, we note that for stars with temperatures below $\sim 3500$~K, the scatter about the fit is much higher. The plotted residuals also show no pattern with respect to the stellar metallicity. This reinforces the fact that we were not able to solve for a metallicity dependent solution for Equation~\ref{eq:temp_lumin_relation} using our data.  

In the top plot, we overlay Dartmouth 5~Gyr isochrones for [Fe/H]$=0, -0.5$ (dash-dotted and dotted lines, respectively). We find that the observed agreement of our data and the Dartmouth models in Figure~\ref{fig:Lumin_VS_Temp} is fairly decent, although much more favorable in the earlier type stars. To evaluate this agreement of our data and the Dartmouth models more quantitatively, we generated Dartmouth model isochrones for each of our stars at its respective metallicity.  Each isochrone was chosen at an age of 5~Gyr (appropriate for non-evolved low-mass field stars) with the default Helium mass fraction ($Y=0.245 + 1.5*Z$), and alpha-enhanced elements scaled to solar ([$\alpha$/Fe]$=0$)\footnote{We use a [$\alpha$/Fe]$=0.2$ for GJ~53A (See discussion in \citealt{boy08}).}.  We then interpolate through the model grid to find the model temperature, $T_{\rm Mod}$, at the observed luminosity of each star.  The results are plotted in the bottom panel of Figure~\ref{fig:Lumin_VS_Temp}.  The results show that for a given luminosity, models overestimate the stellar temperature by a average of 3\% ($\sim 120$~K) for stars with temperatures under 5000~K. For the 7 stars with temperatures greater than 5000~K, we detect a trend that the temperatures of the more metal-rich stars (orange points; [Fe/H]~$\sim 0.0$) are underestimated by the models by a few percent.  On the other hand, the more metal-poor stars (green points; [Fe/H]~$\sim -0.25$) with temperatures greater than 5000~K show that their temperatures are overestimated by the models by a few percent.  This effect is also seen for stars cooler than 5000~K, where the less abundant the star is in iron, the more offset the star's predicted temperature will be with the observed temperature.  This pattern of metallicity within the residuals in the bottom panel of Figure~\ref{fig:Lumin_VS_Temp} is telling that the models could be over-predicting the influence of metallicity on the global stellar parameters.

\subsection{Stellar Luminosity versus Radius}
\label{sec:empirical_LR_relations}

We use the 33 stars in Table~\ref{tab:fund_params_combined} to derive a solve for an solution on the radius-luminosity plane expressed as:

\begin{eqnarray}
\label{eq:radius_lumin_relation}
\log L ({\rm L_{\odot}}) &	= 	&	-3.5822 (\pm 0.0219) + 6.8639 (\pm 0.1562) X \nonumber \\
						&		&	- 7.1850 (\pm 0.3398) X^2 + 4.5169 (\pm 0.2265) X^3  
\end{eqnarray}

\noindent where $X$ is the stellar radius in solar units.  The median absolute deviation about the fit is 0.008~L$_{\odot}$, and the data and solution are plotted in the top panel of Figure~\ref{fig:Lumin_VS_Radius}. The middle panel in Figure~\ref{fig:Lumin_VS_Radius} shows the fractional deviation in luminosity about the fit. Stars with radii between $\sim 0.4$~R$_{\odot}$ and $\sim 0.67$~R$_{\odot}$ show the tightest correlation, deviating from the fit by less than $\pm 10$\%.  Outside this range in radii, the scatter is worse, where on the lower end the scatter increases to double this level and on the upper end the scatter increases to triple. We observe no pattern with metallicity in the fractional residuals compared to our fit, enforcing the fact that a metallicity dependent solution is not appropriate on the luminosity - radius plane. 

We plot Dartmouth 5~Gyr isochrones for [Fe/H]$=0,-0.5$ on the top portion of Figure~\ref{fig:Lumin_VS_Radius}.  By inspection of these curves, the solar metallicity model agrees well with our empirical fit (Equation~\ref{eq:radius_lumin_relation}, while the metal-poor isochrone appears slightly offset from the data, similar to the luminosity - temperature plots in Figure~\ref{fig:Lumin_VS_Temp}.  To make a fair assessment of the comparison to each star's unique composition, we use the custom-tailored Dartmouth isochrones generated for the 33 stars (discussed above in Section~\ref{sec:empirical_LT_relations}) and interpolate through the model grid to determine the model radius $R_{\rm Mod}$ at the observed luminosity for each star.  The fractional deviations of the observed radii from the model radii are presented in the bottom panel of Figure~\ref{fig:Lumin_VS_Radius}.  This shows a clear offset on the order of 5\% for models to under-predict the stellar radii at a given observed luminosity.  

The deviations in observed radii to those predicted by models become prevalent for stars with radii $<0.7$~R$_{\odot}$ (bottom panel of  Figure~\ref{fig:Lumin_VS_Radius}). For the full range of stars in our sample however, we also detect the same pattern in the residuals shown in the bottom panel of Figure~\ref{fig:Lumin_VS_Temp}, where the stars with lower metallicities have radii that are are more deviant than model predictions than the stars with more solar type metallicity values. While the effect is less pronounced on the radius plane than on the temperature plane, the conclusions remain the same: models appear to be over-predicting the impact stellar metallicity has on the physical stellar properties.

Our results are consistent with the findings in the literature (e.g. \citealt{pop97, rib07, lop07, mor08}), where models are shown to predict temperatures too high and radii too small while still being able to correctly reproduce the luminosity. Such references however, are typically referring to binary star properties, and we note that this result is quite novel to detect such a distinct pattern for single stars.  Additionally, a unique contribution to our project's analysis is that we are able to reveal the pattern of the offset in connection with the stellar metallicity, an observable often-times unreachable for binary stars due to the complexity of the blended spectrum from each component. 

\citet{cas08} use the IRFM (infrared-flux method) in order to determine properties of M-dwarfs.  Their conclusions show that there is a radii offset present when comparing their observations of single stars to models.  However, while similar to our own conclusions, they estimate this offset is on the order of 15-20\%, unlike the 5\% we claim from our analysis.  This difference is possibly caused by the errors introduced in their semi-empirical determinations of the stellar properties.  They also identify a discontinuity at the transition from K- to M-dwarfs, most apparent on the luminosity - temperature plane, that is not reproduced by models.  We do not observe this feature, however, this region from 4200 to 4300~K is not well populated within our interferometrically observed sample, as we only have one data point that lies in this range of temperatures. Perhaps what is more interesting is a region not explicitly identified in \citet{cas08}; the point where the deviations of their data compared to models begins.  The data in their figure~12 show this is at $\sim 5000$~K ($M_{\rm BOL} \sim 6.5$). Their figure 13 (a) and (c) show this point at $\sim 0.7$~R$_{\odot}$ and $\sim 0.7$~M$_{\odot}$, the exact positions we identify the beginning of the trend in our data.


\subsection{Stellar Temperature versus Radius}
\label{sec:empirical_TR_relations}	


In order to link effective temperature to stellar radius, we use the 33 stars in Table~\ref{tab:fund_params_combined} to derive a single parameter solution in the form of a 3rd order polynomial, valid for non-evolved objects with temperatures between 3200 and 5500~K:

\begin{eqnarray}
\label{eq:temp_radius_relation}
R ({\rm R_{\odot}}) &	= 	&	- 10.8828 (\pm 0.1355) + 7.18727 (\pm 0.09468) \times 10^{-3} X \nonumber \\
					&		&	- 1.50957 (\pm 0.02155) \times 10^{-6} X^2 + 1.07572 (\pm 0.01599) \times 10^{-10} X^3 
\end{eqnarray}

\noindent where the variable $X$ is the effective temperature.  Like the radius - color plane (\S \ref{sec:empirical_radius_relations}), there is a lot of scatter in the radius of a star for a given temperature (see Figure~\ref{fig:Radius_VS_Temp}).  However, the data on the temperature - radius plane does not show the same spread in the measurements due to differences in metallicity as in the color-metallicity-radius relations, as its solution is independent of metallicity. We find that the median absolute deviation of the observed versus calculated radius for Equation \ref{eq:temp_radius_relation} is 0.031~R$_{\odot}$. This value is similar to ones derived via the abundance dependent color-metallicity-radius relations of Equation~\ref{eq:radius_metallicity_relation} shown in Table~\ref{tab:poly_solutions_radii}. 

Stars more massive than the sample of K- and M-dwarfs (i.e., masses $\gtrsim 0.8$~M$_{\odot}$) are sensitive to evolutionary effects. For example, with increasing age, a more massive star will first begin to evolve off its zero-age main-sequence position to larger radii and hotter temperatures.  We illustrate this in Figure~\ref{fig:Radius_VS_Temp}, showing the expected model predictions of a 0.8~M$_{\odot}$ star (blue $\ast$), a 0.9~M$_{\odot}$ star (red $+$), and a 1.0~M$_{\odot}$ star (black $\times$) at the age of 1~Gyr and 4.5~Gyr from solar-metallicity Dartmouth model isochrones.  Arrows connecting these positions point to the top-left of the plot, i.e., the direction of evolution to larger radii and hotter temperatures. We also show in Figure~\ref{fig:Radius_VS_Temp} the Sun's present position.  To model the data beyond our sample of K and M dwarfs to hotter temperatures, we re-fit the data adding the Sun as a point of reference to more massive stars, effectively making a solar-age calibrated isochrone.  This relation is expressed as:

\begin{eqnarray}
\label{eq:temp_radius_relation_withsun}
R ({\rm R_{\odot}}) &	= 	&	-8.133 (\pm 0.226) + 5.09342 (\pm 0.16745) \times 10^{-3} X \nonumber \\
					&		&	- 9.86602 (\pm 0.40672) \times 10^{-7} X^2 + 6.47963 (\pm 0.32429) \times 10^{-11} X^3 	
\end{eqnarray}

\noindent where $X$ is the effective temperature.  We show this solution in Figure~\ref{fig:Radius_VS_Temp} as a blue line, as well as the Sun's position (farthest left point in Figure~\ref{fig:Radius_VS_Temp}, assuming $T_{\odot} = 5778$~K, $R = 1$~R$_{\odot}$).  The extrapolation of this curve past 5500~K is shown as a dotted line that intersects the Sun's present position.  A comparison of the temperature - radius relation excluding the Sun (Equation~\ref{eq:temp_radius_relation}) and including the Sun (Equation~\ref{eq:temp_radius_relation_withsun}) shows that while the hottest of the K-stars in our sample are modelled better by Equation~\ref{eq:temp_radius_relation}, there is an inflection point to larger radii beyond our data sample range of spectral type K0 to meet the Sun's position at its age today.

What is more intriguing, is whether or not the models are able to reproduce the trend to the temperatures and radii of single stars, with the omission of the luminosity as a constant like structured in the last two Sections of \ref{sec:empirical_LT_relations} and \ref{sec:empirical_LR_relations}.  The bottom-left panel in Figure~\ref{fig:logL_vs_T_and_R_phil} shows that the Dartmouth models predict the sharp drop in stellar radii to happen at hotter temperatures than what we observe.  This is also seen in the temperature-luminosity (Section~\ref{sec:empirical_LT_relations}) and radius-luminosity (Section~\ref{sec:empirical_LR_relations}) relations displayed in Figure~\ref{fig:Lumin_VS_Temp} and \ref{fig:Lumin_VS_Radius}, where compared to our observations, the Dartmouth models predict temperatures too high and radii too low for a given luminosity. 

Similar to the method described in Sections~\ref{sec:empirical_LT_relations} and \ref{sec:empirical_LR_relations}, we use the custom-tailored Dartmouth isochrones for each star to evaluate expected parameters from the model compared to our observations.  First, we assume that the temperature is a constant within the observations and model grid.  For each star, we then interpolate within the model grid for a model radius at the observed temperature.  The second approach is the reverse, and we assume that the radius is constant within the observations and model grid.  A model temperature is then solved for by interpolation through the model grid at the observed radius. We show the results in Figure~\ref{fig:CHARA_VS_DSEP_radiustemp_offset}.   

The top plot in Figure~\ref{fig:CHARA_VS_DSEP_radiustemp_offset} shows the fractional deviation of temperature with the observations to the Dartmouth models as a function of temperature.  The bottom plot shows the fractional deviation of the observed to Dartmouth model radii as a function of radius. The fractional temperature offsets for stars with temperatures less than 5000~K, show that models predict temperatures higher than we observe by an average of 6\%.  The fractional radius offsets for stars with radii less than 0.7R$_{\odot}$, show that models predict radii smaller than we observe by 10\%, where this offset increases with decreasing radii to $\sim 50$\% for stars with radii  $\sim 0.4$R$_{\odot}$.  The three stars with temperatures below 3300~K (GJ~725B, GJ~551, and GJ~699) produce null results when interpolating at the observed temperatures to derive model radii, and thus their results are not plotted in the bottom plot of Figure~\ref{fig:CHARA_VS_DSEP_radiustemp_offset}.  The computed offsets in temperature and radius given here reflect doubly deteriorating conditions compared to the results in Sections~\ref{sec:empirical_LT_relations} and \ref{sec:empirical_LR_relations}, where the stellar luminosities were held as a constant.  These compounded errors should be regarded with caution when using models to relate stellar temperatures and radii for stars $<5000$~K and $<0.7$R$_{\odot}$.


\subsection{Stellar Mass versus Radius}
\label{sec:empirical_MR_relations}


Data from double-lined spectroscopic eclipsing binary (EB) systems are standards for the mass-radius relations available today. For the following discussion, we collect the binary star parameters presented in table~2 of \citet{tor10} together with the low-mass binaries and secondaries in binaries compiled in \citet{lop07}.  We impose the same criteria as for the single stars in our analysis, limiting the sample to only allow stars with mass and radius $< 0.9$~M$_{\odot}$ and R$_{\odot}$, and where the radius is measured to better than a 5\% error, leaving a total of 24 individual stars.  These properties of the single stars are discussed in Section~\ref{sec:stellar_params}, where we have a total of 33 stars in this range of masses and radii.  Note that although we limited the literature measurements of single stars to only include those with angular diameters measured to better than 5\%, the sample remains unchanged if we filter with respect to the error in the linear radii to be better than 5\%, due to the close proximity of the objects (within $\sim 10$~pc) and thus well known distances.

To determine a mass-radius relation, we use a second order polynomial to fit the data for single stars (Equation~\ref{eq:mass_VS_radius}).  We experimented with higher order polynomials and found that they did not did not improve the fit. We solve for a solution for the EB's in the same manner (Equation~\ref{eq:mass_VS_radius_EB}).  Data and fits are displayed in Figure~\ref{fig:M_vs_R_vs_FeH_withEBs}. The resulting relations are:

\begin{eqnarray}
\label{eq:mass_VS_radius}
{\rm R_{\rm SS}} (R_{\odot}) &	= 	&	 0.0906 (\pm 0.0027) + 0.6063 (\pm 0.0153) M_{\ast} \nonumber \\
							&		&	 + 0.3200 (\pm 0.0165) M_{\ast}^2 
\end{eqnarray}

\begin{eqnarray}
\label{eq:mass_VS_radius_EB}
{\rm R_{\rm EB}} (R_{\odot}) &	= 	&	 0.0135 (\pm 0.0070) + 1.0718 (\pm 0.0373) M_{\ast} \nonumber \\
							&		&	- 0.1297 (\pm 0.0367) M_{\ast}^2 
\end{eqnarray}

\noindent where $M_{\ast}$ is the stellar mass in solar units.  

In Figure~\ref{fig:M_vs_R_vs_FeH_withEBs}, we show the data and solutions for single (circles, solid line) and binary stars (squares, dotted line).  These relations are both consistent with a 1:1 relation to the mass and radius of a star (dashed line).  One can see from the figure that the radii of single and binary stars are indistinguishable for a given mass. This shows that tidal influences on binary components radii may not be a concern when viewed on this scale.  In other words, the scatter in the measurements in EB systems may wash out the effect the binary period might have in the radii of the binary star components (for instance, see discussion in \citealt{kra11}).


\subsection{The Endless Discussion About Model Predictions of Late-Type Stellar Masses and Radii Briefly Continues Here}
\label{sec:modelling}


Historically, large discrepancies have been observed when comparing observed radii with radii predicted by models for low-mass stars (for example, see discussion in \S \ref{sec:introduction} and \citealt{lop07}) in the sense that the models tend to underestimate the stellar radii at a given mass. We compare our measured radii to the model radii from the 5~Gyr model isochrones of Padova, Dartmouth, BCAH98, and Yonsei-Yale \citep{gir00, dot08, bar98, dem04} in Figure~\ref{fig:logL_vs_T_and_R_phil}.  Points for the masses and radii from eclipsing binaries are also plotted.

Due to the density of information from showing several model predictions along with the data and errors, the visibility of the claimed radius discrepancy with models in Figure~\ref{fig:logL_vs_T_and_R_phil} is difficult to see clearly.  While it is abundantly documented in the literature that such a discrepancy in the predicted and observed radii exists for binary stars, the intentions of this work is to show its equivalent - if present - for single stars. Already in Sections~\ref{sec:empirical_LT_relations} and \ref{sec:empirical_LR_relations} we have shown proof of this discrepancy in the luminosity - temperature and luminosity - radius plane, where models over-predict temperatures by $\sim 3$\%, and under-predict radii by $\sim 5$\% compared to our observations.  These results presented on the luminosity - temperature and luminosity - radius plane are more robust for single stars because temperatures and radii are properties of single stars we can directly measure to high precision with interferometry.  However, the ability to comprehensively show the radius discrepancy from this data set is quite difficult on the mass - radius plane, since the mass errors for single stars are characteristically large (as shown on the bottom axis of the plot).  To first order however, we find it useful to present Figure~\ref{fig:mass_VS_radius5_allstars}, a series of plots showing the fractional deviation of the observed to theoretical radii for single stars.

The theoretical radii, R$_{\rm Mod}$, are computed by using the mass (calculated from the mass-luminosity relation; Table~\ref{tab:fund_params_combined}) and interpolating through the custom-tailored mass-radius grid of values computed for each star's metallicity in a 5~Gyr Dartmouth model isochrone.  The result is a value for a model radius at the given mass and metallicity.  The difference is then between the observed radius (from interferometry; Table~\ref{tab:fund_params_combined}) and the model radius (from interpolation of the mass from the M-L relation in the model mass-radius grid).  Since the single star mass errors are quite large, the errors in the theoretical radii dominate the y-errors shown in Figure~\ref{fig:mass_VS_radius5_allstars}.  Again, we caution that the absence of any signal in a ($R_{\rm observed} - R_{\rm model}$) plot could be misleading when using single stars as clues into this discrepancy on the mass-radius plane due to the large mass errors. 

On the top panel of Figure~\ref{fig:mass_VS_radius5_allstars}, we show the difference in observed versus model radii as a function of mass.  The two middle and two bottom panels show the fractional deviation of radius as a function of metallicity [Fe/H] and stellar activity level $L_X/L_{\rm BOL}$ (Table~\ref{tab:fund_params_combined}), since previously proposed diagnostics to explain any disagreement with models have been linked to abundances and stellar activity levels for stars in this mass regime.  The bottom four panels are segregated to show stars of all masses $< 0.9$M$_{\odot}$ and only admit stars with mass $< 0.6$M$_{\odot}$ (left panel and right panel, respectively).

The top panel on the plot shows that even with the stated large errors, stars with masses below $0.42$~M$_{\odot}$ (8 total) have bigger radii than those of the Dartmouth models, where 6 of the 8 deviate by several sigma.  We identify these 8 stars with masses below $0.42$~M$_{\odot}$ as blue points within the right hand side plots of Figure~\ref{fig:mass_VS_radius5_allstars}.   Regarding the fractional deviation in radii for the blue points with respect to metallicity, we find that for metallicities [Fe/H]~$<-0.35$, 4 of the 5 blue points deviate by more than 1$\sigma$.  Any pattern with respect to $L_X/L_{\rm BOL}$ is less clear within the data.  The fact that we do not see any trend with activity is not a surprise, due to the low levels of these single field stars compared to their active binary star counterparts (typically a decrease in X-ray flux by two to three orders of magnitude; see \citealt{lop07}).

Our conclusion supports the known discrepancy in observed radii compared to models for stars less than 0.42~M$_{\odot}$, showing a slight dependence on metallicity. This conclusion is different from previous works that focused on interferometric radii of single stars.  For instance, although \citet{ber06} show the presence of the radius discrepancy with models, they observe an opposite trend compared to our own: that stars with the most deviant stellar radii have larger metallicities. Contradictory to the \citet{ber06} conclusion as well as the one presented here is the analysis from \citet{dem09}, where they find that the radii of single stars are consistent with predictions of stellar models.  

The differences in the previously published results and the one presented here are multi-fold.  A primary difference is the sample used in the analysis of each work. \citet{ber06} applied no filter to the diameter precision allowed in their analysis, and at the time of publication, this amounted to 10 interferometric radii of M-dwarfs available for their analysis. Minding our filter of using only data with sizes determined to better than 5\%, this constitutes to only 5 of the 10 stars used in the \citet{ber06}.  In addition to limiting the precision of measurements allowed in our analysis, we also tripled the sample size of M-dwarfs (14), making for better statistics and thus giving more weight to our results. Our work also uses recently defined metallicity calibrations for M-dwarfs (see references in Table~\ref{tab:fund_params_combined}). These new references use techniques to measure metallicity that are more refined than the ones used to derive the metallicities for the stars cited in \citet{ber06}, and thus could explain the different conclusions presented here.  However, by comparing the metallicities used in \citet{ber06} to the ones we use in this work, we find an agreement within $0.02 - 0.14$~dex.  This is less than the 1-$\sigma$ uncertainty in metallicity ($\sim 0.2$~dex) and thus should not be the cause of new trend we observe.

The sample of M-dwarfs at the time of the \citet{dem09} publication was identical to the time of \citet{ber06}\footnote{While \citet{dem09} report on the diameters of low-mass stars, only 2 of the stars they observe are M-dwarfs, and are repeated measurements of stars with diameters already existing in the literature. Furthermore, these points are not averaged, but counted twice in their analysis.}.  In their analysis, they include both K- and M-dwarf measurements, choosing to disregard all but one measurement of the \citet{ber06} data, and limit published measurements to have a precision better than 10\%.

The comparison of the radii offsets examined in \citet{dem09} come from an analysis of an expanded mass range encompassing both K- and M-dwarfs, similar to the one presented here on the LHS panels of our Figure~\ref{fig:mass_VS_radius5_allstars}.  In the full mass range, we come to the same conclusion as \citet{dem09} in their figure 9; that we see no deviation in single star radii with respect to metallicity\footnote{\citet{dem09} quote a 5\% error in the derived mass of the star, and thus errors on the theoretical radius, and  are likely underestimates (see the errors on their figure~9).}.  However, we find that by adding the stellar mass as an additional constraint, the higher-mass stars that do not show offsets from models are more likely to hide signals in the low-mass stars that $do$ contain offsets (for example, see the RHS panels of Figure~\ref{fig:mass_VS_radius5_allstars}). It is thus a misleading claim to pronounce the conclusions in \citet{ber06} void, since they are analysing the wrong mass range (the full span of K- and M-dwarfs, as opposed to just M-dwarfs in \citealt{ber06}), in addition to not bringing any new measurements for interpretation. Lastly, as opposed to the methods in both \citet{ber06} and \citet{dem09} where a solar metallicity model is assumed, we compare the observations to models that are specifically tailored to each star's composition.   


\subsection{Summary of Global [Luminosity, Temperature, Radius, Mass] Relations}
\label{sec:empirical_LTR_relations}


We find that the empirical relations we derive relating the global properties of: luminosity - temperature, luminosity - radius, and temperature - radius show no dependence on the metallicity of a star. This conclusion, while unexpected, challenges the Vogt-Russell theorem, which states that the properties of a star should exhibit a unique solution for a star of given mass and chemical composition (see \citealt{wei04}, and references therein).  Generally speaking, when regarding relations to global properties of stars, models predict that the resulting scatter we observe is caused from varying the stellar metallicity. However, the majority of metallicities used in this work are from a uniform source, and should only have systematic offsets with respect to each other.  So although we do not see this trend for the scatter in our data to be caused by metallicity (e.g., see Figure~\ref{fig:Radius_VS_Temp}), we propose that it could be a residual effect from underestimated errors in metallicity measurements for these types of stars (average metallicity error is $\sim 0.2$~dex).  It is important to note however, that we are only reviewing this lack of dependence with the composition on the stellar physical parameters with respect to iron abundances.  Alternate elemental abundances such as Helium that are not accounted for in the models can have an effect on a star's physical properties.  However, our data not only shows no correlations with differences in iron abundances, but also suggests that models over-predict the dependence on the stellar global parameters with metallicity as illustrated in the pattern of the plotted residuals in the observations versus models of Figures~\ref{fig:Lumin_VS_Temp} and \ref{fig:Lumin_VS_Radius}.

With our data, we are able to develop an empirical method of quantifying the effect a star's metallicity has on the observed color index. For instance, our analysis shows that metallicity does not effect the stellar radius when relating the global properties of stars (luminosity - radius, temperature - radius), while on the other hand, the analysis of the color - metallicity - radius relations in Section~\ref{sec:empirical_radius_relations} (Figure~\ref{fig:radius_VS_colors_b}), identifies a substantial spread in radii for a given color index, noting a dependence on metallicity.  This leads us to deduce that the metallicity only directly affects the observed color index of a star.  The effect can be visualized in the color - metallicity - radius relations (Figure~\ref{fig:radius_VS_colors_b}) as a shift in the horizontal direction for our solutions with different metallicity, as opposed to the vertical direction, where at a fixed color index, the radii changes with respect to metallicity.  We can quantify this shift by defining a zero metallicity color index: we find that for a differential metallicity of 0.25 dex, the ($B-V$), ($V-R$), ($V-I$), ($V-J$), ($V-H$), and ($V-K$) color indices change by $\sim$ 0.05, 0.15, 0.2, 0.3, 0.3, and 0.3 magnitudes, respectively.  This argument, while described with respect to our observed radii, also holds using the stellar luminosities as a proxy.

In this work, the masses for single stars are derived indirectly via the mass-luminosity relation (MLR) in \citet{hen93} (Section~\ref{sec:metallicities}). This relation, as well as others in the literature (e.g. \citealt{del00}), do not incorporate metallicity in their formulation and analysis. Therefore, the application of a MLR may produce spurious results for an object with a different composition than the objects used to develop the MLRs.  To alleviate this concern, we used the $K$-band form of the MLR, which is thought to be insensitive to metallicity: a fortunate by-product to a proper balancing act of equalizing luminosities and flux redistributions in the infrared with changing metallicity (see \citealt{del00} for their discussion and empirical verification). However, unlike luminosity-temperature, luminosity-radius, and temperature-radius planes, we are not able to confirm or deny any metallicity dependence on the mass-radius plane with our data because the mass errors are too large for the single stars in our sample (Section~\ref{sec:modelling}, Figure~\ref{fig:M_vs_R_vs_FeH_withEBs}).

Within Sections~\ref{sec:color_temperature}, \ref{sec:empirical_radius_relations}, and \ref{sec:empirical_lumin_relations}, we use a multi-variable function to relate the observed colors and metallicities to the global properties of temperature, radius, and luminosity. We found that inclusion of metallicity as a variable was essential to properly relate the astrophysical quantities ($T_{\rm EFF}$, $R$, and $L$) to the observed colors of our stars.  Contrary to the color solutions, we find that linking the global properties (temperature, radius, luminosity, mass) of a star (\ref{sec:empirical_LT_relations}, \ref{sec:empirical_LR_relations}, \ref{sec:empirical_TR_relations}, and \ref{sec:empirical_MR_relations}) do not appear to be sensitive to the stellar metallicity within the observational errors used in this work \footnote{Since there is a physical coupling of the temperature, radius, and luminosity through the Stephan-Boltzmann equation, we expect this connection to their properties to exhibit similar attributes.}.  In other words, our data show that the metallicity only affects the observed color-index of a star, and thus using the metallicity-dependent transformations in order to convert colors into the stellar properties of temperatures, radii, and luminosities is essential. However, the analysis clearly shows that for the range of stars that we observe, the metallicity does not impact the physical properties of a star in a way that we are able to measure: {\it throwing a bucketful of metals in a star does not make it expand in size or cool its surface temperature, it simply morphs the observed color index.}


\section{A Need to Validate Binary Star Temperatures}
\label{sec:teff_validation}


Typically, when making comparisons of single and binary star properties, the community has viewed the data on the mass-radius plane, as we present in Section~\ref{sec:modelling}.  While it is true that EBs represent the highest precision in, and largest volume of, the available data for stellar masses and radii, it is still not as fundamental when adding single stars to the analysis because single stars must rely on mass-luminosity relations to derive their masses. However, the data set we introduce here turns the table to the single stars as the most voluminous data set of precise measurements to K- and M-dwarf properties.  Because of this, we compare the fundamental parameters of both the single and binary stars on the temperature-radius plane, as they are the most directly determined measurements for single stars.

While individual component masses and radii can be determined for double-lined EB systems, only the ratio of effective temperatures (and thus luminosities), may be calculated. Solutions for estimating EB temperatures are generally constrained by photometric calibrations, where the results often vary due to differences within analyses \citep{tor10}. Furthermore, the empirical data to calibrate the temperature scales of late-type stars have only recently become available, making any absolute zero point in temperature scales subject to uncertainties. The basic technique of EB star temperatures relies on an assumed temperature of the primary star that is then translated to the properties of the secondary component.  The calculation of the EB stellar luminosities are then based on the Stephan-Boltzmann equation and interstellar extinction (if present). Additionally, a measure of the accurate stellar abundances of the EB system is typically unreachable due to the complex blending of spectral features, which are dependent on the (unconstrained) individual temperature of each component in the system.  Comprehensive reviews of binary star properties can be found in, e.g.,\citet{and91} and \citet{tor10}. 

Using the same collection of data for binary stars as described in Section~\ref{sec:modelling}, we regard the differences on the temperature-radius plane for single stars and binary stars in Figure~\ref{fig:R_vs_T_withEBs_withDSEP} \footnote{Note that not all binary stars used in Section~\ref{sec:modelling} have published values for the temperatures.}.  Although there is considerable scatter, the binary star data are clearly separated from the single stars on the temperature-radius plane.  It appears that either the EB components tend to have larger radii (up to $\sim 50$\% for radii $< 0.6$~M$_{\odot}$) compared to a single star of the same temperature, or the EB components have cooler surface temperatures (up to $\sim 500$~K) compared to single stars of the same radius. A combination of the two effects could also explain the observed disagreement. 

Since our conclusions in the previous section show that mass-radius relationship shows minimal differences between single and binary radii (Section~\ref{sec:empirical_MR_relations}, Figure~\ref{fig:M_vs_R_vs_FeH_withEBs}), we may infer that the systematic differences in Figure~\ref{fig:R_vs_T_withEBs_withDSEP} arise from an offset in temperature, not radius, between single stars and binary stars.  Given an approximate 1:1 relation between mass and radius, the observed offset is also present for the data if viewed on the temperature-mass plane, where for a given mass, the temperatures of EBs are lower than those of single stars by several hundred Kelvin. 

The interpretation of the discrepancy between single and binary star temperatures to be a consequence of the diverse physical nature of the two different sets of stars is premature, as we must first demonstrate that the observed effect is genuine.  As such, even without knowledge of how the binary stars temperatures are derived in Figure~\ref{fig:R_vs_T_withEBs_withDSEP} (technique or calibration), it is straightforward to show that depending on what computational method is adopted to determine $T_{\rm EFF}$, different solutions for $T_{\rm EFF}$ will be obtained. For example, Figure~\ref{fig:R_vs_T_withEBs_withDSEP} shows the temperature-radius curve derived for single stars (Section~\ref{sec:empirical_TR_relations}, Equation~\ref{eq:temp_radius_relation_withsun}), the 5~Gyr solar metallicity Dartmouth model curve, and the tabulated radii and temperature values for main-sequence stars in Allen's Astrophysical Quantities (Allen's AQ; \citealt{cox00}).  We find that the \citet{cox00} temperature values generally agree with the temperatures for the EB systems, having temperatures $\sim 200$ to 300~K cooler than what is predicted by both the empirical relation (Equation~\ref{eq:temp_radius_relation_withsun}) and the Dartmouth model.

Once the method to determine EB temperatures is revisited, we may then proceed to uncover if there is any residual discrepancy seen in the temperatures of single and binary stars.  Any detection of an offset at this point would likely be of astrophysical origin. For instance, both higher magnetic activity and star-spots have been previously postulated by, e.g., \citet{tor02a} and \citet{lop05}, and would result in an overall lower effective temperature for a more active star. Based on the normalized X-ray emission $L_X/L_{\rm BOL}$ results from Table~\ref{tab:fund_params_combined}, and column 5 in table 1 from \citet{lop07}, the stars in binaries are $10^2 - 10^3$ times more active than the single star counterparts. Such a scenario would implicate the elevated activity levels and increase of surface inhomogeneities (spots) of EBs playing an important role in the difference of effective temperatures in single and binary stars.


\section{Conclusion}
\label{sec:conclusion}


In this work, we present 21 interferometrically determined radii of nearby K and M dwarfs.  This survey nearly doubles the number of directly determined radii for single stars in this range of spectral types.  With our measurements of angular diameter and bolometric flux, we are able to empirically determine stellar linear radii, luminosities, and effective temperatures of the entire sample.  Additional properties of the stellar mass and the $L_X/L_{\rm BOL}$ activity indicator are computed for each source.  We develop empirical relations of the broad-band color indices as a functions of temperatures, radii, and luminosities of the objects.  We find that adding metallicity as an additional parameter is necessary to properly constrain our fits, where the data show a notional effect on the color indices due to metallicity among the K-type stars, becoming very strong by spectral type $\sim$~M0 and later.  These relations are accurate to $\sim$2\% (temperature), $\sim$5\% (radius), and $\sim$4\% (luminosity).   

Solutions for relating the global stellar parameters (i.e., temperature-radius, temperature-luminosity, and mass-radius), on the other hand, are found to be independent of metallicity.  For example, a solar metallicity star with an observed radius of $0.6$~R$_{\odot}$ will have the same mass, temperature, and luminosity as a metal-poor counterpart.  This allows us to empirically quantify the influence of metallicity on the color index, which appears to impose the strongest dependence towards the K- to M- star transition. 

\begin{figure}										
  \centering
   \begin{tabular}{cc}
         \epsfig{file=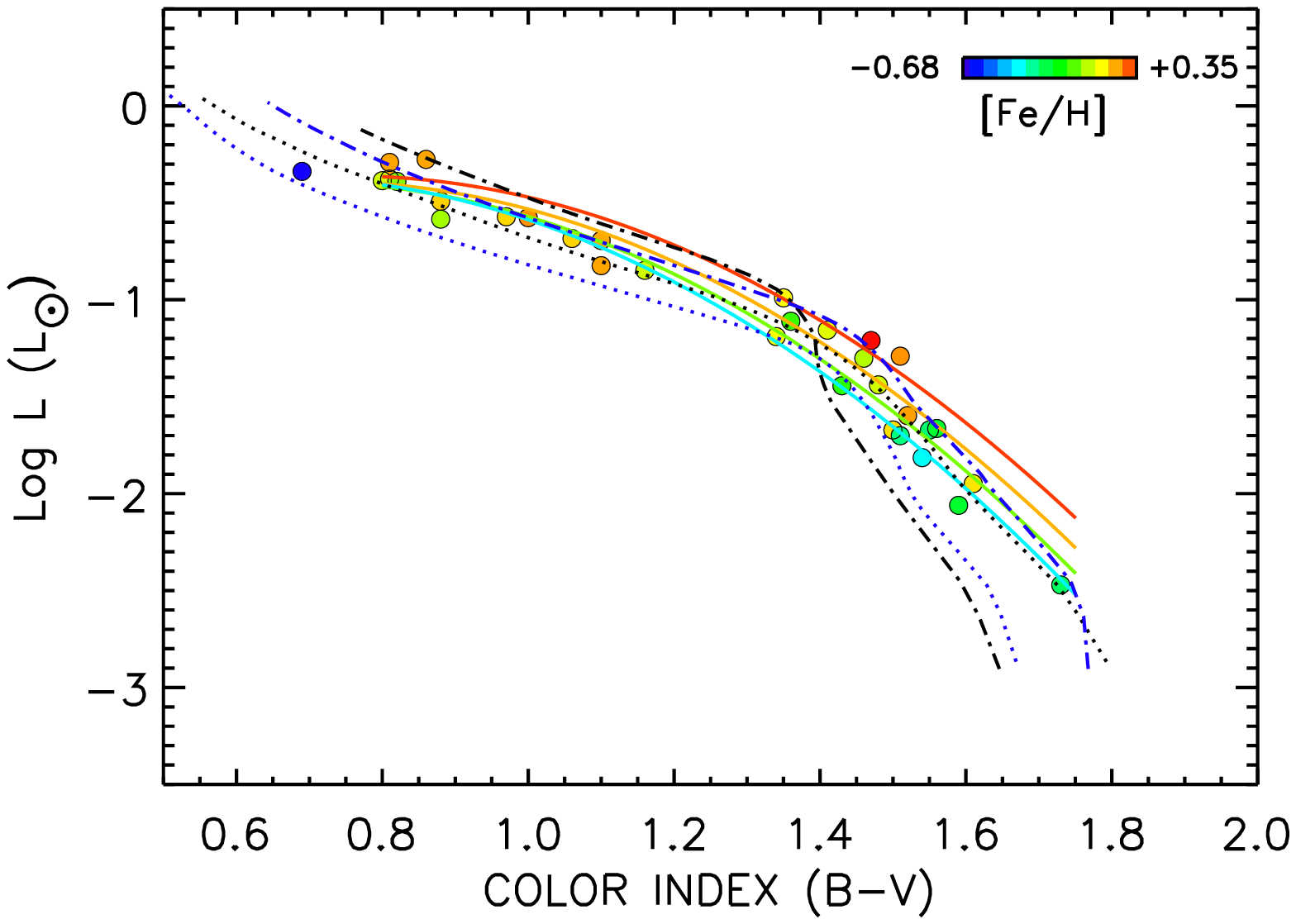, width=0.5\linewidth, clip=}	&
         \epsfig{file=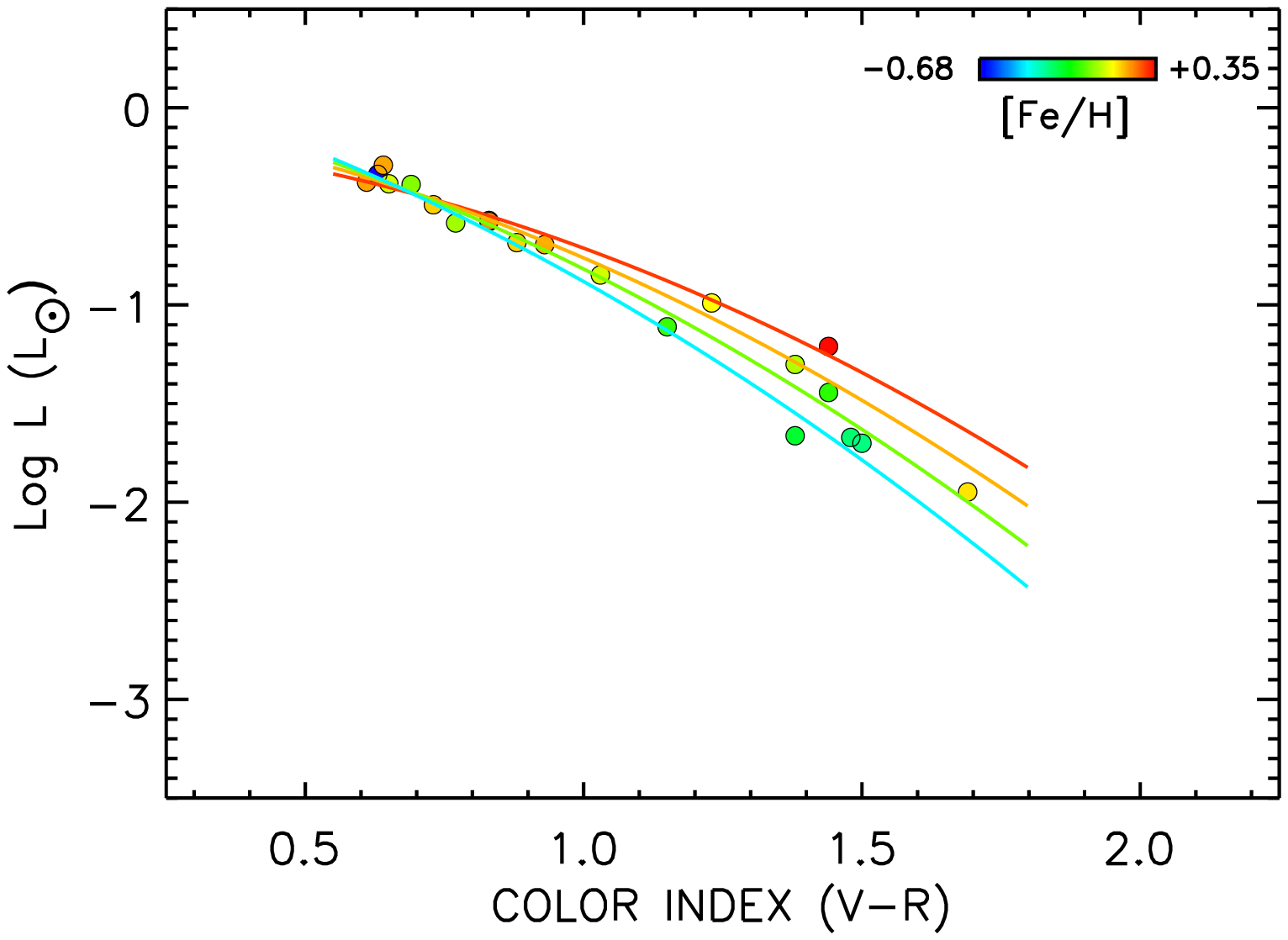, width=0.5\linewidth, clip=} \\
         \epsfig{file=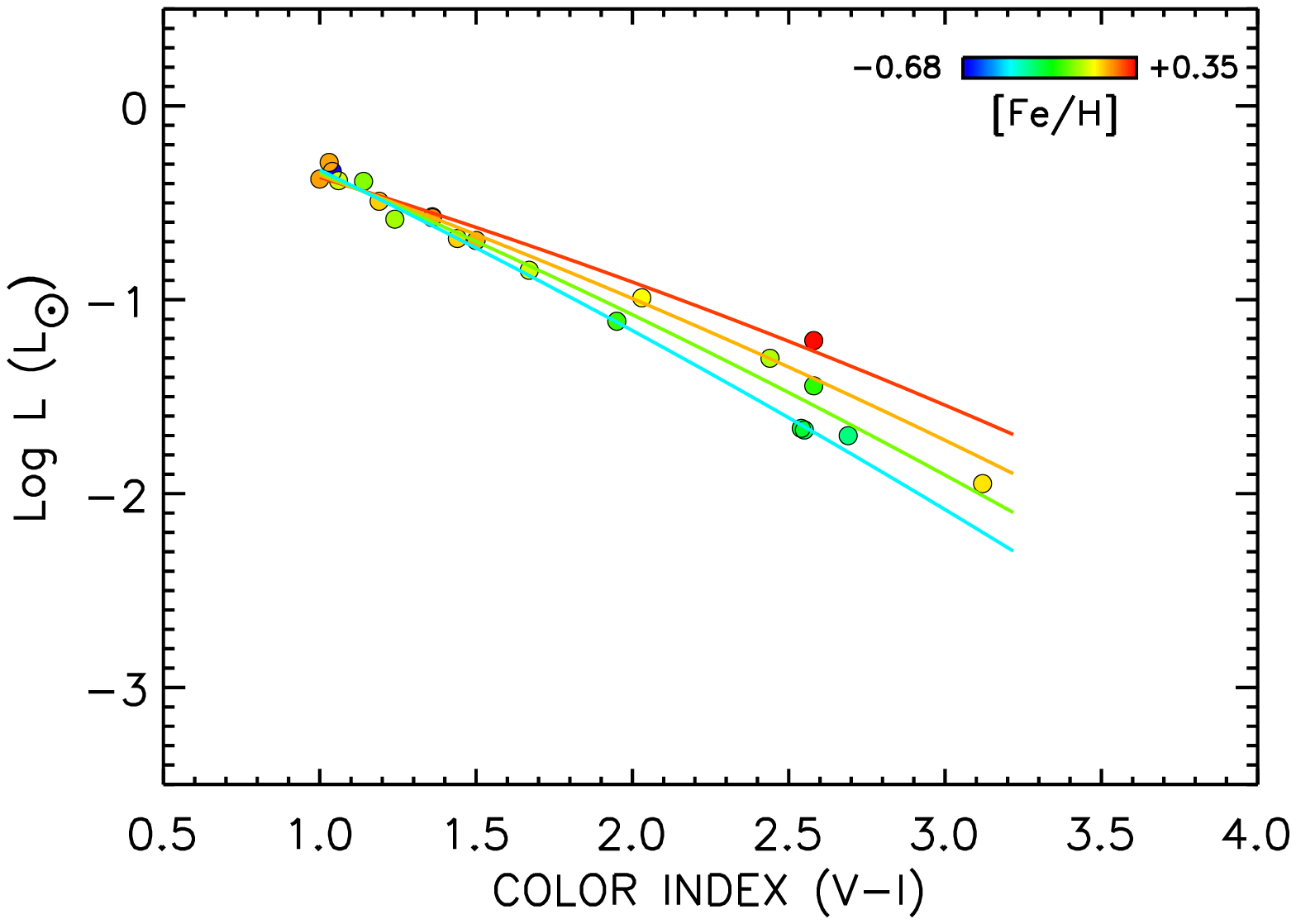, width=0.5\linewidth, clip=} &
         \epsfig{file=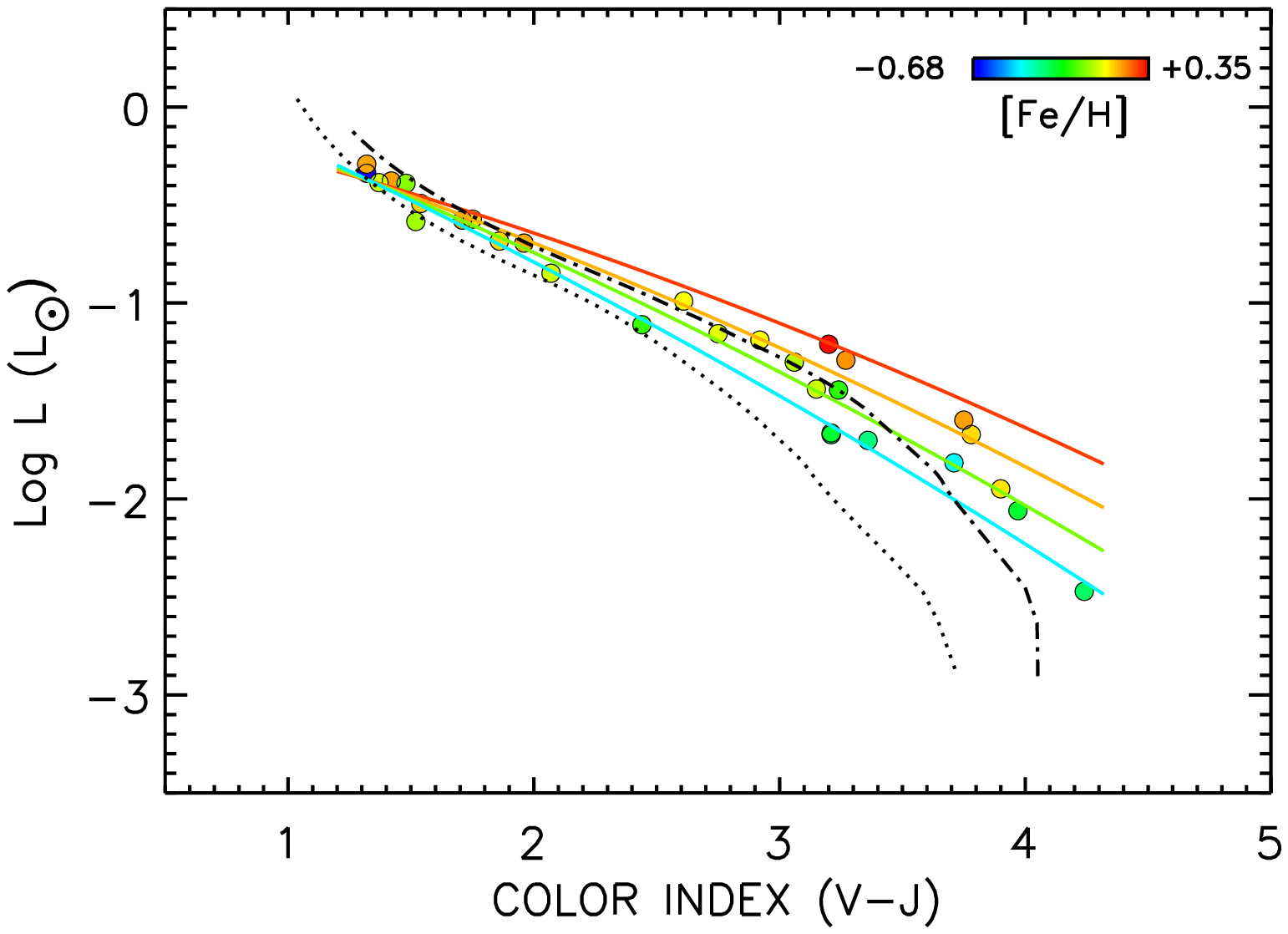, width=0.5\linewidth, clip=} \\
         \epsfig{file=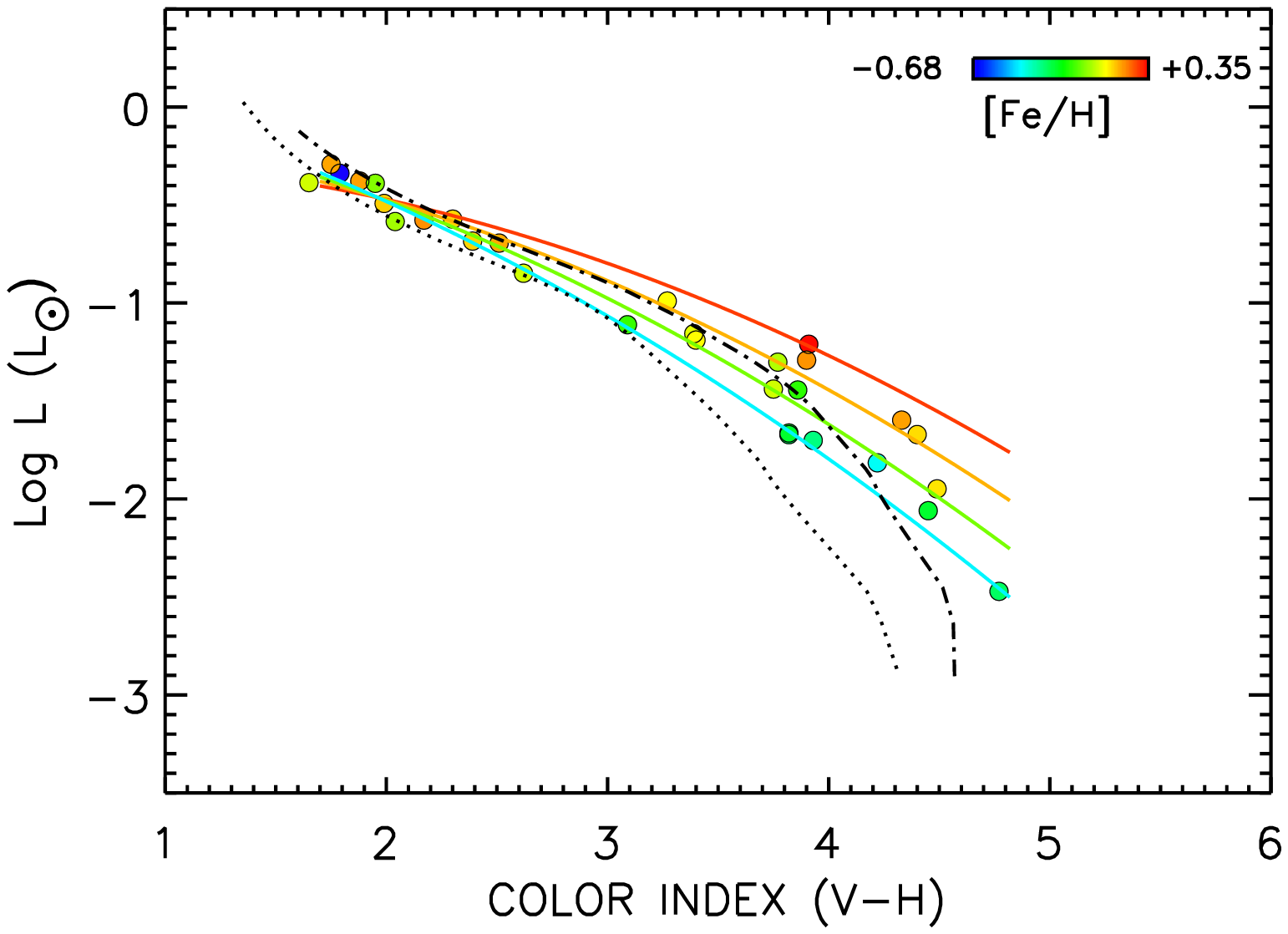, width=0.5\linewidth, clip=} &
         \epsfig{file=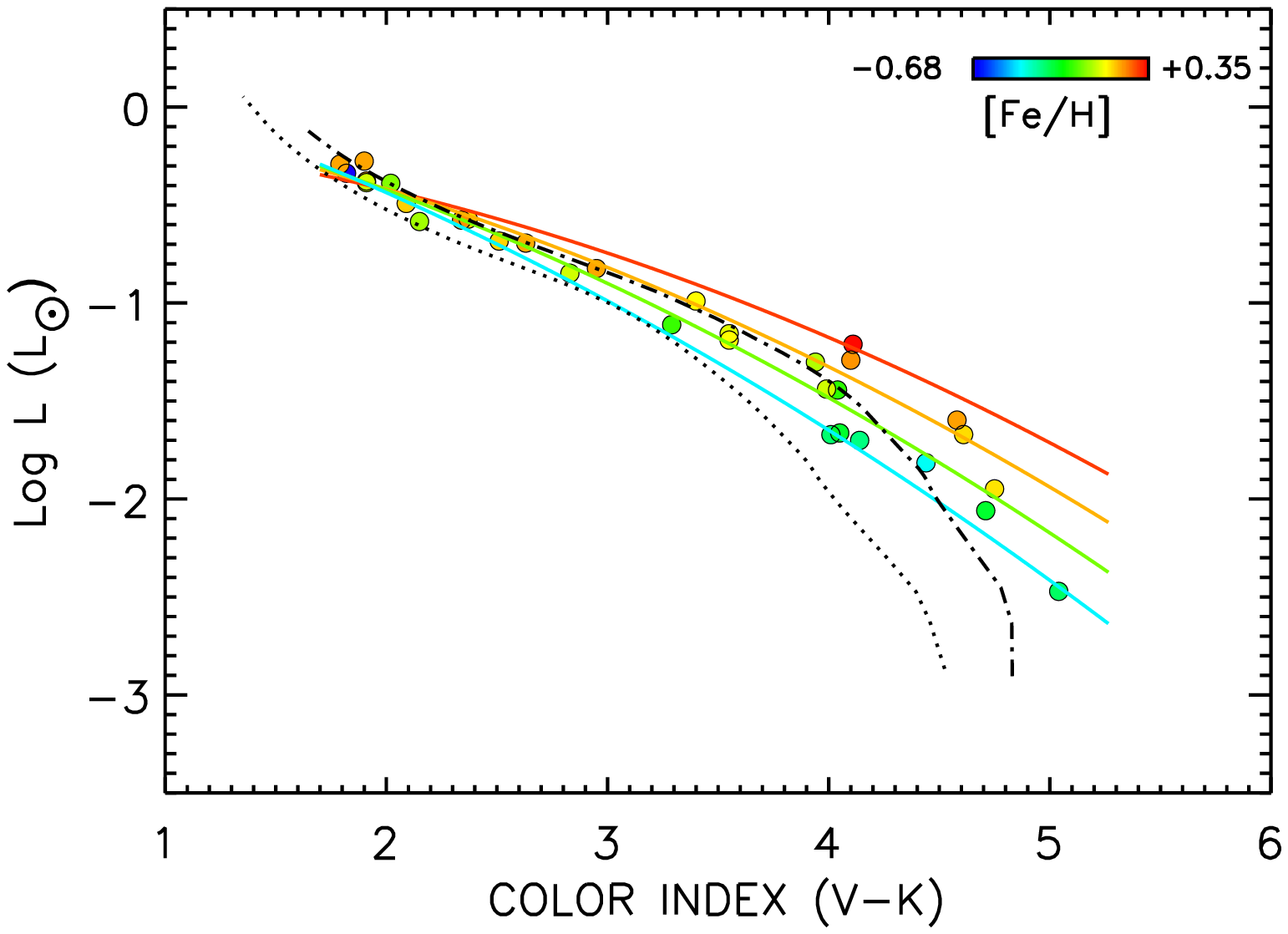, width=0.5\linewidth, clip=} 
    \end{tabular}
  \caption[] {Empirical color-metallicity-luminosity relations presented in Section~\ref{sec:empirical_lumin_relations}.  The color of the data point reflects the metallicity of the star. The colored lines are the solutions for the metallicity dependent fits, where the line color (red, orange, green, teal) represents our solution for an iso-metallicity line to [Fe/H]~=~$+0.25,0.0,-0.25,-0.5$. The panels displaying the ($B-V$), ($V-J$), ($V-H$), and ($V-K$) relations also include Dartmouth 5~Gyr isochrones as dash-dotted ([Fe/H]~$= 0.0$) and dotted lines ([Fe/H]~$=-0.5$).  Within the ($B-V$) panel, we include solutions with the same Dartmouth model specifications, but using the semi-empirical color transformation (blue lines) to derive the ($B-V$) colors.  See Equation~\ref{eq:luminosity_metallicity_relation} for the form of the equation and Table~\ref{tab:poly_solutions_lumin} for the coefficients and statistical overview for each color-metallicity-luminosity solution.}
 \label{fig:lumin_VS_colors_b}
 \end{figure}

\begin{figure}										
  \centering
     \begin{tabular}{cc}
     \epsfig{file=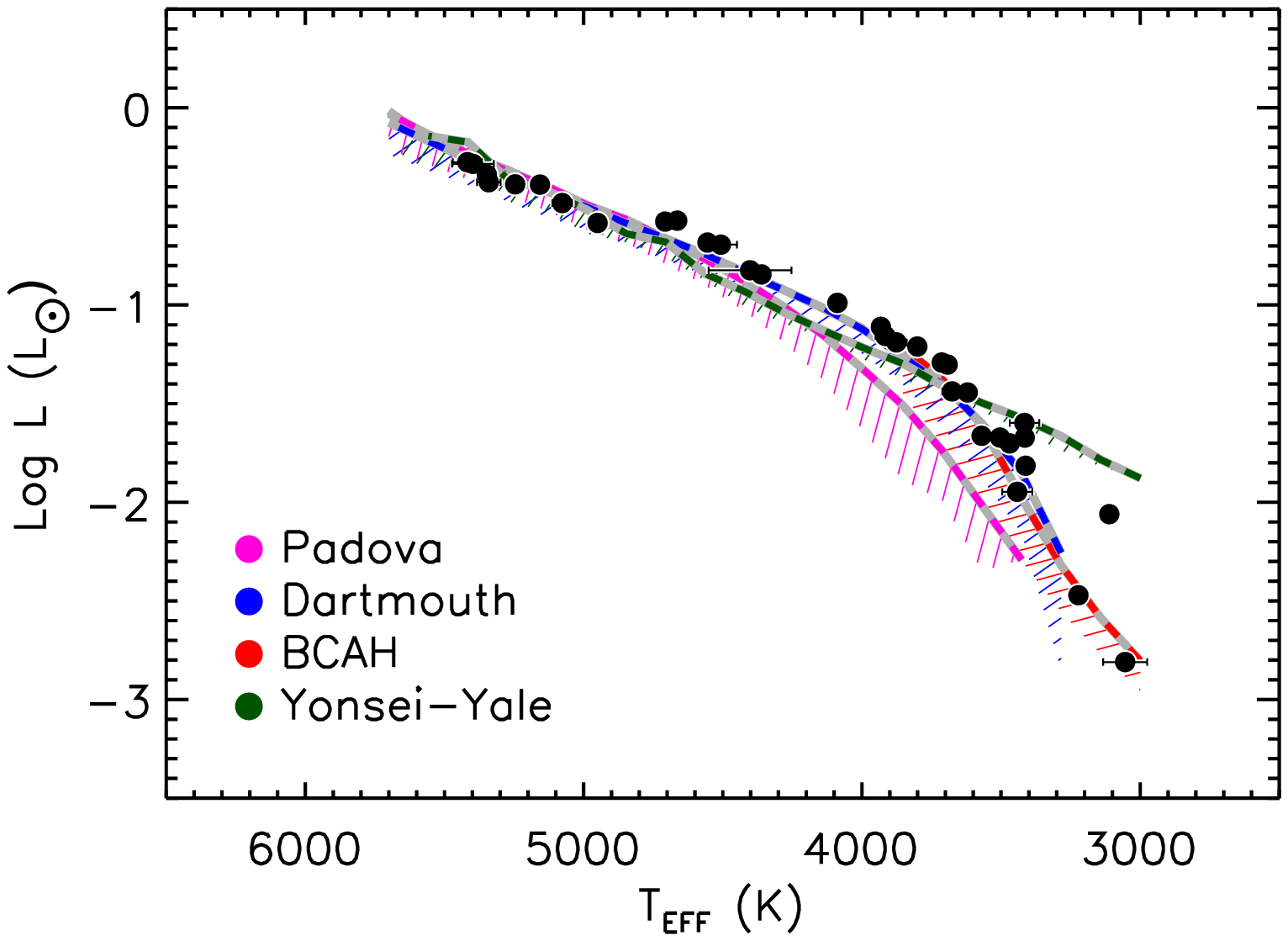, width=0.5\linewidth, clip=} &
\epsfig{file=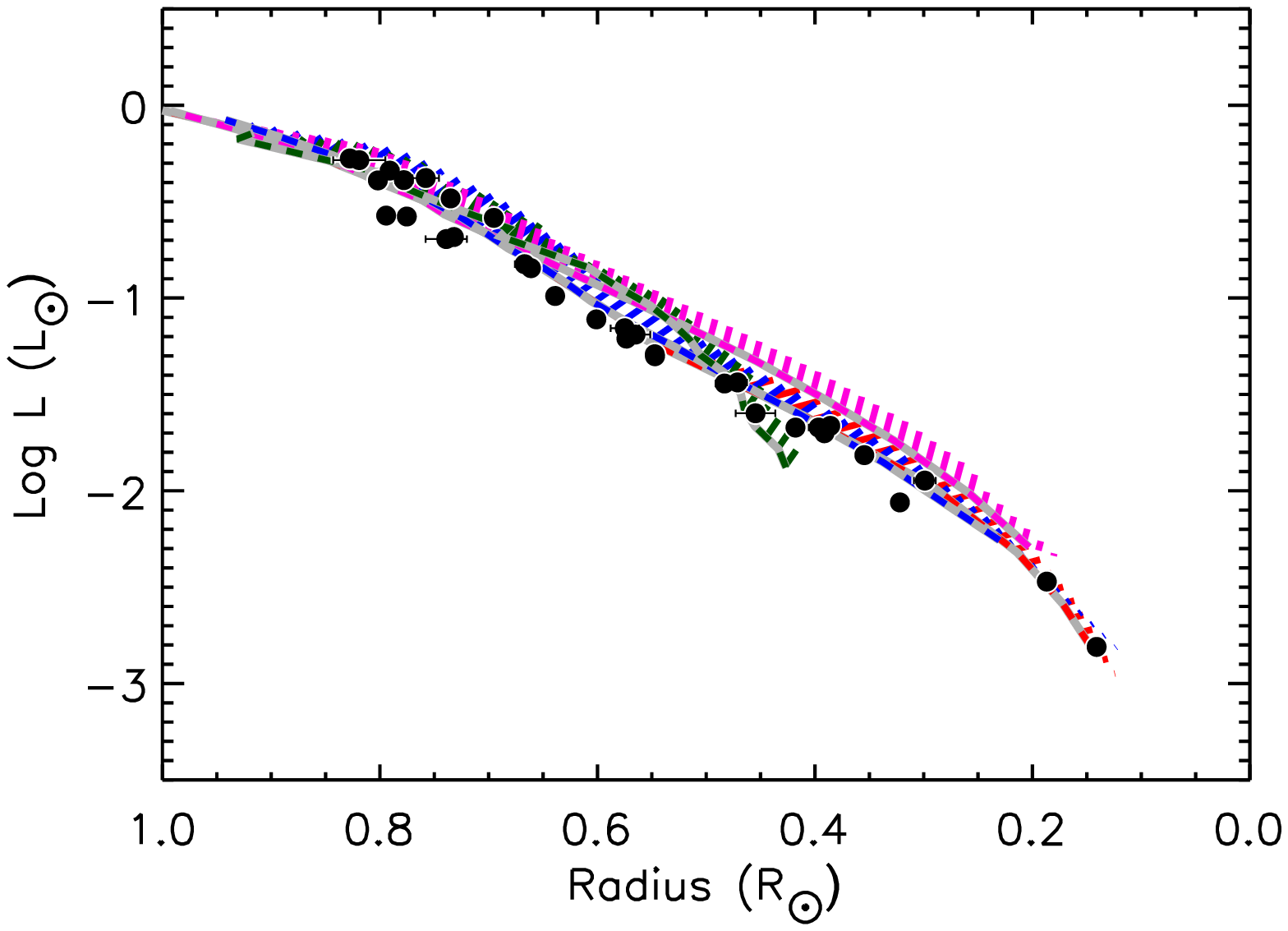, width=0.5\linewidth, clip=} \\	
\epsfig{file=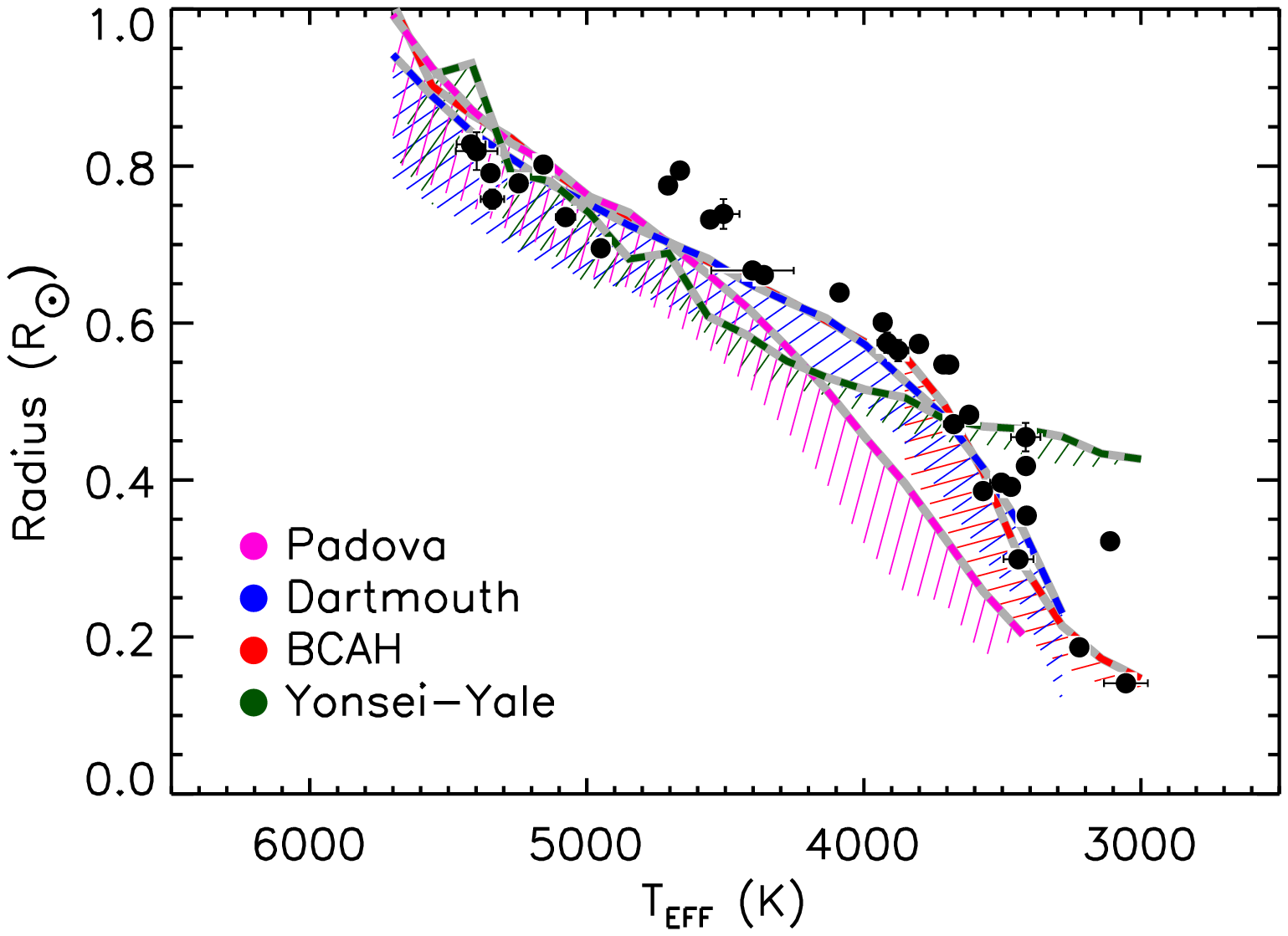, width=.5\linewidth,clip=} &
          \epsfig{file=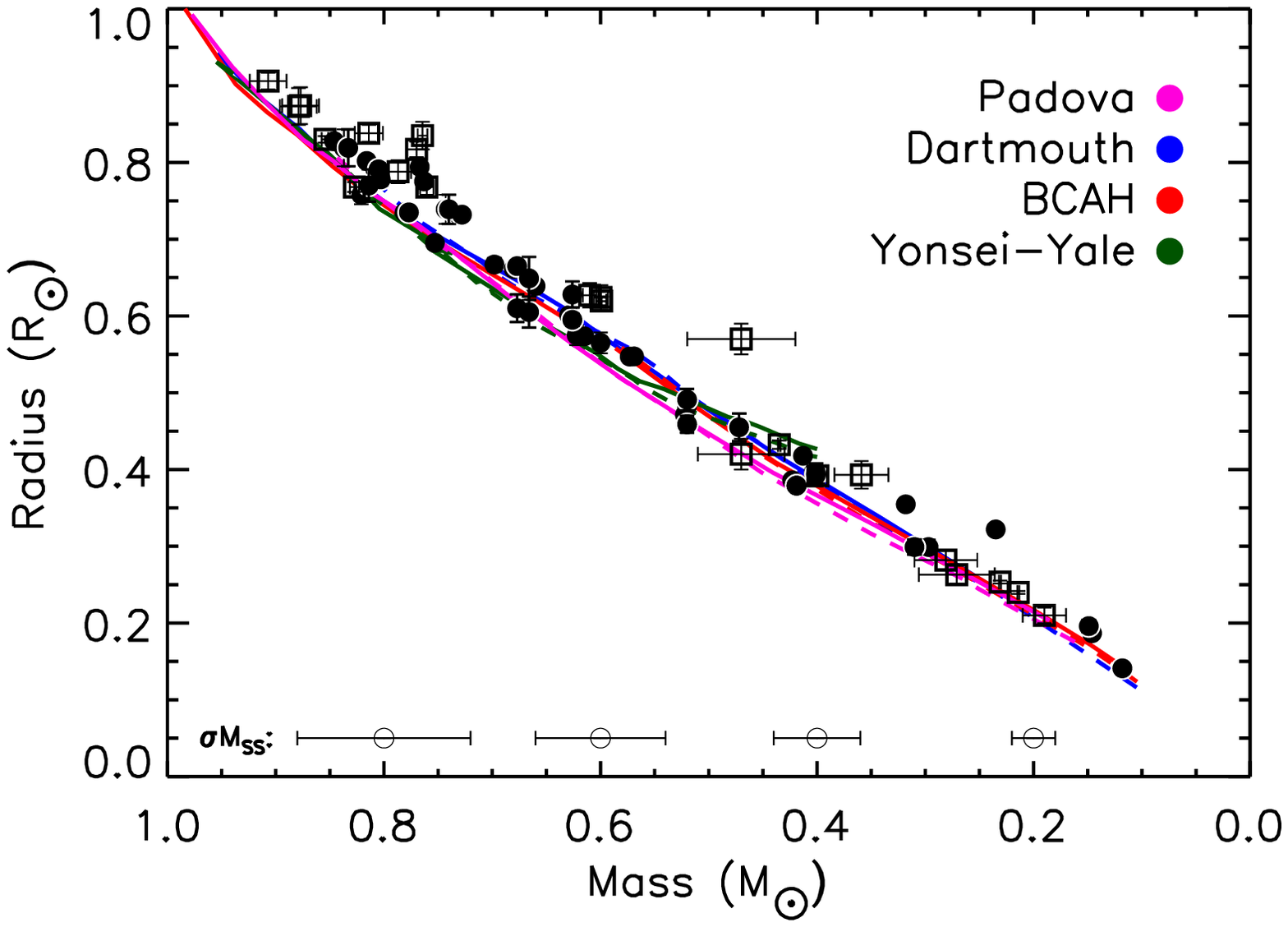, width=0.5\linewidth,clip=} 
\end{tabular}
  \caption[] {The 5~Gyr models isochrones of Padova, Dartmouth, BCAH98, and Yonsei-Yale \citep{gir00, dot08, bar98, dem04} displayed along with the single star measurements presented in Table~\ref{tab:fund_params_combined} (filled circles). In the top-right, top-left, and bottom-left plots, the dashed line corresponds to the solar metallicity model (see color legend within plots), and the hashed region spans a metallicity of 0 to $-0.5$~dex. In the bottom-right plot, we show observations of single stars (filled circles; Table~\ref{tab:fund_params_combined}) and binary stars (open squares, see Section~\ref{sec:empirical_MR_relations}). The solid and dashed lines mark a change in metallicity from [Fe/H]~$=0$ to $-0.5$.  Characteristic one-sigma mass errors for single stars ($\sigma M_{SS}$) are shown close to the bottom axis. See Section~\ref{sec:global_relations} for details.}
   \label{fig:logL_vs_T_and_R_phil}
   \end{figure}

\begin{figure}										
  \centering
     \begin{tabular}{c}
\epsfig{file=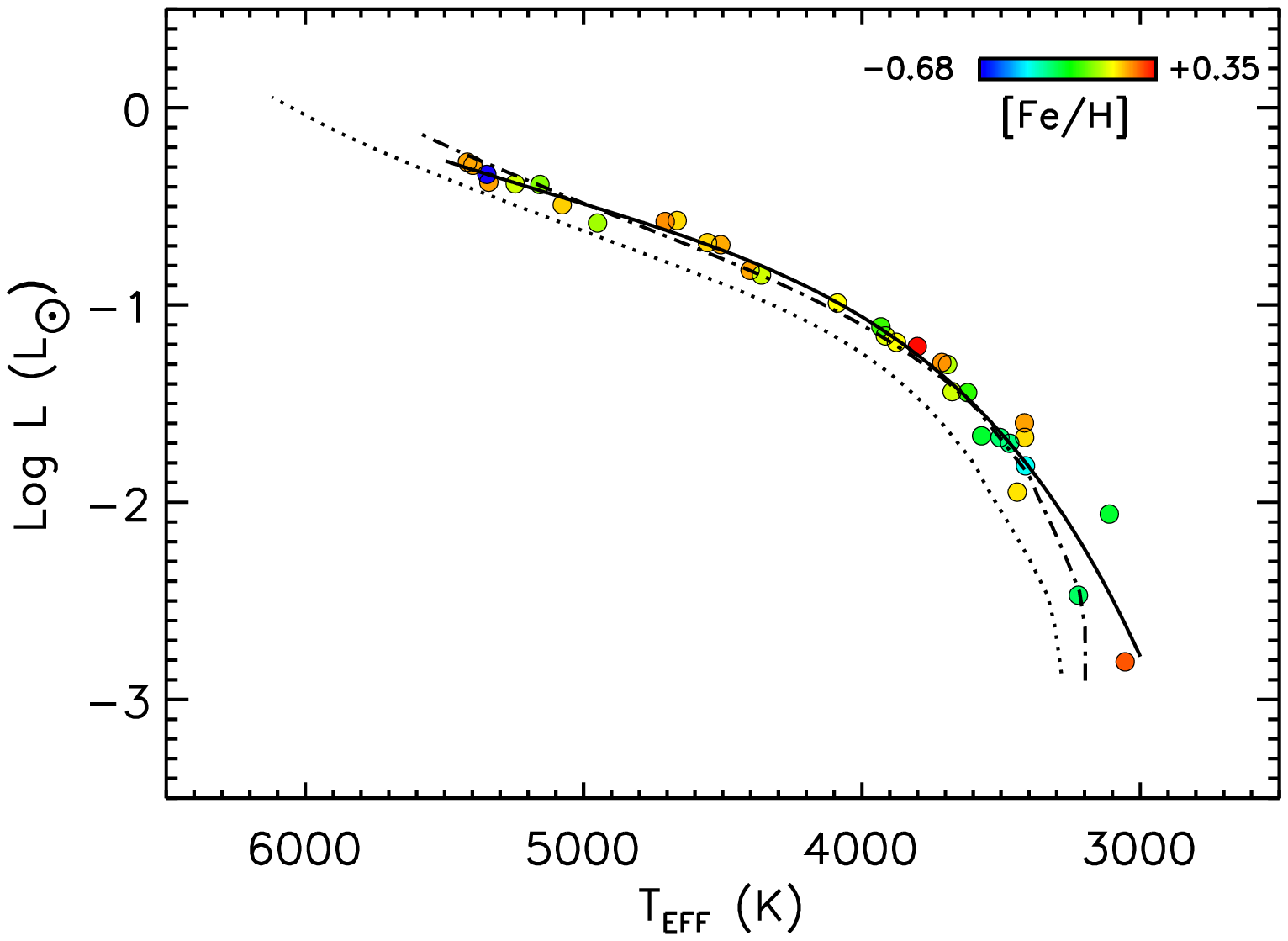, width=0.45\linewidth, clip=} \\
\epsfig{file=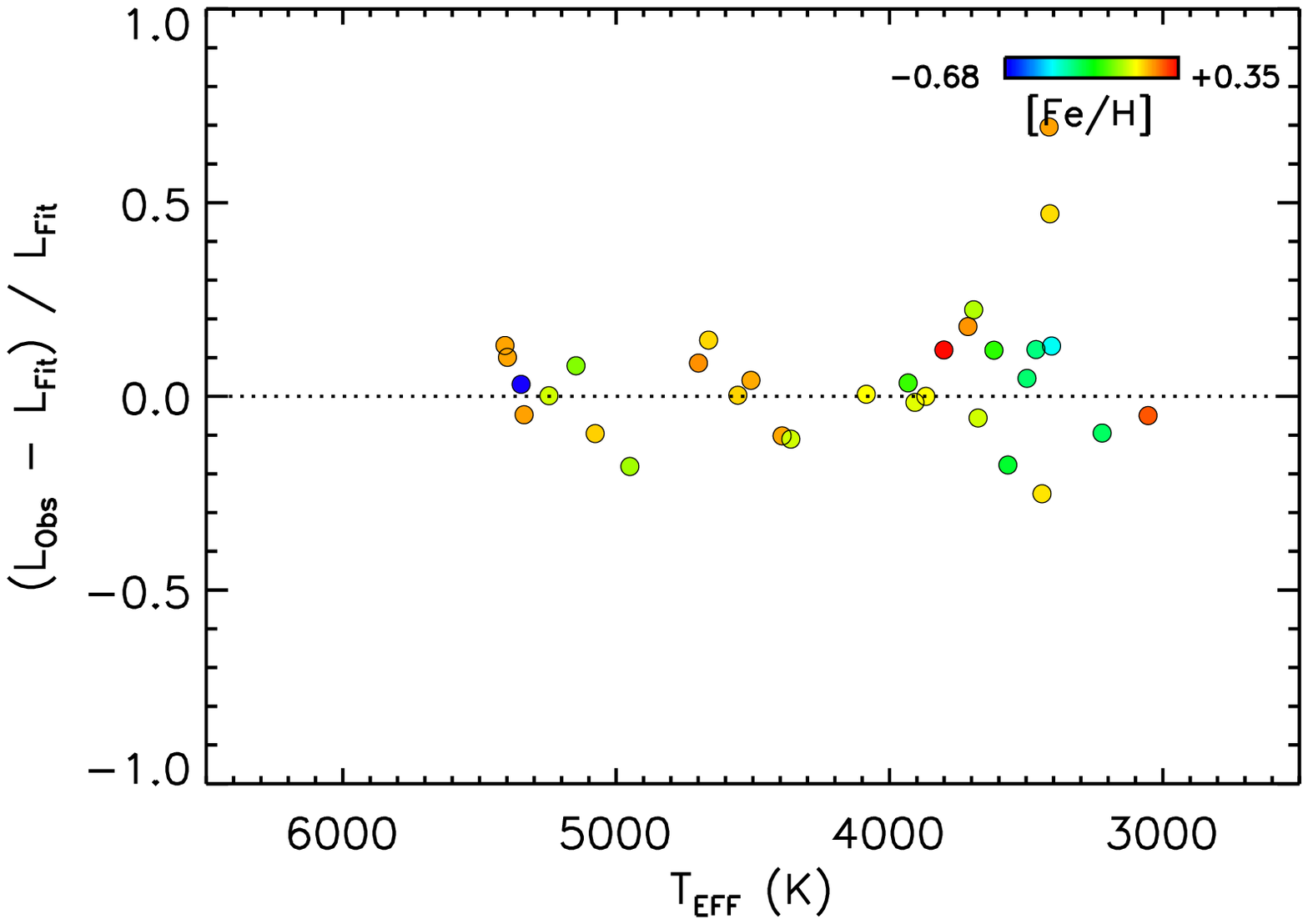, width=0.45\linewidth, clip=} \\
\epsfig{file=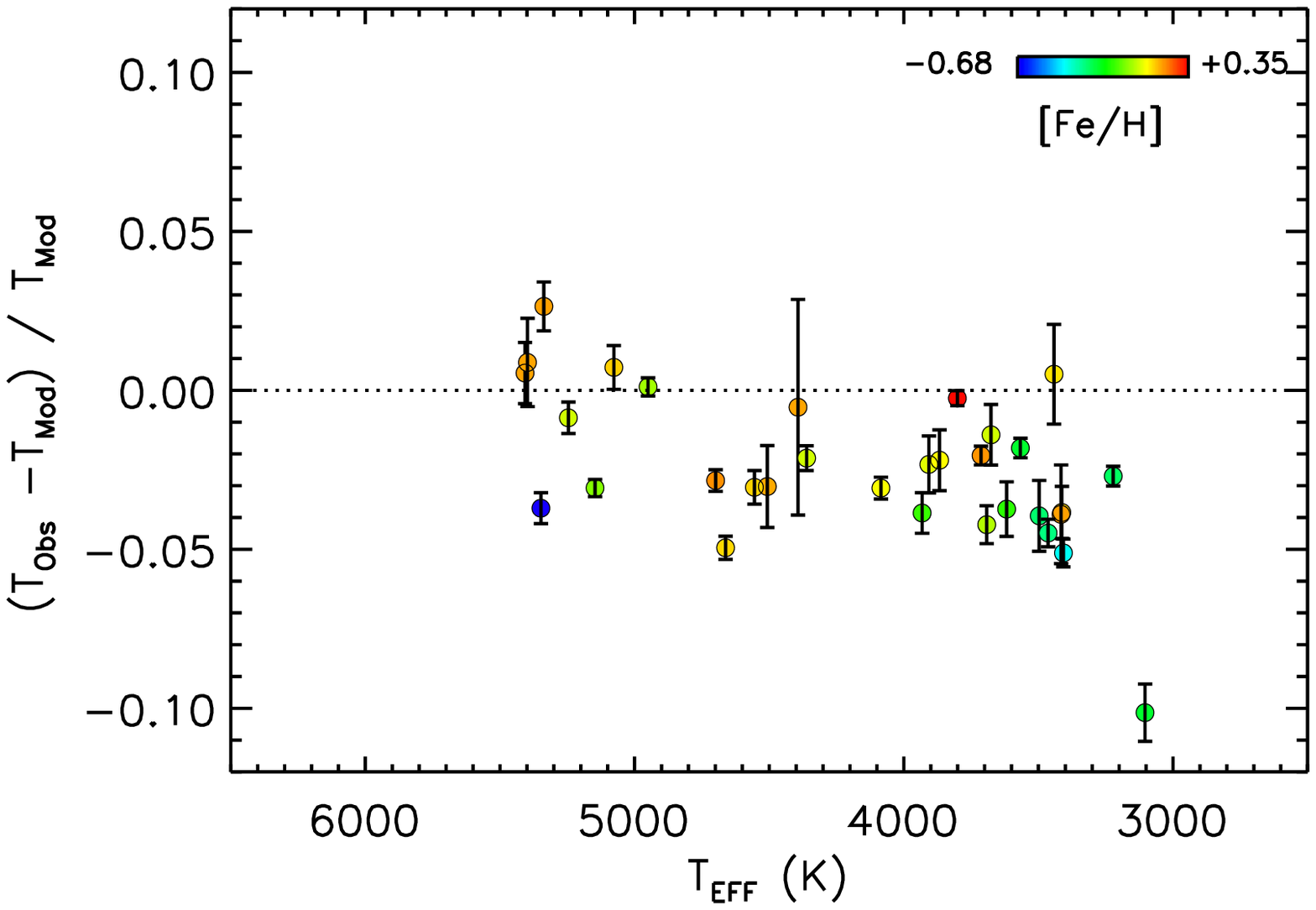, width=0.45\linewidth, clip=} 		
\end{tabular}
  \caption[] {{\it Top:} Temperature-luminosity relation derived here is shown as a solid black line (Equation~\ref{eq:temp_lumin_relation}) for the approximate range the relation holds true.  Dartmouth model 5~Gyr isochrones are also shown as dash-dotted ([Fe/H]~$=0$) and dotted lines ([Fe/H]~$=-0.5$). {\it Middle:} The fractional difference in luminosity (observed versus our fit to the data) is shown versus temperature. The dotted line represents a zero deviation from the fit. In both the top and middle plots, errors are not plotted, but typically smaller than the size of the data points. {\it Bottom:} The observed temperature versus the temperature predicted by interpolating the observed luminosity in the custom-tailored Dartmouth model for each star ($T_{\rm Mod}$). The dotted line shows a 1:1 relation, and the y-errors shown are the scaled observational errors in temperature. In all plots, the color of the data point reflects the metallicity of the star as depicted in the legend. See Section~\ref{sec:empirical_LT_relations} for details. }
   \label{fig:Lumin_VS_Temp}
   \end{figure}

\begin{figure}										
  \centering
     \begin{tabular}{c}
\epsfig{file=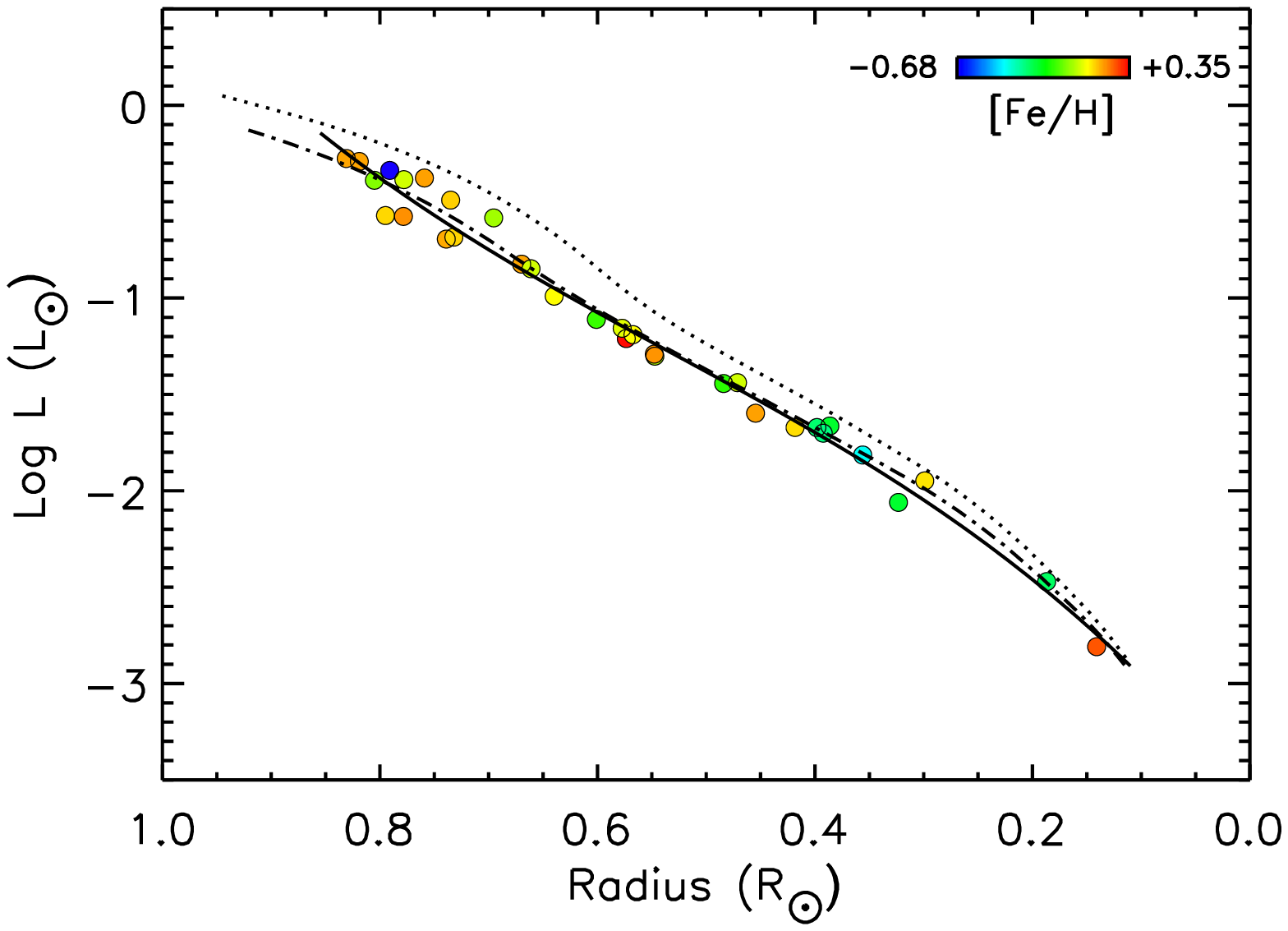, width=0.45\linewidth, clip=} \\	
\epsfig{file=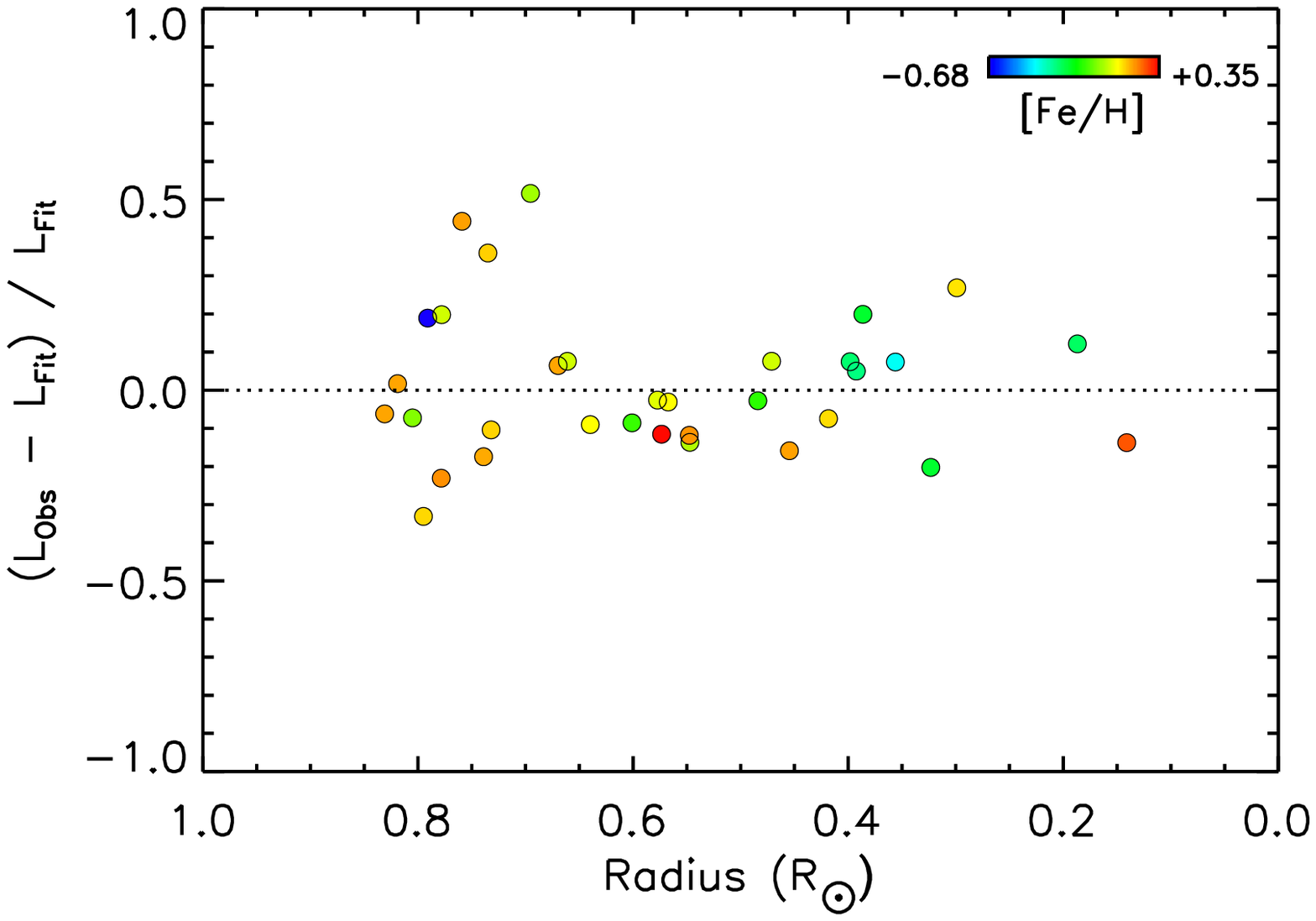, width=0.45\linewidth, clip=} \\
\epsfig{file=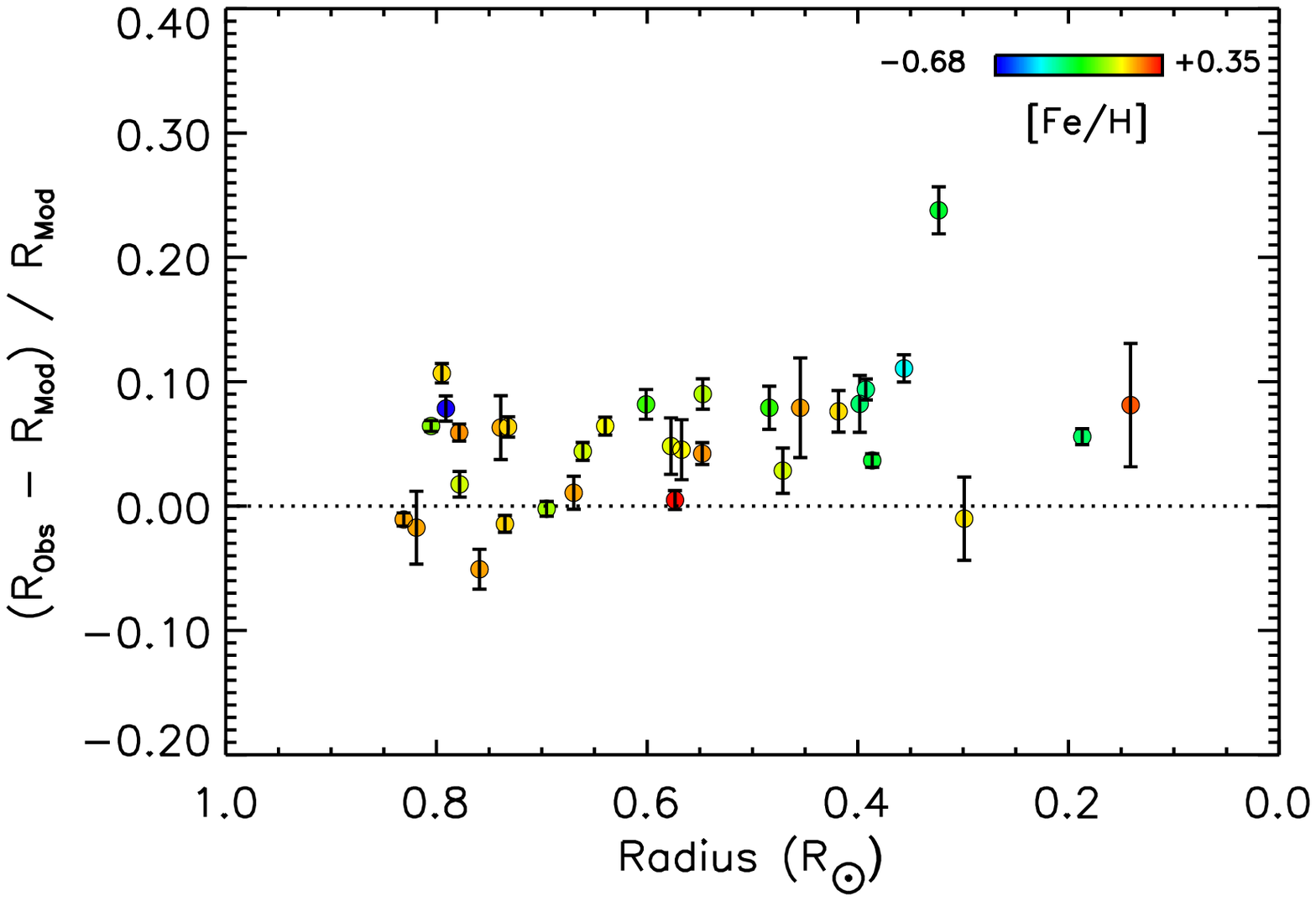, width=0.45\linewidth, clip=} 		
\end{tabular}
  \caption[] {{\it Top:} Radius-luminosity relation derived here is shown as a solid black line (Equation~\ref{eq:radius_lumin_relation}) for the approximate range the relation holds true.  Dartmouth model 5~Gyr isochrones are also shown as dash-dotted ([Fe/H]~$=0$) and dotted lines ([Fe/H]~$=-0.5$). {\it Middle:} The fractional difference in luminosity (observed versus fit to the data) is shown versus radius. The dotted line represents a zero deviation from the fit. In both the top and middle plots, errors are not plotted, but typically smaller than the size of the data points. {\it Bottom:} The observed radius versus the radius predicted by interpolating the observed luminosity in the custom-tailored  Dartmouth model for each star ($R_{\rm Mod}$). The dotted line shows a 1:1 relation, and the y-errors shown are the scaled observational errors in radius. In all plots, the color of the data point reflects the metallicity of the star as depicted in the legend. See Section~\ref{sec:empirical_LR_relations} for details. }
   \label{fig:Lumin_VS_Radius}
   \end{figure}

\begin{figure}										
  \centering 
          \epsfig{file=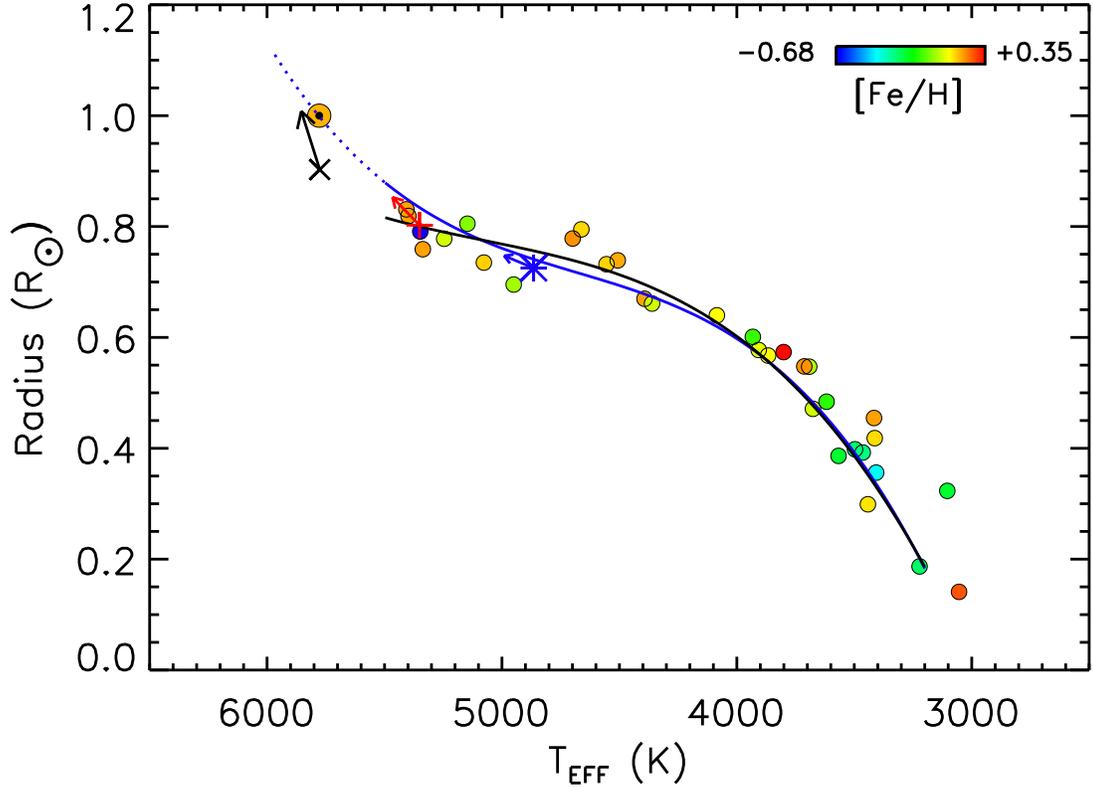, width=1.0\linewidth,clip=} 
  \caption[] {Temperature-radius relation is shown as a solid black line (Equation~\ref{eq:temp_radius_relation}) for the approximate range that the relation holds true. The relation including the Sun (Equation~\ref{eq:temp_radius_relation_withsun}) is shown as a blue line, where the dotted line is an extrapolation of the curve to higher temperatures with the Sun as a reference point.  The color of the data point reflects the metallicity of the star.  Errors are not plotted, but are typically of the sizes of the data points.  The left-most orange dot is the Sun. Solar metallicity Dartmouth models are used to plot the positions at 1~Gyr for stars with masses of 1.0~M$_{\odot}$, 0.9M$_{\odot}$, and 0.8M$_{\odot}$ (black $\times$, red $+$, and blue $\ast$ respectively).  Arrows point to their positions at an age of 4.5~Gyr. See Section~\ref{sec:empirical_TR_relations} for details.} 
 \label{fig:Radius_VS_Temp}
 \end{figure}

\begin{figure}										
  \centering
     \begin{tabular}{c}
\epsfig{file=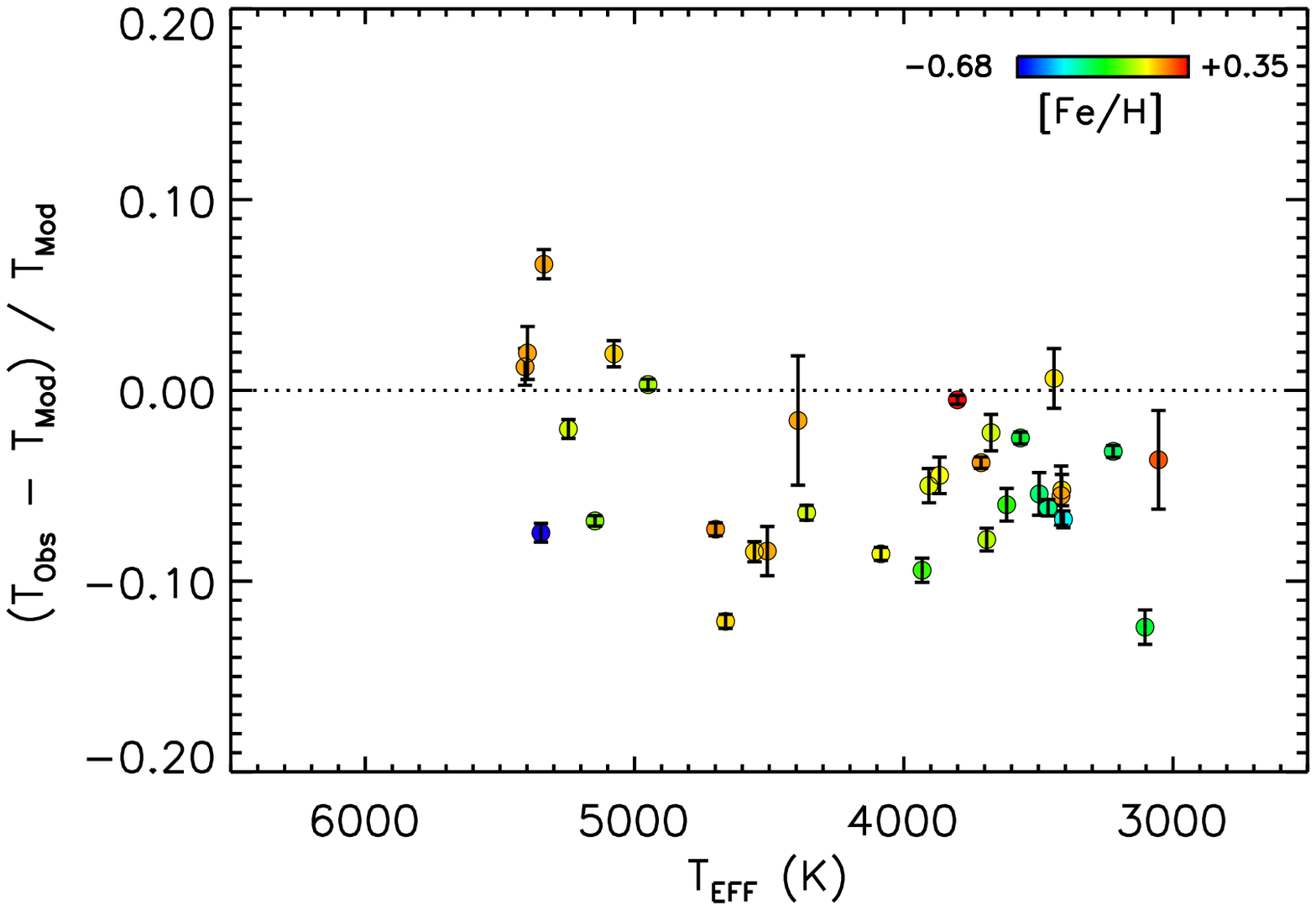, width=0.45\linewidth, clip=} \\	
\epsfig{file=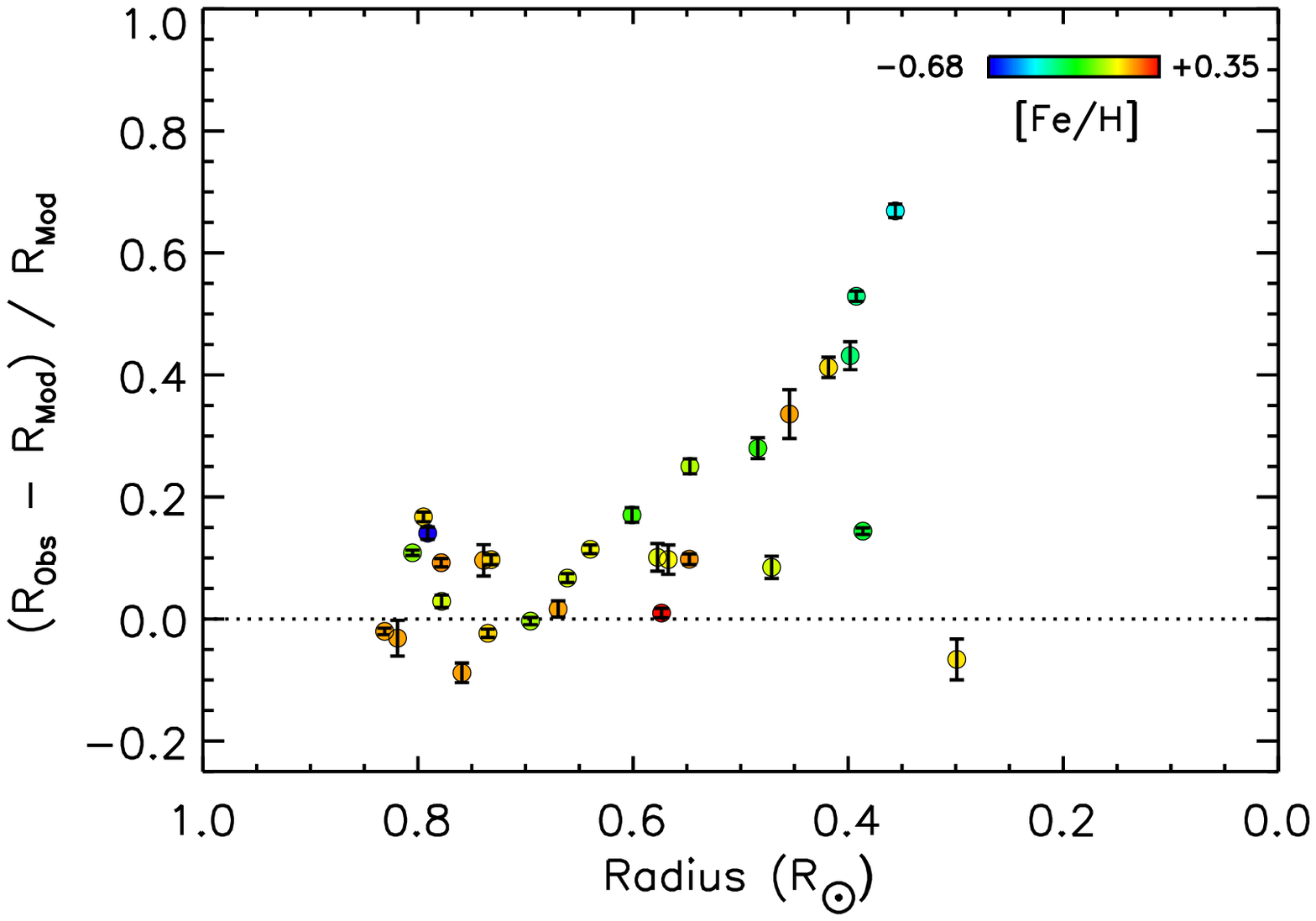, width=0.45\linewidth, clip=} 
\end{tabular}
  \caption[] {{\it Top:} The observed temperature versus the temperature predicted by interpolating the observed radius in the custom-tailored Dartmouth model for each star ($T_{\rm Mod}$). {\it Bottom:} The observed radius versus the radius predicted by interpolating the observed temperature in the custom-tailored Dartmouth model for each star ($R_{\rm Mod}$). The dotted line shows zero deviation from the model, and the y-errors shown are the scaled observational errors in temperature (top plot) and radius (bottom plot). In all plots, the color of the data point reflects the metallicity of the star as depicted in the legend. See Section~\ref{sec:empirical_TR_relations} for details. }
   \label{fig:CHARA_VS_DSEP_radiustemp_offset}
   \end{figure}

\begin{figure}										
  \centering
          \epsfig{file=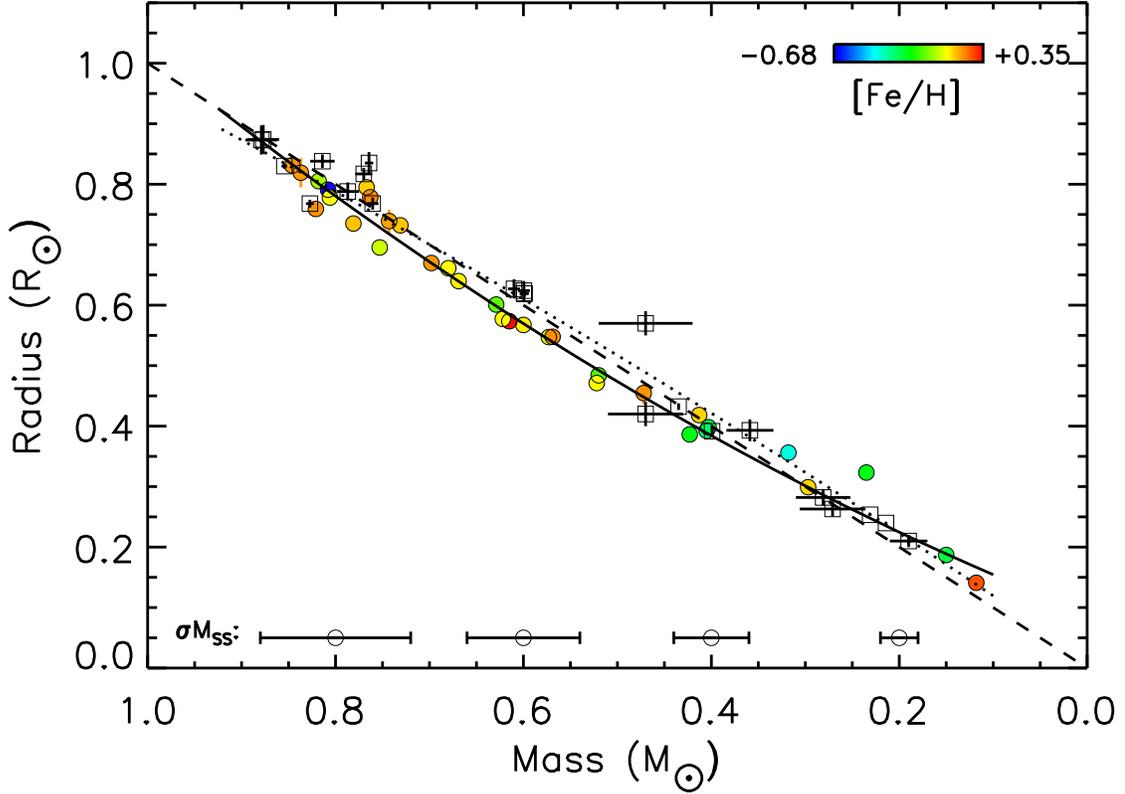,width=1.0\linewidth, clip=} 
  \caption[] {Mass-radius relations for single and binary stars as expressed in Equations~\ref{eq:mass_VS_radius} and \ref{eq:mass_VS_radius_EB}.  The filled circles and solid line are the data and solution for single stars. The open squares and dotted line are for the EB stars. The measured 1-sigma errors are shown for radii, but are typically smaller than the data point. Single star mass errors are not shown for clarity.  We show a typical single star mass error bar for a given mass at the bottom of the plot window indicating a value of $\sigma M_{\rm SS} \sim 10$~\%.  Although the mass errors for single stars are large, we do not detect any metallicity dependence on the mass-radius relation.  See Section~\ref{sec:empirical_MR_relations} for details.}
 \label{fig:M_vs_R_vs_FeH_withEBs}
 \end{figure}

We show that models predict a drop in luminosity at 0.1~L$_{\odot}$ and radius at 0.6~R$_{\odot}$ at colors several tenths of a magnitude too blue than what we observe (Figure~\ref{fig:lumin_VS_colors_b}). It is unclear whether the cause of this is within the model physics itself, or whether it lies within the errors of color tables used in the color transformations \citep{dot08}.  We are able to bypass errors introduced from the color tables by looking directly at the empirically measured luminosity-temperature and luminosity-radius data.  In this, we find that evolutionary models over-predict the temperatures for stars with temperatures $<5000$~K by $\sim 3$\%, and under-predict the radii for stars with radii $<0.7$~R$_{\odot}$ by $\sim 5$\%. This follows suit with similar conclusions in the literature that models have a tendency to underestimate the radius and overestimate the temperature in order for the luminosity to come out right \citep{pop97, mor08}.  Our data also suggest that for the range of stars observed, the influence of metallicity on the global parameters appears to be over accounted for in models (Figure~\ref{fig:Lumin_VS_Temp} and \ref{fig:Lumin_VS_Radius}).

\begin{figure}										
  \centering
  \epsfig{file=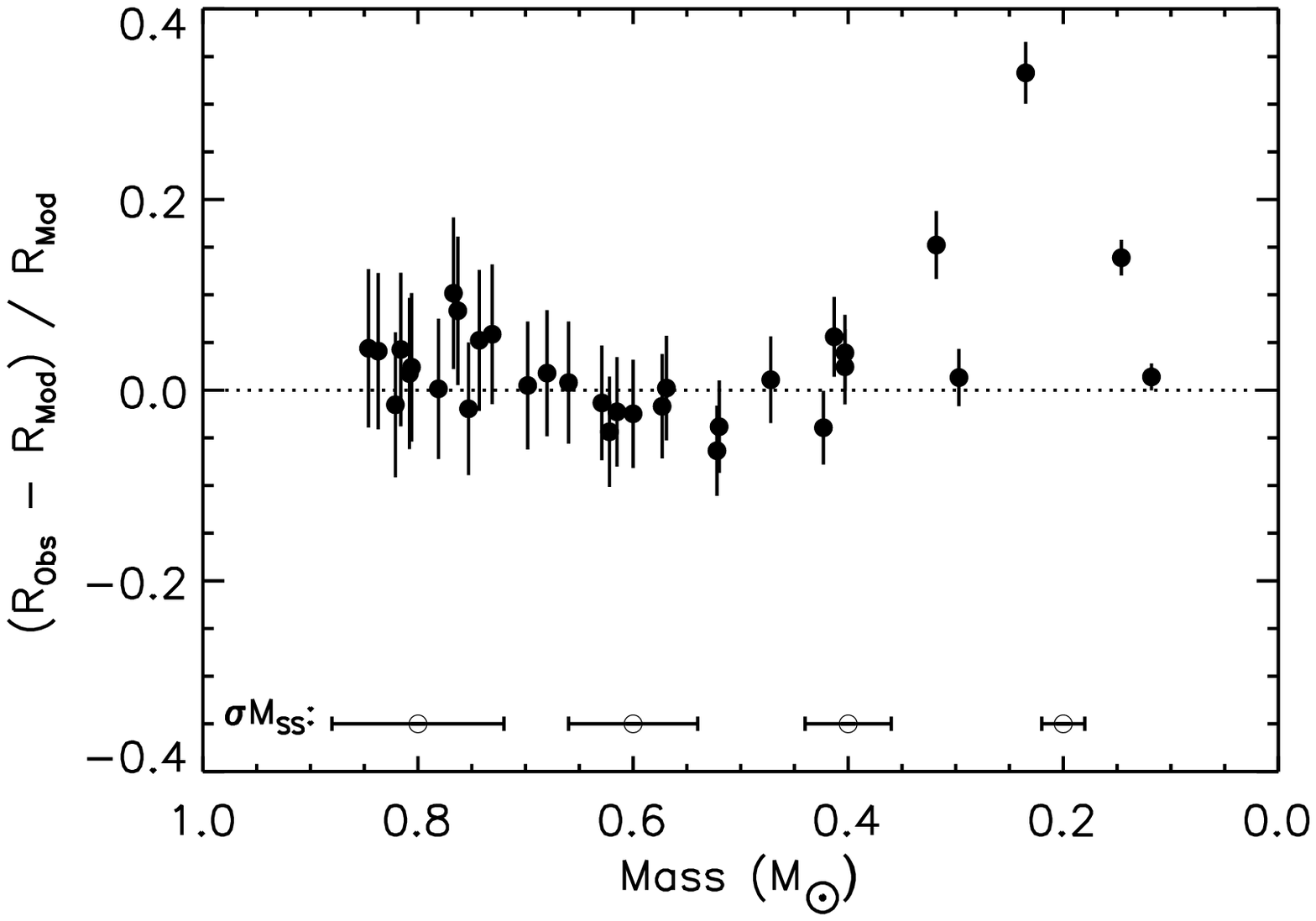, width=.50\linewidth,clip=}	\\
  \begin{tabular}{cc}
          \epsfig{file=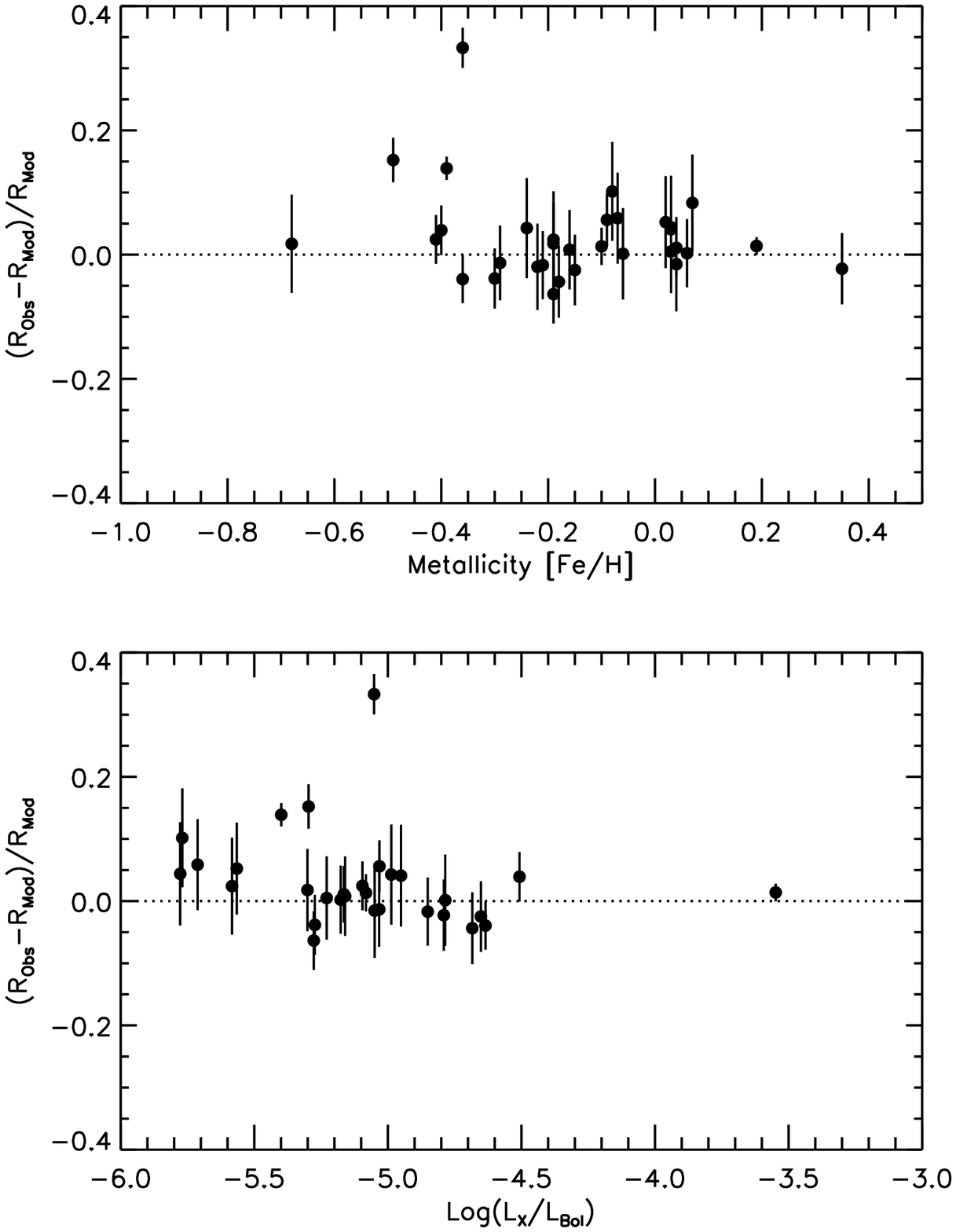, width=.50\linewidth,clip=} &
          \epsfig{file=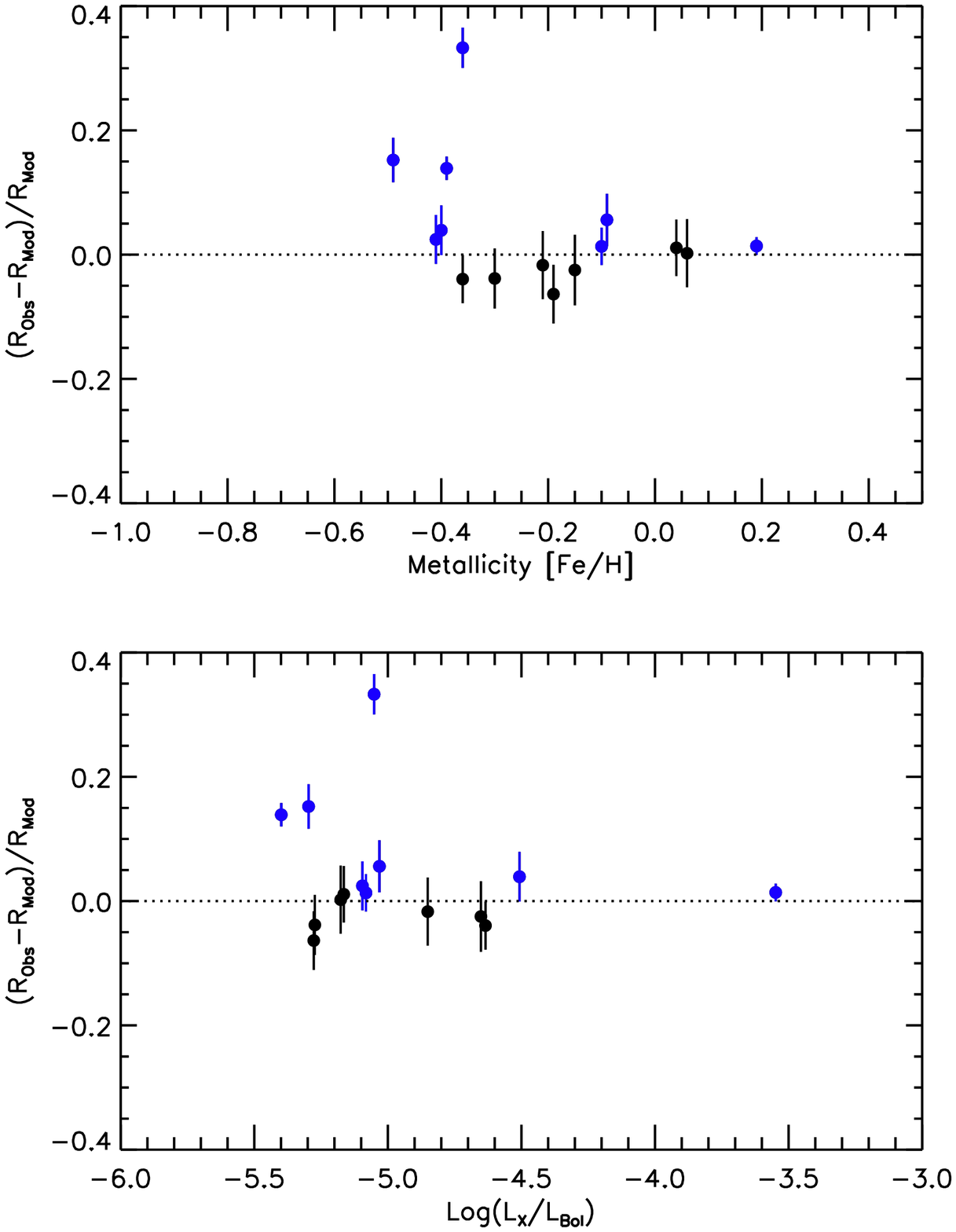, width=.50\linewidth,clip=} 
\end{tabular}
  \caption[] {Fractional offset in radii versus mass ({\it top}), metallicity [Fe/H] ({\it middle}) and normalized X-ray to bolometric luminosity $L_X/L_{\rm BOL}$ ({\it bottom}) for single stars compared to 5~Gyr Dartmouth model isochrones custom-tailored to each star. Characteristic one-sigma mass errors for single stars ($\sigma M_{SS}$) are shown close to the bottom axis in the top plot.  The extent of the y-errors in all plots show a fractional deviation in radius of 10\% (due to errors in mass estimates). The left middle and bottom panels show the full sample of K- and M-dwarfs ($M < 0.9$~M$_{\odot}$).  The right middle and bottom panels show only the M-dwarfs ($M < 0.6$~M$_{\odot}$), where the blue points indicate stars with masses $< 0.42$~M$_{\odot}$. The dotted line indicates a zero deviation.  See Section~\ref{sec:modelling} for details.  }
 \label{fig:mass_VS_radius5_allstars}
 \end{figure}

The data set examined in conjunction with a collection of EBs, shows that the radii for single stars are consistent within errors to those in eclipsing systems for a given mass, both essentially following a 1:1 relation between mass and radius for stars with masses $< 0.9$~M$_{\odot}$ and radii $<0.9$~R$_{\odot}$. On the mass-radius plane, we are able to confirm the previously documented offset between the directly measured radii and the ones based on theory observed in late-type binary stars at or below the full convection limit of $\sim 0.4$~M$_{\odot}$.  We evaluate this discrepancy as a function of mass, metallicity, and activity, revealing a mild correlation to the radius offset to the metallicity of a star, where stars with [Fe/H]~$<-0.35$ all have radii larger than the model predictions (Figure~\ref{fig:mass_VS_radius5_allstars}).

\begin{figure}										
  \centering
          \epsfig{file=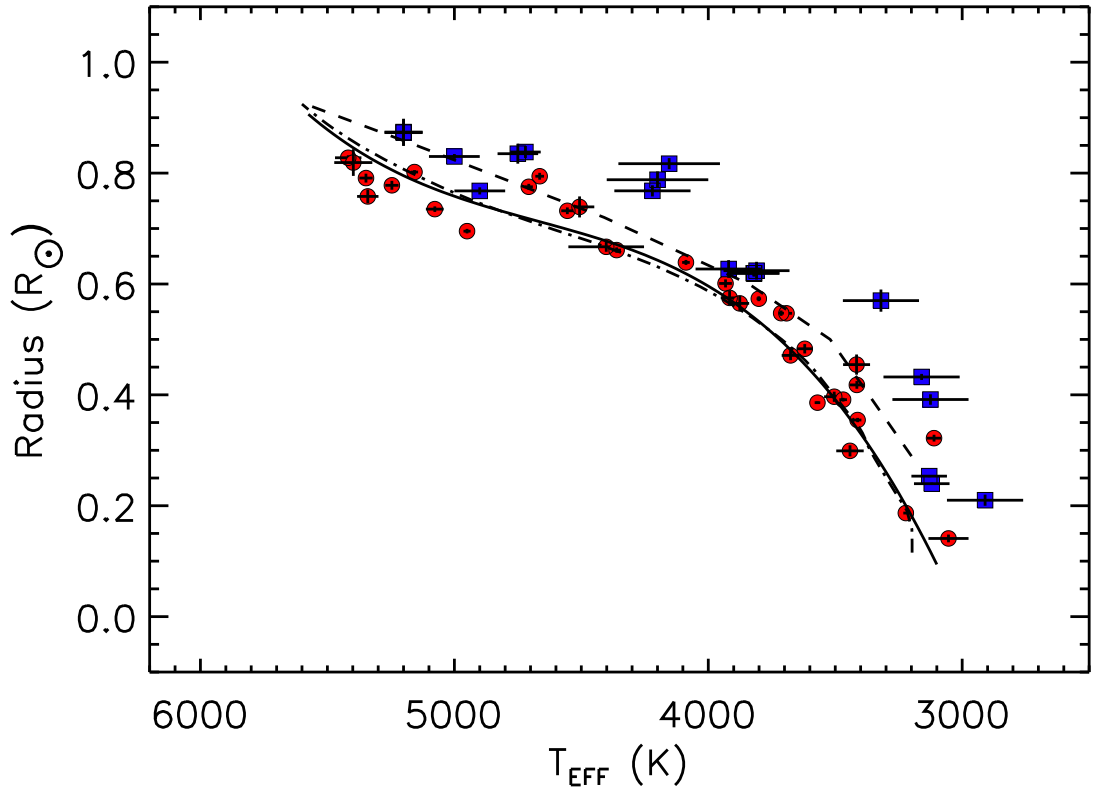, width=1.0\linewidth,clip=} 
  \caption[] {Stellar radius versus temperature for single stars (red circles) and EBs (blue squares). The solid line is the polynomial fit to the empirically determined single star data presented in Equation~\ref{eq:temp_radius_relation_withsun}. The solar metallicity solution from the Dartmouth models is shown as the dash-dot line. The tabulated values for temperatures and radii for main sequence stars in Allen's AQ \citep{cox00} is plotted as a dashed line, predicting a decreased temperature of $\sim$200 to 300~K for a star of given radius. Note the correlations to binary star temperatures and radii show to be much more suited for the Allen's AQ \citet{cox00} temperature scale. See Section~\ref{sec:teff_validation} for details.} 
 \label{fig:R_vs_T_withEBs_withDSEP}
 \end{figure}

Comparing single and binary star properties on the temperature - radius plane, we find that for given stellar radius, binary star effective temperature estimates in the literature are systematically lower than the interferometrically determined temperatures of single stars by $\sim 200 - 300$~K (Figure~\ref{fig:R_vs_T_withEBs_withDSEP}). The major extent of this discrepancy is likely to be caused by differences in measuring techniques as well as insufficiently accurate temperature calibrations.  However, we emphasize that this analysis alone can not exclude the role stellar activity may have on stellar temperature and radius.  We strongly encourage the validation of EB temperatures in order determine to what extent higher activity rates may influence the physical properties of binary stars.

\begin{figure}										
  \centering
         \epsfig{file=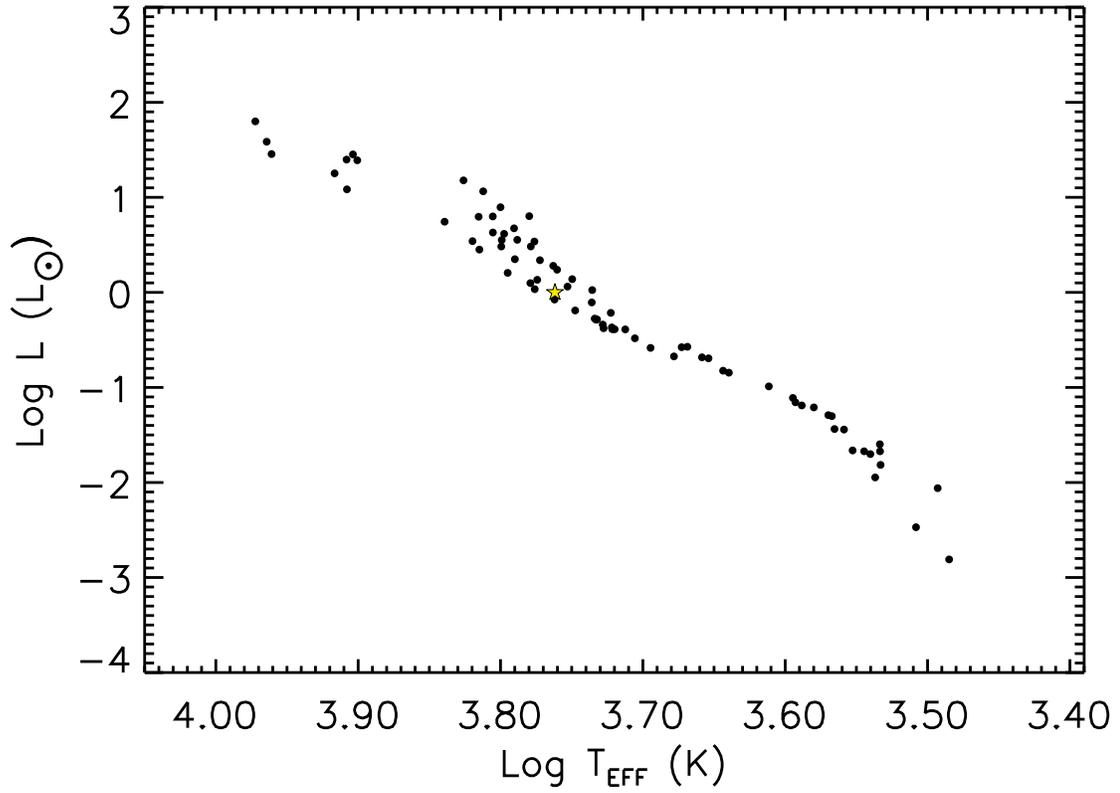, width=1.0\linewidth, clip=} 
  \caption[] {An empirically determined H-R diagram for the collection of stars presented in this paper and in Paper~I \citep{boy12}. The yellow five-pointed star is the Sun.}
   \label{fig:lumin_VS_temp_b}
   \end{figure}

One practical application of measuring these stellar fundamental properties for stars in the solar neighborhood is the construction of an empirically determined HR diagram with astrophysical, as opposed to observable, axes.  This is shown in Figure~\ref{fig:lumin_VS_temp_b}, which comprises the 33 stars with diameters presented in this paper as well as those from the literature (Table~\ref{tab:fund_params_combined}), and the 41 early-type counterparts (A, F, and G-type dwarfs) that were examined in an analogous way in Paper I of this series \citep{boy12}, for a total of 74 stars.  Such a well-populated HR diagram for late-type stars in the theoretical plane, but based only on empirical data, has not existed to-date due to the aforementioned paucity in the astronomical literature of precise, directly determined radii.

At the time of writing, the only two stars from our compiled sample that host exoplanets in orbits around them are GJ~436 and GJ~581 (see \citealt{von12} and \citealt{von11a}, and references therein). However, ongoing radial velocity and transit exoplanet surveys with increasing sensitivity are continuously producing new planet discoveries\footnote{The Exoplanet Encyclopedia (exoplanet.eu; \citealt{sch11}).}. We therefore calculate the boundaries of the system habitable zones (HZs) for the stars in Table~\ref{tab:fund_params_combined} based on our measured stellar astrophysical properties. The HZ boundaries are given in Table~\ref{tab:HZs}, where $R_{\rm inner}$ and $R_{\rm outer}$ correspond to the inner and outer radius of the HZ, respectively, following the formalism in \citet{jon10}. The HZs for the stars in Paper~I are also defined in this Table.

CHARA is currently the only interferometer with the sensitivity and resolution to enable this kind of work to this kind of precision. We are thus actively continuing our survey for stellar angular diameters to populate the rarefied parameter space of fully characterized K- and M- dwarfs in our solar neighborhood. The addition of new data will improve the empirical relations we develop in this paper, delivering the foundation needed to facilitate fundamental understanding of stellar astrophysics. 

Finally, the ongoing surveys to detect extrasolar planets are making constant progress in sensitivity and sophistication, and many of them are concentrating on late-type stars in the solar neighborhood because 1) the systems are bright and can thus be studied in great detail, and 2) the low stellar mass puts the habitable zone closer to the parent star, making the detection of exoplanets in the habitable zones at correspondingly lower periods as compared to, e.g., the Earth - Sun system, much more straightforward. Knowledge of the respective stellar radius, effective temperature, and luminosity characterizes the radiation environment of the orbiting exoplanets and thus gives insight into the first of the crucial conditions that must be fullfilled if the planets were to be able to host liquid water on their surfaces.



\acknowledgments

We would like to thank Greg Feiden for his helpful conversations and advice in the Dartmouth stellar models. We would like to thank the referee, whose comments undoubtedly improved the manuscript in a major way.  TSB acknowledges support provided by NASA through Hubble Fellowship grant \#HST-HF-51252.01 awarded by the Space Telescope Science Institute, which is operated by the Association of Universities for Research in Astronomy, Inc., for NASA, under contract NAS 5-26555.  STR acknowledges partial support from NASA grant NNH09AK731.  The CHARA Array is funded by the National Science Foundation through NSF grant AST-0908253 and by Georgia State University through the College of Arts and Sciences. This research has made use of the SIMBAD literature database, operated at CDS, Strasbourg, France, and of NASA's Astrophysics Data System. This research has made use of the VizieR catalogue access tool, CDS, Strasbourg, France.

\clearpage
\bibliographystyle{apj}            
\bibliography{apj-jour,paper}      

\begin{thebibliography}{98}
\expandafter\ifx\csname natexlab\endcsname\relax\def\natexlab#1{#1}\fi

\bibitem[{{Allard} {et~al.}(2012){Allard}, {Homeier}, \& {Freytag}}]{all12}
{Allard}, F., {Homeier}, D., \& {Freytag}, B. 2012, Royal Society of London
  Philosophical Transactions Series A, 370, 2765

\bibitem[{{Alonso} {et~al.}(1996){Alonso}, {Arribas}, \&
  {Martinez-Roger}}]{alo96}
{Alonso}, A., {Arribas}, S., \& {Martinez-Roger}, C. 1996, \aaps, 117, 227

\bibitem[{{Andersen}(1991)}]{and91}
{Andersen}, J. 1991, \aapr, 3, 91

\bibitem[{{Anderson} \& {Francis}(2011)}]{and11}
{Anderson}, E., \& {Francis}, C. 2011, VizieR Online Data Catalog, 5137, 0

\bibitem[{{Arribas} \& {Martinez Roger}(1989)}]{1989AA...215..305A}
{Arribas}, S., \& {Martinez Roger}, C. 1989, \aap, 215, 305

\bibitem[{{Baraffe} {et~al.}(1998){Baraffe}, {Chabrier}, {Allard}, \&
  {Hauschildt}}]{bar98}
{Baraffe}, I., {Chabrier}, G., {Allard}, F., \& {Hauschildt}, P.~H. 1998, \aap,
  337, 403

\bibitem[{{Berger} {et~al.}(2006){Berger}, {Gies}, {McAlister}, {Brummelaar},
  {Henry}, {Sturmann}, {Sturmann}, {Turner}, {Ridgway}, {Aufdenberg}, \&
  {M{\'e}rand}}]{ber06}
{Berger}, D.~H. {et~al.} 2006, \apj, 644, 475

\bibitem[{{Bessel}(1990)}]{1990AAS...83..357B}
{Bessel}, M.~S. 1990, \aaps, 83, 357

\bibitem[{{Bessell}(1990)}]{bes90}
{Bessell}, M.~S. 1990, \pasp, 102, 1181

\bibitem[{{Blackwell} \& {Petford}(1991)}]{bla91}
{Blackwell}, D.~E., \& {Petford}, A.~D. 1991, \aap, 250, 459

\bibitem[{{Bonfils} {et~al.}(2005){Bonfils}, {Delfosse}, {Udry}, {Santos},
  {Forveille}, \& {S{\'e}gransan}}]{bon05}
{Bonfils}, X., {Delfosse}, X., {Udry}, S., {Santos}, N.~C., {Forveille}, T., \&
  {S{\'e}gransan}, D. 2005, \aap, 442, 635

\bibitem[{{Boyajian}(2009)}]{boy09a}
{Boyajian}, T.~S. 2009, PhD thesis, Georgia State University

\bibitem[{{Boyajian} {et~al.}(2008){Boyajian}, {McAlister}, {Baines}, {Gies},
  {Henry}, {Jao}, {O'Brien}, {Raghavan}, {Touhami}, {ten Brummelaar},
  {Farrington}, {Goldfinger}, {Sturmann}, {Sturmann}, {Turner}, \&
  {Ridgway}}]{boy08}
{Boyajian}, T.~S. {et~al.} 2008, \apj, 683, 424

\bibitem[{{Boyajian} {et~al.}(2012){Boyajian}, {McAlister}, {van Belle},
  {Gies}, {ten Brummelaar}, {von Braun}, {Farrington}, {Goldfinger}, {O'Brien},
  {Parks}, {Richardson}, {Ridgway}, {Schaefer}, {Sturmann}, {Sturmann},
  {Touhami}, {Turner}, \& {White}}]{boy12}
------. 2012, \apj, 746, 101

\bibitem[{{Boyajian} {et~al.}(2010){Boyajian}, {von Braun}, {van Belle}, {ten
  Brummelaar}, {Ciardi}, {Henry}, {Lopez-Morales}, {McAlister}, {Ridgway},
  {Farrington}, {Goldfinger}, {Sturmann}, {Sturmann}, \& {Turner}}]{boy10}
------. 2010, ArXiv e-prints

\bibitem[{{Casagrande} {et~al.}(2008){Casagrande}, {Flynn}, \&
  {Bessell}}]{cas08}
{Casagrande}, L., {Flynn}, C., \& {Bessell}, M. 2008, \mnras, 389, 585

\bibitem[{{Chabrier} \& {Baraffe}(1997)}]{cha97}
{Chabrier}, G., \& {Baraffe}, I. 1997, \aap, 327, 1039

\bibitem[{{Chabrier} {et~al.}(2007){Chabrier}, {Gallardo}, \&
  {Baraffe}}]{cha07}
{Chabrier}, G., {Gallardo}, J., \& {Baraffe}, I. 2007, \aap, 472, L17

\bibitem[{{Christou} \& {Drummond}(2006)}]{2006AJ....131.3100C}
{Christou}, J.~C., \& {Drummond}, J.~D. 2006, \aj, 131, 3100

\bibitem[{{Claret}(2000)}]{cla00}
{Claret}, A. 2000, \aap, 363, 1081

\bibitem[{{Cowley} {et~al.}(1967){Cowley}, {Hiltner}, \&
  {Witt}}]{1967AJ.....72.1334C}
{Cowley}, A.~P., {Hiltner}, W.~A., \& {Witt}, A.~N. 1967, \aj, 72, 1334

\bibitem[{{Cox}(2000)}]{cox00}
{Cox}, A.~N. 2000, {Allen's astrophysical quantities} (Allen's astrophysical
  quantities, 4th ed.~Publisher: New York: AIP Press; Springer, 2000.~ed.
  Arthur N.~Cox.~ ISBN: 0387987460)

\bibitem[{{Cutri} {et~al.}(2003{\natexlab{a}}){Cutri}, {Skrutskie}, {van Dyk},
  {Beichman}, {Carpenter}, {Chester}, {Cambresy}, {Evans}, {Fowler}, {Gizis},
  {Howard}, {Huchra}, {Jarrett}, {Kopan}, {Kirkpatrick}, {Light}, {Marsh},
  {McCallon}, {Schneider}, {Stiening}, {Sykes}, {Weinberg}, {Wheaton},
  {Wheelock}, \& {Zacarias}}]{cut03}
{Cutri}, R.~M. {et~al.} 2003{\natexlab{a}}, {The 2MASS All Sky Catalog of Point
  Sources} (Pasadena: IPAC)

\bibitem[{{Cutri} {et~al.}(2003{\natexlab{b}}){Cutri}, {Skrutskie}, {van Dyk},
  {Beichman}, {Carpenter}, {Chester}, {Cambresy}, {Evans}, {Fowler}, {Gizis},
  {Howard}, {Huchra}, {Jarrett}, {Kopan}, {Kirkpatrick}, {Light}, {Marsh},
  {McCallon}, {Schneider}, {Stiening}, {Sykes}, {Weinberg}, {Wheaton},
  {Wheelock}, \& {Zacarias}}]{2003yCat.2246....0C}
------. 2003{\natexlab{b}}, VizieR Online Data Catalog, 2246, 0

\bibitem[{{Delfosse} {et~al.}(2000){Delfosse}, {Forveille}, {S{\'e}gransan},
  {Beuzit}, {Udry}, {Perrier}, \& {Mayor}}]{del00}
{Delfosse}, X., {Forveille}, T., {S{\'e}gransan}, D., {Beuzit}, J.-L., {Udry},
  S., {Perrier}, C., \& {Mayor}, M. 2000, \aap, 364, 217

\bibitem[{{Demarque} {et~al.}(2004){Demarque}, {Woo}, {Kim}, \& {Yi}}]{dem04}
{Demarque}, P., {Woo}, J.-H., {Kim}, Y.-C., \& {Yi}, S.~K. 2004, \apjs, 155,
  667

\bibitem[{{Demory} {et~al.}(2009){Demory}, {S{\'e}gransan}, {Forveille},
  {Queloz}, {Beuzit}, {Delfosse}, {di Folco}, {Kervella}, {Le Bouquin},
  {Perrier}, {Benisty}, {Duvert}, {Hofmann}, {Lopez}, \& {Petrov}}]{dem09}
{Demory}, B. {et~al.} 2009, \aap, 505, 205

\bibitem[{{di Folco} {et~al.}(2007){di Folco}, {Absil}, {Augereau},
  {M{\'e}rand}, {Coud{\'e} du Foresto}, {Th{\'e}venin}, {Defr{\`e}re},
  {Kervella}, {ten Brummelaar}, {McAlister}, {Ridgway}, {Sturmann}, {Sturmann},
  \& {Turner}}]{dif07}
{di Folco}, E. {et~al.} 2007, \aap, 475, 243

\bibitem[{{Di Folco} {et~al.}(2004){Di Folco}, {Th{\'e}venin}, {Kervella},
  {Domiciano de Souza}, {Coud{\'e} du Foresto}, {S{\'e}gransan}, \&
  {Morel}}]{dif04}
{Di Folco}, E., {Th{\'e}venin}, F., {Kervella}, P., {Domiciano de Souza}, A.,
  {Coud{\'e} du Foresto}, V., {S{\'e}gransan}, D., \& {Morel}, P. 2004, \aap,
  426, 601

\bibitem[{{Dotter} {et~al.}(2008){Dotter}, {Chaboyer}, {Jevremovi{\'c}},
  {Kostov}, {Baron}, \& {Ferguson}}]{dot08}
{Dotter}, A., {Chaboyer}, B., {Jevremovi{\'c}}, D., {Kostov}, V., {Baron}, E.,
  \& {Ferguson}, J.~W. 2008, \apjs, 178, 89

\bibitem[{{Ducati}(2002)}]{2002yCat.2237....0D}
{Ducati}, J.~R. 2002, VizieR Online Data Catalog, 2237, 0

\bibitem[{{Edvardsson} {et~al.}(1993){Edvardsson}, {Andersen}, {Gustafsson},
  {Lambert}, {Nissen}, \& {Tomkin}}]{edv93}
{Edvardsson}, B., {Andersen}, J., {Gustafsson}, B., {Lambert}, D.~L., {Nissen},
  P.~E., \& {Tomkin}, J. 1993, \aap, 275, 101

\bibitem[{{Eggenberger} {et~al.}(2008{\natexlab{a}}){Eggenberger}, {Miglio},
  {Carrier}, {Fernandes}, \& {Santos}}]{egg08}
{Eggenberger}, P., {Miglio}, A., {Carrier}, F., {Fernandes}, J., \& {Santos},
  N.~C. 2008{\natexlab{a}}, \aap, 482, 631

\bibitem[{{Eggenberger} {et~al.}(2008{\natexlab{b}}){Eggenberger}, {Miglio},
  {Carrier}, {Fernandes}, \& {Santos}}]{2008AA...482..631E}
------. 2008{\natexlab{b}}, \aap, 482, 631

\bibitem[{{Elias} {et~al.}(1982){Elias}, {Frogel}, {Matthews}, \&
  {Neugebauer}}]{1982AJ.....87.1029E}
{Elias}, J.~H., {Frogel}, J.~A., {Matthews}, K., \& {Neugebauer}, G. 1982, \aj,
  87, 1029

\bibitem[{{Erro}(1971)}]{1971BITon...6..143E}
{Erro}, B.~I. 1971, Boletin del Instituto de Tonantzintla, 6, 143

\bibitem[{{Frogel} {et~al.}(1972){Frogel}, {Kleinmann}, {Kunkel}, {Ney}, \&
  {Strecker}}]{1972PASP...84..581F}
{Frogel}, J.~A., {Kleinmann}, D.~E., {Kunkel}, W., {Ney}, E.~P., \& {Strecker},
  D.~W. 1972, \pasp, 84, 581

\bibitem[{{Girardi} {et~al.}(2000){Girardi}, {Bressan}, {Bertelli}, \&
  {Chiosi}}]{gir00}
{Girardi}, L., {Bressan}, A., {Bertelli}, G., \& {Chiosi}, C. 2000, \aaps, 141,
  371

\bibitem[{{Glass}(1975)}]{1975MNRAS.171P..19G}
{Glass}, I.~S. 1975, \mnras, 171, 19P

\bibitem[{{Gray}(1992)}]{gra92}
{Gray}, D.~F. 1992, {The Observation and Analysis of Stellar Photospheres} (The
  Observation and Analysis of Stellar Photospheres, by David F.~Gray,
  pp.~470.~ISBN 0521408687.~Cambridge, UK: Cambridge University Press, June
  1992.)

\bibitem[{{Hanbury Brown} {et~al.}(1974){Hanbury Brown}, {Davis}, {Lake}, \&
  {Thompson}}]{han74}
{Hanbury Brown}, R.~H., {Davis}, J., {Lake}, R.~J.~W., \& {Thompson}, R.~J.
  1974, \mnras, 167, 475

\bibitem[{{Hawley} {et~al.}(1996){Hawley}, {Gizis}, \& {Reid}}]{haw96}
{Hawley}, S.~L., {Gizis}, J.~E., \& {Reid}, I.~N. 1996, \aj, 112, 2799

\bibitem[{{Henry} \& {McCarthy}(1993)}]{hen93}
{Henry}, T.~J., \& {McCarthy}, Jr., D.~W. 1993, \aj, 106, 773

\bibitem[{{Irwin} {et~al.}(2011){Irwin}, {Quinn}, {Berta}, {Latham}, {Torres},
  {Burke}, {Charbonneau}, {Dittmann}, {Esquerdo}, {Stefanik}, {Oksanen},
  {Buchhave}, {Nutzman}, {Berlind}, {Calkins}, \& {Falco}}]{irw11}
{Irwin}, J.~M. {et~al.} 2011, \apj, 742, 123

\bibitem[{{Johnson}(1964)}]{1964BOTT....3..305J}
{Johnson}, H.~L. 1964, Boletin de los Observatorios Tonantzintla y Tacubaya, 3,
  305

\bibitem[{{Johnson}(1965)}]{1965ApJ...141..170J}
------. 1965, \apj, 141, 170

\bibitem[{{Johnson} {et~al.}(1968){Johnson}, {MacArthur}, \&
  {Mitchell}}]{1968ApJ...152..465J}
{Johnson}, H.~L., {MacArthur}, J.~W., \& {Mitchell}, R.~I. 1968, \apj, 152, 465

\bibitem[{{Johnson} {et~al.}(1966){Johnson}, {Mitchell}, {Iriarte}, \&
  {Wisniewski}}]{1966CoLPL...4...99J}
{Johnson}, H.~L., {Mitchell}, R.~I., {Iriarte}, B., \& {Wisniewski}, W.~Z.
  1966, Communications of the Lunar and Planetary Laboratory, 4, 99

\bibitem[{{Johnson} \& {Morgan}(1953)}]{1953ApJ...117..313J}
{Johnson}, H.~L., \& {Morgan}, W.~W. 1953, \apj, 117, 313

\bibitem[{{Jones} \& {Sleep}(2010)}]{jon10}
{Jones}, B.~W., \& {Sleep}, P.~N. 2010, \mnras, 407, 1259

\bibitem[{{Kervella} \& {Fouqu{\'e}}(2008)}]{ker08}
{Kervella}, P., \& {Fouqu{\'e}}, P. 2008, \aap, 491, 855

\bibitem[{{Kervella} {et~al.}(2008){Kervella}, {M{\'e}rand}, {Pichon},
  {Th{\'e}venin}, {Heiter}, {Bigot}, {ten Brummelaar}, {McAlister}, {Ridgway},
  {Turner}, {Sturmann}, {Sturmann}, {Goldfinger}, \& {Farrington}}]{ker08a}
{Kervella}, P. {et~al.} 2008, \aap, 488, 667

\bibitem[{{Kraus} {et~al.}(2011){Kraus}, {Tucker}, {Thompson}, {Craine}, \&
  {Hillenbrand}}]{kra11}
{Kraus}, A.~L., {Tucker}, R.~A., {Thompson}, M.~I., {Craine}, E.~R., \&
  {Hillenbrand}, L.~A. 2011, \apj, 728, 48

\bibitem[{{Lane} {et~al.}(2001){Lane}, {Boden}, \& {Kulkarni}}]{lan01}
{Lane}, B.~F., {Boden}, A.~F., \& {Kulkarni}, S.~R. 2001, \apjl, 551, L81

\bibitem[{{Leggett}(1992)}]{leg92}
{Leggett}, S.~K. 1992, \apjs, 82, 351

\bibitem[{{Lejeune} {et~al.}(1998){Lejeune}, {Cuisinier}, \& {Buser}}]{lej98}
{Lejeune}, T., {Cuisinier}, F., \& {Buser}, R. 1998, \aaps, 130, 65

\bibitem[{{L{\'o}pez-Morales}(2007)}]{lop07}
{L{\'o}pez-Morales}, M. 2007, \apj, 660, 732

\bibitem[{{L{\'o}pez-Morales} \& {Ribas}(2005)}]{lop05}
{L{\'o}pez-Morales}, M., \& {Ribas}, I. 2005, \apj, 631, 1120

\bibitem[{{Luck} \& {Heiter}(2005)}]{luc05}
{Luck}, R.~E., \& {Heiter}, U. 2005, \aj, 129, 1063

\bibitem[{{Markwardt}(2009)}]{mar09}
{Markwardt}, C.~B. 2009, in Astronomical Society of the Pacific Conference
  Series, Vol. 411, Astronomical Data Analysis Software and Systems XVIII, ed.
  {D.~A.~Bohlender, D.~Durand, \& P.~Dowler}, 251--+

\bibitem[{{Mermilliod}(1997)}]{1997yCat.2168....0M}
{Mermilliod}, J.~C. 1997, VizieR Online Data Catalog, 2168, 0

\bibitem[{{Michelson} \& {Pease}(1921)}]{mic21}
{Michelson}, A.~A., \& {Pease}, F.~G. 1921, \apj, 53, 249

\bibitem[{{Morales} {et~al.}(2010){Morales}, {Gallardo}, {Ribas}, {Jordi},
  {Baraffe}, \& {Chabrier}}]{mor10}
{Morales}, J.~C., {Gallardo}, J., {Ribas}, I., {Jordi}, C., {Baraffe}, I., \&
  {Chabrier}, G. 2010, \apj, 718, 502

\bibitem[{{Morales} {et~al.}(2008){Morales}, {Ribas}, \& {Jordi}}]{mor08}
{Morales}, J.~C., {Ribas}, I., \& {Jordi}, C. 2008, \aap, 478, 507

\bibitem[{{Mould} \& {Hyland}(1976)}]{1976ApJ...208..399M}
{Mould}, J.~R., \& {Hyland}, A.~R. 1976, \apj, 208, 399

\bibitem[{{Mullan} \& {MacDonald}(2001)}]{mul01}
{Mullan}, D.~J., \& {MacDonald}, J. 2001, \apj, 559, 353

\bibitem[{{Neves} {et~al.}(2011){Neves}, {Bonfils}, {Santos}, {Delfosse},
  {Forveille}, {Allard}, {Nat{\'a}rio}, {Fernandes}, \& {Udry}}]{nev11}
{Neves}, V. {et~al.} 2011, ArXiv e-prints

\bibitem[{{Ochsenbein} {et~al.}(2000){Ochsenbein}, {Bauer}, \&
  {Marcout}}]{och00}
{Ochsenbein}, F., {Bauer}, P., \& {Marcout}, J. 2000, \aaps, 143, 23

\bibitem[{{Persson} {et~al.}(1977){Persson}, {Aaronson}, \&
  {Frogel}}]{1977AJ.....82..729P}
{Persson}, S.~E., {Aaronson}, M., \& {Frogel}, J.~A. 1977, \aj, 82, 729

\bibitem[{{Pickles}(1998)}]{pic98}
{Pickles}, A.~J. 1998, \pasp, 110, 863

\bibitem[{{Popper}(1997)}]{pop97}
{Popper}, D.~M. 1997, \aj, 114, 1195

\bibitem[{{Ram{\'{\i}}rez} \& {Mel{\'e}ndez}(2005)}]{ram05}
{Ram{\'{\i}}rez}, I., \& {Mel{\'e}ndez}, J. 2005, \apj, 626, 465

\bibitem[{{Ribas} {et~al.}(2007){Ribas}, {Morales}, {Jordi}, {Baraffe},
  {Chabrier}, \& {Gallardo}}]{rib07}
{Ribas}, I., {Morales}, J., {Jordi}, C., {Baraffe}, I., {Chabrier}, G., \&
  {Gallardo}, J. 2007, ArXiv e-prints, 711

\bibitem[{{Ribas} {et~al.}(2008){Ribas}, {Morales}, {Jordi}, {Baraffe},
  {Chabrier}, \& {Gallardo}}]{rib08}
{Ribas}, I., {Morales}, J.~C., {Jordi}, C., {Baraffe}, I., {Chabrier}, G., \&
  {Gallardo}, J. 2008, \memsai, 79, 562

\bibitem[{{Rojas-Ayala} {et~al.}(2012){Rojas-Ayala}, {Covey}, {Muirhead}, \&
  {Lloyd}}]{roj12}
{Rojas-Ayala}, B., {Covey}, K.~R., {Muirhead}, P.~S., \& {Lloyd}, J.~P. 2012,
  \apj, 748, 93

\bibitem[{{Schneider} {et~al.}(2011){Schneider}, {Dedieu}, {Le Sidaner},
  {Savalle}, \& {Zolotukhin}}]{sch11}
{Schneider}, J., {Dedieu}, C., {Le Sidaner}, P., {Savalle}, R., \&
  {Zolotukhin}, I. 2011, \aap, 532, A79

\bibitem[{{S{\'e}gransan} {et~al.}(2003){S{\'e}gransan}, {Kervella},
  {Forveille}, \& {Queloz}}]{seg03}
{S{\'e}gransan}, D., {Kervella}, P., {Forveille}, T., \& {Queloz}, D. 2003,
  \aap, 397, L5

\bibitem[{{Siess} {et~al.}(1997){Siess}, {Forestini}, \& {Dougados}}]{sie97}
{Siess}, L., {Forestini}, M., \& {Dougados}, C. 1997, \aap, 324, 556

\bibitem[{{Sturmann} {et~al.}(2010){Sturmann}, {Ten Brummelaar}, {Sturmann}, \&
  {McAlister}}]{stu10}
{Sturmann}, J., {Ten Brummelaar}, T., {Sturmann}, L., \& {McAlister}, H.~A.
  2010, in Society of Photo-Optical Instrumentation Engineers (SPIE) Conference
  Series, Vol. 7734, Society of Photo-Optical Instrumentation Engineers (SPIE)
  Conference Series

\bibitem[{{ten Brummelaar} {et~al.}(2008){ten Brummelaar}, {McAlister},
  {Ridgway}, {Gies}, {Sturmann}, {Sturmann}, {Turner}, {M{\'e}rand},
  {Thompson}, {Farrington}, \& {Goldfinger}}]{ten08}
{ten Brummelaar}, T.~A. {et~al.} 2008, in Society of Photo-Optical
  Instrumentation Engineers (SPIE) Conference Series, Vol. 7013, Society of
  Photo-Optical Instrumentation Engineers (SPIE) Conference Series

\bibitem[{{ten Brummelaar} {et~al.}(2005){ten Brummelaar}, {McAlister},
  {Ridgway}, {Bagnuolo}, {Turner}, {Sturmann}, {Sturmann}, {Berger}, {Ogden},
  {Cadman}, {Hartkopf}, {Hopper}, \& {Shure}}]{ten05}
{ten Brummelaar}, T.~A. {et~al.} 2005, \apj, 628, 453

\bibitem[{{Torres} {et~al.}(2010){Torres}, {Andersen}, \&
  {Gim{\'e}nez}}]{tor10}
{Torres}, G., {Andersen}, J., \& {Gim{\'e}nez}, A. 2010, \aapr, 18, 67

\bibitem[{{Torres} \& {Ribas}(2002)}]{tor02a}
{Torres}, G., \& {Ribas}, I. 2002, \apj, 567, 1140

\bibitem[{{Valenti} \& {Fischer}(2005)}]{val05}
{Valenti}, J.~A., \& {Fischer}, D.~A. 2005, \apjs, 159, 141

\bibitem[{{van Belle} \& {van Belle}(2005)}]{van05}
{van Belle}, G.~T., \& {van Belle}, G. 2005, \pasp, 117, 1263

\bibitem[{{van Belle} {et~al.}(2008){van Belle}, {van Belle}, {Creech-Eakman},
  {Coyne}, {Boden}, {Akeson}, {Ciardi}, {Rykoski}, {Thompson}, {Lane}, \&
  {Collaboration}}]{van08}
{van Belle}, G.~T. {et~al.} 2008, \apjs, 176, 276

\bibitem[{{van Belle} \& {von Braun}(2009)}]{van09}
{van Belle}, G.~T., \& {von Braun}, K. 2009, \apj, 694, 1085

\bibitem[{{van Leeuwen}(2007)}]{van07}
{van Leeuwen}, F. 2007, {Hipparcos, the New Reduction of the Raw Data}
  (Hipparcos, the New Reduction of the Raw Data.~By Floor van Leeuwen,
  Institute of Astronomy, Cambridge University, Cambridge, UK Series:
  Astrophysics and Space Science Library, Vol.~ 350 20 Springer Dordrecht)

\bibitem[{{VandenBerg} {et~al.}(2010){VandenBerg}, {Casagrande}, \&
  {Stetson}}]{van10}
{VandenBerg}, D.~A., {Casagrande}, L., \& {Stetson}, P.~B. 2010, \aj, 140, 1020

\bibitem[{{VandenBerg} \& {Clem}(2003)}]{van03}
{VandenBerg}, D.~A., \& {Clem}, J.~L. 2003, \aj, 126, 778

\bibitem[{{Veeder}(1974)}]{1974AJ.....79.1056V}
{Veeder}, G.~J. 1974, \aj, 79, 1056

\bibitem[{{Voges} {et~al.}(1999){Voges}, {Aschenbach}, {Boller},
  {Br{\"a}uninger}, {Briel}, {Burkert}, {Dennerl}, {Englhauser}, {Gruber},
  {Haberl}, {Hartner}, {Hasinger}, {K{\"u}rster}, {Pfeffermann}, {Pietsch},
  {Predehl}, {Rosso}, {Schmitt}, {Tr{\"u}mper}, \& {Zimmermann}}]{vog99}
{Voges}, W. {et~al.} 1999, \aap, 349, 389

\bibitem[{{von Braun} {et~al.}(2012){von Braun}, {Boyajian}, {Kane}, {Hebb},
  {van Belle}, {Farrington}, {Ciardi}, {Knutson}, {ten Brummelaar},
  {L{\'o}pez-Morales}, {McAlister}, {Schaefer}, {Ridgway}, {Collier Cameron},
  {Goldfinger}, {Turner}, {Sturmann}, \& {Sturmann}}]{von12}
{von Braun}, K. {et~al.} 2012, \apj, 753, 171

\bibitem[{{von Braun} {et~al.}(2011){von Braun}, {Boyajian}, {Kane}, {van
  Belle}, {Ciardi}, {L{\'o}pez-Morales}, {McAlister}, {Henry}, {Jao}, {Riedel},
  {Subasavage}, {Schaefer}, {ten Brummelaar}, {Ridgway}, {Sturmann},
  {Sturmann}, {Mazingue}, {Turner}, {Farrington}, {Goldfinger}, \&
  {Boden}}]{von11a}
------. 2011, \apjl, 729, L26+

\bibitem[{{von Braun} {et~al.}(2008){von Braun}, {van Belle}, {Ciardi},
  {L{\'o}pez-Morales}, {Hoard}, \& {Wachter}}]{von08}
{von Braun}, K., {van Belle}, G.~T., {Ciardi}, D.~R., {L{\'o}pez-Morales}, M.,
  {Hoard}, D.~W., \& {Wachter}, S. 2008, \apj, 677, 545

\bibitem[{{Weis}(1993)}]{1993AJ....105.1962W}
{Weis}, E.~W. 1993, \aj, 105, 1962

\bibitem[{{Weiss} {et~al.}(2004){Weiss}, {Hillebrandt}, {Thomas}, \&
  {Ritter}}]{wei04}
{Weiss}, A., {Hillebrandt}, W., {Thomas}, H.-C., \& {Ritter}, H. 2004, {Cox and
  Giuli's Principles of Stellar Structure}

\bibitem[{{Wright} {et~al.}(2011){Wright}, {Drake}, {Mamajek}, \&
  {Henry}}]{wri11}
{Wright}, N.~J., {Drake}, J.~J., {Mamajek}, E.~E., \& {Henry}, G.~W. 2011,
  \apj, 743, 48

\end{thebibliography}

	\newpage
\begin{deluxetable}{rccccc}
\tabletypesize{\tiny}
\tablewidth{0pt}
\tablecaption{Observation Log\label{tab:observations}}
\tablehead{
\colhead{\textbf{ }} &
\colhead{\textbf{UT}} &
\colhead{\textbf{ }} &
\colhead{\textbf{ }} &
\colhead{\textbf{\# of}} &
\colhead{\textbf{Calibrator}}  \\
\colhead{\textbf{Object}} &
\colhead{\textbf{Date}} &
\colhead{\textbf{Baseline}} &
\colhead{\textbf{Filter}} &
\colhead{\textbf{Brackets}} &
\colhead{\textbf{HD}}	
}

\startdata

GJ 15A	&	2008/09/16	&	E1/W1	&	$K^{\prime}$	&	6	&	6920	\\
		&	2008/09/17	&	S1/E1	&	$K^{\prime}$	&	12	&	905, 3765	\\
		&	2008/10/23	&	E1/W1	&	$K^{\prime}$	&	3	&	905	\\
		&	2009/12/01	&	E1/W2	&	$K^{\prime}$	&	5	&	905	\\
		&	2009/12/02	&	S2/E2	&	$K^{\prime}$	&	5	&	3765, 6920	\\
		&	2010/10/10	&	S1/E2	&	$H$				&	4	&	905	\\
		&	2011/08/21	&	S1/W1	&	$H$				&	5	&	905	\\	\\ 

GJ 33	&	2009/08/20	&	S1/E1	&	$K^{\prime}$	&	18	&	2344, 2454, 6288	\\
		&	2010/09/19	&	S1/E1	&	$H$				&	5	&	4928	\\	\\	
				
GJ 105	&	2009/08/21	&	S1/E1	&	$K^{\prime}$	&	4	&	16647, 16970	\\
		&	2010/09/16	&	E1/W1	&	$H$				&	6	&	16970	\\	
		&	2010/09/18	&	E1/W1	&	$H$				&	6	&	16970	\\	\\

GJ 166A	&	2008/10/23	&	E1/W1	&	$K^{\prime}$	&	3	&	22879	\\
		&	2008/10/24	&	E1/W1	&	$K^{\prime}$	&	6	&	22879, 29391	\\
		&	2009/12/01	&	E1/W2	&	$K^{\prime}$	&	5	&	22879, 29391	\\	\\

GJ 205	&	2008/10/23	&	E1/W1	&	$K^{\prime}$	&	2	&	38858	\\
		&	2009/12/01	&	E1/W2	&	$K^{\prime}$	&	8	&	33608, 38858	\\
		&	2010/09/19	&	S1/E1	&	$H$				&	12	&	33256, 38858	\\	\\

GJ 338A	&	2008/10/22	&	S1/E1	&	$K^{\prime}$	&	7	&	69548, 80290	\\	
		&	2008/10/24	&	E1/W1	&	$K^{\prime}$	&	7	&	69548, 80290	\\	
		&	2009/12/01	&	E1/W2	&	$K^{\prime}$	&	6	&	69548, 80290	\\	
		&	2009/12/02	&	S2/E2	&	$K^{\prime}$	&	2	&	80290	\\
		&	2009/12/03	&	S2/E2	&	$K^{\prime}$	&	2	&	69548	\\
		
GJ 338B	&	2008/10/22	&	S1/E1	&	$K^{\prime}$	&	5	&	69548, 80290	\\	
		&	2008/10/24	&	E1/W1	&	$K^{\prime}$	&	7	&	69548, 80290	\\	
		&	2009/12/01	&	E1/W2	&	$K^{\prime}$	&	6	&	69548, 80290	\\	
		&	2009/12/02	&	S2/E2	&	$K^{\prime}$	&	1	&	80290 	\\
		&	2009/12/03	&	S2/E2	&	$K^{\prime}$	&	2	&	69548	\\ \\
							
GJ 380	&	2009/11/22	&	S1/E1	&	$K^{\prime}$	&	15	&	85795, 90508	\\		
		&	2010/11/10	&	E1/W1	&	$H$				&	3	&	89021, 90839	\\ \\	
			
GJ 411	&	2009/12/02	&	S2/E2	&	$K^{\prime}$	&	6	&	 90277, 99984 	\\	
		&	2009/12/03	&	S2/E2	&	$K^{\prime}$	&	2	&	 90277, 99984 	\\	
		&	2009/12/04	&	S2/E2	&	$K^{\prime}$	&	3	&	  90840			\\	\\						
							
GJ 412A	&	2009/05/28	&	E1/W1	&	$K^{\prime}$	&	5	&	99984	\\
		&	2009/12/02	&	S2/E2	&	$K^{\prime}$	&	9	&	99984, 103799 	\\	
		&	2009/12/03	&	S2/E2	&	$K^{\prime}$	&	13	&	90277, 99984, 103799	\\	
		&	2009/12/04	&	S2/E2	&	$K^{\prime}$	&	11	&	90840, 99984, 103799			\\	\\							

GJ 526	&	2006/04/19	&	S1/E1	&	$K^{\prime}$	&	11	&	119550, 120066	\\
		&	2009/05/27	&	E1/W1	&	$K^{\prime}$	&	2	&	121560	\\
		&	2010/06/30	&	S1/E1	&	$H$	&	4	&	119550	\\
		&	2010/07/01	&	S1/E1	&	$H$	&	3	&	119550	\\ \\
		
GJ 631	&	2009/05/27	&	E1/W1	&	$K^{\prime}$	&	6	&	147449	\\		
		&	2009/05/28	&	E1/W1	&	$K^{\prime}$	&	7	&	150177	\\
		&	2009/06/18	&	S1/E1	&	$K^{\prime}$	&	2	&	147449	\\
		&	2010/04/04	&	S1/E1	&	$H$				&	5	&	147449	\\	\\
		
GJ 687	&	2004/06/29	&	S1/E1	&	$K^{\prime}$	&	1	&	151514	\\
		&	2004/05/27	&	S1/E1	&	$K^{\prime}$	&	7	&	154633, 151514	\\
		&	2004/05/31	&	S1/E1	&	$K^{\prime}$	&	2	&	154633	\\
		&	2010/06/27	&	E1/W1	&	$H$				&	6	&	158633	\\
		&	2010/06/28	&	E1/W1	&	$H$				&	2	&	151514	\\
		&	2010/06/29	&	E1/W1	&	$H$				&	4	&	158633	\\	\\
		
GJ 699	&	2009/05/28	&	E1/W1	&	$K^{\prime}$	&	1	&	164353	\\
		&	2009/08	21	&	S1/E1	&	$K^{\prime}$	&	2	&	161868	\\
		&	2010/06/30	&	S1/E1	&	$H$				&	6	&	164353, 171834		\\
		&	2010/07/01	&	S1/E1	&	$H$				&	4	&	164353	\\	\\

GJ 702A	&	2009/05/26	&	E1/W1	&	$K^{\prime}$	&	3	&	164353	\\	
		&	2009/05/27	&	E1/W1	&	$K^{\prime}$	&	2	&	164353	\\
		&	2009/08/21	&	S1/E1	&	$K^{\prime}$	&	7	&	161868	\\
		&	2010/07/01	&	S1/E1	&	$K^{\prime}$	&	8	&	164353	\\	\\
		
GJ 702B	&	2009/05/27	&	E1/W1	&	$K^{\prime}$	&	1	&	164353	\\
		&	2009/08/21	&	S1/E1	&	$K^{\prime}$	&	1	&	161868	\\	\\

GJ 725A	&	2008/10/22	&	S1/E1	&	$K^{\prime}$	&	3	&	168151, 186760	\\
		&	2009/05/26	&	E1/W1	&	$K^{\prime}$	&	3	&	168151	\\
		&	2009/05/27	&	E1/W1	&	$K^{\prime}$	&	13	&	168151, 173920	\\ \\		

GJ 725B	&	2008/10/22	&	S1/E1	&	$K^{\prime}$	&	5	&	168151, 186760	\\
		&	2009/05/26	&	E1/W1	&	$K^{\prime}$	&	2	&	168151	\\
		&	2009/05/27	&	E1/W1	&	$K^{\prime}$	&	14	&	168151, 173920	\\ \\				

GJ 809	&	2009/11/20	&	S1/E1	&	$K^{\prime}$	&	6	&	197373	\\
		&	2009/11/21	&	S1/E1	&	$K^{\prime}$	&	4	&	197373	\\
		&	2009/11/22	&	S1/E1	&	$K^{\prime}$	&	4	&	197950	\\
		&	2010/09/16	&	E1/W1	&	$H$				&	10	&	197373, 197950	\\
		&	2010/09/20	&	S1/E1	&	$H$				&	6	&	197373,	197950	\\	\\

GJ 880	&	2009/08/21	&	S1/E1	&	$K^{\prime}$	&	10	&	217813, 218261	\\
		&	2009/12/02	&	S2/E2	&	$K^{\prime}$	&	7	&	217813, 218261	\\
		&	2010/07/26	&	S1/E1	&	$K^{\prime}$	&	10	&	217813, 218261	\\	
		&	2010/07/27	&	S1/E1	&	$H$				&	9	&	217813, 218235	\\
		&	2010/07/28	&	S1/E1	&	$H$				&	9	&	218235	\\
		&	2011/08/18	&	S1/E1	&	$H$				&	6	&	217813, 218261	\\	\\
		
GJ 892	&	2008/09/15	&	S2/E1	&	$K^{\prime}$	&	10	&	219623, 221354	\\	
		&	2008/09/16	&	E1/W1	&	$K^{\prime}$	&	9	&	219623, 221354	\\
		&	2008/09/17	&	S1/E1	&	$K^{\prime}$	&	8	&	219623, 221354	\\
		&	2009/11/22	&	S1/E1	&	$K^{\prime}$	&	7	&	219623, 221354	\\	\\

\enddata

\end{deluxetable}
\newpage
\begin{deluxetable}{rccccccc}
\tabletypesize{\tiny}
\tablewidth{0pt}
\tablecaption{Calibrators Observed\label{tab:calibs}}
\tablehead{
\colhead{\textbf{Calibrator}} &
\colhead{\textbf{RA}} &
\colhead{\textbf{DEC}} &
\colhead{\textbf{Spectral}} &
\colhead{\textbf{$V$}} &
\colhead{\textbf{$K$}} &
\colhead{\textbf{$\theta_{\rm SED} \pm \sigma$}}	& 
\colhead{\textbf{Target (s)}}	 \\
\colhead{\textbf{HD}} &
\colhead{\textbf{(hh mm ss)}}	&
\colhead{\textbf{(dd mm ss)}}	&
\colhead{\textbf{Type}} &
\colhead{\textbf{(mag)}}	&
\colhead{\textbf{(mag)}}	&
\colhead{\textbf{(mas)}}	&
\colhead{\textbf{HD}}	 
}
\startdata

905	&	00 13 31	&	41 02 07	&	F0~IV	&	5.72	&	4.84	&	$0.397\pm0.020$	&	GJ~15A	\\
2344	&	00 27 20	&	02 48 51	&	G4~III	&	7.19	&	4.93	&	$0.422\pm0.026$	&	GJ~33	\\
2454	&	00 28 20	&	10 11 23	&	F6~V	&	6.04	&	4.94	&	$0.391\pm0.014$	&	GJ~33	\\
3765	&	00 40 49	&	40 11 14	&	K2~V	&	7.36	&	5.16	&	$0.413\pm0.035$	&	GJ~15A	\\
6288	&	01 03 49	&	01 22 01	&	F1~V	&	6.08	&	5.48	&	$0.290\pm0.016$	&	GJ~33	\\
6920	&	01 10 19	&	42 04 53	&	F8~V	&	5.68	&	4.46	&	$0.508\pm0.015$	&	GJ~15A	\\
16647	&	02 40 16	&	06 06 43	&	F3~V	&	6.26	&	5.24	&	$0.339\pm0.012$	&	GJ~105	\\
16970	&	02 43 18	&	03 14 09	&	A2~V	&	3.47	&	3.08	&	$0.770\pm0.068$	&	GJ~105	\\
22879	&	03 40 22	&	$-$03 13 01\phs	&	F9~V	&	6.74	&	5.18	&	$0.342\pm0.021$	&	GJ~166A	\\
29391	&	04 37 36	&	$-$02 28 25\phs	&	F0~V	&	5.22	&	4.54	&	$0.432\pm0.025$	&	GJ~166A	\\
33608	&	05 11 19	&	$-$02 29 27\phs	&	F5~V	&	5.90	&	4.82	&	$0.415\pm0.014$	&	GJ~205	\\
38858	&	05 48 35	&	$-$04 05 41\phs	&	G4~V	&	5.97	&	4.41	&	$0.526\pm0.029$	&	GJ~205	\\
69548	&	08 20 26	&	57 44 36	&	F4~V	&	5.89	&	4.90	&	$0.401\pm0.016$	&	GJ~338A, GJ~338B	\\
80290	&	09 20 44	&	51 15 58	&	F3~V	&	6.15	&	4.97	&	$0.385\pm0.016$	&	GJ~338A, GJ~338B	\\
85795	&	09 55 43	&	49 49 11	&	A3~III	&	5.28	&	5.08	&	$0.329\pm0.031$	&	GJ~380	\\
89021	&	10 17 06	&	42 54 52	&	A2~IV	&	3.44	&	3.42	&	$0.701\pm0.064$	&	GJ~380	\\
90277	&	10 25 55	&	33 47 46	&	F0~V	&	4.73	&	3.97	&	$0.599\pm0.034$	&	GJ~411, GJ~412A	\\
90508	&	10 28 04	&	48 47 06	&	G1~V	&	6.44	&	4.87	&	$0.442\pm0.022$	&	GJ~380	\\
90839	&	10 30 38	&	55 58 50	&	F8~V	&	4.83	&	3.64	&	$0.731\pm0.025$	&	GJ~380	\\
90840	&	10 30 06	&	38 55 30	&	A4~V	&	5.79	&	5.48	&	$0.283\pm0.011$	&	GJ~411, GJ~412A	\\
99984	&	11 30 31	&	43 10 24	&	F4~V	&	5.95	&	4.59	&	$0.483\pm0.020$	&	GJ~411, GJ~412A	\\
103799	&	11 57 15	&	40 20 37	&	F6~V	&	6.61	&	5.34	&	$0.343\pm0.013$	&	GJ~412A	\\
119550	&	13 43 36	&	14 21 56	&	G2~V	&	6.94	&	5.33	&	$0.345\pm0.011$	&	GJ~526	\\
120066	&	13 46 57	&	06 21 01	&	G0~V	&	6.30	&	4.85	&	$0.428\pm0.013$	&	GJ~526	\\
121560	&	13 55 50	&	14 03 23	&	F6~V	&	6.10	&	4.84	&	$0.422\pm0.018$	&	GJ~526	\\
147449	&	16 22 04	&	01 01 44	&	F0~V	&	4.82	&	4.09	&	$0.564\pm0.022$	&	GJ~631	\\
150177	&	16 39 39	&	$-$09 33 16\phs	&	F0~V	&	6.34	&	4.98	&	$0.391\pm0.019$	&	GJ~631	\\
151541	&	16 42 39	&	68 06 08	&	K1~V	&	7.56	&	5.69	&	$0.307\pm0.023$	&	GJ~687	\\
154633	&	17 02 16	&	64 36 03	&	G5~V	&	6.10	&	3.89	&	$0.761\pm0.057$	&	GJ~687	\\
158633	&	17 25 00	&	67 18 24	&	K0~V	&	6.43	&	4.52	&	$0.542\pm0.043$	&	GJ~687	\\
161868	&	17 47 54	&	02 42 26	&	A0~V	&	3.75	&	3.62	&	$0.638\pm0.031$	&	GJ~699, GJ~702A, GJ~702B	\\
164353	&	18 00 39	&	02 55 54	&	B5~I	&	3.97	&	4.00	&	$0.417\pm0.030$	&	GJ~699, GJ~702A, GJ~702B	\\
168151	&	18 13 54	&	64 23 50	&	F5~V	&	5.03	&	3.94	&	$0.618\pm0.023$	&	GJ~725A, GJ~725B	\\
170073	&	18 23 55	&	58 48 03	&	A2~V	&	4.99	&	4.78	&	$0.376\pm0.034$	&	GJ~725A, GJ~725B	\\
171834	&	18 36 39	&	06 40 18	&	F3~V	&	5.44	&	4.46	&	$0.473\pm0.020$	&	GJ~699	\\
173920	&	18 44 55	&	54 53 50	&	G5~III	&	6.26	&	4.38	&	$0.602\pm0.011$	&	GJ~725A, GJ~725B	\\
178207	&	19 04 55	&	53 23 48	&	A0~V	&	5.38	&	5.41	&	$0.276\pm0.023$	&	GJ~725A, GJ~725B	\\
186760	&	19 43 14	&	58 00 60	&	G0~V	&	6.30	&	4.91	&	$0.431\pm0.018$	&	GJ~725A, GJ~725B	\\
197373	&	20 40 18	&	60 30 19	&	F6~V	&	6.03	&	4.94	&	$0.404\pm0.013$	&	GJ~809	\\
197950	&	20 43 11	&	66 39 27	&	A8~V	&	5.60	&	5.06	&	$0.349\pm0.012$	&	GJ~809	\\
217813	&	23 03 05	&	20 55 07	&	G1~V	&	6.66	&	5.15	&	$0.377\pm0.013$	&	GJ~880	\\
218235	&	23 06 18	&	18 31 04	&	F6~V	&	6.13	&	5.07	&	$0.379\pm0.010$	&	GJ~880	\\
218261	&	23 06 32	&	19 54 39	&	F7~V	&	6.30	&	5.14	&	$0.380\pm0.012$	&	GJ~880	\\
219623	&	23 16 42	&	53 12 49	&	F7~V	&	5.60	&	4.31	&	$0.518\pm0.027$	&	GJ~892	\\
221354	&	23 31 22	&	59 09 56	&	K2~V	&	6.74	&	4.80	&	$0.451\pm0.020$	&	GJ~892	\\

\enddata

\end{deluxetable}
\newpage
\begin{deluxetable}{lcccccc}
\tabletypesize{\tiny}
\tablewidth{0pt}
\tablecaption{Angular Diameters of K and M Dwarfs\label{tab:diameters}}
\tablehead{
\colhead{\textbf{Star}} &
\colhead{\textbf{\# of}} &
\colhead{\textbf{Reduced}} &
\colhead{\textbf{$\theta_{\rm UD} \pm \sigma$}} &
\colhead{\textbf{ }} &
\colhead{\textbf{$\theta_{\rm LD} \pm \sigma$}} &
\colhead{\textbf{$\theta_{\rm LD}$}} \\
\colhead{\textbf{Name}} &
\colhead{\textbf{Obs.}}	&
\colhead{\textbf{$\chi^{2}$}}	&
\colhead{\textbf{(mas)}} &
\colhead{\textbf{$\mu_{\lambda}$}}	&
\colhead{\textbf{(mas)}}	&
\colhead{\textbf{\% err}}
}
\startdata

GJ 15A	&	37	&	0.93	&	$0.970\pm0.005$	&	0.398	&	$1.005\pm0.005$	&	0.5	\\
GJ 33	&	23	&	0.86	&	$0.836\pm0.004$	&	0.425	&	$0.868\pm0.004$	&	0.4	\\
GJ 105	&	16	&	0.76	&	$0.989\pm0.007$	&	0.443	&	$1.030\pm0.007$	&	0.7	\\
GJ 166A	&	14	&	1.01	&	$1.444\pm0.006$	&	0.410	&	$1.504\pm0.006$	&	0.4	\\
GJ 205	&	22	&	1.40	&	$0.904\pm0.003$	&	0.453	&	$0.943\pm0.004$	&	0.4	\\
GJ 338A	&	23	&	0.74	&	$0.834\pm0.014$	&	0.491	&	$0.871\pm0.015$	&	1.7	\\
GJ 338B	&	21	&	0.76	&	$0.823\pm0.016$	&	0.453	&	$0.856\pm0.016$	&	1.9	\\
GJ 380	&	18	&	1.00	&	$1.169\pm0.008$	&	0.491	&	$1.225\pm0.009$	&	0.7	\\
GJ 411	&	11	&	1.75	&	$1.380\pm0.013$	&	0.391	&	$1.432\pm0.013$	&	0.9	\\
GJ 412A	&	35	&	1.25	&	$0.739\pm0.016$	&	0.398	&	$0.764\pm0.017$	&	2.2	\\
GJ 526	&	29	&	2.60	&	$0.807\pm0.013$	&	0.398	&	$0.835\pm0.014$	&	1.6	\\
GJ 631	&	20	&	0.54	&	$0.701\pm0.011$	&	0.396	&	$0.724\pm0.011$	&	1.6	\\
GJ 687	&	22	&	1.63	&	$0.830\pm0.013$	&	0.391	&	$0.859\pm0.014$	&	1.6	\\
GJ 699	&	13	&	0.66	&	$0.917\pm0.005$	&	0.408	&	$0.952\pm0.005$	&	0.6	\\
GJ 702A	&	20	&	1.51	&	$1.460\pm0.004$	&	0.396	&	$1.515\pm0.005$	&	0.3	\\
GJ 702B	&	2	&	0.35	&	$1.169\pm0.015$	&	0.476	&	$1.221\pm0.015$	&	1.3	\\
GJ 725A	&	19	&	0.58	&	$0.907\pm0.008$	&	0.391	&	$0.937\pm0.008$	&	0.9	\\
GJ 725B	&	21	&	1.42	&	$0.822\pm0.015$	&	0.408	&	$0.851\pm0.015$	&	1.8	\\
GJ 809	&	31	&	1.33	&	$0.698\pm0.008$	&	0.398	&	$0.722\pm0.008$	&	1.1	\\
GJ 880	&	51	&	0.94	&	$0.716\pm0.004$	&	0.453	&	$0.744\pm0.004$	&	0.5	\\
GJ 892	&	34	&	0.69	&	$1.063\pm0.007$	&	0.443	&	$1.106\pm0.007$	&	0.7	\\

\enddata

\end{deluxetable}
\newpage
\begin{deluxetable}{lcccccc}
\tabletypesize{\tiny}
\tablewidth{0pt}
\tablecaption{Comparison of Angular Diameters\label{tab:comp_stars}}
\tablehead{
\colhead{\textbf{Star}} &
\colhead{\textbf{$\theta_{\rm LD} \pm \sigma$}} &
\colhead{\textbf{ }} &
\colhead{\textbf{ }} &
\colhead{\textbf{Spectral}} &
\colhead{\textbf{$\theta_{\rm LD}$}} &
\colhead{\textbf{ }}  \\
\colhead{\textbf{Name}} &
\colhead{\textbf{(mas)}}	&
\colhead{\textbf{Reference}}	&
\colhead{\textbf{Instrument}} &
\colhead{\textbf{Type}}	&
\colhead{\textbf{\% err}} &
\colhead{\textbf{$\Delta\theta_{\rm LD}/\sigma_{\rm C}$\tablenotemark{a}}}
}
\startdata

GJ   15A	&	1.005	$\pm$	0.005	&	This work	&	CHARA	&	M1.5 V	&	0.5	&	0.0	\\
	&	0.988	$\pm$	0.016	&	\citet{ber06}	&	CHARA	&		&	1.6	&	1.0	\\
	&	1.027	$\pm$	0.059	&	\citet{van09}	&	PTI	&		&	5.7\tablenotemark{c}	&	$-0.4$\phs	\\
															
GJ  33	&	0.868	$\pm$	0.004	&	This work	&	CHARA	&	K2   V	&	0.5	&	0.0	\\
	&	0.933	$\pm$	0.064	&	\citet{van09}	&	PTI	&		&	6.9\tablenotemark{c}	&	$-1.0$\phs	\\
															
GJ 105	&	1.03	$\pm$	0.007	&	This work	&	CHARA	&	K3   V	&	0.7	&	0.0	\\
	&	0.936	$\pm$	0.070	&	\citet{lan01}	&	PTI	&		&	7.5\tablenotemark{c}	&	1.3	\\
															
GJ  205	&	0.943	$\pm$	0.004	&	This work	&	CHARA	&	M1.5 V	&	0.4	&	0.0	\\
	&	1.149	$\pm$	0.110	&	\citet{seg03}	&	VLTI	&		&	9.6\tablenotemark{c}	&	$-1.9$\phs	\\
															
GJ  166A	&	1.504	$\pm$	0.006	&	This work	&	CHARA	&	K1   Ve	&	0.4	&	0.0	\\
	&	1.437	$\pm$	0.039	&	\citet{dem09}	&	VLTI	&		&	2.7	&	1.7	\\
															
GJ  380	&	1.225	$\pm$	0.008	&	This work	&	CHARA	&	K7.0 V	&	0.7	&	0.0	\\
	&	1.155	$\pm$	0.040\tablenotemark{b}	&	\citet{lan01}	&	PTI	&		&	3.5	&	1.7	\\
	&	1.238	$\pm$	0.053	&	\citet{van09}	&	PTI	&		&	4.3	&	$-0.2$\phs	\\
															
GJ  411	&	1.432	$\pm$	0.013	&	This work	&	CHARA	&	M2.0 V	&	0.9	&	0.0	\\
	&	1.436	$\pm$	0.030	&	\citet{lan01}	&	PTI	&		&	2.1	&	$-0.1$\phs	\\
	&	1.439	$\pm$	0.048	&	\citet{van09}	&	PTI	&		&	3.3	&	$-0.1$\phs	\\
															
GJ  526	&	0.835	$\pm$	0.014	&	This work	&	CHARA	&	M1.5 V	&	1.7	&	0.0	\\
	&	0.845	$\pm$	0.057	&	\citet{ber06}	&	CHARA	&		&	6.7\tablenotemark{c}	&	$-0.2$\phs	\\
															
															
GJ  631	&	0.724	$\pm$	0.011	&	This work	&	CHARA	&	K0 V	&	1.5	&	0.0	\\
	&	0.888	$\pm$	0.066	&	\citet{van09}	&	PTI	&		&	7.4\tablenotemark{c}	&	$-2.5$\phs	\\
															
GJ  687	&	0.859	$\pm$	0.014	&	This work	&	CHARA	&	M3.0 V	&	1.6	&	0.0	\\
	&	1.009	$\pm$	0.077	&	\citet{ber06}	&	CHARA	&		&	7.6\tablenotemark{c}	&	$-1.9$\phs	\\
															
GJ  699	&	0.952	$\pm$	0.005	&	This work	&	CHARA	&	M4.0 V	&	0.5	&	0.0	\\
	&	1.004	$\pm$	0.040	&	\citet{lan01}	&	PTI	&		&	4.0	&	$-1.3$\phs	\\
															
GJ  880	&	0.744	$\pm$	0.004	&	This work	&	CHARA	&	M1.5 V	&	0.5	&	0.0	\\
	&	0.934	$\pm$	0.059	&	\citet{ber06}	&	CHARA	&		&	6.3\tablenotemark{c}	&	$-3.2$\phs

\enddata

\tablenotetext{a}{We define the combined error as $\sigma_{\rm C} = (\sigma_{\rm This work}^2 + \sigma_{\rm Reference}^2)^{0.5}$.}
\tablenotetext{b}{We caution the reader that the value of $\theta_{\rm LD}$ cited in  \citet{lan01} is less than $\theta_{\rm UD}$.}
\tablenotetext{c}{These measurements have errors $> 5$\%. See Section~\ref{sec:discussion} for details.} 

\end{deluxetable}

\newpage
\begin{landscape}
\begin{deluxetable}{lcccccccc}
\tabletypesize{\tiny}
\tablewidth{0pt}
\tablecaption{Object Photometry used in Relations \label{tab:Object_phot_forfits}}
\tablehead{
\colhead{\textbf{ }} &
\colhead{\textbf{$B$}} &
\colhead{\textbf{$V$}} &
\colhead{\textbf{$R$}} &
\colhead{\textbf{$I$}} &
\colhead{\textbf{$J$}} &
\colhead{\textbf{$H$}} &
\colhead{\textbf{$K$}} &
\colhead{\textbf{ }} \\
\colhead{\textbf{Star}} &
\colhead{\textbf{(mag)}} &
\colhead{\textbf{(mag)}} &
\colhead{\textbf{(mag)}} &
\colhead{\textbf{(mag)}} &
\colhead{\textbf{(mag)}} &
\colhead{\textbf{(mag)}} &
\colhead{\textbf{(mag)}} &
\colhead{\textbf{Reference$_{BVRI}$, Reference$_{JHK}$}}
}
\startdata

GJ   15A	\dotfill &	9.63	&	8.07	&	6.69	&	5.53	&	4.86	&	4.25	&	4.02	&	\citet{1965ApJ...141..170J},\citet{1975MNRAS.171P..19G}	\\
GJ  33	\dotfill &	6.64	&	5.76	&	4.99	&	4.52	&	4.24	&	3.72	&	3.61	&	\citet{1966CoLPL...4...99J},\citet{1968ApJ...152..465J}	\\
GJ 53A	\dotfill &	5.87	&	5.18	&	4.55	&	4.14	&	3.86	&	3.39	&	3.36	&  \citet{1966CoLPL...4...99J},\citet{1968ApJ...152..465J}	\\
GJ 75	\dotfill &	6.44	&	5.63	&	4.99	&	4.60	&	4.31	&	3.88	&	3.84	&  \citet{1966CoLPL...4...99J},\citet{1989AA...215..305A}	\\
GJ 105	\dotfill &	6.79	&	5.82	&	4.99	&	4.46	&	4.07	&	3.52	&	3.45	&	\citet{1966CoLPL...4...99J},\citet{1975MNRAS.171P..19G}	\\
GJ 144	\dotfill &	4.62	&	3.74	&	3.01	&	2.55	&	2.20	&	1.75	&	1.65	&  \citet{1966CoLPL...4...99J},\citet{1975MNRAS.171P..19G}	\\
GJ  166A	\dotfill &	5.25	&	4.43	&	3.74	&	3.29	&	2.95	&	2.48	&	2.41	&	\citet{1966CoLPL...4...99J},\citet{1968ApJ...152..465J}	\\
GJ  205	\dotfill &	9.44	&	7.97	&	6.53	&	5.39	&	4.77	&	4.06	&	3.86	&	\citet{1965ApJ...141..170J},\citet{1976ApJ...208..399M}	\\
GJ  338A	\dotfill &	9.05	&	7.64	&	\nodata	&	\nodata	&	4.89	&	4.25	&	4.09	&	\citet{1967AJ.....72.1334C},\citet{1974AJ.....79.1056V}	\\
GJ  338B	\dotfill &	9.04	&	7.70	&	\nodata	&	\nodata	&	4.78	&	4.30	&	4.15	&	\citet{1967AJ.....72.1334C},\citet{1974AJ.....79.1056V}	\\
GJ  380	\dotfill &	7.94	&	6.59	&	5.36	&	4.56	&	3.98	&	3.32	&	3.19	&	\citet{1965ApJ...141..170J},\citet{1975MNRAS.171P..19G}	\\
GJ  411	\dotfill &	9.00	&	7.49	&	5.99	&	4.80	&	4.13	&	3.56	&	3.35	&	\citet{1964BOTT....3..305J},\citet{1975MNRAS.171P..19G}	\\
GJ  412A	\dotfill &	10.32	&	8.77	&	7.29	&	6.22	&	5.56	&	4.95	&	4.76	&	\citet{1965ApJ...141..170J},\citet{1975MNRAS.171P..19G}	\\
GJ 436	\dotfill &	12.17	&	10.65	&	\nodata	&	\nodata	&	6.90	&	6.32	&	6.07	&  \citet{1993AJ....105.1962W},\citet{2003yCat.2246....0C}	\\
GJ  526	\dotfill &	9.93	&	8.50	&	7.06	&	5.92	&	5.26	&	4.64	&	4.46	&	\citet{1965ApJ...141..170J},\citet{1975MNRAS.171P..19G}	\\
GJ 551	\dotfill &	13.02	&	11.05	&	8.68	&	6.42	&	5.33	&	4.73	&	4.37	&	\citet{1972PASP...84..581F},\citet{1976ApJ...208..399M}	\\
GJ 570A	\dotfill &	6.88	&	5.78	&	4.85	&	4.28	&	3.82	&	3.27	&	3.15	&  \citet{1965ApJ...141..170J},\citet{1976ApJ...208..399M}	\\
GJ 581	\dotfill &	12.19	&	10.58	&	8.89	&	7.46	&	6.68	&	6.09	&	5.83	&  \citet{1965ApJ...141..170J},\citet{1976ApJ...208..399M}	\\
GJ  631	\dotfill &	6.55	&	5.74	&	5.13	&	4.74	&	4.32	&	3.86	&	3.83	&	\citet{1966CoLPL...4...99J},\citet{1968ApJ...152..465J}	\\
GJ  687	\dotfill &	10.65	&	9.15	&	\nodata	&	\nodata	&	5.37	&	4.75	&	4.54	&	\citet{1953ApJ...117..313J},\citet{1977AJ.....82..729P}	\\
GJ  699	\dotfill &	11.27	&	9.54	&	\nodata	&	\nodata	&	5.30	&	4.77	&	4.50	&	\citet{1997yCat.2168....0M},\citet{1976ApJ...208..399M}	\\
GJ  702A	\dotfill &	5.06	&	4.20	&	\nodata	&	\nodata	&	\nodata	&	\nodata	&	2.30	&	\citet{2008AA...482..631E},\citet{2006AJ....131.3100C}	\\
GJ  702B	\dotfill &	7.15	&	6.05	&	\nodata	&	\nodata	&	\nodata	&	\nodata	&	3.10	&	\citet{2008AA...482..631E},\citet{2006AJ....131.3100C}	\\
GJ  725A	\dotfill &	10.44	&	8.90	&	\nodata	&	\nodata	&	5.19	&	4.68	&	4.46	&	\citet{1953ApJ...117..313J},\citet{1974AJ.....79.1056V}	\\
GJ  725B	\dotfill &	11.28	&	9.69	&	\nodata	&	\nodata	&	5.72	&	5.24	&	4.98	&	\citet{1953ApJ...117..313J},\citet{1974AJ.....79.1056V}	\\
GJ 764	\dotfill &	5.49	&	4.69	&	4.04	&	3.63	&	3.32	&	3.04	&	2.78	&  \citet{1966CoLPL...4...99J},\citet{1966CoLPL...4...99J}	\\
GJ 809	\dotfill &	10.04	&	8.58	&	7.20	&	6.14	&	5.52	&	4.81	&	4.64	&	\citet{1971BITon...6..143E},\citet{1977AJ.....82..729P}	\\
GJ 820A	\dotfill &	6.39	&	5.23	&	4.20	&	3.56	&	3.16	&	2.61	&	2.40	&  \citet{1966CoLPL...4...99J},\citet{1968ApJ...152..465J}	\\
GJ 820B	\dotfill &	7.38	&	6.02	&	4.87	&	4.07	&	3.58	&	2.93	&	2.73	&  \citet{1966CoLPL...4...99J},\citet{1968ApJ...152..465J}	\\
GJ 845	\dotfill &	5.75	&	4.69	&	3.81	&	3.25	&	2.83	&	2.30	&	2.18	&  \citet{1966CoLPL...4...99J},\citet{1976ApJ...208..399M}	\\
GJ 880	\dotfill &	10.19	&	8.68	&	\nodata	&	\nodata	&	5.41	&	4.78	&	4.58	&	\citet{1971BITon...6..143E},\citet{1971BITon...6..143E}	\\
GJ 887	\dotfill &	8.83	&	7.35	&	\nodata	&	\nodata	&	4.20	&	3.60	&	3.36	&  \citet{1990AAS...83..357B},\citet{1976ApJ...208..399M}	\\
GJ 892	\dotfill &	6.57	&	5.57	&	4.74	&	4.21	&	3.86	&	3.40	&	3.23	&	\citet{1966CoLPL...4...99J},\citet{1968ApJ...152..465J}	


\enddata

\tablecomments{Refer to Section~\ref{sec:metallicities} for details.}
\end{deluxetable}

\end{landscape}

\newpage
\begin{landscape}
\begin{deluxetable}{lccccccccccc}
\tabletypesize{\tiny}
\tablewidth{0pt}
\tablecaption{Fundamental Parameters\label{tab:fund_params_combined}}
\tablehead{
\colhead{\textbf{Star}} &
\colhead{\textbf{Spectral}} &
\colhead{\textbf{Metallicity}} &
\colhead{\textbf{Metallicity}} &
\colhead{\textbf{$R \pm \sigma$}} &
\colhead{\textbf{Radius}} &
\colhead{\textbf{$F_{\rm BOL} \pm \sigma$}} &
\colhead{\textbf{$L \pm \sigma$}} &
\colhead{\textbf{$T_{\rm EFF} \pm \sigma$}} &
\colhead{\textbf{Mass}} &
\colhead{\textbf{ }}	&
\colhead{\textbf{ }} \\
\colhead{\textbf{Name}} &
\colhead{\textbf{Type}}	&
\colhead{\textbf{[Fe/H]}} &
\colhead{\textbf{Reference\tablenotemark{a}}} &
\colhead{\textbf{($R_{\rm \odot}$)}}	&
\colhead{\textbf{Reference\tablenotemark{b}}} &
\colhead{\textbf{($10^{-8}$~erg/cm/s$^{2}$)}} &
\colhead{\textbf{($L_{\rm \odot}$)}}	&
\colhead{\textbf{(K)}}	&
\colhead{\textbf{($M_{\rm \odot}$)}}	&
\colhead{\textbf{$L_{\rm X}$/$L_{\rm BOL}$}}	&
\colhead{\textbf{ }}
}
\startdata


GJ   15A$^{\dagger}$	&	M1.5 V	&		$-0.36$\phs	&		1	&		0.3874	$\pm$	0.0023	&	this work	&	5.420	$\pm$	0.044	&	0.02173	$\pm$	0.00021	&	3563	$\pm$	11	&	0.423	&	2.32E-05	& \\
GJ  33	&	K2   V	&		$-0.22$\phs	&		2	&		0.6954	$\pm$	0.0041	&	this work	&	15.060	$\pm$	0.100	&	0.26073	$\pm$	0.00263	&	4950	$\pm$	14	&	0.753	&	8.59E-07	& \\
GJ 105	&	K3   V	&		$-0.08$\phs	&		2	&		0.7949	$\pm$	0.0062	&	this work	&	16.680	$\pm$	0.103	&	0.26790	$\pm$	0.00239	&	4662	$\pm$	17	&	0.767	&	1.70E-06	& \\
GJ  166A$^{\dagger}$	&	K1   Ve	&		$-0.24$\phs	&		2	&		0.8061	$\pm$	0.0036	&	this work	&	52.690	$\pm$	0.394	&	0.40782	$\pm$	0.00319	&	5143	$\pm$	14	&	0.816	&	1.03E-05	& \\
GJ  205	&	M1.5 V	&		$0.35$	&		3	&		0.5735	$\pm$	0.0044	&	this work	&	6.182	$\pm$	0.032	&	0.06163	$\pm$	0.00088	&	3801	$\pm$	9\phn	&	0.615	&	1.62E-05	& \\
GJ  338A	&	M0.0 V	&		$-0.18$\phs	&		3	&		0.5773	$\pm$	0.0131	&	this work	&	5.885	$\pm$	0.051	&	0.06974	$\pm$	0.00213	&	3907	$\pm$	35	&	0.622	&	2.07E-05	& \\
GJ  338B	&	K7.0 V	&		$-0.15$\phs	&		3	&		0.5673	$\pm$	0.0137	&	this work	&	5.455	$\pm$	0.037	&	0.06465	$\pm$	0.00194	&	3867	$\pm$	37	&	0.600	&	2.23E-05	& \\
GJ  380$^{\dagger}$	&	K7.0 V	&		$-0.16$\phs	&		2	&		0.6415	$\pm$	0.0048	&	this work	&	13.860	$\pm$	0.093	&	0.10253	$\pm$	0.00088	&	4081	$\pm$	15	&	0.660	&	6.92E-06	& \\
GJ  411$^{\dagger}$	&	M2.0 V	&		$-0.41$\phs	&		3	&		0.3921	$\pm$	0.0037	&	this work	&	9.842	$\pm$	0.060	&	0.01989	$\pm$	0.00014	&	3465	$\pm$	17	&	0.403	&	8.03E-06	& \\
GJ  412A	&	M1.0 V	&		$-0.40$\phs	&		3	&		0.3982	$\pm$	0.0091	&	this work	&	2.908	$\pm$	0.022	&	0.02129	$\pm$	0.00026	&	3497	$\pm$	39	&	0.403	&	3.11E-05	& \\
GJ  526	&	M1.5 V	&		$-0.30$\phs	&		3	&		0.4840	$\pm$	0.0084	&	this work	&	3.979	$\pm$	0.030	&	0.03603	$\pm$	0.00051	&	3618	$\pm$	31	&	0.520	&	5.33E-06	& \\
GJ  631	&	K0 V	&		$0.04$	&		2	&		0.7591	$\pm$	0.0122	&	this work	&	14.160	$\pm$	0.090	&	0.41945	$\pm$	0.00422	&	5337	$\pm$	41	&	0.821	&	8.92E-06	& \\
GJ  687	&	M3.0 V	&		$-0.09$\phs	&		3	&		0.4183	$\pm$	0.0070	&	this work	&	3.332	$\pm$	0.022	&	0.02128	$\pm$	0.00023	&	3413	$\pm$	28	&	0.413	&	9.30E-06	& \\
GJ  699$^{\dagger}$	&	M4.0 V	&		$-0.39$\phs	&		3	&		0.1867	$\pm$	0.0012	&	this work	&	3.262	$\pm$	0.022	&	0.00338	$\pm$	0.00003	&	3224	$\pm$	10	&	0.146	&	3.99E-06	& \\
GJ  702A	&	K0   Ve	&		$0.03$	&		2	&		0.8310	$\pm$	0.0044	&	this work	&	\nodata			&	0.53	$\pm$	0.02\tablenotemark{c}\tablenotemark{c}	&	5407	$\pm$	52	&	0.846	&	1.67E-06	& \\
GJ  702B	&	K5   Ve	&		$0.03$	&		2	&		0.6697	$\pm$	0.0089	&	this work	&	\nodata			&	0.15	$\pm$	0.02\tablenotemark{c}\tablenotemark{c}	&	\phn4393	$\pm$	149	&	0.698	&	5.90E-06	& \\
GJ  725A	&	M3.0 V	&		$-0.49$\phs	&		3	&		0.3561	$\pm$	0.0039	&	this work	&	3.937	$\pm$	0.004	&	0.01531	$\pm$	0.00018	&	3407	$\pm$	15	&	0.318	&	5.05E-06	& \\
GJ  725B	&	M3.5 V	&		$-0.36$\phs	&		3	&		0.3232	$\pm$	0.0061	&	this work	&	2.238	$\pm$	0.013	&	0.00871	$\pm$	0.00012	&	3104	$\pm$	28	&	0.235	&	8.88E-06	& \\
GJ  809	&	M0.5	&		$-0.21$\phs	&		3	&		0.5472	$\pm$	0.0067	&	this work	&	3.224	$\pm$	0.027	&	0.04990	$\pm$	0.00062	&	3692	$\pm$	22	&	0.573	&	1.41E-05	& \\
GJ  880	&	M1.5 V	&		$0.06$	&		1	&		0.5477	$\pm$	0.0048	&	this work	&	3.502	$\pm$	0.017	&	0.05112	$\pm$	0.00074	&	3713	$\pm$	11	&	0.569	&	6.66E-06	& \\
GJ  892	&	K3   V	&		$0.07$	&		2	&		0.7784	$\pm$	0.0053	&	this work	&	19.850	$\pm$	0.086	&	0.26499	$\pm$	0.00152	&	4699	$\pm$	16	&	0.763	&	3.73E-07	& \\
\hline

GJ 15A$^{\dagger}$	&	M1.5V	&	$-$0.36\phs	&	1	&	0.3790	$\pm$	0.0060	&	1	&	5.420	$\pm$	0.044	&	0.02173	$\pm$	0.00017	&	3594	$\pm$	30	&	0.423	&	2.32E-05	& \\
GJ 53A	&	G5Vp	&	$-$0.68\phs	&	4	&	0.7910	$\pm$	0.0080	&	3	&	25.790	$\pm$	0.147\phn	&	0.45845	$\pm$	0.00260	&	5348	$\pm$	26	&	0.808	&	3.73E-07	& \\
GJ 75	&	K0V	&	0.03	&	2	&	0.8190	$\pm$	0.0240	&	3	&	16.460	$\pm$	0.126\phn	&	0.51971	$\pm$	0.00396	&	5398	$\pm$	75	&	0.837	&	1.12E-05	& \\
GJ 144	&	K2V	&	$-$0.06\phs	&	2	&	0.7350	$\pm$	0.0050	&	5	&	102.100	$\pm$	0.457\phn\phn	&	0.32897	$\pm$	0.00147	&	5077	$\pm$	35	&	0.781	&	1.64E-05	& \\
GJ 166A$^{\dagger}$	&	K1V	&	$-$0.24\phs	&	2	&	0.7700	$\pm$	0.0210	&	6	&	52.690	$\pm$	0.394\phn	&	0.40782	$\pm$	0.00304	&	5261	$\pm$	72	&	0.818	&	1.03E-05	& \\
GJ 380$^{\dagger}$	&	K7V	&	$-$0.16\phs	&	2	&	0.6050	$\pm$	0.0200	&	4	&	13.860	$\pm$	0.093\phn	&	0.10253	$\pm$	0.00069	&	4203	$\pm$	73	&	0.669	&	6.92E-06	& \\
GJ 380$^{\dagger}$	&	K7V	&	$-$0.16\phs	&	2	&	0.6490	$\pm$	0.0280	&	2	&	13.860	$\pm$	0.093\phn	&	0.10253	$\pm$	0.00069	&	4060	$\pm$	87	&	0.669	&	6.92E-06	& \\
GJ 411$^{\dagger}$	&	M2V	&	$-$0.41\phs	&	3	&	0.3930	$\pm$	0.0080	&	4	&	9.842	$\pm$	0.060	&	0.01989	$\pm$	0.00012	&	3460	$\pm$	37	&	0.405	&	8.04E-06	& \\
GJ 411$^{\dagger}$	&	M2V	&	$-$0.41\phs	&	3	&	0.3950	$\pm$	0.0130	&	2	&	9.842	$\pm$	0.060	&	0.01989	$\pm$	0.00012	&	3457	$\pm$	58	&	0.405	&	8.04E-06	& \\
GJ 436	&	M3V	&	0.04	&	3	&	0.4546	$\pm$	0.0182	&	10	&	0.788	$\pm$	0.004	&	0.02525	$\pm$	0.00012	&	3416	$\pm$	53	&	0.472	&	6.84E-06	& \\
GJ 551	&	M5.5V	&	0.19	&	5	&	0.1410	$\pm$	0.0070	&	6	&	2.961	$\pm$	0.037	&	0.00155	$\pm$	0.00002	&	3054	$\pm$	79	&	0.118	&	2.83E-04	& \\
GJ 570A	&	K4V	&	0.02	&	2	&	0.7390	$\pm$	0.0190	&	6	&	19.040	$\pm$	0.154\phn	&	0.20232	$\pm$	0.00163	&	4507	$\pm$	58	&	0.743	&	2.72E-06	& \\
GJ 581	&	M2.5V	&	$-$0.10\phs	&	3	&	0.2990	$\pm$	0.0100	&	9	&	0.930	$\pm$	0.006	&	0.01130	$\pm$	0.00008	&	3442	$\pm$	54	&	0.297	&	8.28E-06	& \\	
GJ 699$^{\dagger}$	&	M4Ve	&	$-$0.39\phs	&	3	&	0.1960	$\pm$	0.0080	&	4	&	3.262	$\pm$	0.022	&	0.00338	$\pm$	0.00002	&	3140	$\pm$	63	&	0.150	&	3.99E-06	& \\
GJ 764	&	K0V	&	$-$0.19\phs	&	2	&	0.7780	$\pm$	0.0080	&	3	&	39.670	$\pm$	0.194\phn	&	0.40926	$\pm$	0.00199	&	5246	$\pm$	26	&	0.806	&	2.61E-06	& \\
GJ 820A$^{\dagger\dagger}$	&	K5V	&	$-$0.19\phs	&	2	&	0.6650	$\pm$	0.0050	&	8	&	37.750	$\pm$	0.188\phn	&	0.14295	$\pm$	0.00071	&	4355	$\pm$	17	&	0.680	&	5.00E-06	& \\
GJ 820A$^{\dagger\dagger}$	&	K5V	&	$-$0.19\phs	&	2	&	0.6100	$\pm$	0.0180	&	2	&	37.750	$\pm$	0.188\phn	&	0.14295	$\pm$	0.00071	&	4548	$\pm$	64	&	0.680	&	5.00E-06	& \\
GJ 820B$^{\dagger\dagger}$	&	K7V	&	$-$0.29\phs	&	2	&	0.5950	$\pm$	0.0080	&	8	&	20.340	$\pm$	0.107\phn	&	0.07753	$\pm$	0.00041	&	3954	$\pm$	28	&	0.629	&	9.28E-06	& \\
GJ 820B$^{\dagger\dagger}$	&	K7V	&	$-$0.29\phs	&	2	&	0.6280	$\pm$	0.0170	&	2	&	20.340	$\pm$	0.107\phn	&	0.07753	$\pm$	0.00041	&	3852	$\pm$	53	&	0.629	&	9.28E-06	& \\
GJ 845	&	K5V	&	$-$0.07\phs	&	2	&	0.7320	$\pm$	0.0060	&	6	&	50.730	$\pm$	0.505\phn	&	0.20737	$\pm$	0.00206	&	4555	$\pm$	24	&	0.731	&	1.94E-06	& \\
GJ 887$^{\dagger\dagger}$	&	M0.5V	&	$-$0.19\phs	&	2	&	0.4910	$\pm$	0.0140	&	7	&	10.920	$\pm$	0.127\phn	&	0.03651	$\pm$	0.00042	&	3612	$\pm$	53	&	0.522	&	5.28E-06	& \\
GJ 887$^{\dagger\dagger}$	&	M0.5V	&	$-$0.19\phs	&	2	&	0.4590	$\pm$	0.0110	&	6	&	10.920	$\pm$	0.127\phn	&	0.03651	$\pm$	0.00042	&	3727	$\pm$	47	&	0.522	&	5.28E-06	& \\

\hline
\hline	\\ [-.5mm]


Star$^{\dagger,\dagger\dagger}$	&		\multicolumn{3}{c}{ }	&	$<R> \pm \sigma$ (R$_{\odot}$)	&	\multicolumn{3}{c}{ }	& $<T_{\rm EFF}> \pm \sigma$ (K)	&	&	&	Average from	\\

\hline \\ [-.5mm]

GJ 15A	&	\multicolumn{3}{c}{ }	&	0.3863$\pm$0.0021	&	\multicolumn{3}{c}{ }	&	3567$\pm$11	&	&	&	this work, 1	\\
GJ 166A	&	\multicolumn{3}{c}{ }	&	0.8051$\pm$0.0035	&	\multicolumn{3}{c}{ }	&	5147$\pm$14	&	&	&	this work, 2	\\
GJ 380	&	\multicolumn{3}{c}{ }	&	0.6398$\pm$0.0046	&	\multicolumn{3}{c}{ }	&	4085$\pm$14	&	&	&	this work, 2, 4	\\
GJ 411	&	\multicolumn{3}{c}{ }	&	0.3924$\pm$0.0033	&	\multicolumn{3}{c}{ }	&	3464$\pm$15	&	&	&	this work, 2, 4	\\
GJ 699	&	\multicolumn{3}{c}{ }	&	0.1869$\pm$0.0012	&	\multicolumn{3}{c}{ }	&	3222$\pm$10	&	&	&	this work, 4	\\
	
GJ 820A	&	\multicolumn{3}{c}{ }	&	0.6611$\pm$0.0048	&	\multicolumn{3}{c}{ }	&	4361$\pm$17	&	&	&	2, 8	\\
GJ 820B	&	\multicolumn{3}{c}{ }	&	0.6010$\pm$0.0072	&	\multicolumn{3}{c}{ }	&	3932$\pm$25	&	&	&	2, 8	\\
GJ 887	&	\multicolumn{3}{c}{ }	&	0.4712$\pm$0.0086	&	\multicolumn{3}{c}{ }	&	3676$\pm$35	&	&	&	6, 7	

\enddata

\tablenotetext{a}{Metallicity references are 1) \citet{nev11}, 2) \citet{and11}, 3) \citet{roj12}, 4) \citet{boy08}, 5) \citet{edv93}, and 6) \citet{bon05}.  See Section~\ref{sec:metallicities} for details.}

\tablenotetext{b}{Interferometric references for measured radii of stars 1)~\citet{ber06}, 2)~\citet{van09}, 3)~\citet{boy08}, 4)~\citet{lan01}, 5)~\citet{dif07}, 6)~\citet{dem09}, 7)~\citet{seg03}, 8)~\citet{ker08a}, 9)~\citet{von11a}, 10)~\citet{von12}.}

\tablenotetext{c}{The photometry for these two sources is too blended to perform this measurement for this work.  This value if from \citet{egg08}.}

\tablecomments{The top portion are new measurements made in this work.  The middle portion lists measurements of stellar radii found in the literature, with precision of better than 5\%. Stars with multiple measurements are marked with a $^{\dagger}$ or a $^{\dagger\dagger}$. Stars are marked with a $^\dagger$ if the mean is includes a measurement from this work, and a $^{\dagger\dagger}$ if the mean is from two literature measurements.  The bottom portion of the table lists the stars with multiple measurements, and the weighted mean for their radii and temperatures, all other parameters remain unaffected when combining the multiple sources for measured radii.  All bolometric flux, luminosity, temperature, mass, and $L_X/L_{\rm BOL}$ values are computed/measured in this work.  Refer to Sections~\ref{sec:stellar_params}, \ref{sec:literature}, and \ref{sec:discussion} for details.}
 
\end{deluxetable}
\end{landscape}



				

\newpage
\begin{landscape}
\begin{deluxetable}{lcccccccccc}
\tabletypesize{\tiny}
\tablewidth{0pt}
\tablecaption{G-type stars from \citet{boy12} \label{tab:DT1_stars}}
\tablehead{
\colhead{\textbf{Star}} &
\colhead{\textbf{$B$}} &
\colhead{\textbf{$V$}} &
\colhead{\textbf{$R$}} &
\colhead{\textbf{$I$}} &
\colhead{\textbf{$J$}} &
\colhead{\textbf{$H$}} &
\colhead{\textbf{$K$}} &
\colhead{\textbf{$T_{\rm EFF} \pm \sigma$}} &
\colhead{\textbf{Metallicity}} &
\colhead{\textbf{ }} \\
\colhead{\textbf{HD}} &
\colhead{\textbf{(mag)}} &
\colhead{\textbf{(mag)}} &
\colhead{\textbf{(mag)}} &
\colhead{\textbf{(mag)}} &
\colhead{\textbf{(mag)}} &
\colhead{\textbf{(mag)}} &
\colhead{\textbf{(mag)}} &
\colhead{\textbf{(K)}} &
\colhead{\textbf{[Fe/H]}} &
\colhead{\textbf{Reference$_{BVRI}$, Reference$_{JHK}$}}
}
\startdata

19373	&	4.65	&	4.05	&	3.52	&	3.23	&	3.06	&	2.73	&	2.69	&	5915$\pm$29	&	$0.09$	&		\citet{1964BOTT....3..305J,1968ApJ...152..465J}	\\

34411	&	5.33	&	4.71	&	4.18	&	3.86	&	3.62	&	3.33	&	3.28	&	5749$\pm$48	&	$0.05$	&		\citet{1964BOTT....3..305J,1968ApJ...152..465J}	\\

39587	&	5.00	&	4.41	&	3.90	&	3.59	&	3.34	&	3.04	&	2.97	&	5961$\pm$36	&	$-0.16$\phs	&		\citet{1964BOTT....3..305J,1968ApJ...152..465J}	\\

82885	&	6.18	&	5.41	&	4.79	&	4.42	&	4.14	&	3.77	&	3.70	&	5434$\pm$45	&	$0.06$	&		\citet{1964BOTT....3..305J,2002yCat.2237....0D}	\\

101501	&	6.08	&	5.34	&	4.73	&	4.37	&	4.02	&	3.61	&	3.60	&	5270$\pm$32	&	$-0.12$\phs	&		\citet{1964BOTT....3..305J,1968ApJ...152..465J}	\\

109358	&	4.86	&	4.27	&	3.73	&	3.42	&	3.20	&	2.88	&	2.83	&	5653$\pm$72	&	$-0.30$\phs	&		\citet{1964BOTT....3..305J,2002yCat.2237....0D}	\\

114710	&	4.84	&	4.26	&	3.77	&	3.47	&	3.22	&	2.95	&	2.89	&	5936$\pm$33	&	$-0.06$\phs	&		\citet{1964BOTT....3..305J,1968ApJ...152..465J}	\\

\enddata

\tablecomments{Photometry, temperatures, and metallicities of G dwarfs in \citet{boy12}. Refer to Section~\ref{sec:color_temperature} for details.}
\end{deluxetable}

\end{landscape}

\newpage
\begin{landscape}
\begin{deluxetable}{lcccccc}
\tabletypesize{\tiny}
\tablewidth{0pt}
\tablecaption{Solutions to $T_{\rm EFF}$ Relations \label{tab:poly_solutions}}
\tablehead{
\colhead{\textbf{Coefficient}}  & 
\colhead{\textbf{($B-V$) - [Fe/H]}}  & 
\colhead{\textbf{($V-R$) - [Fe/H]}}  & 
\colhead{\textbf{($V-I$) - [Fe/H]}}  & 
\colhead{\textbf{($V-J$) - [Fe/H]}} 	 & 
\colhead{\textbf{($V-H$) - [Fe/H]}} 	 & 
\colhead{\textbf{($V-K$) - [Fe/H]}} 		
}
\startdata

$a_0$\dotfill&$8010\pm52$ &$7646\pm42$ &$7325\pm33$ &$7308\pm26$ &$7641\pm33$ &$7643\pm32$ \\
$a_1$\dotfill&$-4095\pm94$\phs&$-4295\pm80$\phs&$-2262\pm35$\phs&$-1775\pm21$\phs&$-1611\pm22$\phs&$-1523\pm21$\phs\\
$a_2$\dotfill&$819\pm41$ &$1058\pm37$ &$313\pm9$ &$198\pm4$ &$151\pm4$ &$134\pm3$ \\
$a_3$\dotfill&$133\pm57$ &$304\pm66$ &$-15\pm34$\phs&$71\pm20$ &$177\pm19$ &$137\pm17$ \\
$a_4$\dotfill&$39\pm85$ &$-77\pm92$\phs&$393\pm84$ &$100\pm64$ &$-319\pm73$\phs&$-202\pm72$\phs\\
$a_5$\dotfill&$-362\pm57$\phs&$102\pm78$ &$733\pm77$ &$317\pm54$ &$185\pm55$ &$157\pm56$ \\
\hline
n\dotfill&GKM=40; KM=33&GKM=29; KM=22&GKM=29; KM=22&GKM=38; KM=31&GKM=38; KM=31&GKM=40; KM=33\\
Range\dotfill&$0.7-1.7$&$0.6-1.7$&$1.0-3.0$&$1.2-4.0$&$1.5-4.5$&$1.5-5.0$\\
Median d$T$ (K)\dotfill&GKM=70; KM=68&GKM=63; KM=62&GKM=73; KM=68&GKM=75; KM=70&GKM=49; KM=48&GKM=49; KM=43\\

\enddata 


\tablecomments{The values $a_0, a_1, a_2, a_3, a_4, a_5$ are the coefficients to the color-metallicity-temperature relations in the form of Equation~\ref{eq:temp_relation}.  The bottom three rows show $n$, the number of data points used for the combined G, K, and M star fit as well as just the K and M star fit, Range, the range in color index where each relation holds true, and Median d$T$, the median absolute deviation for the combined G, K, and M star fit as well as just the K and M star fit. See Figure~\ref{fig:temp_VS_colors_a} for a graphical representation of the solutions, and refer to Section~\ref{sec:color_temperature} for discussion.}

\end{deluxetable}

\end{landscape}


\newpage
\begin{landscape}
\begin{deluxetable}{lcccccc}
\tabletypesize{\tiny}
\tablewidth{0pt}
\tablecaption{Solutions to Radius Relations\label{tab:poly_solutions_radii}}
\tablehead{
\colhead{\textbf{Coefficient}} &
\colhead{\textbf{($B-V$) - [Fe/H]}} &
\colhead{\textbf{($V-R$) - [Fe/H]}} 	&
\colhead{\textbf{($V-I$) - [Fe/H]}} 	&
\colhead{\textbf{($V-J$) - [Fe/H]}} 	&
\colhead{\textbf{($V-H$) - [Fe/H]}} 	&
\colhead{\textbf{($V-K$) - [Fe/H]}} 	
}
\startdata

$a_0$\dotfill	&$0.3830\pm0.0174$ 	&	$0.9122\pm0.0218$ 	&	$0.8632\pm0.0180$ 	&	$0.9647\pm0.0082$ 	&	$0.8843\pm0.0107$ 	&	$0.9191\pm0.0097$ 	\\
$a_1$\dotfill	&$0.9907\pm0.0297$ 	&	$-0.0936\pm0.0420$\phs	&	$-0.0198\pm0.0199$\phs	&	$-0.1018\pm0.0058$\phs	&	$-0.0106\pm0.0067$\phs	&	$-0.0230\pm0.0060$\phs	\\
$a_2$\dotfill	&$-0.6038\pm0.0124$\phs	&	$-0.1274\pm0.0196$\phs	&	$-0.0472\pm0.0052$\phs	&	$-0.0125\pm0.0011$\phs	&	$-0.0222\pm0.0011$\phs	&	$-0.0191\pm0.0009$\phs	\\
$a_3$\dotfill	&$0.1470\pm0.0236$ 	&	$0.3682\pm0.0250$ 	&	$0.1566\pm0.0129$ 	&	$0.0786\pm0.0073$ 	&	$0.1064\pm0.0068$ 	&	$0.0789\pm0.0058$ 	\\
$a_4$\dotfill	&$-0.0451\pm0.0340$\phs	&	$-0.2662\pm0.0356$\phs	&	$-0.1531\pm0.0328$\phs	&	$-0.0179\pm0.0240$\phs	&	$-0.1533\pm0.0271$\phs	&	$-0.0643\pm0.0239$\phs	\\
$a_5$\dotfill	&$-0.1340\pm0.0222$\phs	&	$-0.1652\pm0.0279$\phs	&	$-0.0321\pm0.0277$\phs	&	$0.0286\pm0.0187$ 	&	$-0.0127\pm0.0193$\phs	&	$0.0089\pm0.0189$ 	\\
\hline
n	\dotfill	&	31	&	21	&	21	&	30	&	30	&	32	\\
Range	\dotfill	&	$0.8-1.6$	&	$0.6-1.6$	&	$1.0-3.2$	&	$1.3-4.3$	&	$1.6-4.8$	&	$1.7-5.0$		\\
Median d$R$ (R$_{\odot}$)	\dotfill	&	0.030	&	0.041	&	0.026	&	0.036	&	0.027	&	0.027

\enddata 

\tablecomments{The values $a_0, a_1, a_2, a_3, a_4, a_5$ are the coefficients to the color-metallicity-radius relations in the form of Equation~\ref{eq:radius_metallicity_relation}.  The bottom three rows show $n$, the number of data points used in the fit, Range, the range in color index where each relation holds true, and Median d$R$, the median absolute deviation of the fit. See Figure~\ref{fig:radius_VS_colors_b} for a graphical representation of the solutions, and refer to Section~\ref{sec:empirical_radius_relations} for details.}

\end{deluxetable}

\end{landscape}

\newpage
\begin{landscape}
\begin{deluxetable}{lcccccc}
\tabletypesize{\tiny}
\tablewidth{0pt}
\tablecaption{Solutions to Luminosity Relations \label{tab:poly_solutions_lumin}}
\tablehead{
\colhead{\textbf{Coefficient}} &
\colhead{\textbf{($B-V$) - [FeH]}} &
\colhead{\textbf{($V-R$) - [FeH]}} &
\colhead{\textbf{($V-I$) - [FeH]}} &
\colhead{\textbf{($V-J$) - [FeH]}} 	&
\colhead{\textbf{($V-H$) - [FeH]}} 	&
\colhead{\textbf{($V-K$) - [FeH]}} 	
}
\startdata

$a_0$\dotfill&	$-1.2895\pm0.0379$\phs&	$0.0132\pm0.0257$ &	$0.1838\pm0.0208$ &	$0.1571\pm0.0157$ &	$-0.0851\pm0.0208$\phs&	$0.0613\pm0.0196$ 	\\
$a_1$\dotfill&	$2.5206\pm0.0667$ &	$-0.3282\pm0.0498$\phs&	$-0.4935\pm0.0232$\phs&	$-0.3525\pm0.0129$\phs&	$-0.0494\pm0.0143$\phs&	$-0.1329\pm0.0129$\phs	\\
$a_2$\dotfill&	$-1.7633\pm0.0282$\phs&	$-0.4462\pm0.0225$\phs&	$-0.0476\pm0.0059$\phs&	$-0.0364\pm0.0025$\phs&	$-0.0726\pm0.0023$\phs&	$-0.0535\pm0.0020$\phs	\\
$a_3$\dotfill&	$0.4953\pm0.0390$ &	$0.7308\pm0.0340$ &	$0.3842\pm0.0178$ &	$0.2967\pm0.0093$ &	$0.3445\pm0.0088$ &	$0.3048\pm0.0080$ 	\\
$a_4$\dotfill&	$-0.2946\pm0.0550$\phs&	$-0.5193\pm0.0465$\phs&	$-0.4313\pm0.0432$\phs&	$-0.3921\pm0.0313$\phs&	$-0.6759\pm0.0355$\phs&	$-0.6038\pm0.0339$\phs	\\
$a_5$\dotfill&	$0.1827\pm0.0385$ &	$-0.0565\pm0.0374$\phs&	$0.0128\pm0.0370$ &	$0.0129\pm0.0285$ &	$-0.0042\pm0.0293$\phs&	$-0.0614\pm0.0294$\phs	\\
\hline
n\dotfill	&	31	&	21	&	21	&	30	&	30	&	32	\\
Range\dotfill	&	$0.8-1.6$	&	$0.6-1.8$	&	$1.0-3.5$	&	$1.3-4.3$	&	$1.6-4.8$	&	$1.7-5.0$		\\
Median d$L$ (L$_{\odot}$)\dotfill&	0.0180	&	0.0100	&	0.0110	&	0.0089	&	0.0120	&	0.0066

\enddata 

\tablecomments{The values $a_0, a_1, a_2, a_3, a_4, a_5$ are the coefficients to the color-metallicity-luminosity relations in the form of Equation~\ref{eq:luminosity_metallicity_relation}.  The bottom three rows show $n$, the number of data points used in the fit, Range, the range in color index where each relation holds true, and Median d$L$, the median absolute deviation of the fit. See Figure~\ref{fig:lumin_VS_colors_b} for a graphical representation of the solutions, and refer to Section~\ref{sec:empirical_lumin_relations} for details.}

\end{deluxetable}

\end{landscape}
\newpage
\begin{deluxetable}{lccc}
\tabletypesize{\tiny}
\tablewidth{0pt}
\tablecaption{Spectral Type Lookup Table \label{tab:spectral_type_table}}
\tablehead{
\colhead{\textbf{Spectral}} &
\colhead{\textbf{ }} &
\colhead{\textbf{$T_{\rm EFF} \pm \sigma$}} &
\colhead{\textbf{$R \pm \sigma$}} \\
\colhead{\textbf{Type}} &
\colhead{\textbf{$n$}} &
\colhead{\textbf{(K)}}	&
\colhead{\textbf{($R_{\odot}$)}}	\\
}
\startdata

K0	\dotfill	&	5	&	5347	$\pm$	20	&	0.7956	$\pm$	0.0076	\\
K1	\dotfill	&	1	&	5147	$\pm$	14	&	0.8051	$\pm$	0.0035	\\
K2	\dotfill	&	2	&	5013	$\pm$	14	&	0.7152	$\pm$	0.0006	\\
K3	\dotfill	&	2	&	4680	$\pm$	15	&	0.7867	$\pm$	0.0006	\\
K4	\dotfill	&	1	&	4507	$\pm$	58	&	0.7390	$\pm$	0.0190	\\
K5	\dotfill	&	3	&	4436	$\pm$	74	&	0.6876	$\pm$	0.0021	\\
K7	\dotfill	&	3	&	3961	$\pm$	11	&	0.6027	$\pm$	0.0047	\\
M0	\dotfill	&	1	&	3907	$\pm$	35	&	0.5773	$\pm$	0.0131	\\
M0.5	\dotfill	&	2	&	3684	$\pm$	9\phn	&	0.5092	$\pm$	0.0013	\\
M1	\dotfill	&	1	&	3497	$\pm$	39	&	0.3982	$\pm$	0.0091	\\
M1.5	\dotfill	&	4	&	3674	$\pm$	10	&	0.4979	$\pm$	0.0026	\\
M2	\dotfill	&	1	&	3464	$\pm$	15	&	0.3924	$\pm$	0.0033	\\
M2.5	\dotfill	&	1	&	3442	$\pm$	54	&	0.2990	$\pm$	0.0100	\\
M3	\dotfill	&	3	&	3412	$\pm$	19	&	0.4097	$\pm$	0.0075	\\
M3.5	\dotfill	&	1	&	3104	$\pm$	28	&	0.3232	$\pm$	0.0061	\\
M4	\dotfill	&	1	&	3222	$\pm$	10	&	0.1869	$\pm$	0.0012	\\
M5.5	\dotfill	&	1	&	3054	$\pm$	79	&	0.1410	$\pm$	0.0070	

\enddata

\tablecomments{See Section~\ref{sec:spectral_type_relations} for details.}
 
\end{deluxetable}
\newpage
\begin{deluxetable}{lcc}
\tabletypesize{\tiny}
\tablewidth{0pt}
\tablecaption{Habitable Zone Regions\label{tab:HZs}}
\tablehead{
\colhead{\textbf{Star}} &
\colhead{\textbf{$R_{\rm inner}$}} &
\colhead{\textbf{$R_{\rm outer}$}} \\
\colhead{\textbf{Name}} &
\colhead{\textbf{(AU)}} &
\colhead{\textbf{(AU)}}	
}
\startdata

GJ 15A 	&	0.143	&	0.286	\\
GJ 33 	&	0.454	&	0.902	\\
GJ 53A 	&	0.583	&	1.160	\\
GJ 75 	&	0.618	&	1.230	\\
GJ 105 	&	0.471	&	0.934	\\
GJ 144 	&	0.505	&	1.003	\\
GJ 166A 	&	0.559	&	1.111	\\
GJ 205 	&	0.238	&	0.475	\\
GJ 338A 	&	0.252	&	0.502	\\
GJ 338B 	&	0.243	&	0.484	\\
GJ 380 	&	0.302	&	0.601	\\
GJ 411 	&	0.137	&	0.275	\\
GJ 412A 	&	0.142	&	0.284	\\
GJ 436 	&	0.155	&	0.311	\\
GJ 526 	&	0.183	&	0.367	\\
GJ 551 	&	0.039	&	0.078	\\
GJ 570A 	&	0.414	&	0.821	\\
GJ 581 	&	0.103	&	0.208	\\
GJ 631 	&	0.558	&	1.111	\\
GJ 687 	&	0.142	&	0.285	\\
GJ 699 	&	0.057	&	0.115	\\
GJ 702A 	&	0.623	&	1.242	\\
GJ 702B 	&	0.359	&	0.712	\\
GJ 725A 	&	0.121	&	0.242	\\
GJ 725B 	&	0.092	&	0.186	\\
GJ 764 	&	0.555	&	1.105	\\
GJ 809 	&	0.215	&	0.430	\\
GJ 820A 	&	0.351	&	0.697	\\
GJ 820B 	&	0.265	&	0.528	\\
GJ 845 	&	0.417	&	0.828	\\
GJ 880 	&	0.218	&	0.435	\\
GJ 887 	&	0.184	&	0.368	\\
GJ 892 	&	0.467	&	0.926	\\

\hline

HD 4614	&	0.907	&	1.820	\\

HD 5015	&	1.507	&	3.023	\\



HD 16895	&	1.194	&	2.401	\\

HD 19373	&	1.207	&	2.419	\\

HD 20630	&	0.759	&	1.519	\\

HD 22484	&	1.415	&	2.838	\\

HD 30652	&	1.296	&	2.620	\\

HD 34411	&	1.092	&	2.185	\\

HD 39587	&	0.846	&	1.697	\\

HD 48737	&	2.633	&	5.323	\\

HD 56537	&	3.532	&	7.340	\\

HD 58946	&	1.748	&	3.558	\\

HD 81937	&	2.943	&	5.971	\\

HD 82328	&	2.210	&	4.454	\\

HD 82885	&	0.756	&	1.507	\\

HD 86728	&	0.987	&	1.970	\\

HD 90839	&	1.005	&	2.023	\\

HD 95418	&	4.603	&	9.811	\\

HD 97603	&	3.291	&	6.851	\\

HD 101501	&	0.676	&	1.345	\\

HD 102870	&	1.513	&	3.042	\\

HD 103095	&	0.416	&	0.825	\\

HD 109358	&	0.898	&	1.795	\\

HD 114710	&	0.950	&	1.905	\\

HD 118098	&	2.740	&	5.722	\\

HD 126660	&	1.607	&	3.236	\\

HD 128167	&	1.424	&	2.883	\\

HD 131156	&	0.677	&	1.351	\\

HD 141795	&	2.294	&	4.776	\\

HD 142860	&	1.374	&	2.769	\\

HD 146233	&	0.879	&	1.751	\\

HD 162003	&	2.039	&	4.093	\\

HD 164259	&	1.926	&	3.896	\\

HD 173667	&	1.962	&	3.959	\\

HD 177724	&	3.658	&	7.772	\\

HD 182572	&	1.141	&	2.284	\\


HD 185395	&	1.614	&	3.257	\\

HD 210418	&	3.307	&	6.867	\\

HD 213558	&	3.173	&	6.733	\\

HD 215648	&	1.734	&	3.488	\\

HD 222368	&	1.487	&	2.997	

\enddata

\tablecomments{Habitable zone boundaries for the K- and M-stars in this work and A-, F-, and G-stars from \citet{boy12} (top and bottom portion of the table, respectively).  See Section~\ref{sec:conclusion} for details.}
 
\end{deluxetable}
\newpage

\newpage
\newpage
 \newpage
 \newpage
\newpage

 \newpage
\clearpage
 \newpage

\newpage

\newpage

\newpage
\clearpage

\newpage

\newpage
       
\newpage

\newpage

\newpage

\newpage

\clearpage

\newpage

\newpage

\clearpage
\newpage


\end{document}